\documentclass[12pt]{article}
\usepackage{amssymb}
\usepackage{amsmath}
\usepackage{mathtools}
\usepackage{bm}
\usepackage{physics}
\usepackage{latexsym}
\usepackage{graphicx}
\usepackage{float}
\usepackage{booktabs}
\usepackage{tabularx}
\newcolumntype{C}{>{\centering\arraybackslash}X}
\usepackage{caption}
\usepackage{subcaption}
\usepackage{hyperref}
\usepackage{pdflscape}
\title{\bf Thermodynamic topology of AdS black holes with $F^{\alpha\beta}F^{\gamma\lambda}R_{\alpha\gamma}R_{\beta\lambda}$ coupling in the grand canonical ensemble}
\author{Faramarz Rahmani\thanks{Corresponding author: faramarz.rahmani@abru.ac.ir}\,\, and \,\,Mehdi Sadeghi\thanks{Email: mehdi.sadeghi@abru.ac.ir}\hspace{2mm}\\
{\small {\em Department of Physics, Faculty of Basic Sciences,}}\\
{\small {\em Ayatollah Boroujerdi University, Boroujerd, Iran}}
}
\date{\today}
\begin{document}

\maketitle

\begin{abstract}

We investigate the thermodynamic and topological phase structure of charged AdS black holes with a nonminimal gauge--curvature coupling in the grand canonical ensemble. In the minimal-coupling limit $\epsilon=0$, the system exhibits only Hawking--Page-like behavior and does not possess van der Waals criticality. Working perturbatively in the nonminimal coupling, we find that the $\epsilon\neq0$ interaction generates a rich phase structure as the electric potential is varied, including several distinct regimes of critical behavior, a double-critical region, and a regime where conventional criticality disappears. We complement the conventional thermodynamic analysis with a topological description based on the winding number of the thermodynamic vector field. The resulting $r_h$--$\tau$ diagrams reveal that the ordering and morphology of the black-hole branches depend sensitively on both the pressure and the off-shell inverse temperature, with monotonic, disconnected, S-shaped, and cusp-like structures appearing in different parameter ranges. Remarkably, these substantial rearrangements of the solution branches can occur without changing the global winding number. In particular, the system remains in the $W=+1$ topological class beyond the range of conventional criticality, before eventually entering a $W=0$ sector. Our results demonstrate that thermodynamic topology provides complementary global information that is not captured by conventional local criticality criteria alone.

\end{abstract}

%%%%%%%%%%%%%%%%%%%%%%%%%%%%%%%%%%%%%%%%%%%%%%%%%%%%%
\noindent PACS numbers: 11.25.Tq, 04.70.Dy, 04.70.Bw, 04.70.Dy ,04.50.Kd

\noindent \textbf{Keywords:} AdS black holes, Black hole phase transitions, Thermodynamic topology, Non-minimal coupling
%%%%%%%%%%%%%%%%%%%%%%%%%%%%%%%%%%%%%%%%%%%%%%%%%%%%%

\section{Introduction} \label{sec1}

The discovery that black holes possess thermodynamic properties made a deep connection between gravitation and thermodynamics that continues to inspire novel theoretical developments \cite{Bardeen:1973gs, Hawking:1975vcx,Hawking:1982dh}. This connection has been further enriched by the AdS/CFT correspondence, which provides a holographic dictionary relating gravitational dynamics in anti-de Sitter spacetimes to quantum field theories living on their boundaries. Within this framework, black hole thermodynamics acquires additional significance, as phase transitions in the bulk correspond to critical phenomena in the dual field theory, offering new insights into strongly coupled quantum systems~\cite{Witten:1998qj}.

The thermodynamic phase structure of AdS black holes has proven remarkably rich and continues to reveal unexpected features. In the extended phase space, where the cosmological constant is treated as a thermodynamic pressure, these systems exhibit behavior strikingly analogous to ordinary thermodynamic systems. Van der Waals-type liquid-gas phase transitions, reentrant phase transitions, and triple points have all been identified in various black hole systems. This analog behavior has established black hole thermodynamics as a valuable arena for studying critical phenomena~\cite{Kastor:2009wy,Dolan2011,Kubiznak:2012wp, Belhaj:2012bg}.

Nonlinear extensions of Einstein-Maxwell theory have played a central role in revealing this rich thermodynamic landscape. Models incorporating Born-Infeld electrodynamics introduce maximal field strength effects that regularize divergences and give rise to novel critical phenomena\cite{Chemissany:2008fy,Banerjee:2011cz,Fernando:2006gh}.
 Higher-curvature corrections, such as Gauss-Bonnet and Lovelock gravities, modify the thermodynamic phase space in ways that can produce non-mean-field critical exponents and multiple horizon branches while preserving the second-order nature of the field equations~\cite{Cai:2001dz, Frassino:2014pha}.
 Massive gravity theories allow the graviton mass to act as an additional thermodynamic variable, enabling even more complex phase structures \cite{Cai:2014znn,Hendi:2017fxp}. 
Scalar-tensor theories and Horndeski gravity introduce additional degrees of freedom that can significantly alter the thermodynamic behavior of black holes~\cite{Anabalon:2013oea, Miao:2016aol}.

Among these extensions, the thermodynamic study of nonminimal couplings between gauge fields and spacetime curvature occupy a particularly important position. Such interactions arise naturally from effective field theory considerations, representing the leading corrections to the Einstein-Maxwell action when higher-dimensional or quantum degrees of freedom are integrated out. They also appear in the low-energy effective actions of string theory and in curvature-induced corrections to electrodynamics \cite{Balakin:2005fu,Balakin2010,Guo:2021ere}.

The study of black hole thermodynamics has traditionally proceeded through conventional equilibrium thermodynamic analysis. This approach involves computing the thermodynamic quantities like temperature, entropy, free energy, and heat capacity from the gravitational solution and examining their behavior to identify phase transitions and critical phenomena. While remarkably successful, this method often requires case-by-case analysis of specific models and can obscure deeper structural patterns governing thermodynamic behavior across different theories. The conventional approach excels at identifying the quantitative details of phase transitions, such as critical temperatures and pressures, but may not capture the universal features that characterize entire classes of thermodynamic systems.

In recent years, a complementary perspective has emerged through the application of topological methods to black hole thermodynamics. This approach, rooted in Duan's $\phi$-mapping topological current theory, treats thermodynamic critical points as topological defects in parameter space~\cite{Duan1984,Duan1979}. By constructing a vector field from the generalized free energy and analyzing its winding numbers around critical points, one can assign topological charges to different black hole branches. Stable branches typically carry positive winding numbers, unstable branches negative winding numbers, and phase transitions correspond to creations or annihilations of pairs of defects with opposite charges. The sum of these winding numbers defines a global topological invariant that characterizes the entire thermodynamic system~\cite{Wei:2021vdx,WeiLiu2026,Ali:2023jox,Gogoi:2025ied,Yerra:2022eov,EslamPanah:2024fls,Hazarika:2024xar,WeiLiuMann2022,Wei:2024gfz,Wu:2022whe,ZhuWu2024,WuGu2024,Chen:2024sow,Zhang:2023uay,Liu:2022aqt,Ali:2024,Sadeghi:2026jic,Rahmani:2025iks}.

This topological framework offers several distinct advantages over conventional analysis. It provides a model-independent classification scheme that groups black hole systems into a small number of universal classes based on the asymptotic behavior of the inverse temperature. The topological invariant is robust under continuous deformations of the thermodynamic potential, making it insensitive to details of the equation of state and dependent only on global properties. This enables comparisons across different gravitational theories and identifies which features of the phase structure are universal and which are model-dependent.

The choice of thermodynamic ensemble plays a crucial role in determining the phase structure and stability of charged AdS black holes. While different ensembles can lead to distinct phase behaviors, this dependence can become more intricate in nonminimal models, where a variety of thermodynamic and topological behaviors may emerge. As we shall see, our model exhibits such diverse behavior in the grand canonical ensemble. We derive the thermodynamic quantities in this ensemble, analyze the phase behavior through conventional thermodynamic methods, and apply the topological framework to determine the global topological invariant and its implications for the phase structure. 

The paper is organized as follows. In Sec.~\ref{sec2}, we briefly review the perturbative black hole solution for the nonminimally coupled system. In Sec.~\ref{sec3}, we first derive the essential thermodynamic quantities and establish the first law in the extended phase space in the canonical ensemble, and then reformulate the thermodynamic relations in the grand canonical ensemble. In Sec.~\ref{sec4}, we analyze the critical behavior and thermodynamic phase structure, identifying six distinct regions of parameter space within two dominant regimes. In Sec.~\ref{sec5}, we briefly introduce the topological formalism based on Duan's $\phi$-mapping theory. In Sec.~\ref{sec6}, we apply this framework to determine the topological classes and winding numbers in each region. Finally, in Sec.~\ref{sec7}, we summarize our findings and discuss the complementary roles of conventional thermodynamics and topological methods.

%%%%%%%%%%%%%%%%%%%%%%%%%%%%%%%%%%%%%%%%%%%%%%%%%%%%%%%
\section{Perturbative Solution for a $F^{\alpha\beta}F^{\gamma\lambda}R_{\alpha\gamma }R_{\beta\lambda}$-AdS Black Hole}\label{sec2}

We consider a four-dimensional asymptotically AdS black hole described by the action
\begin{equation}\label{action}
	S = \int d^{4} x \sqrt{-g} \bigg[ \frac{1}{\kappa}(R - 2\Lambda) - \tfrac{1}{4} \alpha  F_{\mu \nu }F^{ \mu \nu  } + \epsilon F^{\alpha \beta } F^{\gamma \lambda } R_{\alpha \gamma } R_{\beta \lambda }  \bigg],
\end{equation}
where \(R\) is the Ricci scalar, \(\Lambda = -3/l^{2}\) is the cosmological constant (with \(l\) the AdS radius), and \(\kappa = 16\pi G\) in standard units. The Maxwell strength tensor field is  defined as
\begin{align} \label{YM}
	F_{\mu \nu } =\partial _{\mu } A_{\nu } -\partial _{\nu } A_{\mu } ,
\end{align}
and \(A_{\mu}\) the gauge potential. The Ricci tensor is denoted by \(R_{\mu\nu}\).

The parameter $\epsilon$ controls the strength of the nonminimal interaction between the electromagnetic field and spacetime curvature and has dimensions of $L^4$. Such higher-dimensional gauge--curvature operators are naturally motivated within effective field theory, where they are expected to appear as symmetry-allowed corrections to the Einstein--Maxwell action after integrating out heavy degrees of freedom \cite{Drummond:1979pp}.

To construct static, spherically symmetric black-hole solutions, we consider the metric ansatz
\begin{equation}\label{metric}
ds^{2}=-f(r)e^{-2H(r)}dt^{2}
+\frac{dr^{2}}{f(r)}
+r^{2}\left(d\theta^{2}+\sin^{2}\theta\,d\phi^{2}\right),
\end{equation}
where the redshift function \(H(r)\) encodes the effects of the nonminimal matter--curvature interaction. For a purely electric configuration, the gauge potential is taken to be
\begin{equation}\label{background}
A_{\mu}dx^{\mu}=h(r)\,dt,
\end{equation}
which gives rise to the field-strength tensor
\begin{equation}
\label{YM1}
F_{\mu\nu}
=
\left(
\begin{array}{cccc}
0 & -h'(r) & 0 & 0\\
h'(r) & 0 & 0 & 0\\
0 & 0 & 0 & 0\\
0 & 0 & 0 & 0
\end{array}
\right).
\end{equation}
Consequently, the Maxwell invariant becomes
\begin{equation}
\mathcal{F}
=
F_{\mu\nu}F^{\mu\nu}
=
-2e^{2H(r)}h'(r)^2.
\end{equation}

Variation of the action (\ref{action}) with respect to the metric yields the gravitational field equations
\begin{equation}\label{EOM1}
R_{\mu\nu}
-\frac12g_{\mu\nu}R
+\Lambda g_{\mu\nu}
=
\kappa T_{\mu\nu}^{(\mathrm{eff})},
\end{equation}
where
\begin{equation}
T_{\mu\nu}^{(\mathrm{eff})}
=
\alpha T_{\mu\nu}^{(\mathrm{M})}
+
\epsilon T_{\mu\nu}^{(\mathrm{I})}.
\end{equation}
Here,
\begin{equation}
T_{\mu\nu}^{(\mathrm{M})}
=
\frac12F_{\mu}{}^{\alpha}F_{\nu\alpha}
-
\frac18g_{\mu\nu}
F_{\alpha\beta}F^{\alpha\beta}
\end{equation}
is the standard Maxwell energy-momentum tensor, whereas \(T_{\mu\nu}^{(\mathrm{I})}\) denotes the contribution arising from the nonminimal interaction. Owing to its length, the explicit form of \(T_{\mu\nu}^{(\mathrm{I})}\) is presented in Appendix~A.

The Maxwell equations obtained from the action are
\begin{equation}\label{EOM-YM}
\nabla_{\nu}\left(
-\frac{\alpha}{2}F^{\mu\nu}
+
2\epsilon
F^{\alpha\beta}
R_{\alpha}{}^{\mu}
R_{\beta}{}^{\nu}
\right)=0.
\end{equation}

Owing to the nonlinear nature of the nonminimal interaction, an exact analytical solution appears to be intractable. We therefore employ a perturbative expansion in the coupling parameter \(\epsilon\), which is assumed to be small in the sense that \(\epsilon << \ell^4\).  The range of validity of the perturbative expansion in $\epsilon$ has been investigated in~\cite{Sadeghi:2026fyu}. 

 Accordingly, the metric functions and gauge potential are expanded as
\begin{align}
f(r)&=f_0(r)+\epsilon f_1(r)+\mathcal{O}(\epsilon^2),\\
H(r)&=H_0(r)+\epsilon H_1(r)+\mathcal{O}(\epsilon^2),\\
h(r)&=h_0(r)+\epsilon h_1(r)+\mathcal{O}(\epsilon^2).
\end{align}
At zeroth order, corresponding to \(\epsilon=0\), the theory reduces to Einstein--Maxwell gravity. The \(tt\) and \(rr\) components of the field equations (Apeendix B) at this order become
\begin{align}
4rf_0'+4f_0+\kappa\alpha r^2e^{2H_0}h_0'^2
+4\Lambda r^2&=0,\\
4rf_0'+4f_0-8rf_0H_0'
+\kappa\alpha r^2e^{2H_0}h_0'^2
+4\Lambda r^2&=0.
\end{align}
Subtracting these equations yields \(H_0'(r)=0\). The corresponding integration constant can be absorbed by a rescaling of the time coordinate, allowing one to set
\begin{equation}
H_0(r)=0.
\end{equation}
The remaining equations then give the familiar Reissner--Nordstr\"om--AdS solution,
\begin{equation}\label{f0}
f_0(r)=1-\frac{2m_0}{r}
-\frac{\Lambda r^2}{3}+\frac{\alpha\kappa Q^2}{4r^2},
\qquad
m_0=-\frac{\Lambda r_h^3}{6}+\frac{\kappa\alpha Q^2}{8r_h}
+\frac{r_h}{2},
\end{equation}
together with
\begin{equation}\label{h0}
h_0(r)=Q\left(\frac{1}{r}-\frac{1}{r_h}\right),
\end{equation}
where \(r_h\) denotes the radius of the outer event horizon.

Proceeding to first order in the coupling parameter \(\epsilon\), we substitute the perturbative expansions into the complete field equations. Subtracting the \(tt\) and \(rr\) components of the Einstein field equations at \(\mathcal{O}(\epsilon)\) yields
\begin{equation}
64\kappa r f_0 h_0' h_0'' + 32\kappa r^2 f_0 h_0^{(3)} h_0' + 32\kappa r^2 f_0 (h_0'')^2 - 8r f_0 H_1' = 0.
\end{equation}
Integrating this equation gives
\begin{equation}\label{H1}
H_1(r)=C_1-\frac{3}{2}\kappa f_0''h_0'^2
-2\kappa f_0'h_0'h_0''
-\kappa r f_0''h_0'h_0''
-\frac12\kappa r f_0^{(3)}h_0'^2.
\end{equation}
Substituting the zeroth-order solutions (\ref{f0}) and (\ref{h0}) into Eq.~(\ref{H1}), one finds
\begin{equation}\label{HH}
H_1(r)
=
C_1
-\frac{3\Lambda\kappa Q^2}{r^4}
+\frac{7\kappa^2\alpha Q^4}{4r^8}.
\end{equation}
The integration constant \(C_1\) is fixed by requiring the boundary speed of light to be unity, which implies \(C_1=0\)\cite{Balakin2010}.

At first order in \(\epsilon\), the Maxwell equation yields
\begin{equation}\label{h1}
h_1(r)
=
-\frac{\kappa^2\alpha Q^5}{6r^9}
+\frac{\kappa Q^3\Lambda}{5r^5}
+\frac{4\Lambda^2Q}{\alpha r}
+C_2.
\end{equation}

The \(rr\) component of the Einstein field equations at \(\mathcal{O}(\epsilon)\) determines the metric correction,
\begin{align}\label{f1}
f_1(r)
&=
\frac{109\alpha^2\kappa^3Q^6}{144r^{10}}
-\frac{13m_0\alpha\kappa^2Q^4}{2r^9}
+\frac{7\alpha\kappa^2Q^4}{2r^8}
\nonumber\\
&
\quad
-\frac{73\alpha\kappa^2Q^4\Lambda}{30r^6}
+\frac{10m_0\kappa Q^2\Lambda}{r^5}
-\frac{6\kappa Q^2\Lambda}{r^4}
+\frac{11\kappa Q^2\Lambda^2}{3r^2}
-\frac{C_3}{r}.
\end{align}

To determine the remaining integration constants and maintain consistency of the perturbative expansion, we introduce a perturbed horizon radius,
\begin{equation}
r_h'=r_h+\epsilon r_h^{(1)},
\end{equation}
where \(r_h\) denotes the unperturbed horizon radius and \(\epsilon r_h^{(1)}\) its first-order correction.

Expanding the condition \(h(r_h')=0\) gives
\begin{align}
h(r_h+\epsilon r_h^{(1)})
=
h_0(r_h)
+
\epsilon
\left[
h_0'(r_h)r_h^{(1)}
+h_1(r_h)
\right]
+
\mathcal O(\epsilon^2)
=
0,
\end{align}
which leads to
\begin{equation}
r_h^{(1)}
=
-\frac{h_1(r_h)}{h_0'(r_h)}.
\end{equation}
Hence, the perturbed horizon radius becomes
\begin{equation}\label{ph}
r_h'
=
r_h
-
\left(
\frac{h_1(r_h)}{h_0'(r_h)}
\right)\epsilon.
\end{equation}

Imposing the horizon condition \(h(r_h')=0\) at first order fixes $ C_2 $ and the gauge potential takes the form
\begin{equation}\label{hfinal}
\begin{aligned}
h(r) = \frac{Q}{r} - \frac{Q}{r_h} 
+ \epsilon \Bigg[
&-\frac{\kappa^2 \alpha Q^5}{6 r^9}
+ \frac{\kappa Q^3 \Lambda}{5 r^5}
+ \frac{4 \Lambda^2 Q}{\alpha r} \\
&- \frac{4 \Lambda^2 Q}{\alpha r_h}
- \frac{\kappa Q^3 \Lambda}{5 r_h^5}
+ \frac{\kappa^2 \alpha Q^5}{6 r_h^9}
\Bigg]=h_0+h_1 \epsilon,
\end{aligned}
\end{equation}

For consistency of the perturbative description, the horizon displacement inferred from the gauge potential should coincide with that obtained from the metric function, leading to the matching condition
\begin{equation}\label{con}
\frac{h_1(r_h)}{h_0'(r_h)}
=
\frac{f_1(r_h)}{f_0'(r_h)}.
\end{equation}

Since, \(h_1(r_h)=0\), Eq.~(\ref{con}) further requires \(f_1(r_h)=0\), thereby fixing the integration constant \(C_3\) leads to the following form for $f(r)$ up to first order in $\epsilon$:

\begin{equation}\label{f_metric}
\begin{aligned}
f(r) = &\; 1 - \frac{\Lambda r^2}{3} - \frac{2M}{r} + \frac{\alpha \kappa Q^2}{4 r^2} \\
&+ \epsilon \Bigg[
\frac{11}{3} \frac{\kappa Q^2 \Lambda^2}{r^2}
- \frac{6 \kappa Q^2 \Lambda}{r^4}
+ \frac{5 \kappa Q^2 \Lambda}{r^5}\left(r_h - \frac{\Lambda r_h^3}{3} + \frac{\kappa \alpha Q^2}{4 r_h}\right) \\
&\quad - \frac{73 \alpha \kappa^2 Q^4 \Lambda}{30 r^6}
+ \frac{7 \alpha \kappa^2 Q^4}{2 r^8}
+ \frac{13 \alpha \kappa^2 Q^4}{r^9}\left(\frac{\Lambda r_h^3}{12} - \frac{r_h}{4} - \frac{\kappa \alpha Q^2}{16 r_h}\right) \\
&\quad + \frac{109 \alpha^2 \kappa^3 Q^6}{144 r^{10}}
\Bigg] 
\end{aligned}
\end{equation}

where $M$ is corresponding mass parameter and is given by:
\begin{equation}\label{mass}
\begin{aligned}
M
&=
\left(
-\frac{\Lambda r_h^3}{6}
+\frac{r_h}{2}
+\frac{\kappa\alpha Q^2}{8r_h}
\right)
\\
&\quad
+
\epsilon
\left(
\frac{\kappa Q^2\Lambda^2}{r_h}
-
\frac{\kappa Q^2\Lambda}{2r_h^3}
-
\frac{\kappa^2Q^4\alpha\Lambda}{20r_h^5}
+
\frac{\kappa^2Q^4\alpha}{8r_h^7}
-
\frac{\kappa^3Q^6\alpha^2}{36r_h^9}
\right)
= m_0 + m_1 \epsilon.
\end{aligned}
\end{equation}

The physical interpretation of \(M\) can be examined through the asymptotic behavior of
\(g_{tt}=-f(r)e^{-2H(r)}\).
Using the perturbative expansions \(f=f_0+\epsilon f_1\) and \(H=\epsilon H_1\), one finds
\[
g_{tt}
=
-\left(
1-\frac{2m_0}{r}
-\frac{\Lambda r^2}{3}
+\frac{\kappa\alpha Q^2}{4r^2}
\right)
-\epsilon
\bigl(
f_1(r)-2f_0(r)H_1(r)
\bigr)
+\mathcal O(\epsilon^2).
\]
This leads to,
\[
g_{tt}
=
-
\left(
1
-
\frac{\Lambda r^2}{3}
-
\frac{2(m_0+\epsilon m_1)}{r}
\right)
+
\mathcal O\!\left(\frac1{r^4}\right).
\]

Therefore, the conserved mass parameter of the black hole to first order in \(\epsilon\) is identified as
\[
M=m_0+\epsilon m_1,
\]
in agreement with the first law of black-hole thermodynamics, as will be demonstrated in Sec.~\ref{sec3}.

\section{Essential Thermodynamic Relations and the First Law of Thermodynamics}\label{sec3}

To study the thermodynamics of the system in the extended phase space, we identify the cosmological constant with the thermodynamic pressure through the standard relation $\Lambda=-8\pi GP$ and set $G=1$ for simplicity. In this framework, the mass of the black hole is naturally interpreted as the thermodynamic enthalpy. 
Substituting $\Lambda=-8\pi P$ into Eq.~(\ref{mass}) gives the thermodynamic enthalpy,

\begin{equation}\label{entha}
\mathcal{H}
=
\left(
\frac{r_h}{2}
+\frac{Q^2}{8r_h}
+\frac{4\pi P r_h^3}{3}
\right)
+
\left(
-\frac{Q^6}{36r_h^9}
+\frac{Q^4}{8r_h^7}
+\frac{2\pi P Q^4}{5r_h^5}
+\frac{4\pi P Q^2}{r_h^3}
+\frac{64\pi^2P^2Q^2}{r_h}
\right)\epsilon.
\end{equation}

The Hawking temperature, obtained from the surface gravity, is given to first order in $\epsilon$ by

\begin{equation}\label{Temp}
\begin{split}
T &= \frac{1}{2\pi} \left[ \frac{1}{\sqrt{g_{rr}}} \frac{d}{dr} \sqrt{-g_{tt}} \right] \Bigg|_{r = r_h}
   = \frac{e^{-H(r_h)} f'(r_h)}{4 \pi} \\
  &= 2Pr_h + \frac{1}{4\pi r_h} - \frac{Q^2}{16\pi r_h^3}
  + \left(
  -\frac{2PQ^2}{r_h^5}
  + \frac{PQ^4}{r_h^7}
  - \frac{Q^4}{16\pi r_h^9}
  + \frac{Q^6}{32\pi r_h^{11}}
  \right)\epsilon.
\end{split}
\end{equation}

Assuming the validity of the first law, the entropy is obtained by integrating
\[
dS=\frac{1}{T}\left(\frac{\partial \mathcal{H}}{\partial r_h}\right)_{P,Q}dr_h,
\]
which yields

\begin{equation}
S=\pi r_h^2
+\left(
\frac{16\pi^2PQ^2}{r_h^2}
+\frac{\pi Q^4}{2r_h^6}
\right)\epsilon.
\end{equation}
One can use the Wald formula to determine the entropy and observe the consistency with this relation.

Solving Eq.~(\ref{Temp}) for the pressure yields the equation of state,

\begin{equation}
P=
\frac{
32\pi T r_h^{11}+2Q^2r_h^8-8r_h^{10}-Q^6\epsilon
+2Q^4\epsilon r_h^2
}{
32\pi r_h^4
\left(
2r_h^8+Q^4\epsilon-2Q^2\epsilon r_h^2
\right)
}.
\end{equation}

The thermodynamic volume conjugate to the pressure and the electric potential conjugate to the charge are given, respectively, by

\begin{equation}\label{volume}
\begin{aligned}
V
&=
\left(
\frac{\partial\mathcal{H}}{\partial P}
\right)_{Q,S}=\left(
\frac{\partial\mathcal{H}}{\partial P}
\right)_{Q,r_h}
-
\frac{
\left(
\frac{\partial\mathcal{H}}{\partial r_h}
\right)_{P,Q}
\left(
\frac{\partial S}{\partial P}
\right)_{Q,r_h}
}{
\left(
\frac{\partial S}{\partial r_h}
\right)_{P,Q}
}
\\
&=
\frac{4\pi r_h^3}{3}
+
\epsilon
\left(
\frac{96\pi^2PQ^2}{r_h}
+\frac{7\pi Q^4}{5r_h^5}
\right).
\end{aligned}
\end{equation}

and

\begin{equation}\label{potential}
\begin{aligned}
\psi
&=
\left(
\frac{\partial\mathcal{H}}{\partial Q}
\right)_{P,S}=
\left(
\frac{\partial\mathcal{H}}{\partial Q}
\right)_{P,r_h}
-
\frac{
\left(
\frac{\partial\mathcal{H}}{\partial r_h}
\right)_{P,Q}
\left(
\frac{\partial S}{\partial Q}
\right)_{P,r_h}
}{
\left(
\frac{\partial S}{\partial r_h}
\right)_{P,Q}
}
\\
&=
\frac{Q}{4r_h}
+
\epsilon
\left(
\frac{64\pi^2P^2Q}{r_h}
-\frac{2\pi PQ^3}{5r_h^5}
-\frac{Q^5}{24r_h^9}
\right).
\end{aligned}
\end{equation}

To have a correct Smarr relation, the first law of thermodynamics in the canonical ensemble in the extended phase space must take the form

\begin{equation}
d\mathcal{H}
=
T\,dS
+
V\,dP
+
\psi\,dQ
+
\Xi\,d\epsilon,
\end{equation}

where $\Xi$ denotes the thermodynamic quantity conjugate to the nonminimal coupling parameter $\epsilon$,

\begin{equation}\label{Xi}
\begin{aligned}
\Xi
&=
\left(
\frac{\partial\mathcal{H}}{\partial\epsilon}
\right)_{S,P,Q}=\left(
\frac{\partial\mathcal{H}}{\partial\epsilon}
\right)_{P,Q,r_h}
-
\frac{
\left(
\frac{\partial\mathcal{H}}{\partial r_h}
\right)_{P,Q,\epsilon}
\left(
\frac{\partial S}{\partial\epsilon}
\right)_{P,Q,r_h}
}{
\left(
\frac{\partial S}{\partial r_h}
\right)_{P,Q,\epsilon}
}
\\
&=
\frac{32\pi^2P^2Q^2}{r_h}
+\frac{2\pi PQ^4}{5r_h^5}
+\frac{Q^6}{288r_h^9}
\\
&\quad
+\epsilon
\left(
\frac{256\pi^3P^2Q^4}{r_h^5}
+\frac{64\pi^2PQ^6}{5r_h^9}
+\frac{11\pi Q^8}{72r_h^{13}}
\right)
\\
&=
\Xi_0+\epsilon\Xi_1.
\end{aligned}
\end{equation}

Then, the corresponding Smarr relation is satisfied to first order in $\epsilon$,

\begin{equation}
\mathcal{H}
=
2TS
-
2PV
+
\psi Q
+
4\Xi_0\epsilon.
\end{equation}

To investigate the thermodynamic behavior in the grand canonical ensemble, we use the relation between the electric potential $\psi$ and the charge $Q$ given in Eq.~(\ref{potential}). This allows all thermodynamic quantities to be expressed in terms of $\psi$ rather than $Q$, thereby treating the electric potential as the independent thermodynamic variable. In this ensemble, the appropriate thermodynamic potential is the Gibbs free energy,
\begin{equation}\label{G0}
G=\mathcal{H}-TS-\psi Q,
\end{equation}
which governs the phase structure under conditions of fixed temperature, pressure, and electric potential.

\section{Critical Behavior and Thermodynamic Phase Structure in the grand canonical ensemble}\label{sec4}

Using Eq.~(\ref{potential}), the electric charge can be expressed in terms of the electric potential by selecting the real physical branch of the solution. Consequently, all thermodynamic quantities can be rewritten in terms of the electric potential $\psi$, which serves as the independent thermodynamic variable in the grand canonical ensemble. Using relation \ref{potential} and solving for $Q$ the enthalpy for our model in this ensemble is given by
\begin{equation}\label{Eg}
\mathcal{M}
=
\frac{4 \pi P r_h^3}{3}
+2\psi^2r_h
+\frac{r_h}{2}
+\epsilon
\left(
\frac{1024\pi P\psi^4}{5r_h}
+\frac{512\psi^6}{9r_h^3}
+\frac{64\pi P\psi^2}{r_h}
+\frac{32\psi^4}{r_h^3}
\right).
\end{equation}

Since the Gibbs free energy is the relevant thermodynamic potential in the grand canonical ensemble, it is obtained as
\begin{equation}\label{Gibbs}
\begin{split}
G
&=
\mathcal{M}
-
T S
-
\psi Q
\\
&=
-\frac{2\pi Pr_h^3}{3}
-3\psi^2r_h
+\frac{r_h}{4}
\\
&\quad
+\epsilon
\left(
-\frac{2816\pi P\psi^4}{5r_h}
-\frac{2560\psi^6}{9r_h^3}
+\frac{32\pi P\psi^2}{r_h}
+\frac{16\psi^4}{r_h^3}
+512\pi^2P^2\psi^2r_h
\right).
\end{split}
\end{equation}

The entropy in the grand canonical ensemble becomes
\begin{equation}\label{entg}
S
=
\pi r_h^2
+
\epsilon
\left(
256\pi^2P\psi^2
+\frac{128\pi\psi^4}{r_h^2}
\right).
\end{equation}

The temperature relation is
\begin{equation}\label{tempg}
T
=
\left(
\frac{\partial\mathcal{M}}{\partial r_h}
\Big/
\frac{\partial S}{\partial r_h}
\right)_{P,\psi}
=
2Pr_h
+\frac{\psi^2}{\pi r_h}
+\frac{1}{4\pi r_h}
+\epsilon
\left(
\frac{768\psi^4P}{5r_h^3}
+\frac{128\psi^6}{3\pi r_h^5}
-\frac{32\psi^2P}{r_h^3}
-\frac{16\psi^4}{\pi r_h^5}
\right).
\end{equation}

These thermodynamic relations form the basis for our investigation of the critical behavior and phase structure in the grand canonical ensemble. Further, using the temperature relation (\ref{tempg}), the pressure in the grand canonical ensemble is obtained as follows:
\begin{equation}\label{PG}
P=\frac{5\left(12\pi T r_h^5 - 512\epsilon\psi^6 - 12\psi^2 r_h^4 + 192\epsilon\psi^4 - 3r_h^4\right)}{24\pi r_h^2\left(384\epsilon\psi^4 + 5r_h^4 - 80\epsilon\psi^2\right)}.
\end{equation}

\subsection{Classification of the Critical Regions}

To investigate the critical behavior of the system, we first examine the standard criticality conditions,
\begin{equation}\label{crit_t}
\left(\frac{\partial T}{\partial r_h}\right)_P
=
\left(\frac{\partial^2 T}{\partial r_h^2}\right)_P
=
0.
\end{equation}

By analyzing the criticality conditions as the electric potential $\psi$ varies, we identify two broad thermodynamic regimes. In the first regime, the criticality conditions admit real solutions for the critical quantities $(r_c,P_c,T_c)$. A detailed numerical investigation shows that this regime can be further classified into five distinct regions, each exhibiting qualitatively different evolution of the critical parameters as the electric potential increases. These regions include the van der Waals type or small-large black hole phase transition behavior from a local critically perspective, several anomalous critical regimes, and a double-critical region in which two branches of critical points coexist, although only one corresponds to a physically admissible solution.

In the second regime, the criticality conditions no longer admit real solutions. As the electric potential exceeds a threshold value, the oscillatory behavior of the temperature disappears, and the temperature profile evolves into a curve with a single minimum. Consequently, the system no longer exhibits a thermodynamic critical point.

Overall, the parameter space is therefore classified into six distinct regions according to the behavior of the critical quantities and the existence of critical points. To systematically investigate the evolution of the phase structure, we vary the electric potential from $\psi=0.001$ in increments of $\Delta\psi=0.01$. The thermodynamic properties and phase structure of each region are discussed in detail in the following subsections. We shall see in the topological study that these classification differences are more obvious in their details.

\paragraph{Region I: Usual critical Behavior}
We begin our analysis with the low-potential regime, starting from $\psi=0.001$. Throughout this region, the thermodynamic behavior is the  van der Waals-like, characterized by a single critical point satisfying Eq.~(\ref{crit_t}). The region extends approximately from $\psi=0.001$ to $\psi=0.271$. These boundaries should be regarded as numerical estimates rather than exact values, since they are determined from a discrete scan of the parameter space. 

As the electric potential increases within this region, the critical horizon radius $r_c$ increases monotonically, whereas the critical pressure $P_c$ and the critical temperature $T_c$ both decrease monotonically. 
The corresponding critical quantities are listed in Table~\ref{tab:critical_points_part1}.

\begin{table}[H]
\centering
\caption{Critical points in the Region I: $\psi \in [0.001,\,0.271]$. As the electric potential increases, the critical horizon radius increases, while both the critical pressure and the critical temperature decrease monotonically.}
\label{tab:critical_points_part1}
\begin{tabular*}{\textwidth}{@{\extracolsep{\fill}} c c c c}
\toprule
$\psi$ & $r_c$ & $P_c$ & $T_c$ \\
\midrule
0.001  & 0.014802 & 90.804926 & 7.168358 \\
0.011  & 0.049124 & 8.254738  & 2.161999 \\
0.021  & 0.067991 & 4.323601  & 1.566014 \\
0.031  & 0.082835 & 2.928645  & 1.290632 \\
0.041  & 0.095614 & 2.214227  & 1.124341 \\
0.051  & 0.107119 & 1.780133  & 1.010525 \\
0.061  & 0.117763 & 1.488601  & 0.926728 \\
0.071  & 0.127789 & 1.279490  & 0.862032 \\
0.081  & 0.137349 & 1.122365  & 0.810411 \\
0.091  & 0.146545 & 1.000169  & 0.768229 \\
0.101  & 0.155446 & 0.902603  & 0.733149 \\
0.111  & 0.164098 & 0.823078  & 0.703590 \\
0.121  & 0.172533 & 0.757181  & 0.678439 \\
0.131  & 0.180772 & 0.701845  & 0.656891 \\
0.141  & 0.188830 & 0.654871  & 0.638343 \\
0.151  & 0.196715 & 0.614644  & 0.622338 \\
0.161  & 0.204430 & 0.579945  & 0.608520 \\
0.171  & 0.211978 & 0.549844  & 0.596609 \\
0.181  & 0.219356 & 0.523613  & 0.586381 \\
0.191  & 0.226563 & 0.500678  & 0.577655 \\
0.201  & 0.233594 & 0.480581  & 0.570284 \\
0.211  & 0.240443 & 0.462951  & 0.564149 \\
0.221  & 0.247104 & 0.447486  & 0.559151 \\
0.231  & 0.253570 & 0.433939  & 0.555209 \\
0.241  & 0.259834 & 0.422107  & 0.552257 \\
0.251  & 0.265887 & 0.411822  & 0.550241 \\
0.261  & 0.271720 & 0.402947  & 0.549119 \\
0.271  & 0.277323 & 0.395368  & 0.548859\\
\bottomrule
\end{tabular*}
\end{table}

\paragraph{Region II: Anomalous Critical Region}

In this region, which extends approximately from $\psi=0.281$ to $\psi=0.341$, the thermodynamic behavior differs from that of Region I. As the electric potential increases, the critical horizon radius $r_c$ continues to increase, while the critical pressure $P_c$ decreases. In contrast, the critical temperature $T_c$ increases monotonically. Consequently, unlike the first region, the critical pressure and critical temperature no longer exhibit the same monotonic trend.

\begin{table}[H]
\centering
\caption{Critical points in the Region II: $\psi \in [0.281,\,0.341]$. In this region, the critical horizon radius increases and the critical pressure decreases, whereas the critical temperature increases with increasing electric potential.}
\label{tab:critical_points_part2}
\begin{tabular*}{\textwidth}{@{\extracolsep{\fill}} c c c c}
\toprule
$\psi$ & $r_c$ & $P_c$ & $T_c$ \\
\midrule
0.281 & 0.282685 & 0.388994 & 0.549437 \\
0.291 & 0.287796 & 0.383752 & 0.550837 \\
0.301 & 0.292643 & 0.379589 & 0.553055 \\
0.311 & 0.297212 & 0.376463 & 0.556090 \\
0.321 & 0.301490 & 0.374351 & 0.559955 \\
0.331 & 0.305460 & 0.373245 & 0.564670 \\
0.341 & 0.309105 & 0.373152 & 0.570265 \\
\bottomrule
\end{tabular*}
\end{table}

\paragraph{Region III: Positive Critical Scaling}

Region III extends approximately from $\psi=0.351$ to $\psi=0.411$. Unlike the previous regions, all three critical quantities increase monotonically with increasing electric potential. In particular, the critical horizon radius $r_c$, the critical pressure $P_c$, and the critical temperature $T_c$ all increase throughout this interval. This simultaneous increase distinguishes Region III from Regions I and II and represents a qualitatively different critical behavior.

The corresponding critical quantities are summarized in Table~\ref{tab:critical_points_part3}.

\begin{table}[H]
\centering
\caption{Critical points in the Region III: $\psi \in [0.351,\,0.411]$. In this region, the critical horizon radius, critical pressure, and critical temperature all increase monotonically with increasing electric potential.}
\label{tab:critical_points_part3}
\begin{tabular*}{\textwidth}{@{\extracolsep{\fill}} c c c c}
\toprule
$\psi$ & $r_c$ & $P_c$ & $T_c$ \\
\midrule
0.351 & 0.312404 & 0.374096 & 0.576784 \\
0.361 & 0.315334 & 0.376121 & 0.584284 \\
0.371 & 0.317871 & 0.379296 & 0.592843 \\
0.381 & 0.319983 & 0.383717 & 0.602556 \\
0.391 & 0.321637 & 0.389516 & 0.613552 \\
0.401 & 0.322790 & 0.396873 & 0.625994 \\
0.411 & 0.323392 & 0.406033 & 0.640098 \\
\bottomrule
\end{tabular*}
\end{table}

\paragraph{Region IV: Reversed Critical Radius Scaling}

Region IV extends approximately from $\psi=0.421$ to $\psi=0.451$. In contrast to Region III, the critical horizon radius $r_c$ decreases as the electric potential increases, whereas both the critical pressure $P_c$ and the critical temperature $T_c$ continue to increase monotonically. 

The corresponding critical quantities are listed in Table~\ref{tab:critical_points_part4}.

\begin{table}[H]
\centering
\caption{Critical points in the Region IV: $\psi \in [0.421,\,0.451]$. In this region, the critical horizon radius decreases, while the critical pressure and critical temperature increase with increasing electric potential.}
\label{tab:critical_points_part4}
\begin{tabular*}{\textwidth}{@{\extracolsep{\fill}} c c c c}
\toprule
$\psi$ & $r_c$ & $P_c$ & $T_c$ \\
\midrule
0.421 & 0.323381 & 0.417332 & 0.656152 \\
0.431 & 0.322682 & 0.431238 & 0.674546 \\
0.441 & 0.321193 & 0.448424 & 0.695827 \\
0.451 & 0.318782 & 0.469888 & 0.720790 \\
\bottomrule
\end{tabular*}
\end{table}

\paragraph{Region V: Double-Critical Region}

Region V extends approximately from $\psi=0.461$ to $\psi=0.511$. Unlike the previous regions, the criticality conditions (\ref{crit_t}) admit two distinct real solutions for each value of the electric potential. Consequently, two branches of critical points coexist in this interval. Their corresponding critical quantities are listed in Table~\ref{tab:critical_branches}.

The first branch is characterized by an increasing critical horizon radius $r_{c1}$ accompanied by decreasing critical pressure $P_{c1}$ and critical temperature $T_{c1}$ as the electric potential increases. In contrast, the second branch exhibits the opposite behavior: the critical horizon radius $r_{c2}$ decreases, whereas both the critical pressure $P_{c2}$ and the critical temperature $T_{c2}$ increase monotonically.

Although the criticality conditions admit two mathematical solutions, the subsequent thermodynamic analysis demonstrates that only the second branch corresponds to a physically admissible critical point. In particular, the temperature and the Gibbs free energy presented below show that the first branch does not describe a physically realizable thermodynamic phase. Therefore, only the second branch contributes to the physical phase structure of the system.

\begin{table}[H]
\centering
\caption{Critical points in the Region V: $\psi\in[0.461,\,0.511]$. Two distinct mathematical branches of critical points exist in this interval. Subsequent thermodynamic analysis shows that only Branch~2 corresponds to a physically admissible critical point.}
\label{tab:critical_branches}
\begin{tabular*}{\textwidth}{@{\extracolsep{\fill}} c c c c c c c}
\toprule
& \multicolumn{3}{c}{Branch 1} & \multicolumn{3}{c}{Branch 2} \\
\cmidrule(lr){2-4}\cmidrule(lr){5-7}
$\psi$ & $r_{c1}$ & $P_{c1}$ & $T_{c1}$ & $r_{c2}$ & $P_{c2}$ & $T_{c2}$ \\
\midrule
0.461 & 0.091552 & 215.955007 & 64.125250 & 0.315267 & 0.497187 & 0.750647 \\
0.471 & 0.126187 & 34.053206 & 14.385394 & 0.310374 & 0.532919 & 0.787376 \\
0.481 & 0.148926 & 13.691231 & 7.073449 & 0.303666 & 0.581870 & 0.834527 \\
0.491 & 0.169194 & 6.997785 & 4.281478 & 0.294334 & 0.654311 & 0.899520 \\
0.501 & 0.190866 & 3.837540 & 2.789471 & 0.280436 & 0.779604 & 1.002899 \\
0.511 & 0.226313 & 1.775237 & 1.669041 & 0.250276 & 1.185707 & 1.296313 \\
\bottomrule
\end{tabular*}
\end{table}

\paragraph{Region VI: No-Critical-Point Region}

Region VI corresponds to the range $\psi \gtrsim 0.521$, where the criticality conditions given in Eq.~(\ref{crit_t}) no longer admit real solutions. Consequently, the system does not possess a thermodynamic critical point in this regime.

As the electric potential increases beyond the upper bound of Region V, the two critical branches gradually approach one another and eventually disappear. We shall see further details in the topological investigation section. The disappearance of the critical point marks the end of the rich critical behavior observed in the preceding regions. Although the system continues to exhibit nontrivial thermodynamic properties, its phase structure is no longer governed by thermodynamic criticality. The detailed thermodynamic behavior in this regime will be examined in the following sections through the temperature diagrams, Gibbs free energy and topological analysis.

\begin{table}[H]
\centering
\caption{Classification of the thermodynamic regions in the grand canonical ensemble according to the behavior of critical points.}
\label{tab:summary_regions}
\begin{tabular}{c c c l}
\toprule
Region & $\psi$ Range & Critical Points & Characteristics \\
\midrule
I   & $0.001$--$0.271$ & One & vdW-like behavior \\

II  & $0.281$--$0.341$ & One & $P_c\!\downarrow,\;T_c\!\uparrow$ \\

III & $0.351$--$0.411$ & One & $r_c,P_c,T_c$ all increase \\

IV  & $0.421$--$0.451$ & One & $r_c\!\downarrow,\;P_c,T_c\!\uparrow$ \\

V   & $0.461$--$0.511$ & Two & One physical, one unphysical \\

VI  & $\ge0.521$ & None & No criticality \\
\bottomrule
\end{tabular}
\end{table}

For convenience, the six regions identified from the criticality analysis are summarized in Table~\ref{tab:summary_regions}.

\subsection{Phase Structure in Region I}

\begin{figure}[H]
    \centering
    \begin{subfigure}[b]{0.35\textwidth}
        \centering
        \includegraphics[width=\textwidth]{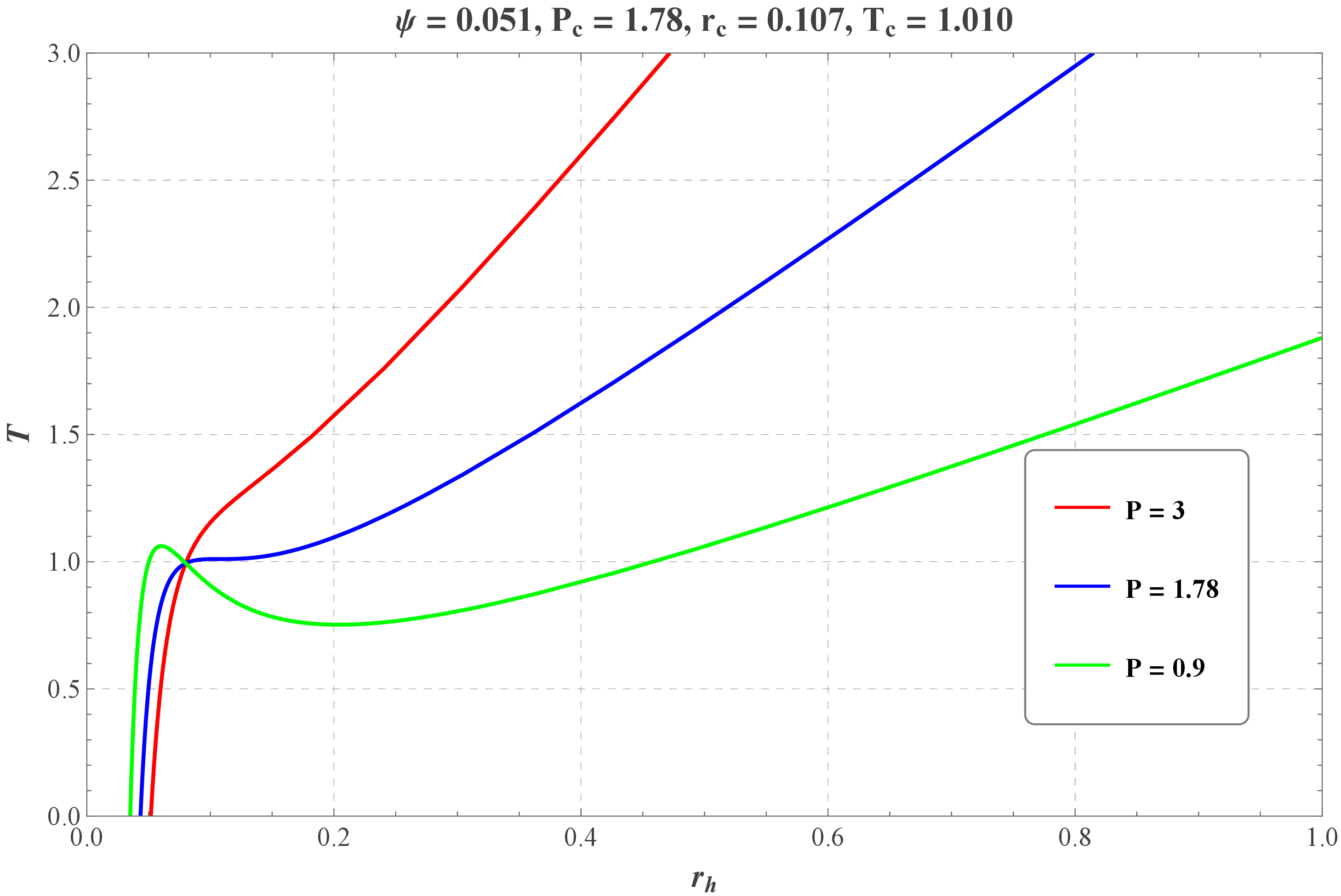}
        \caption{Isobars for $\psi=0.051$ with $(P_c,r_c,T_c)=(1.780,0.107,1.010)$.}
        \label{fig:t1}
    \end{subfigure}
    \quad
    \begin{subfigure}[b]{0.35\textwidth}
        \centering
        \includegraphics[width=\textwidth]{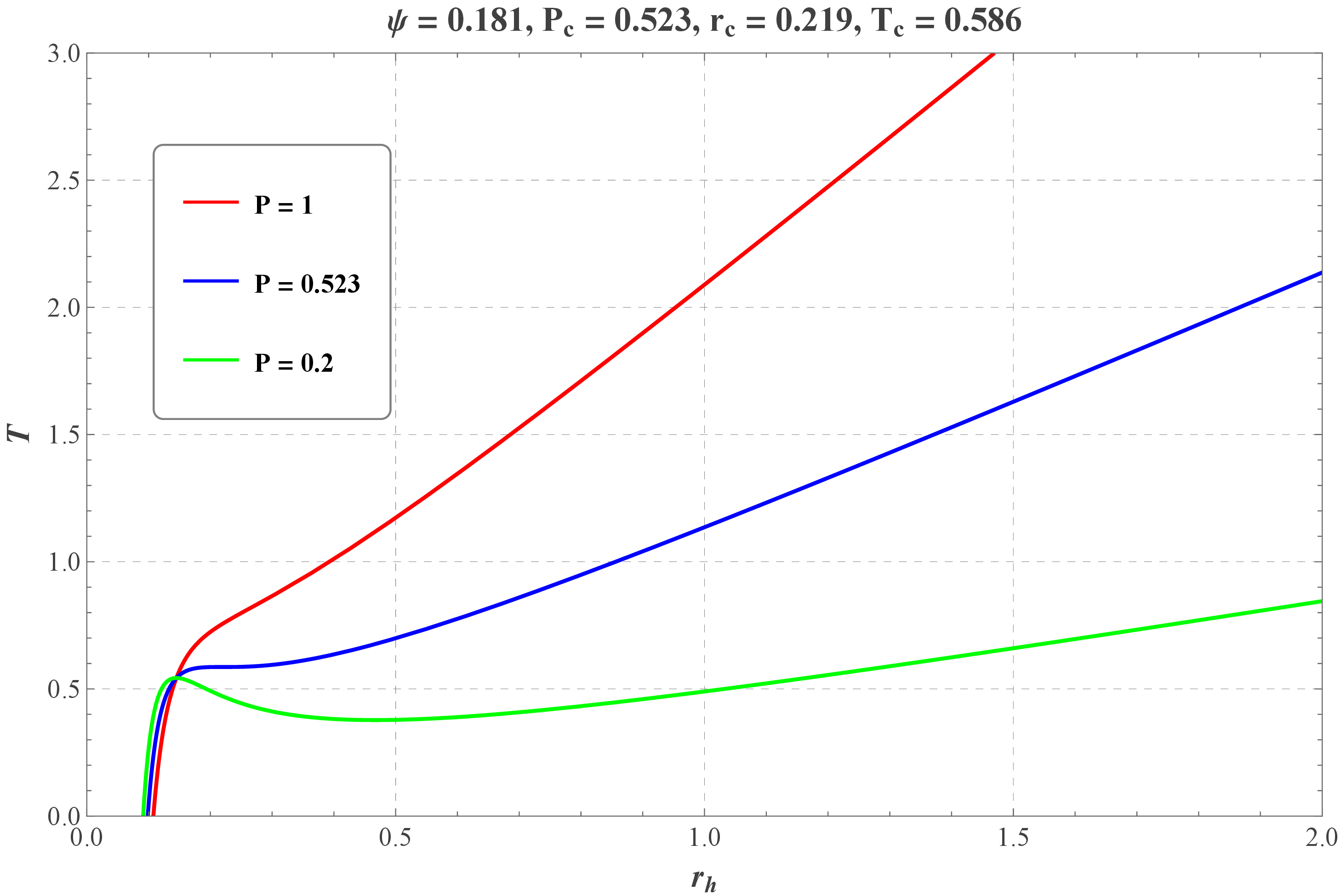}
        \caption{Isobars for $\psi=0.181$ with $(P_c,r_c,T_c)=(0.523,0.219,0.586)$.}
        \label{fig:t2}
    \end{subfigure}

    \par\medskip

    \begin{subfigure}[b]{0.35\textwidth}
        \centering
        \includegraphics[width=\textwidth]{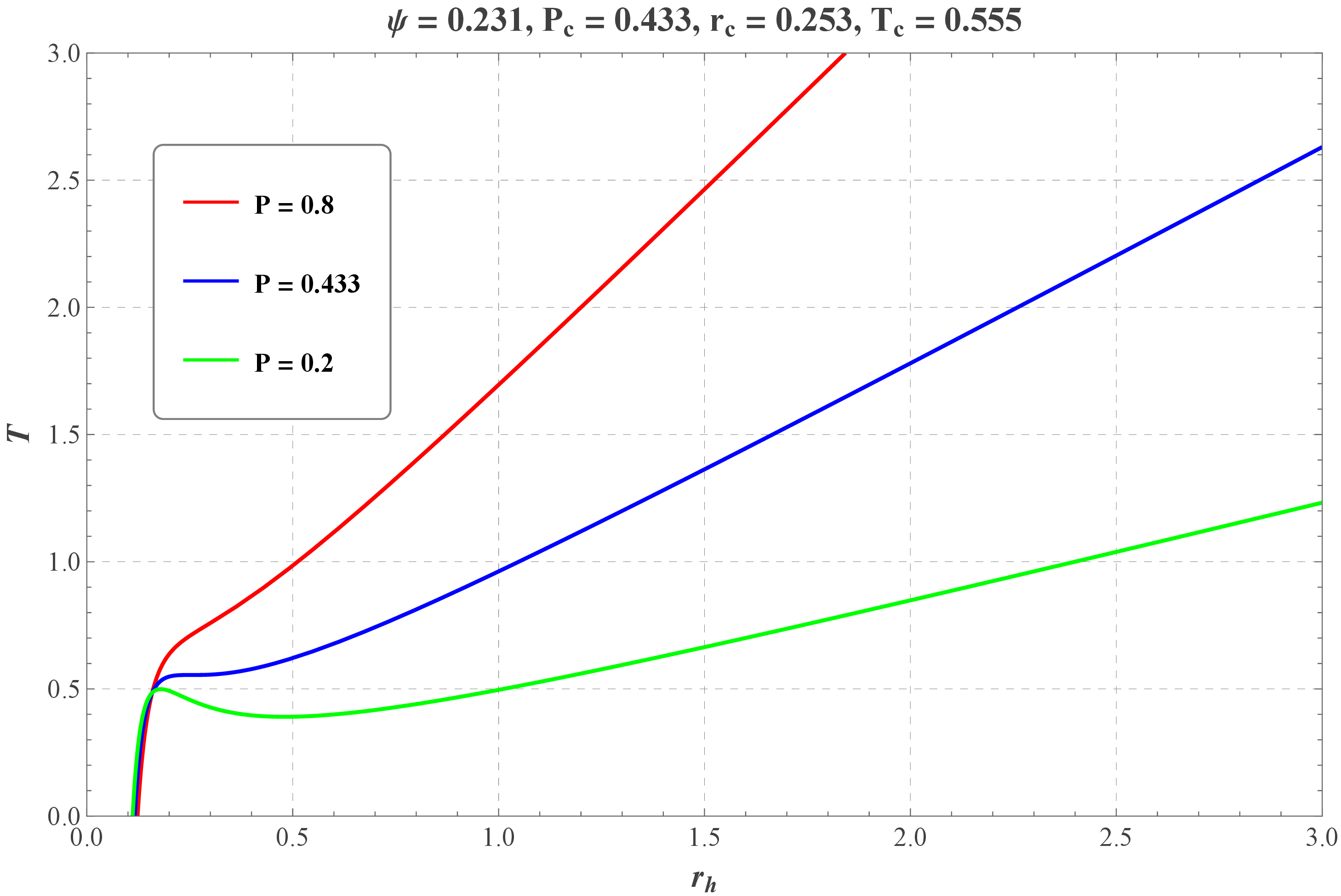}
        \caption{Isobars for $\psi=0.231$ with $(P_c,r_c,T_c)=(0.433,0.253,0.555)$.}
        \label{fig:t3}
    \end{subfigure}
    \quad
    \begin{subfigure}[b]{0.35\textwidth}
        \centering
        \includegraphics[width=\textwidth]{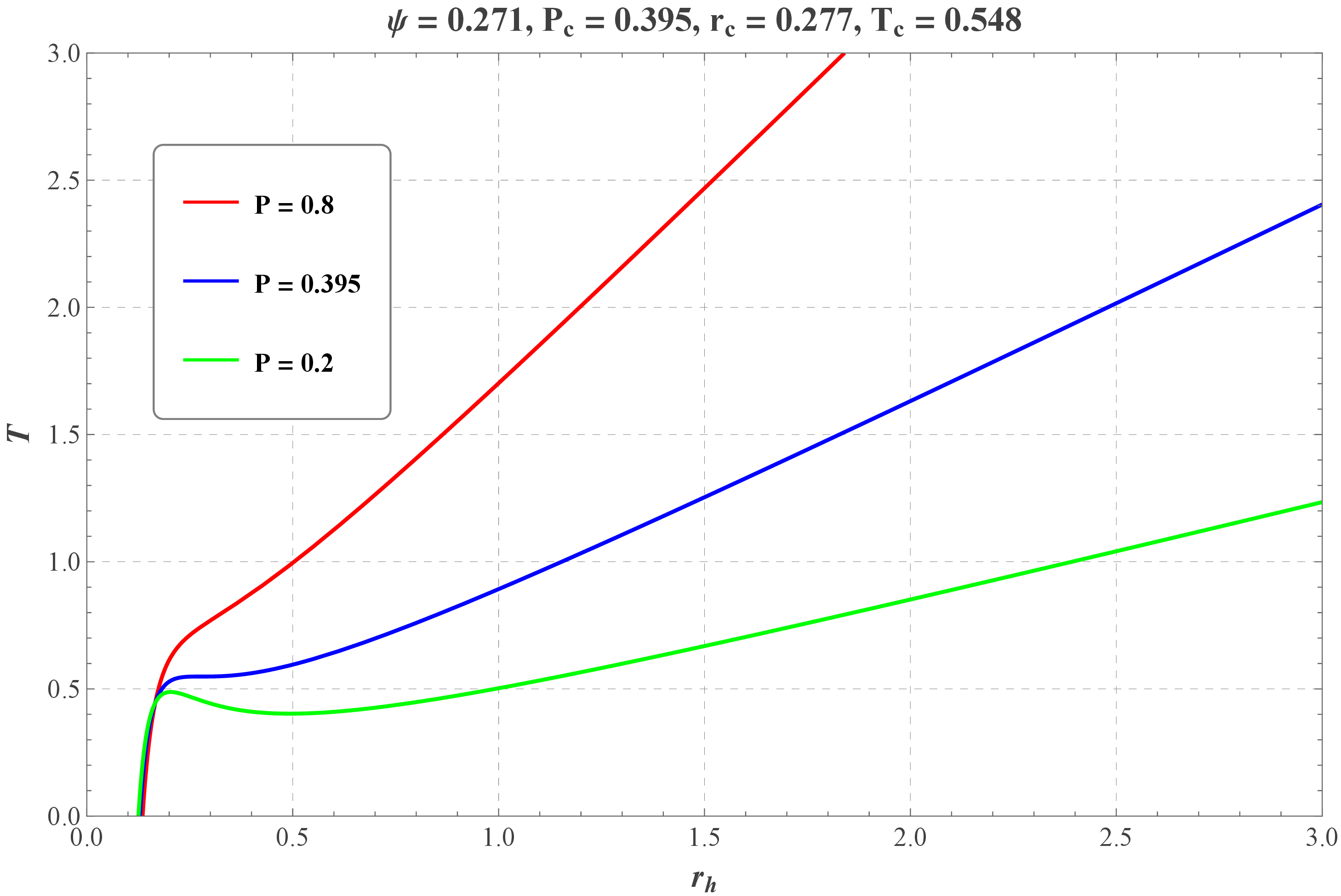}
        \caption{Isobars for $\psi=0.271$ with $(P_c,r_c,T_c)=(0.395,0.277,0.549)$.}
        \label{fig:t44}
    \end{subfigure}

    \caption{Temperature as a function of the horizon radius for representative values of the electric potential in Region~I.}
    \label{fig:temp12344}
\end{figure}

Figure~\ref{fig:temp12344} illustrates the temperature as a function of the horizon radius for representative values of the electric potential in Region~I. The thermodynamic behavior resembles of a  van der Waals-like system.  For pressures above the critical pressure $P_c$, the isobars are almost smooth and monotonic. At the critical pressure, the isobar includes a critical point satisfying the critically condition while for $P<P_c$ the familiar oscillatory behavior characteristic of a first-order phase transition emerges.

The four panels correspond to the critical points $(\psi,P_c,r_c,T_c)=(0.051,1.780,0.107,1.010)$, $(0.181,0.523,0.219,0.586)$, $(0.231,0.433,0.253,0.555)$, and $(0.271,0.395,0.277,0.549)$, respectively. As the electric potential increases within Region~I, the critical pressure decreases monotonically, whereas the critical horizon radius increases monotonically. The critical temperature also decreases throughout this region, consistent with the classification presented in Table~\ref{tab:summary_regions}.

\begin{figure}[H]
    \centering
    \begin{subfigure}[b]{0.30\textwidth}
        \centering
        \includegraphics[width=\textwidth]{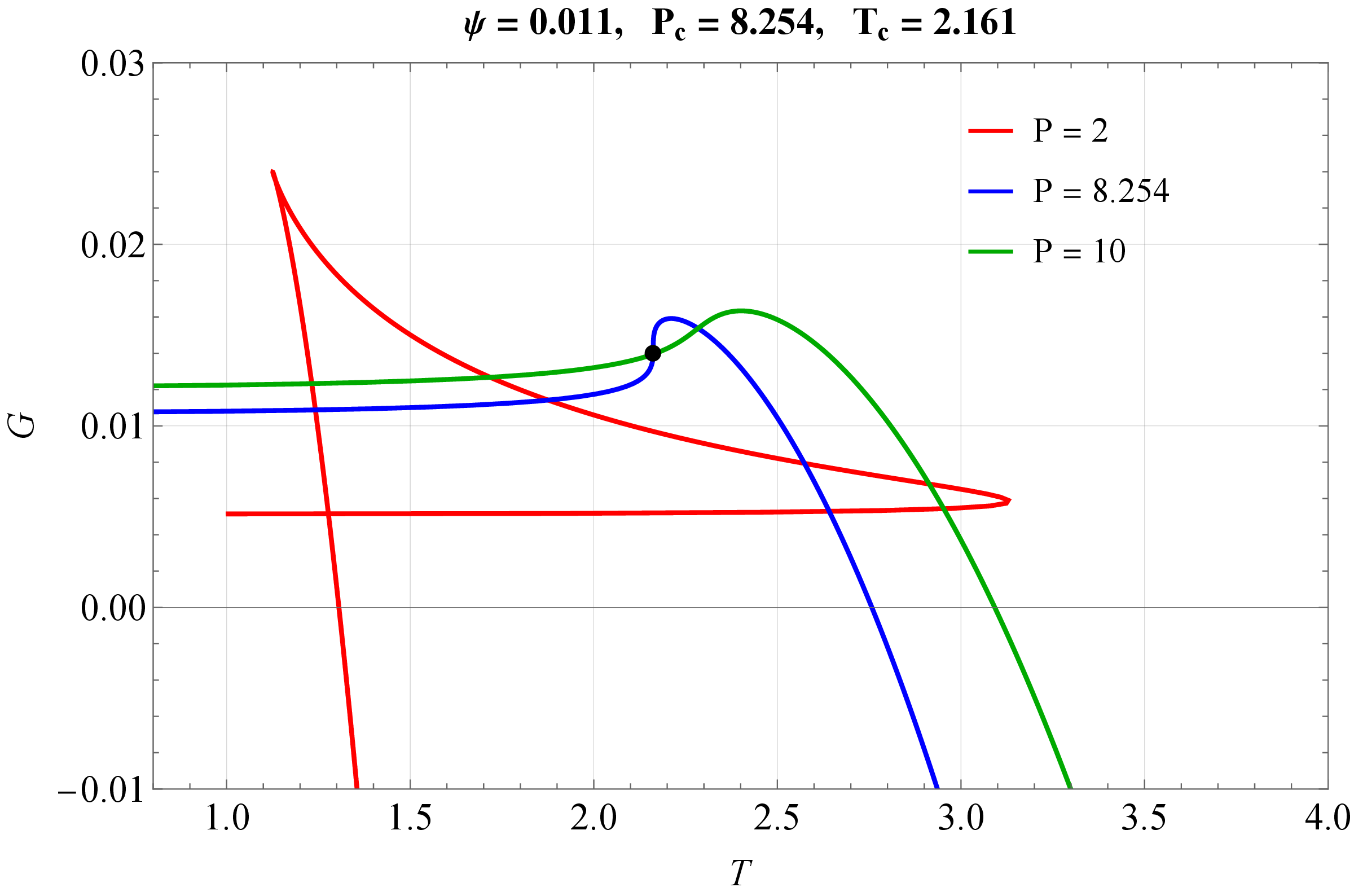}
        \caption{$\psi=0.011$}
        \label{fig:gt00}
    \end{subfigure}
    \hfill
    \begin{subfigure}[b]{0.30\textwidth}
        \centering
        \includegraphics[width=\textwidth]{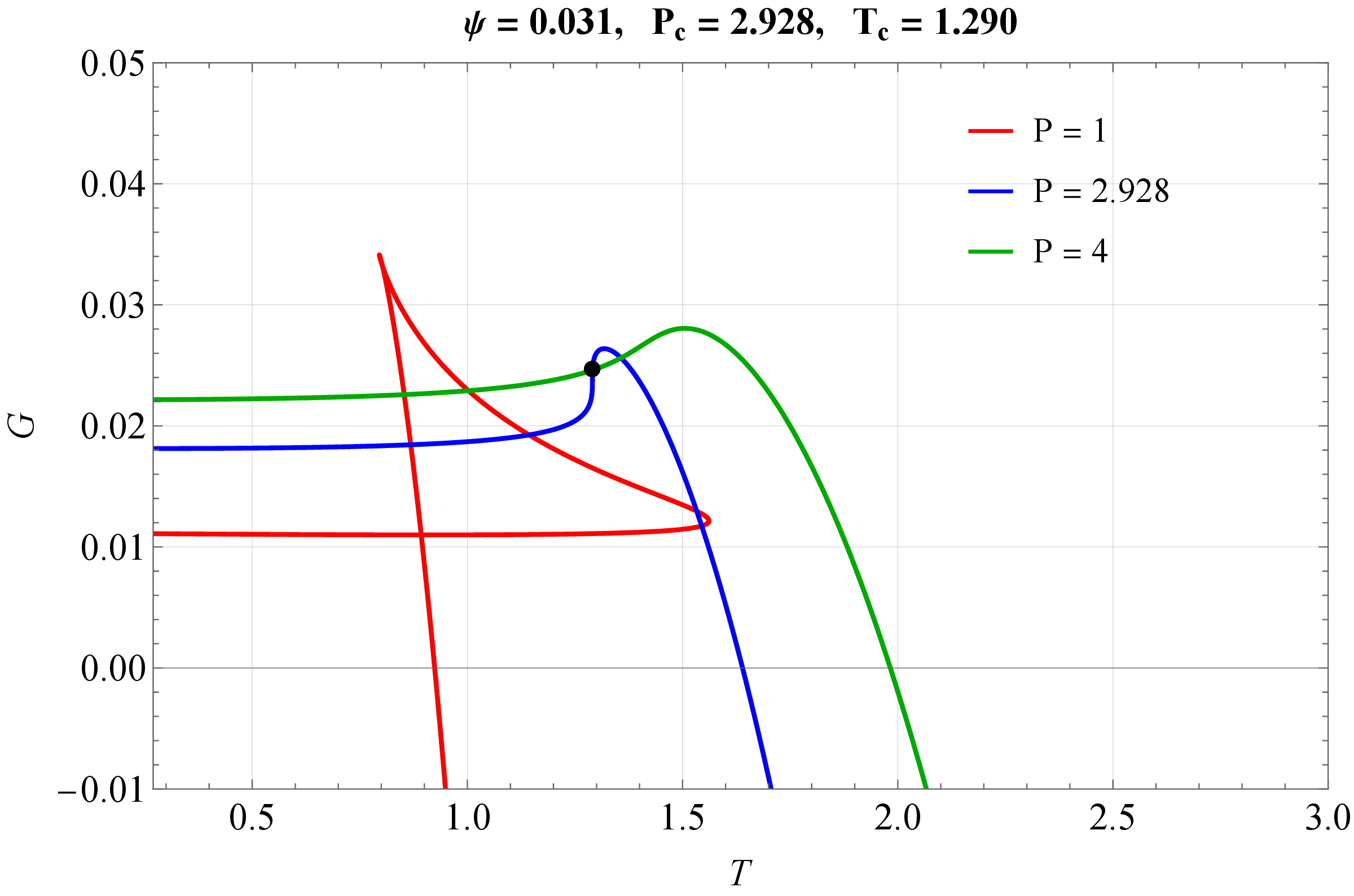}
        \caption{$\psi=0.031$}
        \label{fig:gt0}
    \end{subfigure}
    \hfill
    \begin{subfigure}[b]{0.30\textwidth}
        \centering
        \includegraphics[width=\textwidth]{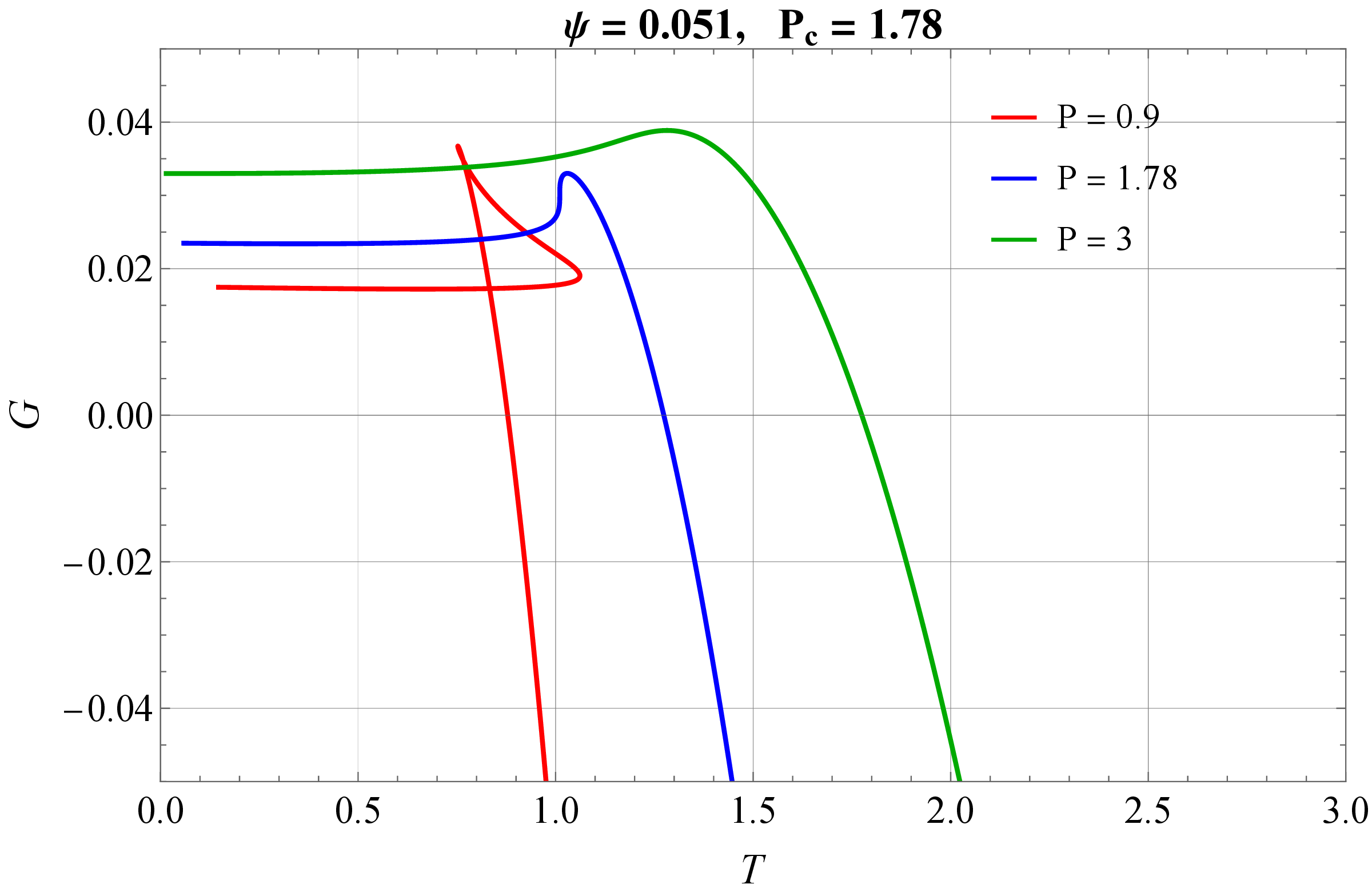}
        \caption{$\psi=0.051$}
        \label{fig:gt1}
    \end{subfigure}

    \par\medskip

    \begin{subfigure}[b]{0.30\textwidth}
        \centering
        \includegraphics[width=\textwidth]{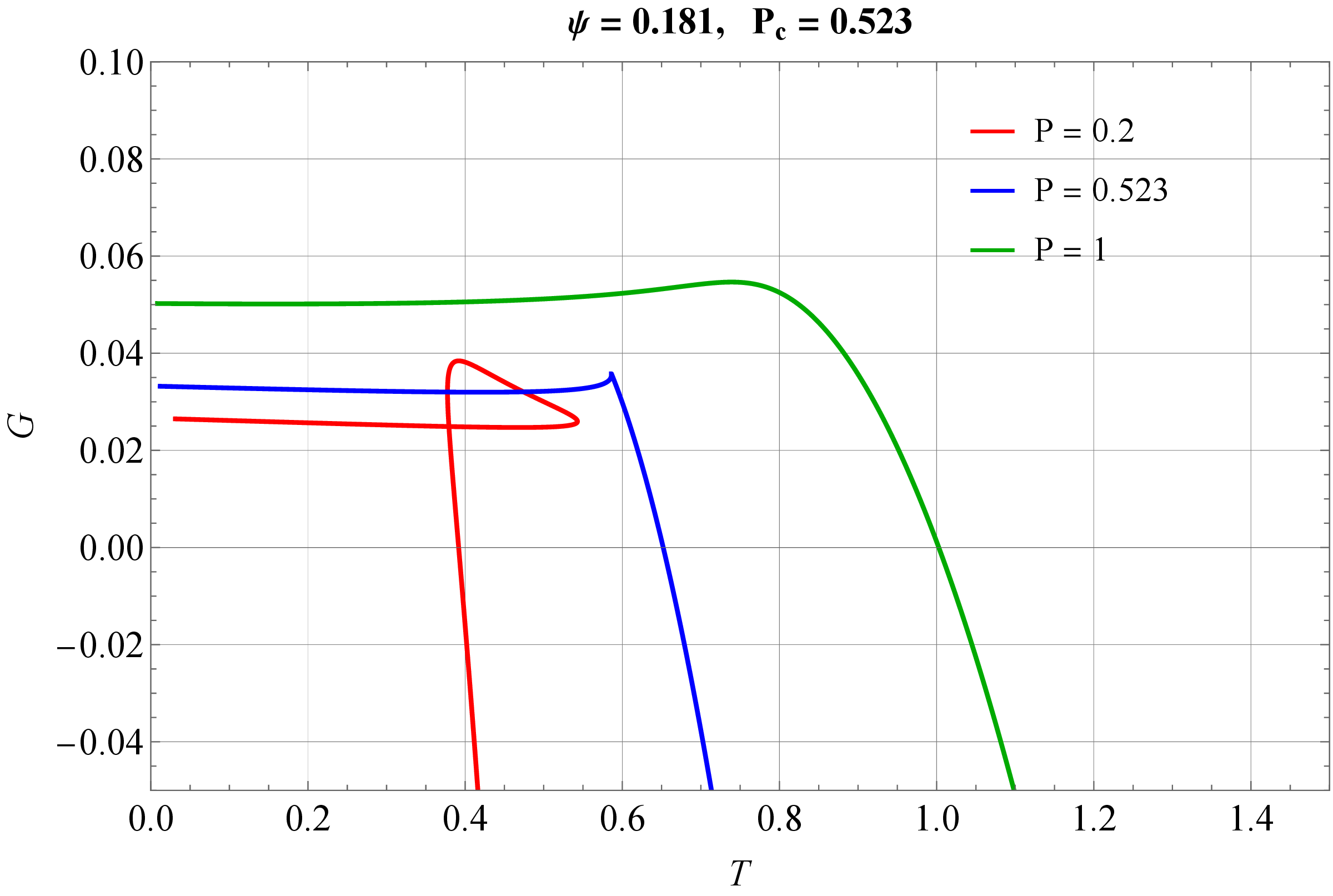}
        \caption{$\psi=0.181$}
        \label{fig:gt2}
    \end{subfigure}
    \hfill
    \begin{subfigure}[b]{0.30\textwidth}
        \centering
        \includegraphics[width=\textwidth]{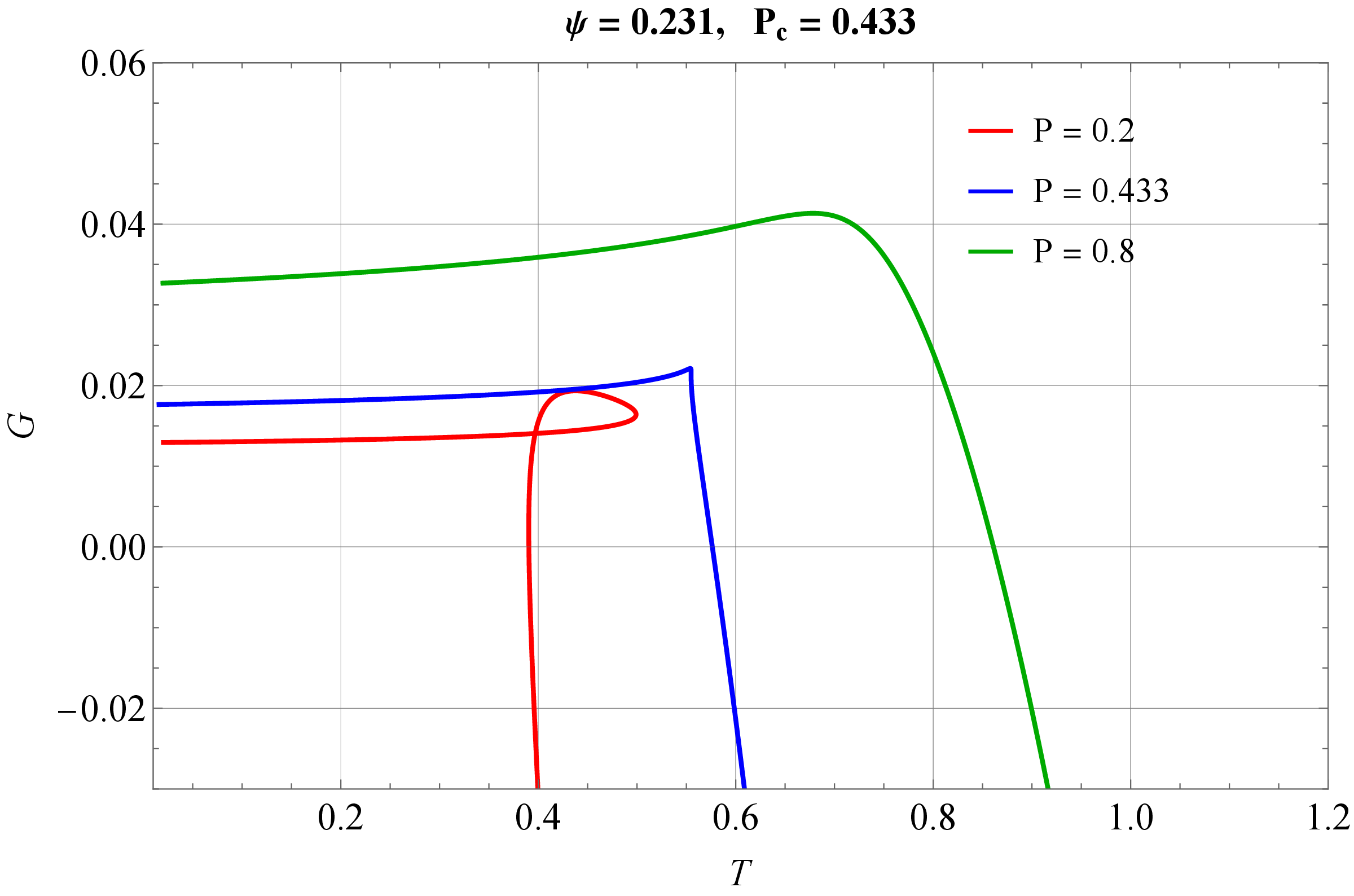}
        \caption{$\psi=0.231$}
        \label{fig:gt3}
    \end{subfigure}
    \hfill
    \begin{subfigure}[b]{0.30\textwidth}
        \centering
        \includegraphics[width=\textwidth]{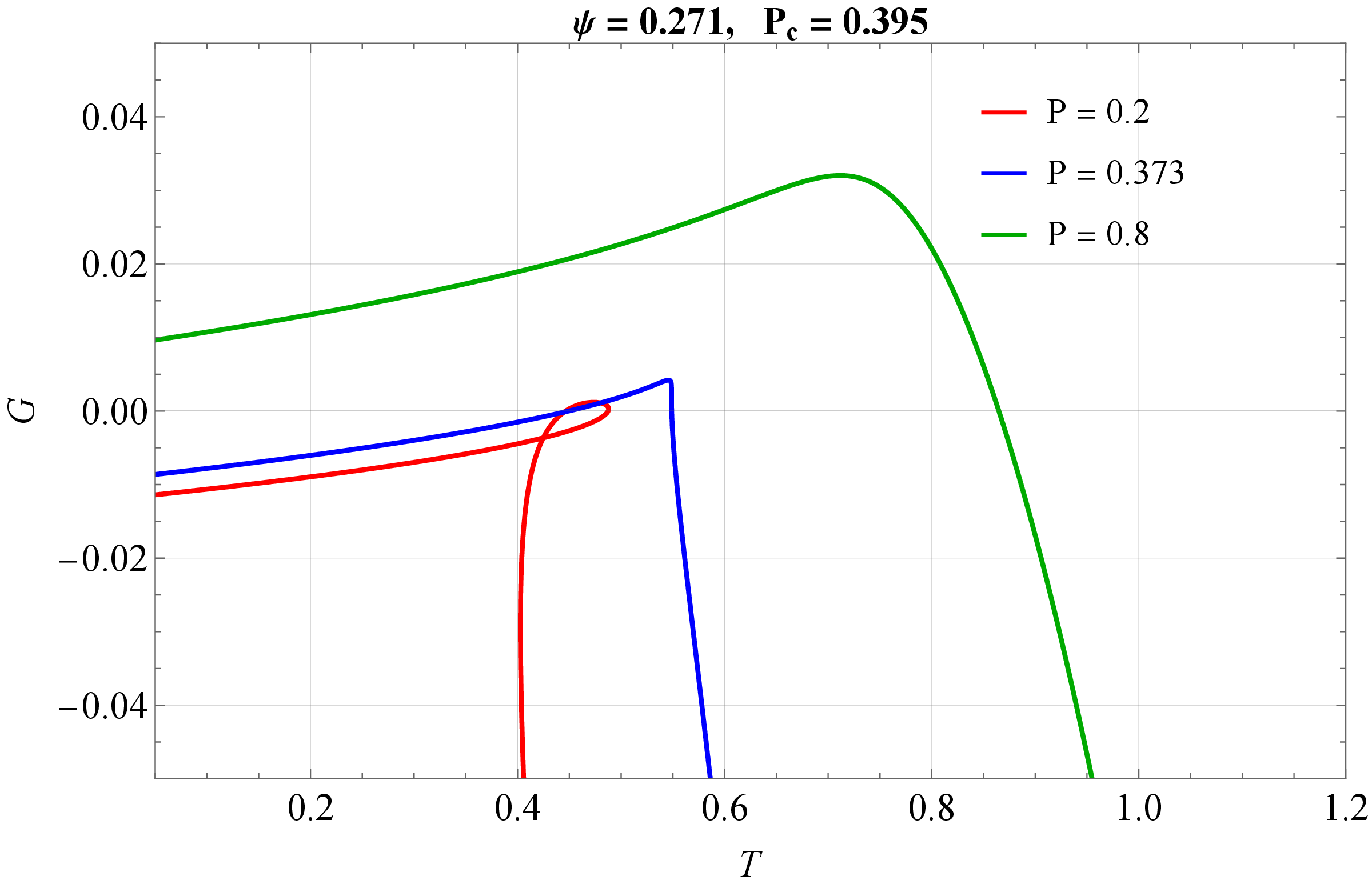}
        \caption{$\psi=0.271$}
        \label{fig:gt44}
    \end{subfigure}

    \caption{Gibbs free energy $G$ as a function of temperature $T$ for representative values of the electric potential in Region~I. The corresponding critical pressures are (a) $P_c=8.254$, (b) $P_c=2.928$, (c) $P_c=1.780$, (d) $P_c=0.523$, (e) $P_c=0.433$, and (f) $P_c=0.395$. For $P<P_c$, the Gibbs free energy exhibits the characteristic swallowtail structure associated with a first-order small/large black hole phase transition. As $P\rightarrow P_c$, the swallowtail shrinks continuously and terminates at the critical point. For $P>P_c$, the Gibbs free energy becomes a smooth single-valued function of temperature. As the electric potential increases, the swallowtail gradually evolves from a sharp to a more rounded structure, while the entire Gibbs free energy curve shifts toward lower values, indicating that the black hole phase becomes increasingly thermodynamically favored.}
    \label{fig:gibss12344}
\end{figure}

The Gibbs free energy as a function of temperature provides a direct characterization of the global thermodynamic phase structure. The corresponding results for representative values of the electric potential are shown in Fig.~\ref{fig:gibss12344}. Throughout Region~I, the system exhibits the characteristic small-large phase transition. For pressures below the critical pressure, the Gibbs free energy develops the familiar swallowtail structure, signaling a first-order small/large black hole phase transition. As the pressure approaches the critical value, the swallowtail continuously shrinks and terminates at the critical point, where the first-order phase transition ends. For pressures above the critical pressure, the Gibbs free energy becomes a smooth single-valued function of temperature, indicating the absence of a first order thermodynamic phase transitions.

Figure~\ref{fig:gibss12344} also illustrates how the Gibbs free energy evolves as the electric potential increases. At relatively small values of $\psi$, the swallowtail possesses sharp cusps with well-defined intersections between the competing thermodynamic branches. As $\psi$ increases, these cusps gradually become smoother and the swallowtail becomes increasingly rounded, reflecting the continuous modification of the phase structure induced by the nonminimal gauge--curvature coupling. Meanwhile, the entire Gibbs free energy curve shifts toward lower values, indicating that increasing the electric potential thermodynamically favors the black hole phase. 

\subsection{Phase Structure in Region II}

\begin{figure}[H]
    \centering

    \begin{subfigure}[b]{0.35\textwidth}
        \centering
        \includegraphics[width=\textwidth]{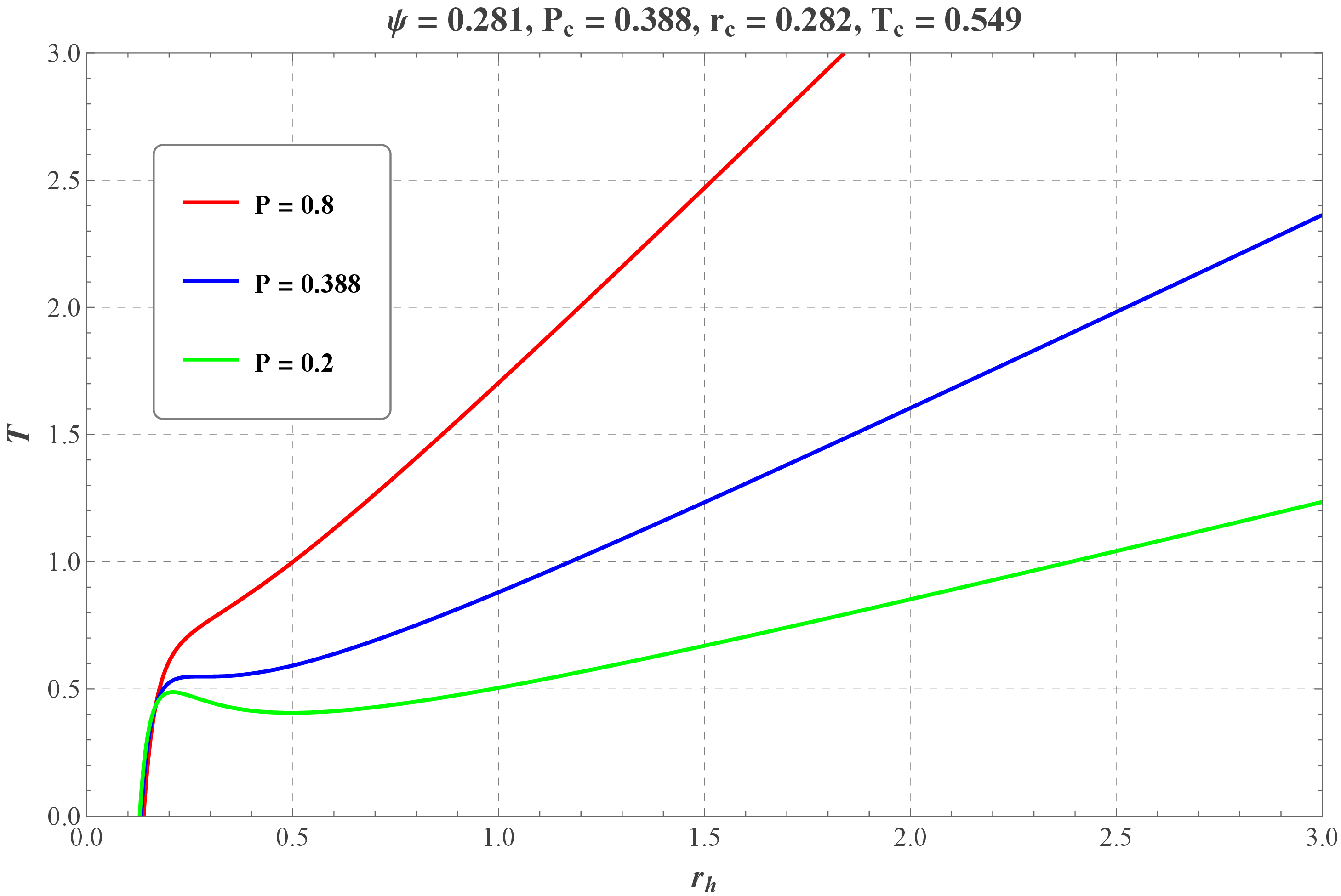}
        \caption{$\psi=0.281$.\\
        $P_c=0.388,\ r_c=0.282,\ T_c=0.549$.}
        \label{fig:t21}
    \end{subfigure}
    \quad
    \begin{subfigure}[b]{0.35\textwidth}
        \centering
        \includegraphics[width=\textwidth]{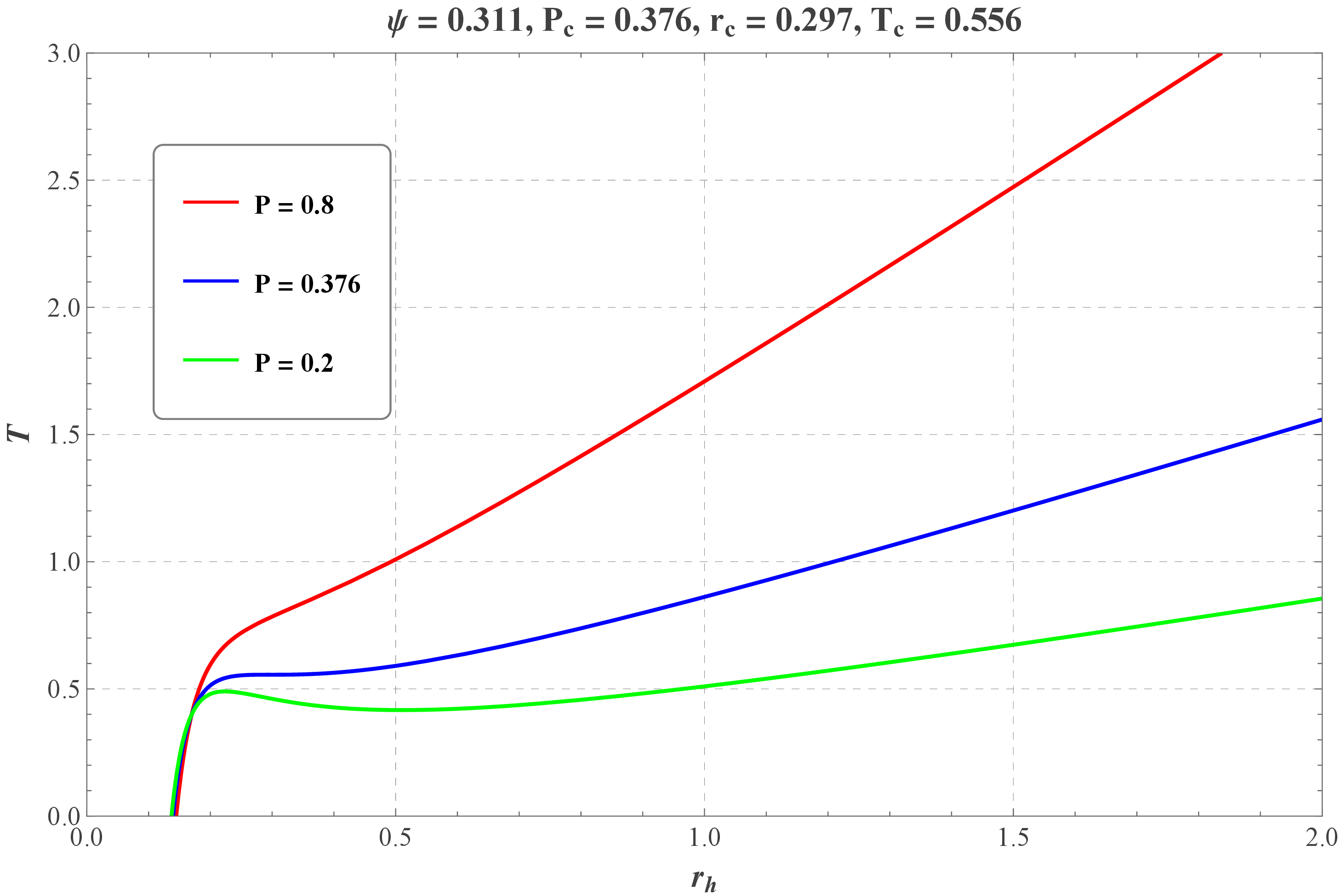}
        \caption{$\psi=0.311$.\\
        $P_c=0.376,\ r_c=0.297,\ T_c=0.556$.}
        \label{fig:t22}
    \end{subfigure}

    \begin{subfigure}[b]{0.35\textwidth}
        \centering
        \includegraphics[width=\textwidth]{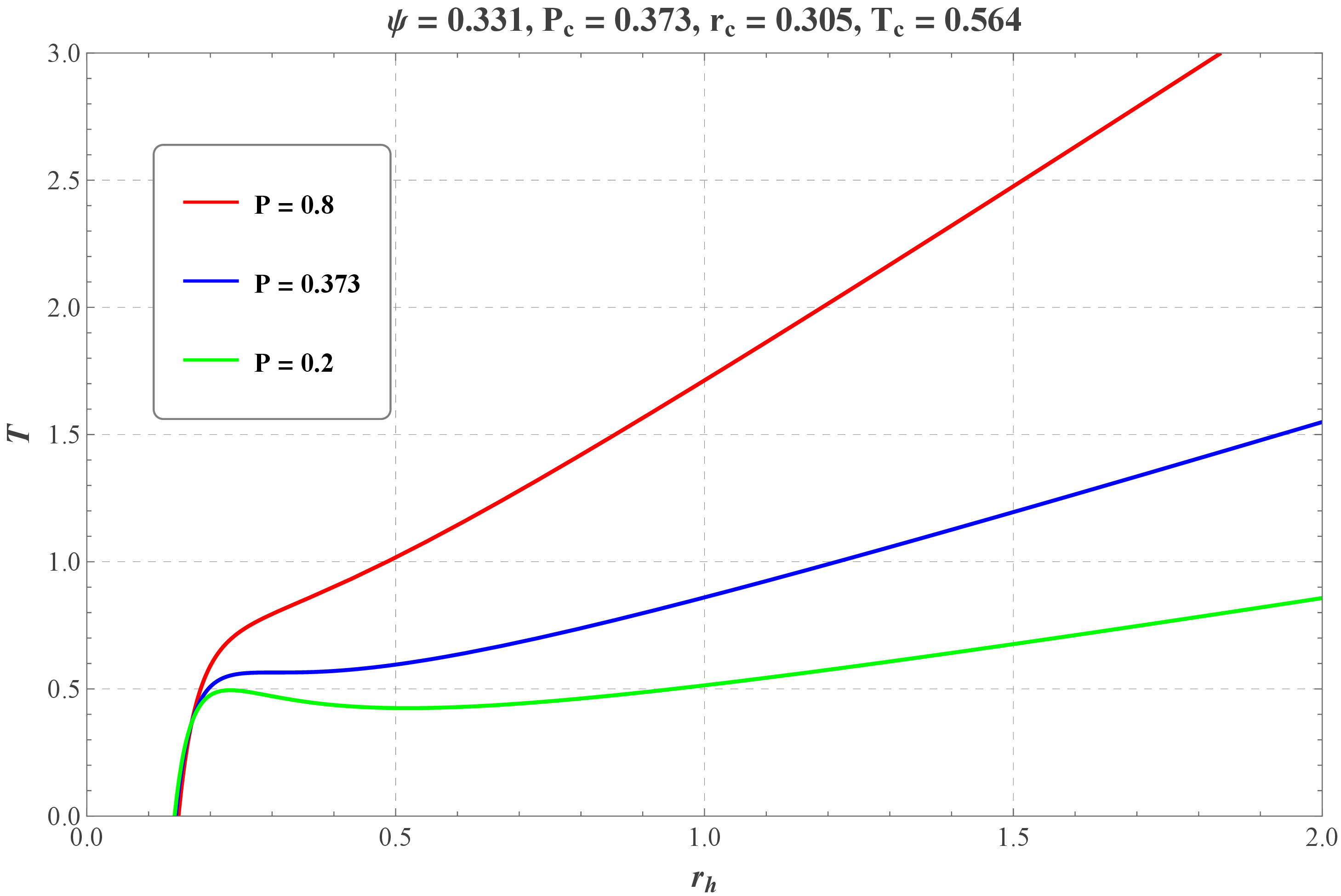}
        \caption{$\psi=0.331$.\\
        $P_c=0.373,\ r_c=0.305,\ T_c=0.564$.}
        \label{fig:t23}
    \end{subfigure}
    \quad
    \begin{subfigure}[b]{0.35\textwidth}
        \centering
        \includegraphics[width=\textwidth]{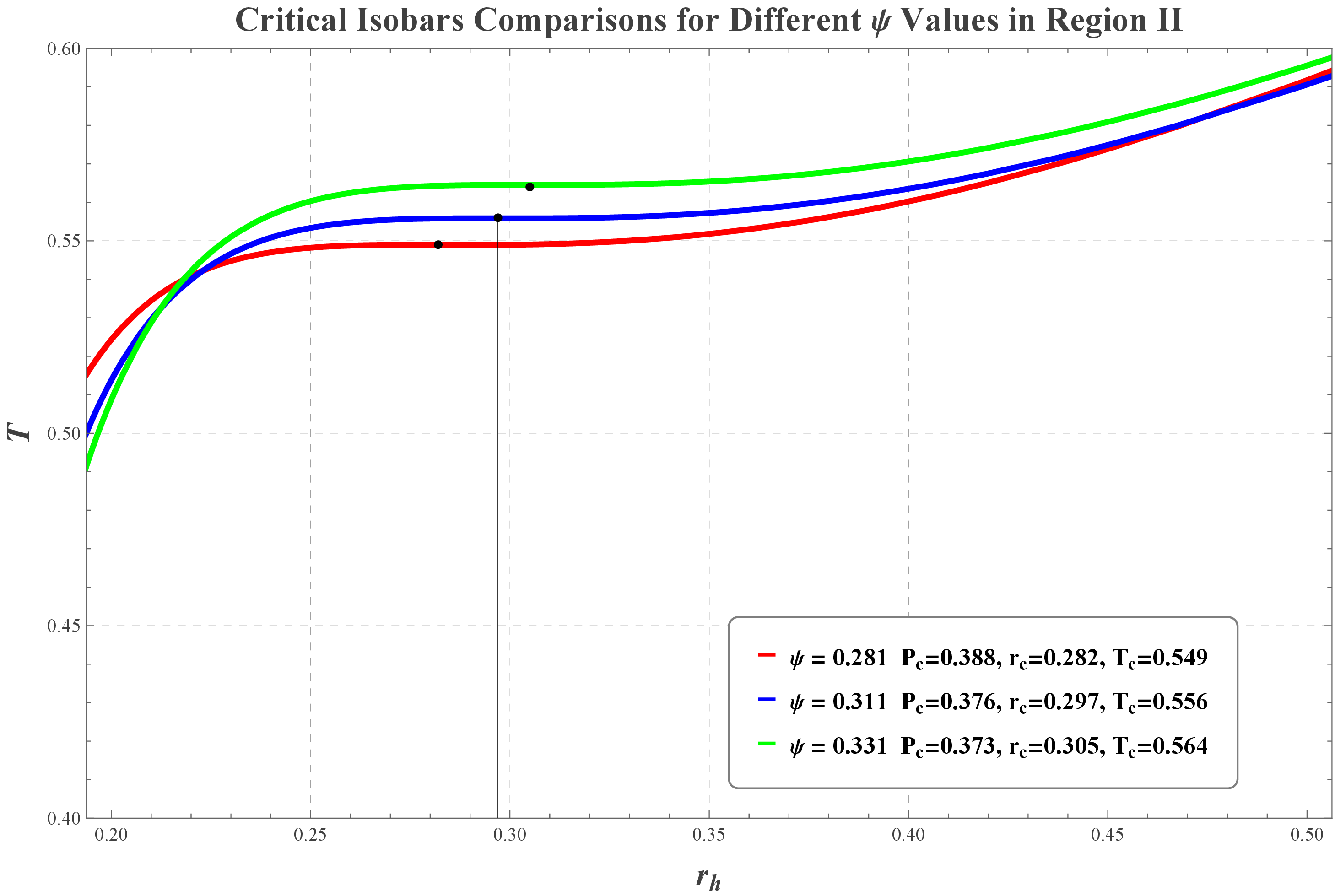}
        \caption{Comparison of\\
        critical isobars.}
        \label{fig:t24}
    \end{subfigure}

    \caption{Temperature as a function of the horizon radius for representative values of the electric potential in Region~II. The first three panels display the temperature profiles at and around the corresponding critical pressure for $\psi=0.281$, $0.311$, and $0.331$. The fourth panel compares the critical isobars, illustrating the monotonic increase of the critical temperature with increasing electric potential.}
    \label{fig:t21to24}
\end{figure}

We next examine the thermodynamic behavior in Region~II for the representative electric potentials $\psi=0.281$, $0.311$, and $0.331$. The corresponding temperature profiles are shown in Fig.~\ref{fig:t21to24}. The qualitative phase structure remains identical to that of a small-large phase transition profile. 

Although the overall thermodynamic behavior is unchanged from Region~I, the evolution of the critical parameters differs. As the electric potential increases throughout Region~II, the critical pressure continues to decrease, while both the critical horizon radius and the critical temperature increase. The first three panels of Fig.~\ref{fig:t21to24} display the temperature profiles for the selected values of $\psi$, whereas the fourth panel compares the critical isobars. The comparison clearly demonstrates the monotonic increase of the critical temperature with increasing electric potential, in agreement with the classification summarized in Table~\ref{tab:summary_regions}.

\begin{figure}[H]
    \centering
    \begin{subfigure}[b]{0.30\textwidth}
        \centering
        \includegraphics[width=\textwidth]{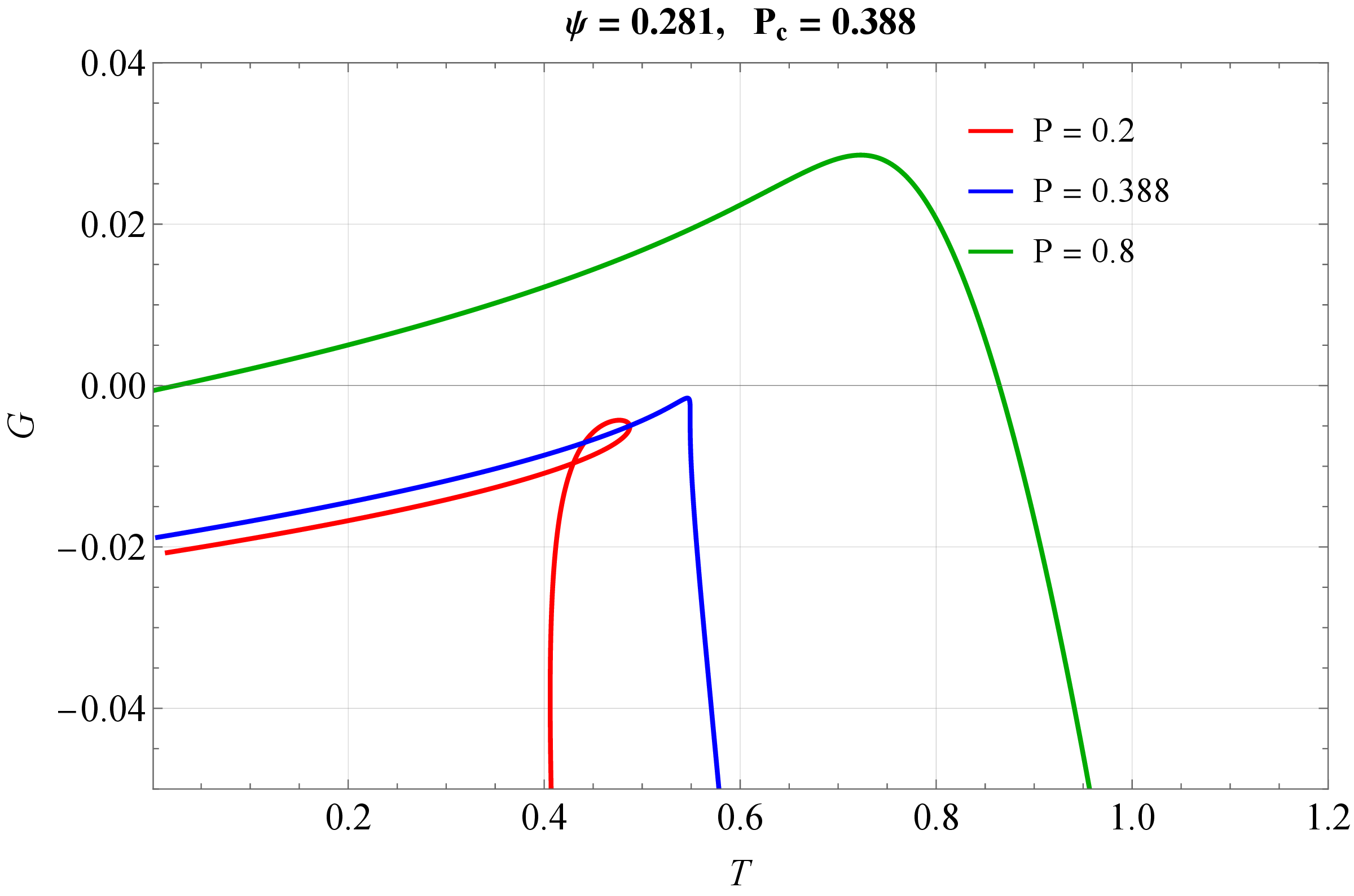}
        \caption{$\psi=0.281$}
        \label{fig:gt21}
    \end{subfigure}
    \hfill
    \begin{subfigure}[b]{0.30\textwidth}
        \centering
        \includegraphics[width=\textwidth]{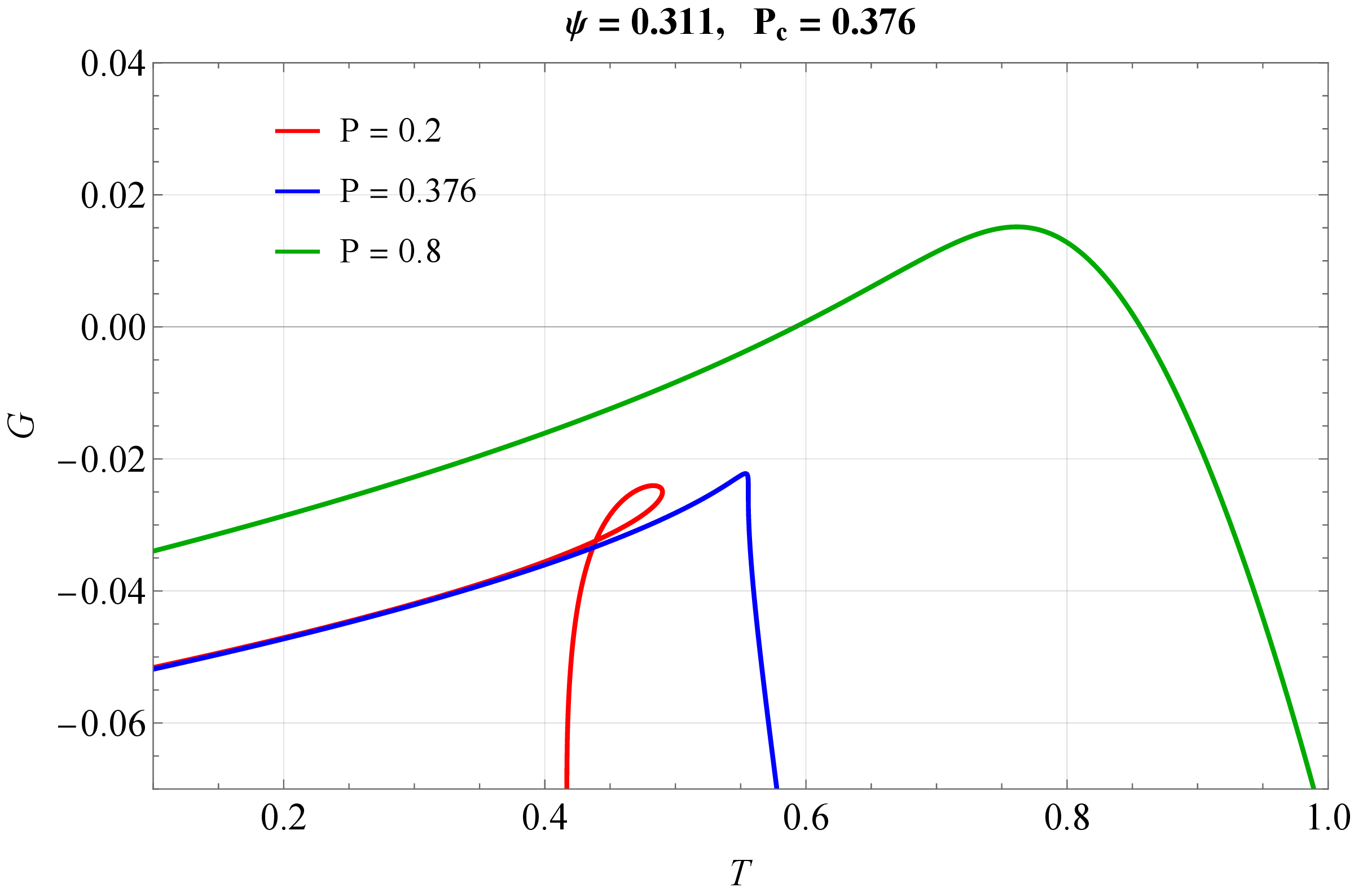}
        \caption{$\psi=0.311$}
        \label{fig:gt22}
    \end{subfigure}
    \hfill
    \begin{subfigure}[b]{0.30\textwidth}
        \centering
        \includegraphics[width=\textwidth]{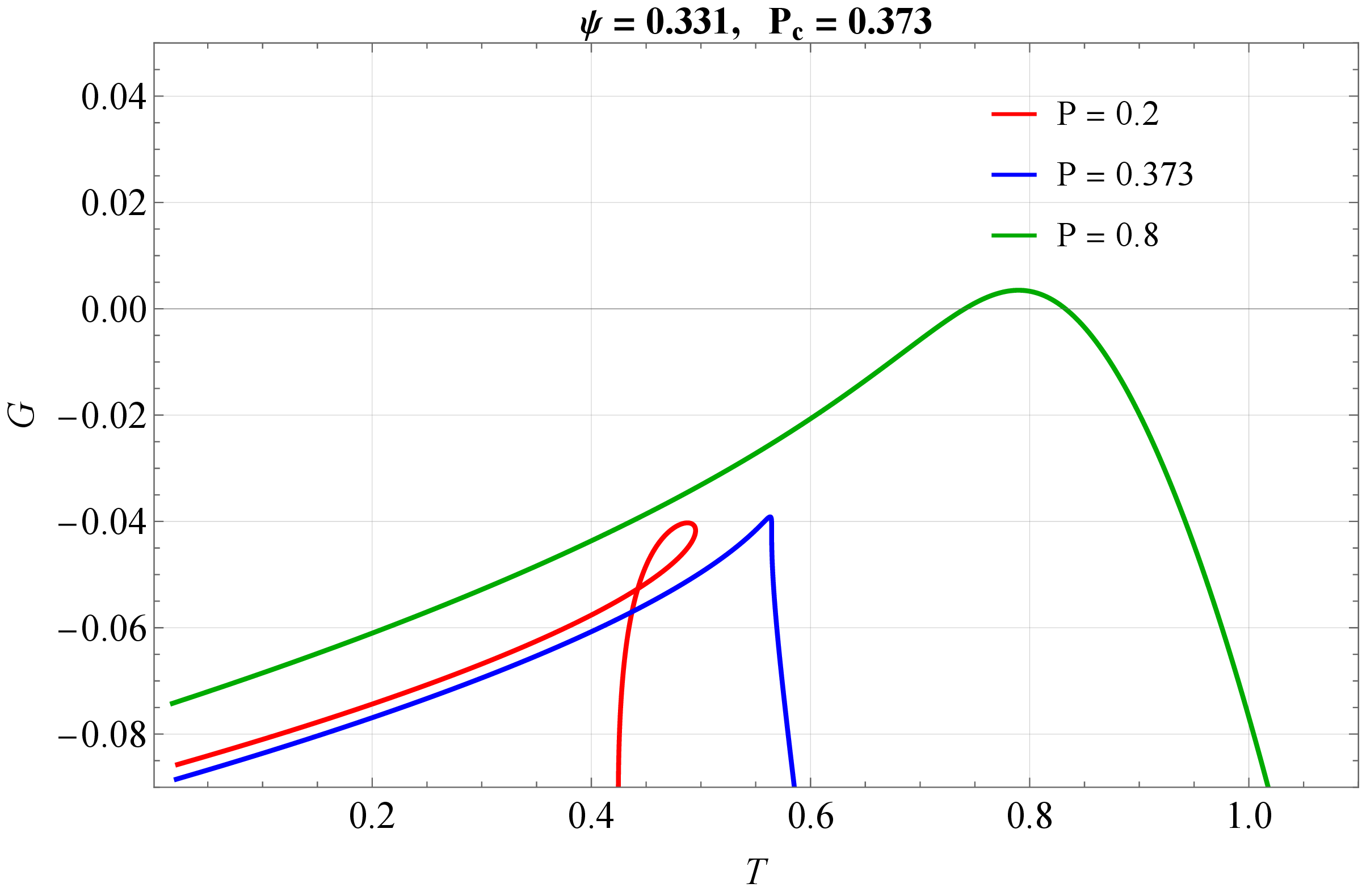}
        \caption{$\psi=0.331$}
        \label{fig:gt23}
    \end{subfigure}

    \par\medskip

    \begin{subfigure}[b]{0.45\textwidth}
        \centering
        \includegraphics[width=\textwidth]{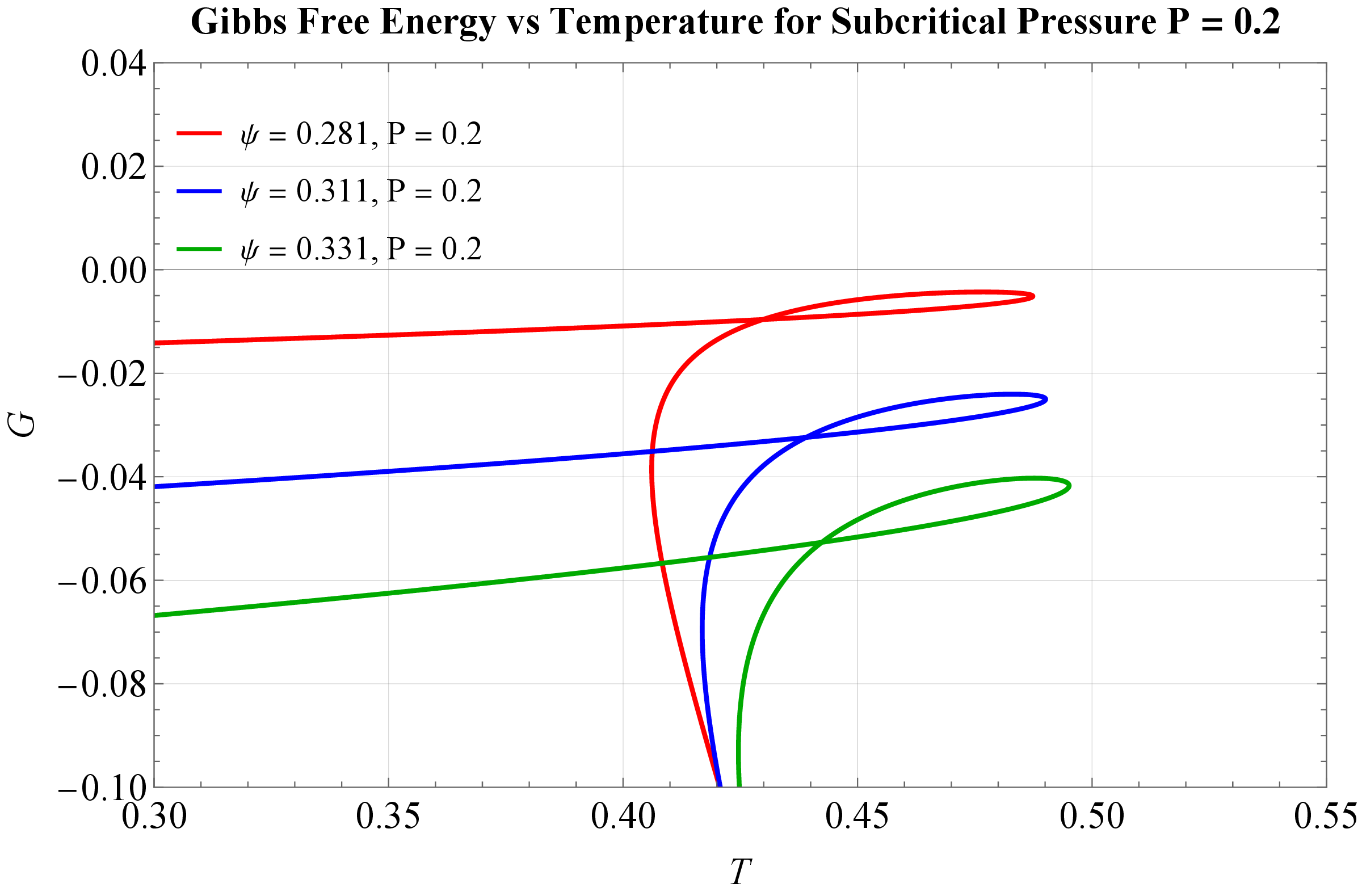}
        \caption{Comparison at $P=0.2$.}
        \label{fig:gt24}
    \end{subfigure}

    \caption{Gibbs free energy $G$ as a function of temperature $T$ for representative values of the electric potential in Region~II. Panels (a)--(c) show the Gibbs free energy for $\psi=0.281$, $0.311$, and $0.331$, respectively. For $P<P_c$, the Gibbs free energy exhibits a rounded swallowtail (ribbon-like) structure characteristic of a first-order phase transition. At $P=P_c$, the swallowtail terminates at the critical point, whereas for $P>P_c$ the Gibbs free energy becomes a smooth single-valued function. Panel (d) compares the subcritical Gibbs free energy curves at the common pressure $P=0.2$, illustrating that the entire swallowtail shifts toward more negative values of $G$ as the electric potential increases.}
    \label{fig:gt21-24}
\end{figure}

The Gibbs free energy as a function of temperature for the representative values of the electric potential in Region~II is shown in Fig.~\ref{fig:gt21-24}. As expected, the thermodynamic behavior resembles a first order small-large phase transition. For pressures below the critical pressure, the Gibbs free energy develops the familiar swallowtail structure associated with a first-order small/large black hole phase transition. Compared with Region~I, however, the swallowtail becomes increasingly rounded, resembling a ribbon-like structure rather than the sharp cusps observed at smaller values of the electric potential. At the critical pressure, the swallowtail terminates at the critical point, while for $P>P_c$ the Gibbs free energy evolves into a smooth single-valued function of temperature.

The first three panels of Fig.~\ref{fig:gt21-24} correspond to $\psi=0.281$, $0.311$, and $0.331$, respectively. The fourth panel compares the Gibbs free energy for these three values at the common subcritical pressure $P=0.2$. As the electric potential increases, the ribbon-like swallowtail gradually shifts toward more negative values of the Gibbs free energy while preserving its overall topology. This downward shift indicates that increasing the electric potential progressively favors the black hole phase thermodynamically.

\subsection{Phase Structure in Region III}

Region~III is characterized by the simultaneous increase of the critical horizon radius $r_c$, critical pressure $P_c$, and critical temperature $T_c$ as the electric potential $\psi$ increases. Despite this distinct scaling behavior of the critical quantities, the qualitative thermodynamic behavior remains similar to that observed in Region~II. In particular, the Gibbs free energy exhibits the characteristic swallowtail structure below the critical pressure, which terminates at the critical point and evolves into a smooth single-valued curve for pressures above $P_c$.

\begin{figure}[H]
    \centering
    \begin{subfigure}[b]{0.40\textwidth}
        \centering
        \includegraphics[width=\textwidth]{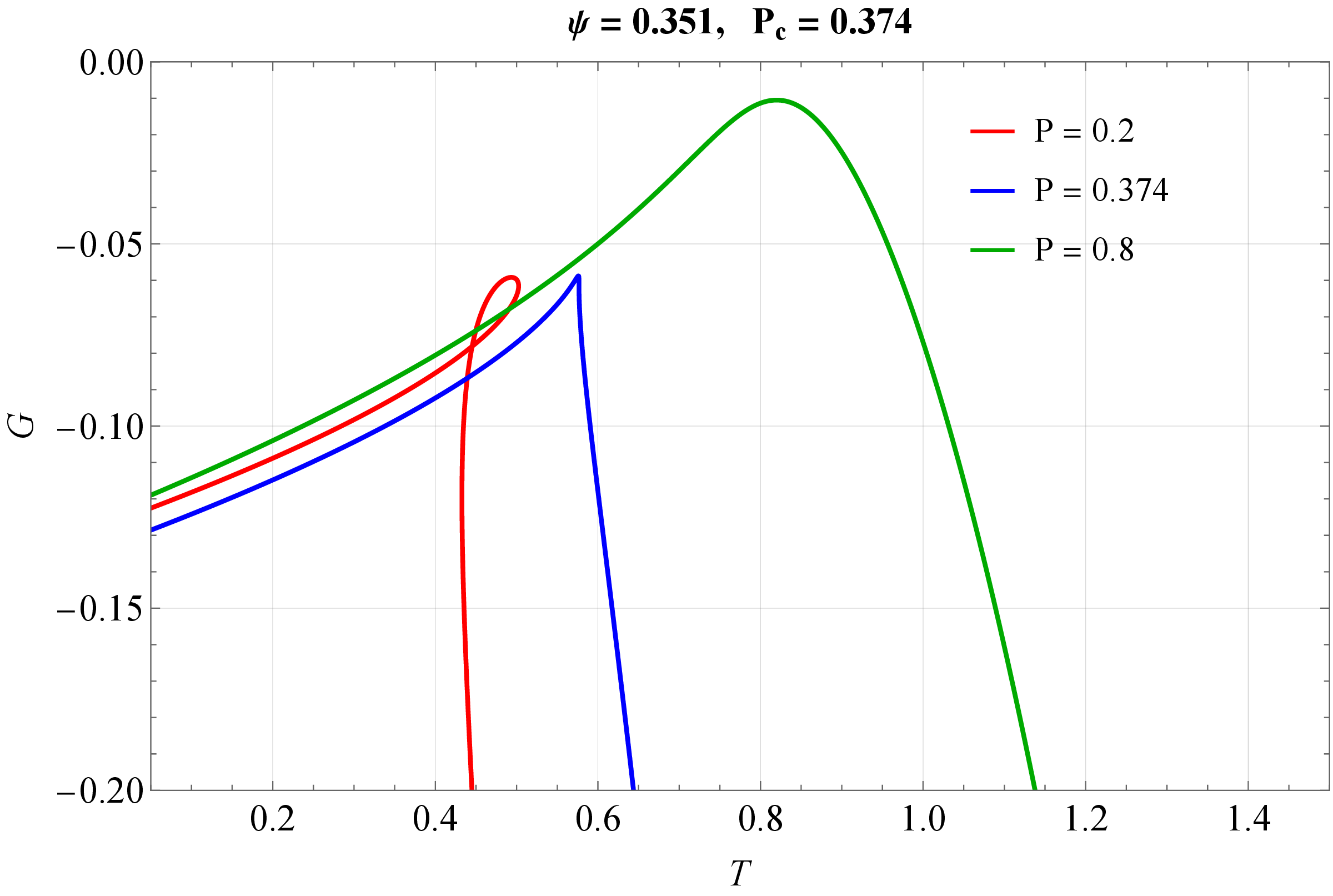}
        \caption{$\psi=0.351$ with $P_c=0.374$. The swallowtail structure is shown for $P=0.2<P_c$, while the curve becomes smooth for $P=0.8>P_c$.}
        \label{fig:gt31}
    \end{subfigure}
    \quad
    \begin{subfigure}[b]{0.40\textwidth}
        \centering
        \includegraphics[width=\textwidth]{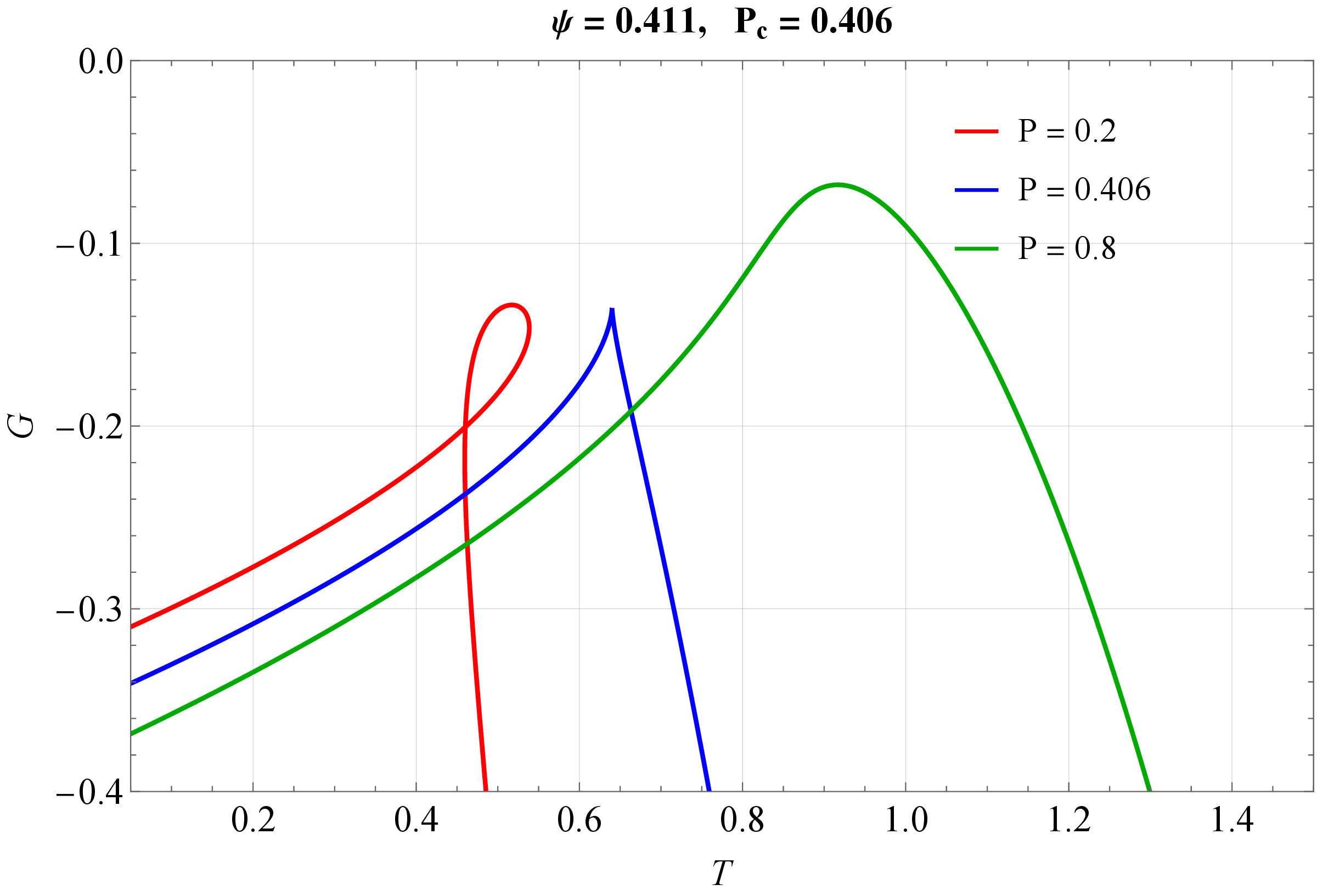}
        \caption{$\psi=0.411$ with $P_c=0.406$. The swallowtail persists for $P=0.2<P_c$ and disappears for $P=0.8>P_c$.}
        \label{fig:gt32}
    \end{subfigure}
    \caption{Gibbs free energy $G$ as a function of temperature $T$ for representative values of the electric potential in Region~III. Although the critical quantities $(r_c,P_c,T_c)$ all increase throughout this region, the qualitative thermodynamic behavior remains unchanged. Below the critical pressure, the Gibbs free energy exhibits the characteristic swallowtail structure associated with a first-order phase transition, whereas above the critical pressure it becomes smooth, indicating a single thermodynamically stable phase.}
    \label{fig:gt31-32}
\end{figure}

Since the overall phase structure is unchanged, we present only two representative Gibbs free energy diagrams corresponding to $\psi=0.351$ and $\psi=0.411$, as shown in Fig.~\ref{fig:gt31-32}.

In contrast to the conventional RN--AdS black hole, for which the van der Waals-like phase transition is characteristic of the canonical ensemble with fixed electric charge, while the grand canonical ensemble with fixed electric potential exhibits a Hawking--Page transition, our nonminimally coupled model displays a qualitatively different thermodynamic behavior. In particular, the grand canonical ensemble can exhibit a van der Waals-like phase structure for appropriate values of the electric potential. This indicates that the nonminimal coupling significantly modifies the ensemble-dependent phase structure and can induce critical behavior that is absent in the conventional RN--AdS case.

\subsection{Phase Structure in Region IV}

\begin{figure}[H]
    \centering
    \begin{subfigure}[b]{0.40\textwidth}
        \centering
        \includegraphics[width=\textwidth]{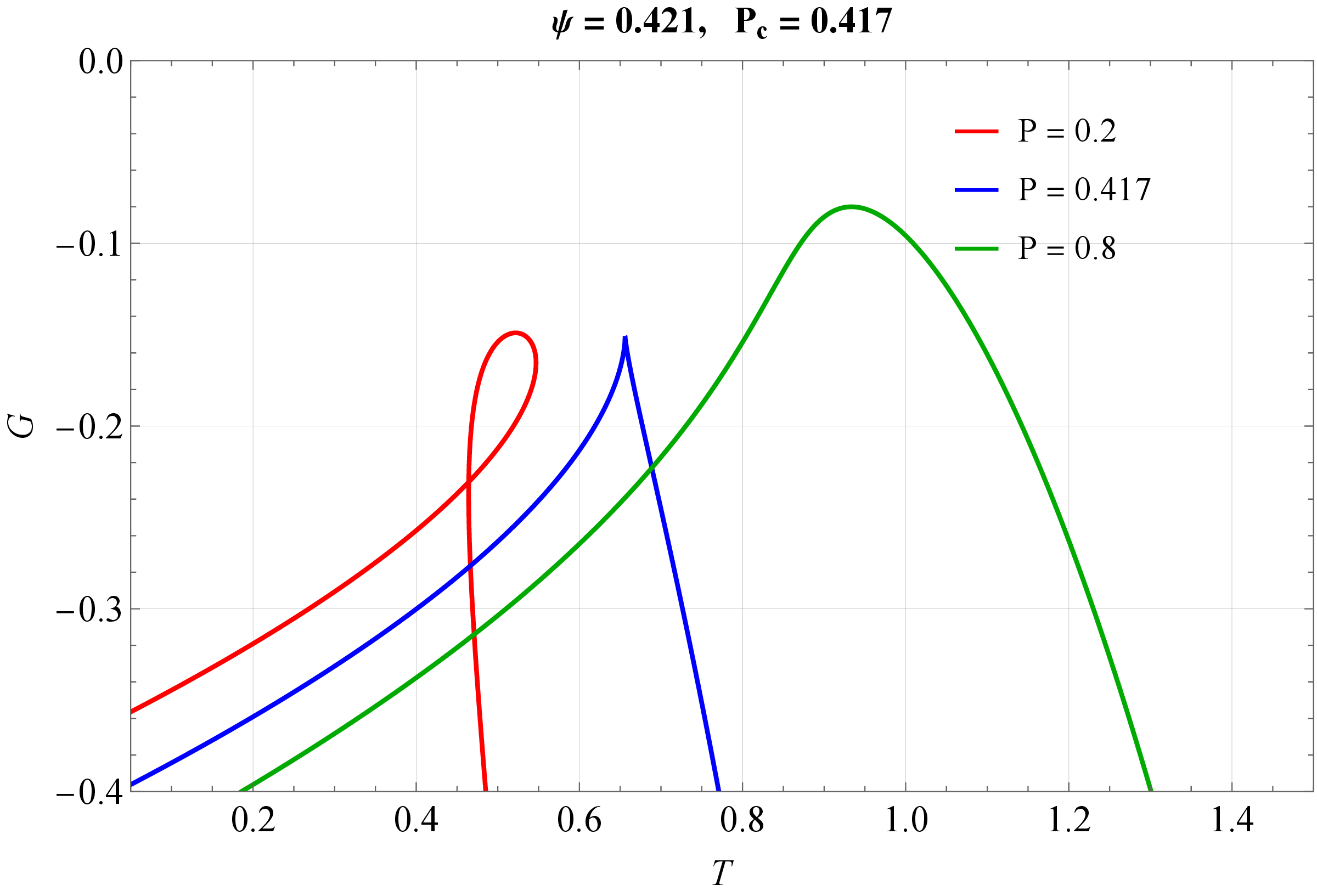}
        \caption{$\psi=0.421$ with $P_c=0.417$. The swallowtail is present for $P=0.2<P_c$ and disappears for $P=0.8>P_c$.}
        \label{fig:gt41}
    \end{subfigure}
    \quad
    \begin{subfigure}[b]{0.40\textwidth}
        \centering
        \includegraphics[width=\textwidth]{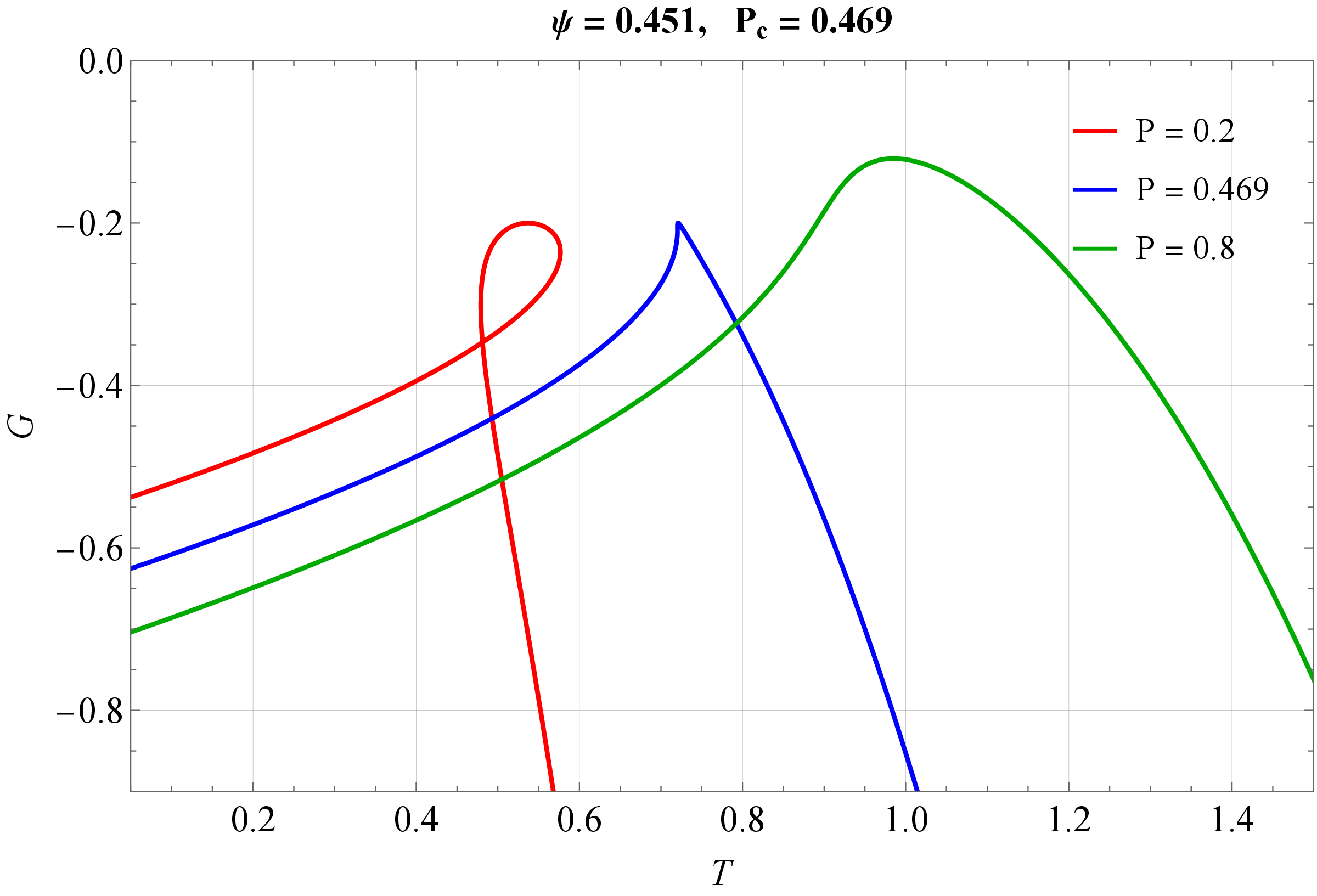}
        \caption{$\psi=0.451$ with $P_c=0.469$. The swallowtail persists for $P=0.2<P_c$ and evolves into a smooth curve for $P=0.8>P_c$.}
        \label{fig:gt42}
    \end{subfigure}

    \caption{Gibbs free energy $G$ as a function of temperature $T$ for representative values of the electric potential in Region~IV. The qualitative phase structure remains of the small/large phase transition. Below the critical pressure, the Gibbs free energy exhibits a swallowtail structure characteristic of a first-order small/large phase transition, while above the critical pressure it becomes smooth, indicating a single thermodynamically stable phase. As the electric potential increases, the critical point shifts toward higher temperatures, whereas the Gibbs free energy is displaced to more negative values.}
    \label{fig:gt41-42}
\end{figure}

The thermodynamic behavior in Region~IV remains qualitatively identical to that of the previous regions. The temperature profiles continue to exhibit the characteristic van der Waals oscillation below the critical pressure, while the Gibbs free energy displays the familiar swallowtail structure associated with a first-order phase transition. Since the temperature profiles do not introduce any new qualitative features, we present only the Gibbs free energy as a function of temperature for two representative values of the electric potential, $\psi=0.421$ and $\psi=0.451$, as shown in Fig.~\ref{fig:gt41-42}.

Unlike Region~III, however, the critical scaling is reversed. As the electric potential increases, both the critical pressure and the critical temperature continue to increase, whereas the critical horizon radius decreases. Thus, the critical point shifts toward smaller black holes while simultaneously requiring higher pressure and higher temperature to reach the critical state.

A notable feature of this region is the opposite scaling of the critical horizon radius compared with Region~III. Although the critical pressure and critical temperature continue to increase with increasing electric potential, the critical horizon radius decreases. This behavior indicates that the nonminimal gauge--curvature coupling drives the system toward criticality through progressively smaller black holes, while simultaneously increasing the thermodynamic conditions required to reach the critical point. Furthermore, the entire Gibbs free-energy curve shifts toward more negative values as $\psi$ increases, implying that the black-hole phase becomes globally more thermodynamically favorable relative to thermal AdS.

\subsection{Phase Structure in Region V}

In this interval of the electric potential, the criticality conditions (\ref{crit_t}) admit two distinct branches of solutions. The corresponding critical points are listed in Table~\ref{tab:critical_branches}. Branch~1 is characterized by a very small critical horizon radius, beginning at approximately $r_{c1}\simeq0.09$, together with unusually large values of the critical pressure and critical temperature. In contrast, Branch~2 displays much smoother variations of the critical quantities and continues the trends observed in Region~IV.

\begin{figure}[H]
    \centering
    \begin{subfigure}[b]{0.35\textwidth}
        \centering
        \includegraphics[width=\textwidth]{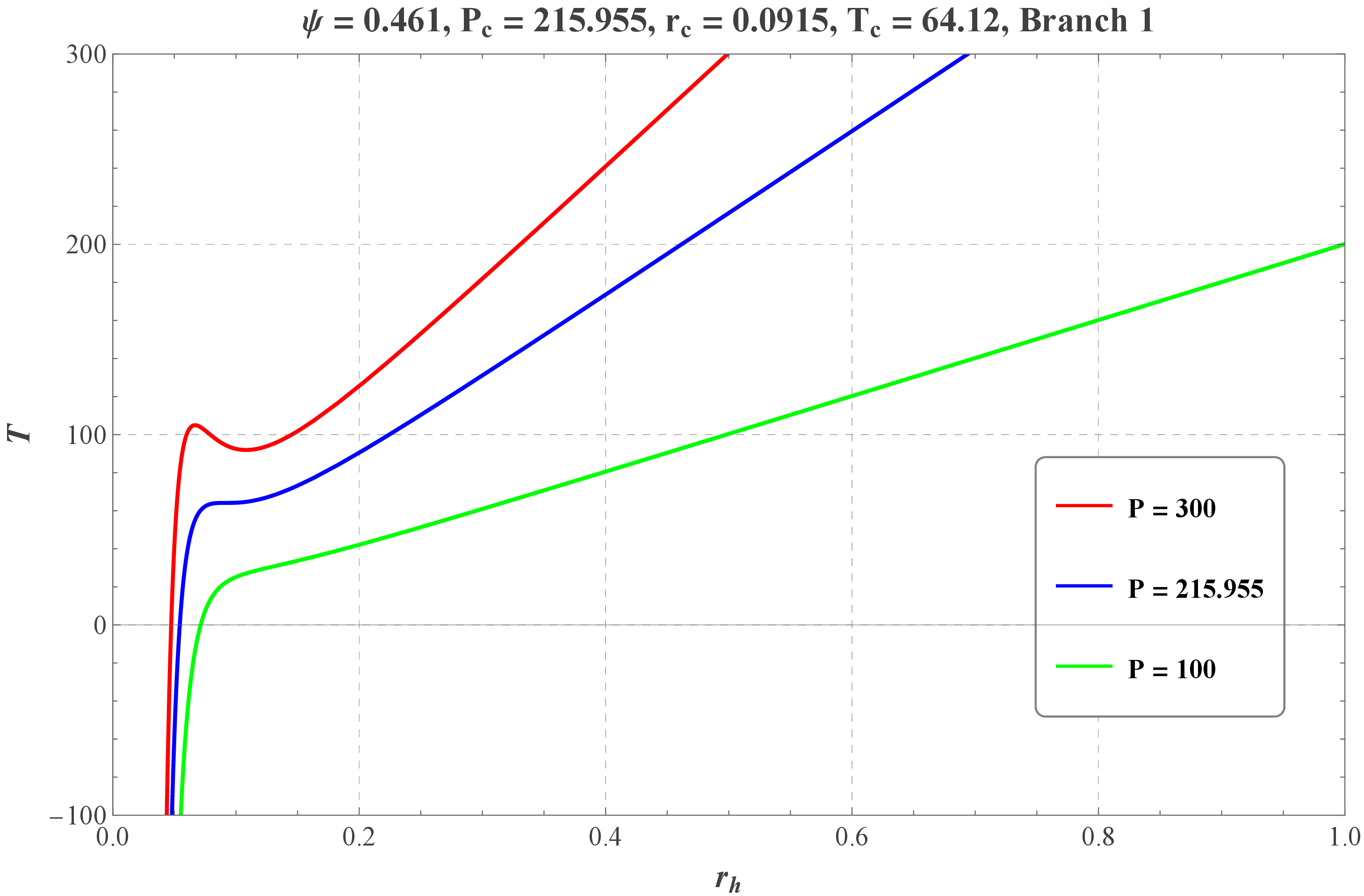}
        \caption{Temperature profile for $\psi=0.461$ with $(P_c,r_c,T_c)=(215.955,\;0.0915,\;64.125)$ (Branch~1).}
        \label{fig:t1b1}
    \end{subfigure}
    \quad
    \begin{subfigure}[b]{0.35\textwidth}
        \centering
        \includegraphics[width=\textwidth]{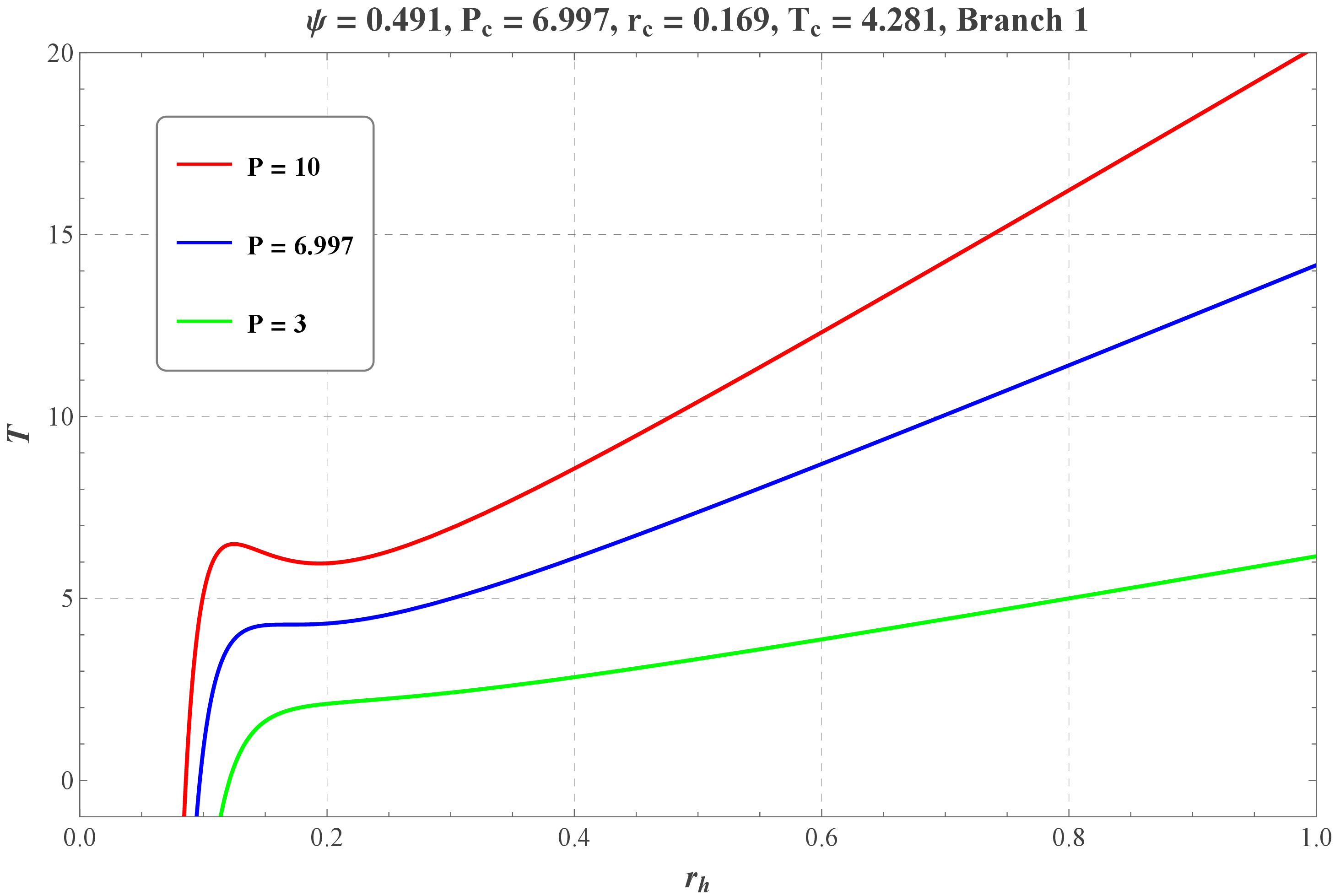}
        \caption{Temperature profile for $\psi=0.491$ with $(P_c,r_c,T_c)=(6.998,\;0.169,\;4.281)$ (Branch~1).}
        \label{fig:t2b1}
    \end{subfigure}

    \caption{Temperature as a function of the horizon radius for representative values of the electric potential in Region~V (Branch~1). The oscillatory structure develops for pressures above the critical pressure, while the subcritical isobars become comparatively smooth. This reversed thermodynamic behavior suggests that, although Branch~1 satisfies the mathematical criticality conditions, it is unlikely to correspond to the physically relevant thermodynamic critical point.}
    \label{fig:t1b1t2b1}
\end{figure}

To determine which branch corresponds to the physically relevant critical behavior, we investigate both the temperature profiles and the Gibbs free energy. Representative temperature curves for Branch~1 at $\psi=0.461$ and $\psi=0.491$ are shown in Fig.~\ref{fig:t1b1t2b1}. Surprisingly,  the isobars become monotonic for $P>P_c$, possess an inflection point at $P=P_c$, and develop an oscillatory region for $P<P_c$. In Branch~1, however, the opposite behavior is observed: the oscillatory structure becomes more pronounced above the critical pressure, whereas the subcritical isobars become comparatively smooth.

\begin{figure}[H]
    \centering
    \begin{subfigure}[b]{0.40\textwidth}
        \centering
        \includegraphics[width=\textwidth]{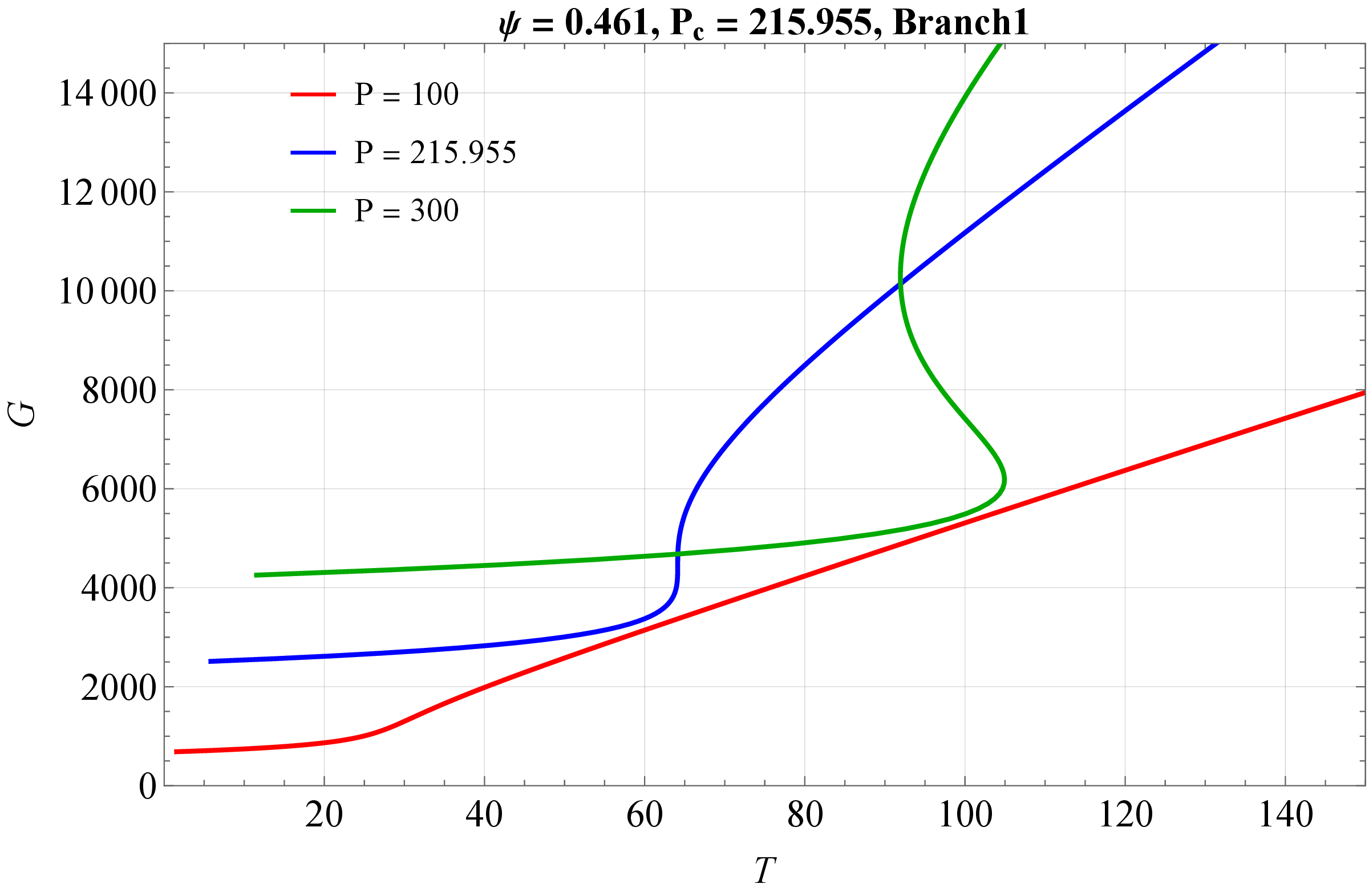}
        \caption{Gibbs free energy for $\psi=0.461$ (Branch~1). The conventional pressure ordering is reversed: the non-monotic structure develops for $P>P_c$, whereas the subcritical curve remains smooth.}
        \label{fig:gt1b1}
    \end{subfigure}
    \quad
    \begin{subfigure}[b]{0.40\textwidth}
        \centering
        \includegraphics[width=\textwidth]{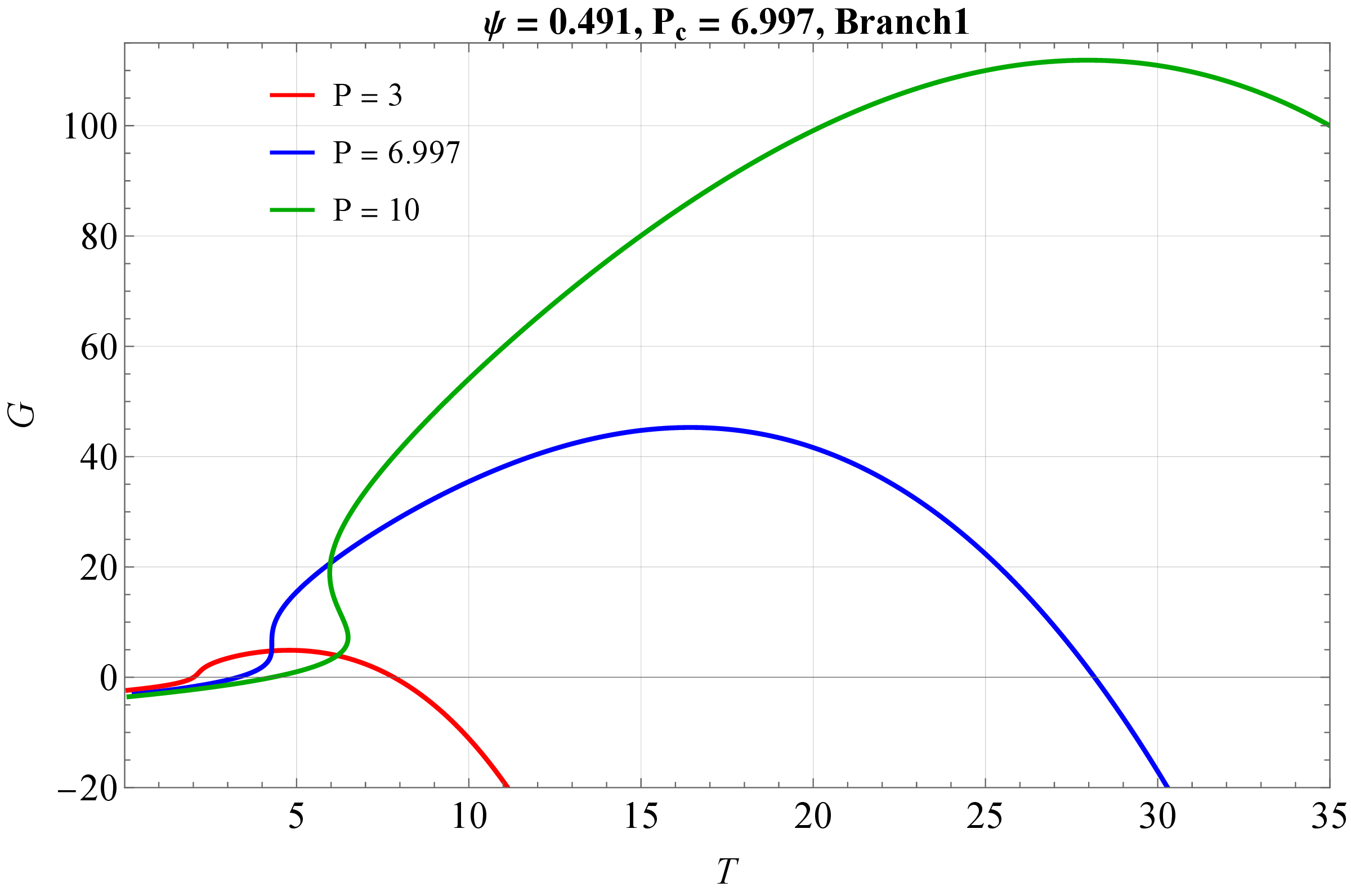}
        \caption{Gibbs free energy for $\psi=0.491$ (Branch~1), showing the same inverse critical behavior observed in panel (a).}
        \label{fig:gt2b1}
    \end{subfigure}
    \caption{Gibbs free energy $G$ as a function of temperature $T$ for representative values of the electric potential in Region~V (Branch~1). The non-monotic curve appears for supercritical pressures ($P>P_c$), while the subcritical curves remain smooth. This inverse pressure dependence, together with the anomalously large values of $(P_c,T_c)$ and the very small critical horizon radius, indicates that Branch~1 is thermodynamically distinct from the conventional RN--AdS critical branch.}
    \label{fig:gt1b1gt2b1}
\end{figure}

The anomalous nature of Branch~1 is further confirmed by the Gibbs free energy in figure \ref{fig:gt1b1gt2b1}. Instead of displaying the characteristic swallowtail structure below the critical pressure and a smooth single-valued curve above it, the $G$--$T$ diagrams exhibit the opposite behavior. Therefore, both the temperature profiles and the Gibbs free-energy analysis consistently indicate that, although Branch~1 satisfies the mathematical criticality conditions, it is unlikely to represent the physically relevant thermodynamic critical point. By contrast, Branch~2 reproduces all the expected thermodynamic signatures of a van der Waals-like system, including the oscillatory temperature profile below the critical pressure and the conventional swallowtail structure of the Gibbs free energy. Consequently, in the following discussion we regard Branch~2 as the physical critical branch of the system.

In the following analysis, we therefore regard Branch~2 as the physically relevant branch for describing the standard thermodynamic phase structure.

\begin{figure}[H]
    \centering
    \begin{subfigure}[b]{0.35\textwidth}
        \centering
        \includegraphics[width=\textwidth]{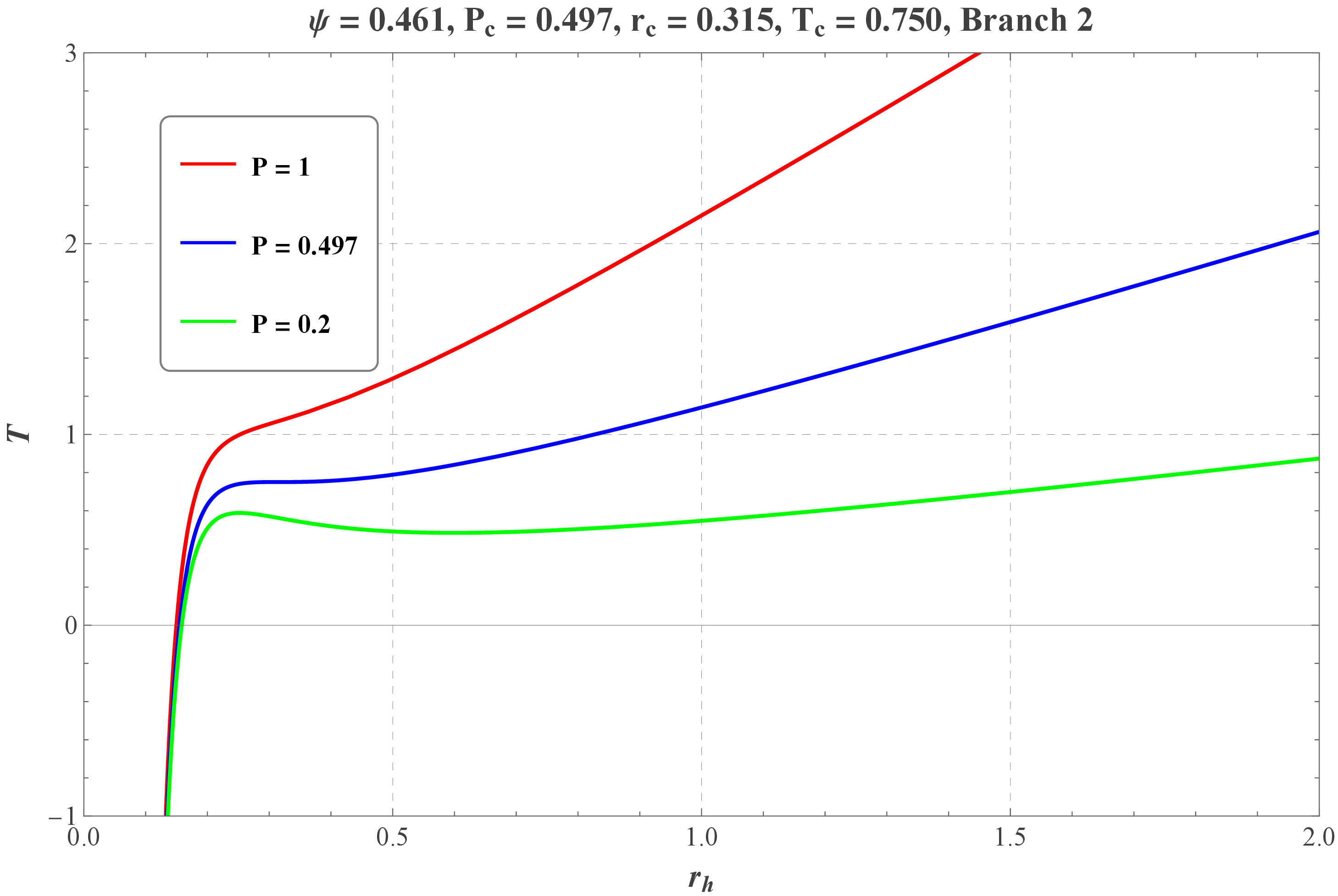}
        \caption{Temperature profile for $\psi=0.461$ with $(P_c,r_c,T_c)=(0.497,\;0.315,\;0.751)$ (Branch~2).}
        \label{fig:t1b2}
    \end{subfigure}
    \quad
    \begin{subfigure}[b]{0.35\textwidth}
        \centering
        \includegraphics[width=\textwidth]{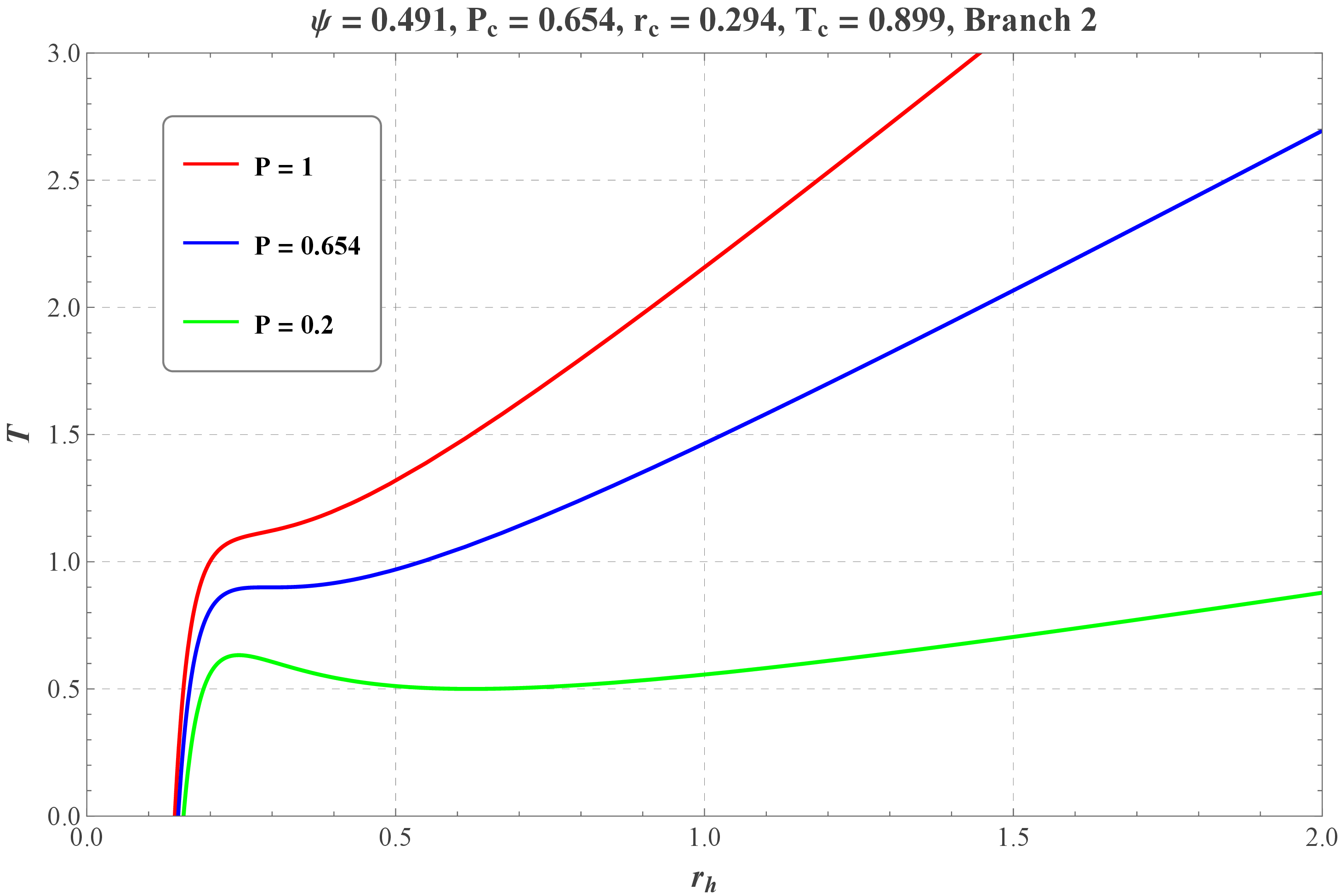}
        \caption{Temperature profile for $\psi=0.491$ with $(P_c,r_c,T_c)=(0.654,\;0.294,\;0.900)$ (Branch~2).}
        \label{fig:t2b2}
    \end{subfigure}

    \caption{Temperature as a function of the horizon radius for representative values of the electric potential in Region~V (Branch~2). Above the critical pressure, the isobars are smooth and monotonic. At the critical pressure, the curve develops an inflection point satisfying Eq.~(\ref{crit_t}), while below the critical pressure the characteristic van der Waals-like oscillation appears, signaling a first-order phase transition between small and large black holes.}
    \label{fig:t1b2t2b2}
\end{figure}

\begin{figure}[H]
    \centering
    \begin{subfigure}[b]{0.40\textwidth}
        \centering
        \includegraphics[width=\textwidth]{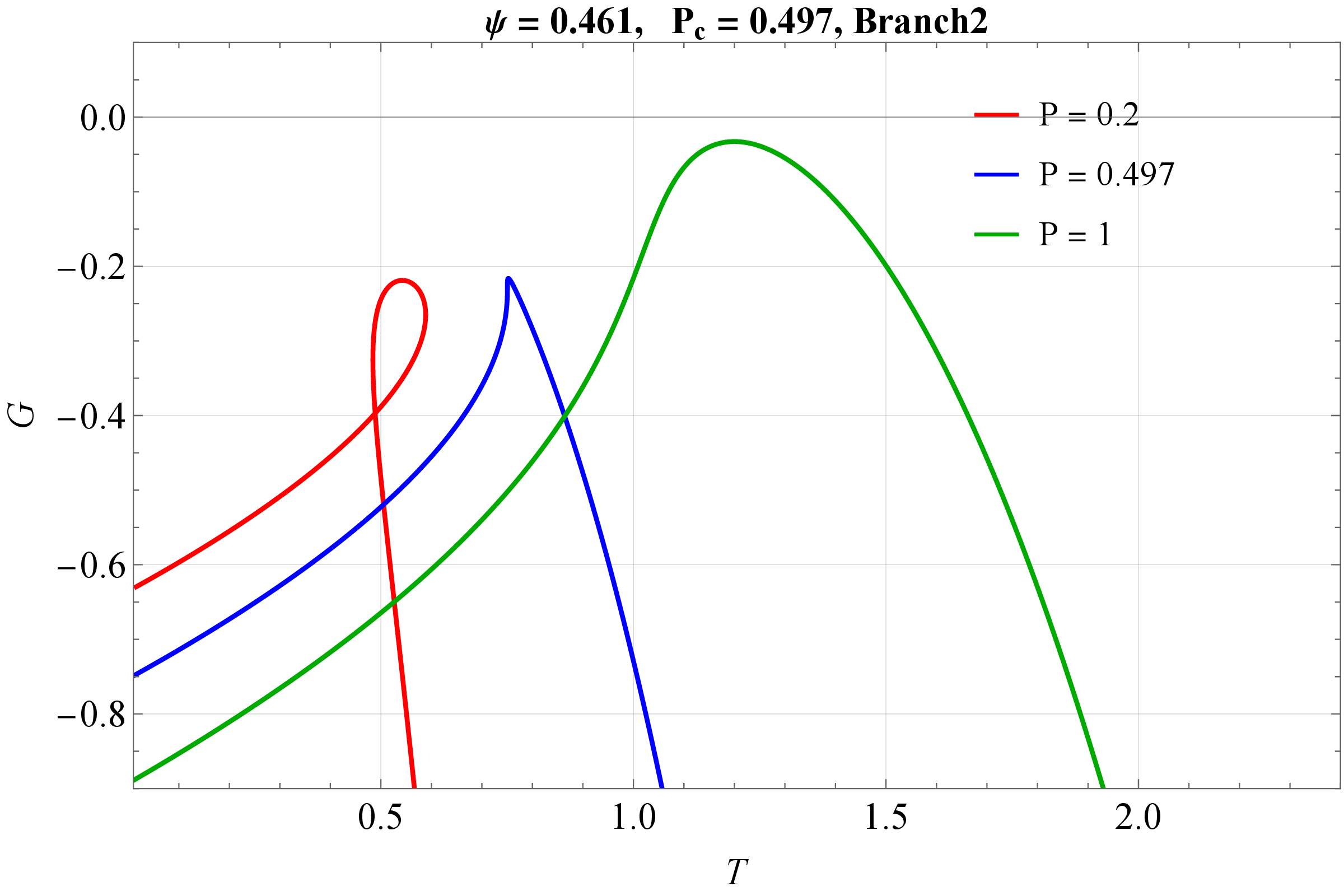}
        \caption{Gibbs free energy for $\psi=0.461$ (Branch~2). The swallowtail appears for $P<P_c$ and disappears for $P>P_c$, in agreement with the standard van der Waals phase transition.}
        \label{fig:gt1b2}
    \end{subfigure}
    \quad
    \begin{subfigure}[b]{0.40\textwidth}
        \centering
        \includegraphics[width=\textwidth]{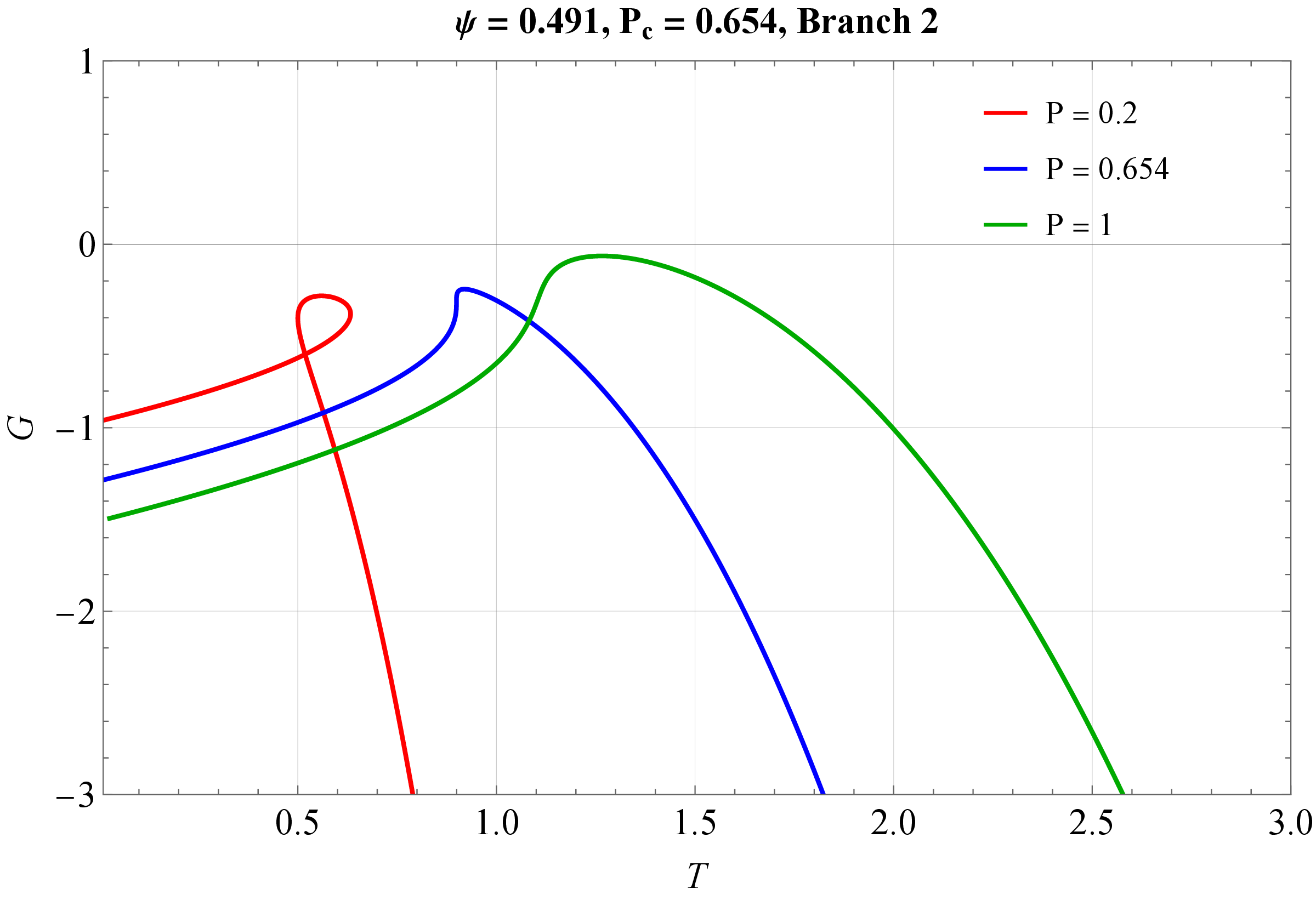}
        \caption{Gibbs free energy for $\psi=0.491$ (Branch~2), showing the same conventional thermodynamic behavior.}
        \label{fig:gt2b2}
    \end{subfigure}

    \caption{Gibbs free energy $G$ as a function of temperature $T$ for representative values of the electric potential in Region~V (Branch~2). Unlike Branch~1, the Gibbs free energy follows the conventional van der Waals-like pattern: a swallowtail structure is present below the critical pressure, collapses to a critical point at $P=P_c$, and evolves into a smooth single-valued curve for $P>P_c$. This confirms that Branch~2 is the physically relevant critical branch governing the thermodynamic phase structure of the system.}
    \label{fig:gt1b2gt2b2}
\end{figure}

For Branch~2, both the temperature and Gibbs free energy exhibit the small/large phase transition.  
The temperature isobars display the standard pattern: above the critical pressure, the temperature varies monotonically with the horizon radius, whereas below the critical pressure an oscillatory region develops, indicating the coexistence of small and large black hole phases. Likewise, the Gibbs free energy exhibits the characteristic swallowtail structure for subcritical pressures, which disappears at and above the critical pressure. These features demonstrate that, despite the unusual evolution of the critical parameters in this region, Branch~2 belongs to the standard thermodynamic universality class. Representative examples for $\psi=0.461$ and $\psi=0.491$ are shown in Figs.~\ref{fig:t1b2t2b2} and \ref{fig:gt1b2gt2b2}.
 
%%%%%%%%%%%

\subsection{Phase Structure in Region VI}

Region~VI begins approximately at $\psi \simeq 0.521$, beyond which the criticality conditions (\ref{crit_t}) no longer admit real solutions for the critical horizon radius. Instead, all solutions become complex, indicating that no physical critical point exists. For example, for $\epsilon=0.001$ and $\psi=0.521$, the solutions of Eq.~(\ref{crit_t}) form four complex-conjugate pairs, demonstrating the complete disappearance of real thermodynamic criticality.

\begin{figure}[H]
    \centering

    \begin{subfigure}[b]{0.30\textwidth}
        \centering
        \includegraphics[width=\textwidth]{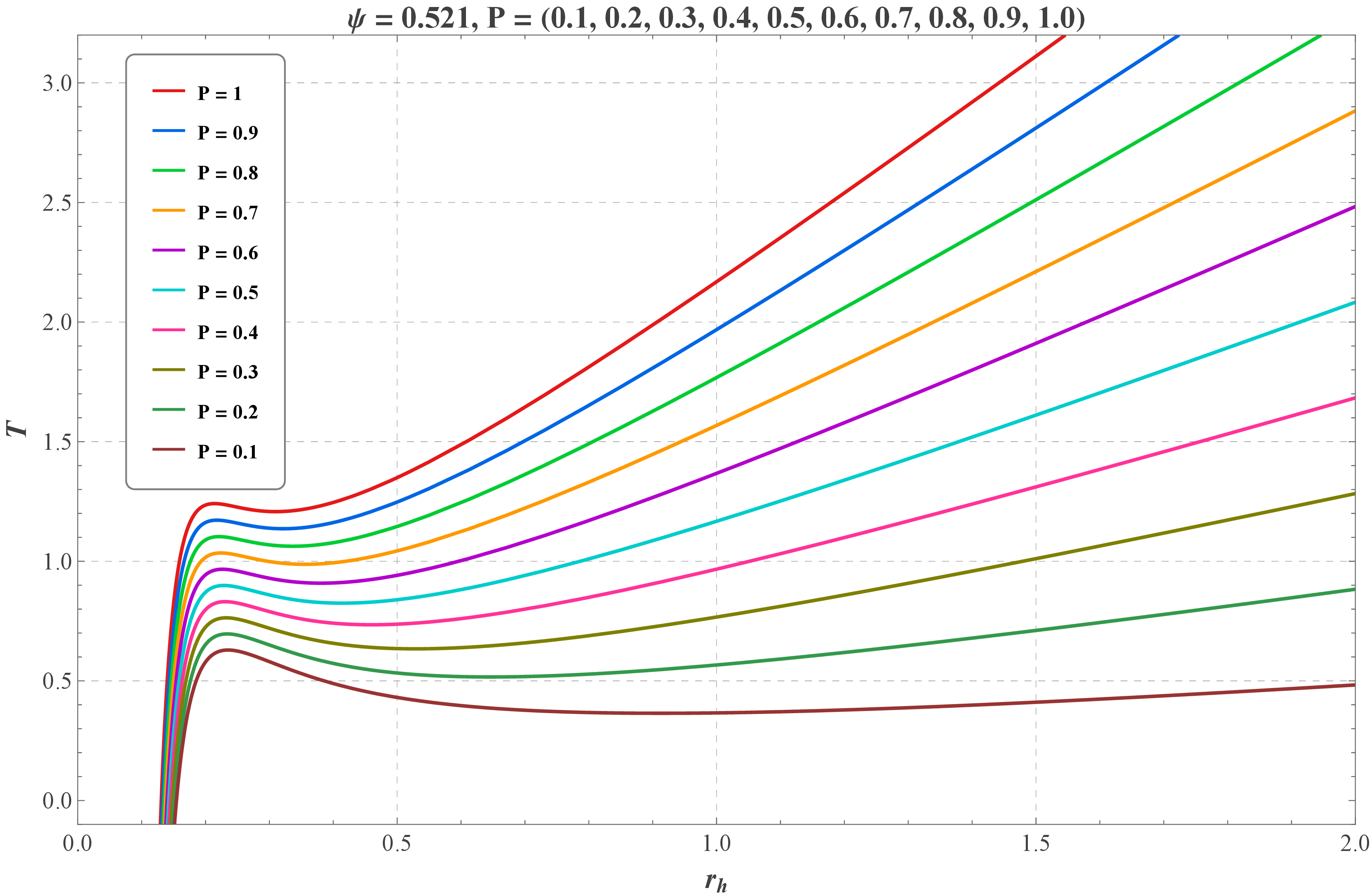}
        \caption{$\psi=0.521$}
        \label{fig:t9}
    \end{subfigure}
    \hfill
    \begin{subfigure}[b]{0.30\textwidth}
        \centering
        \includegraphics[width=\textwidth]{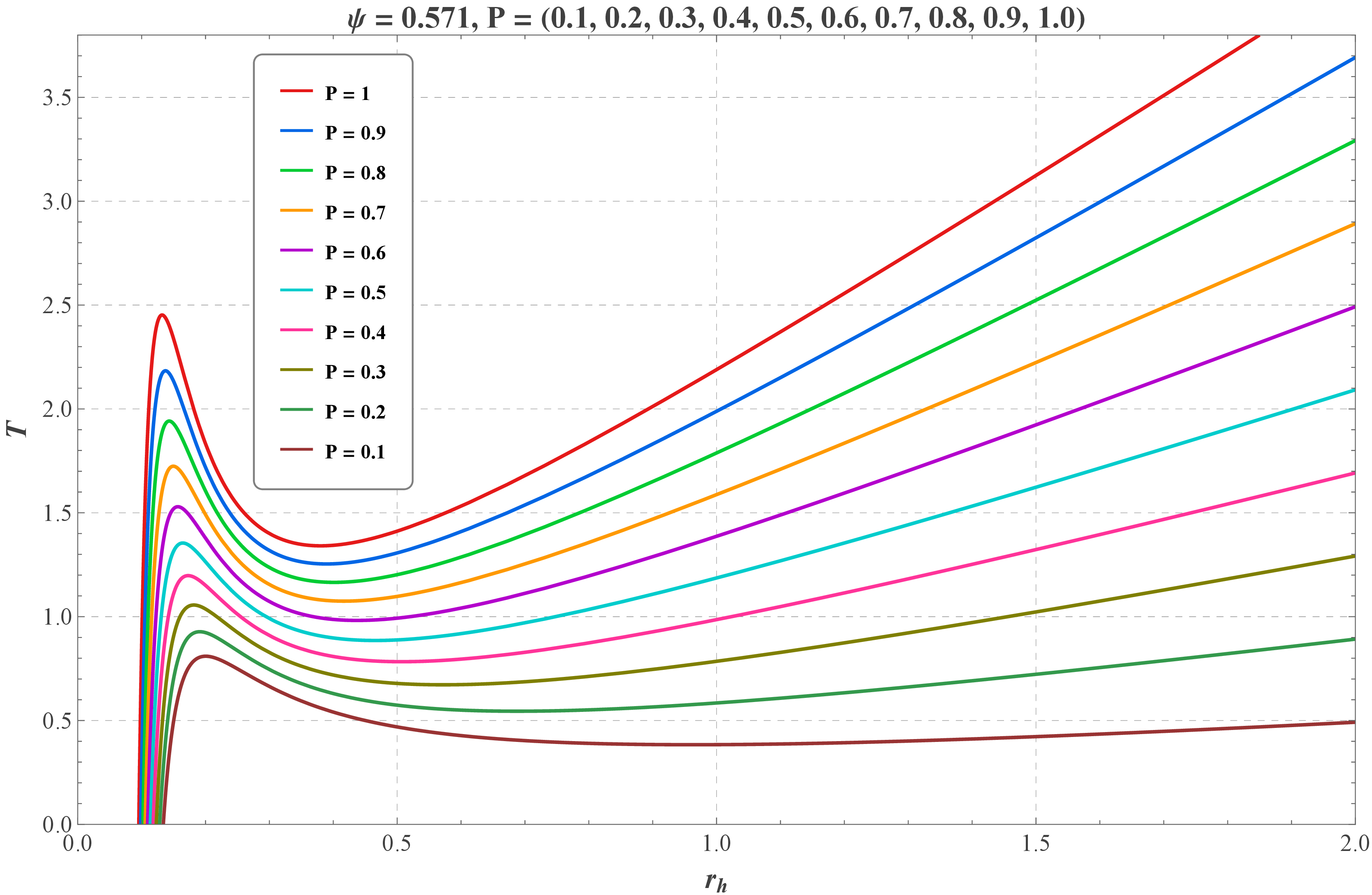}
        \caption{$\psi=0.571$}
        \label{fig:t10}
    \end{subfigure}
    \hfill
    \begin{subfigure}[b]{0.30\textwidth}
        \centering
        \includegraphics[width=\textwidth]{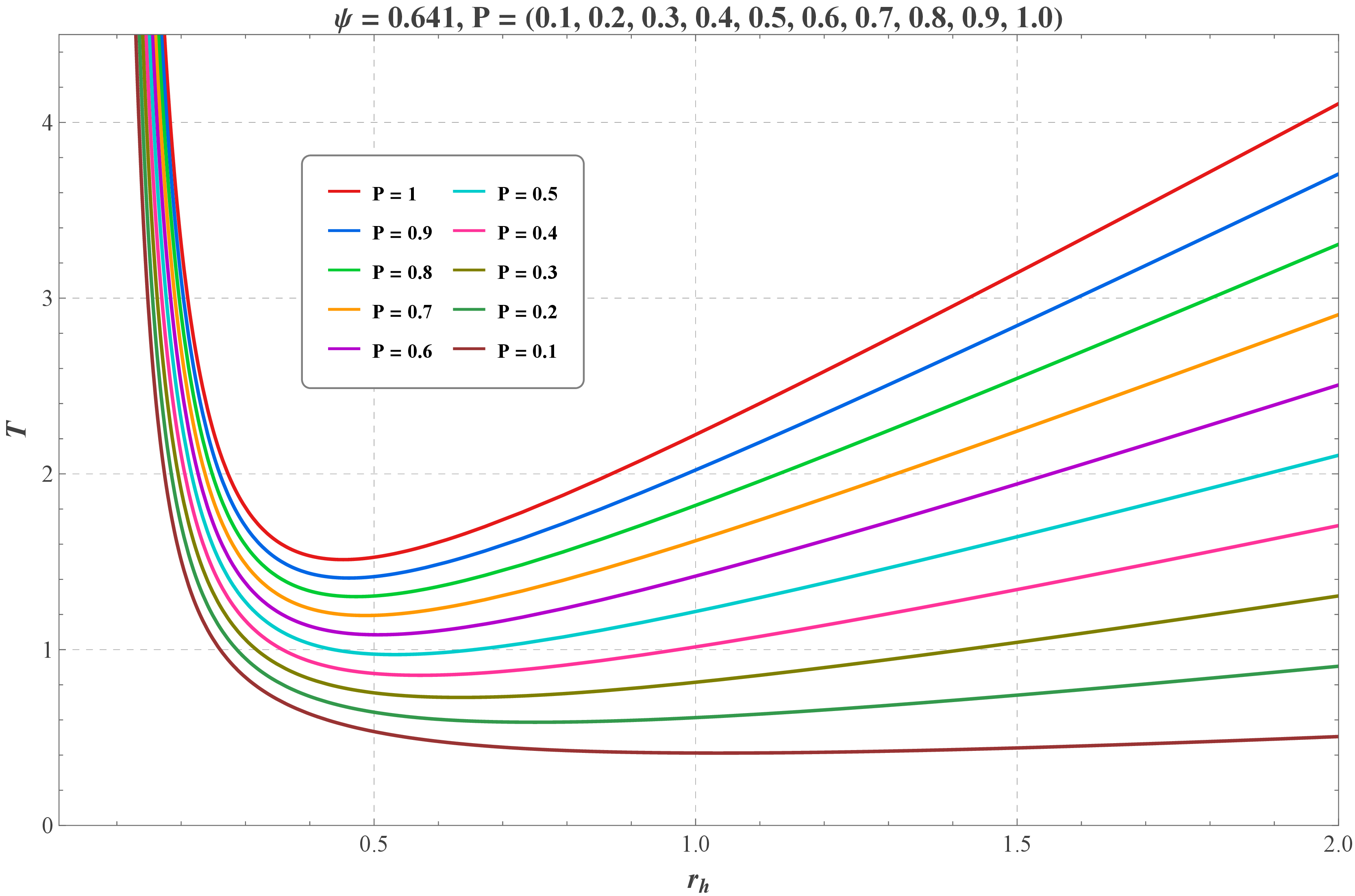}
        \caption{$\psi=0.641$}
        \label{fig:t11}
    \end{subfigure}

    \par\medskip

    \begin{subfigure}[b]{0.30\textwidth}
        \centering
        \includegraphics[width=\textwidth]{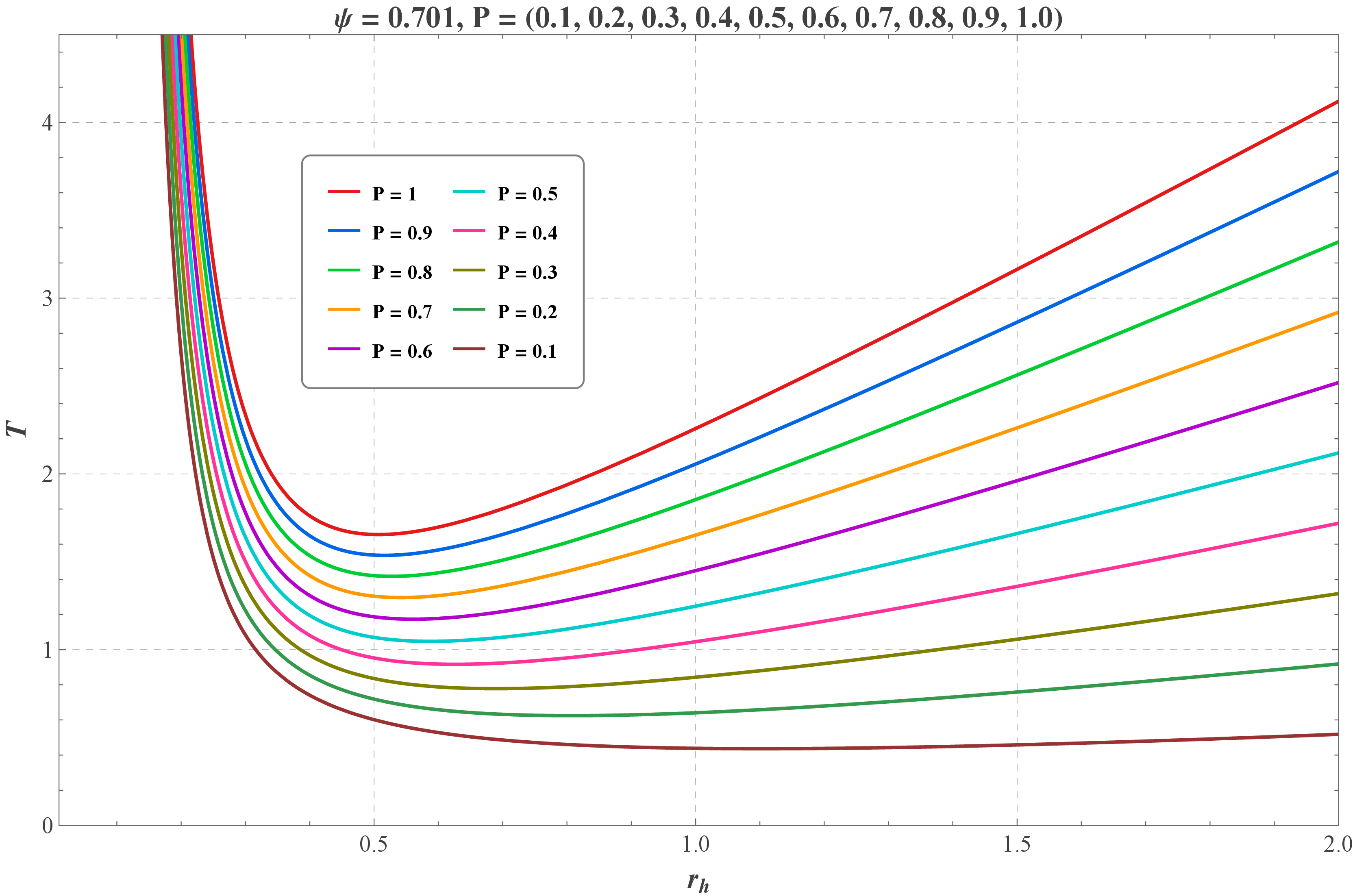}
        \caption{$\psi=0.701$}
        \label{fig:t12}
    \end{subfigure}
    \hfill
    \begin{subfigure}[b]{0.30\textwidth}
        \centering
        \includegraphics[width=\textwidth]{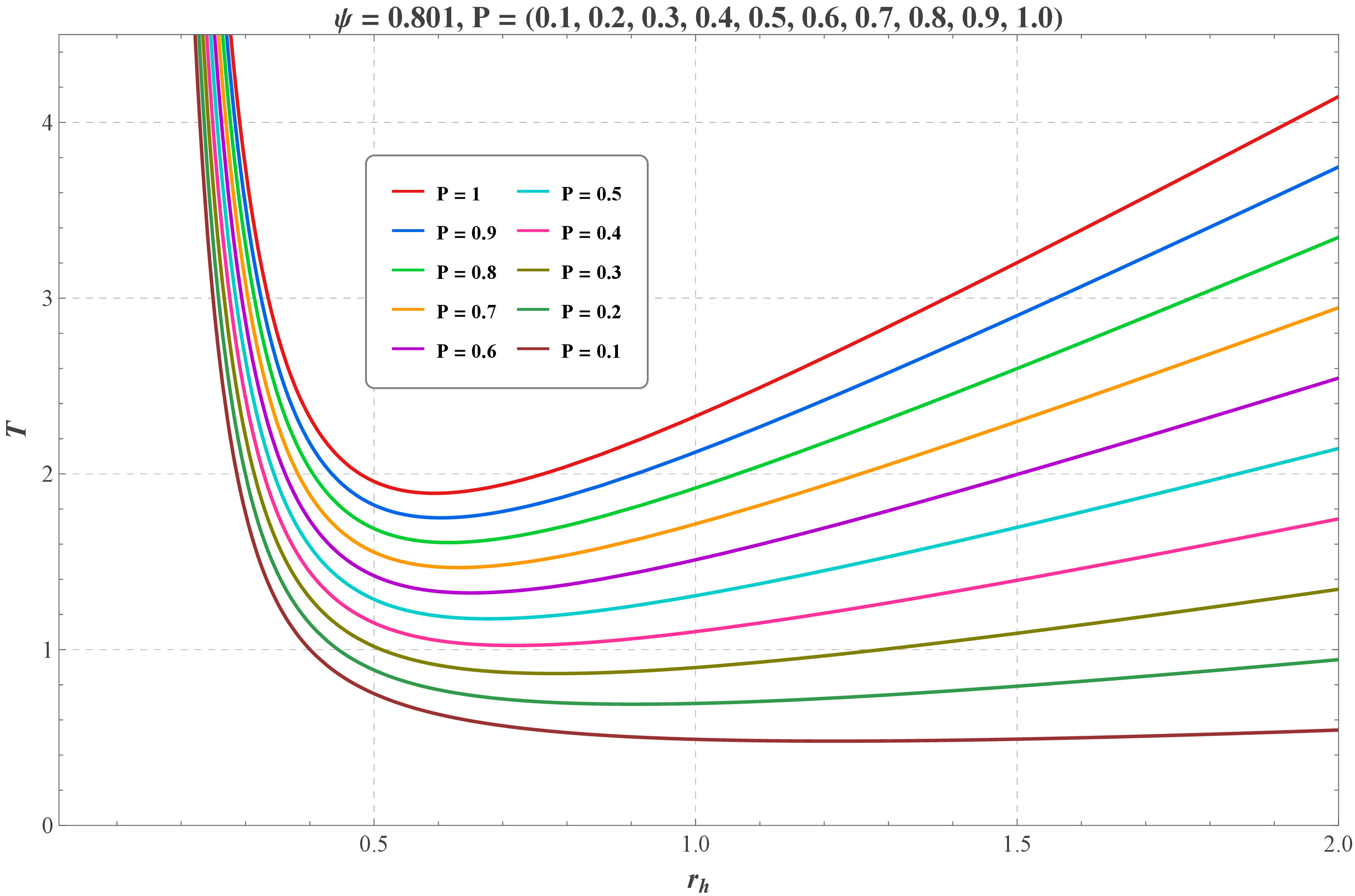}
        \caption{$\psi=0.801$}
        \label{fig:t13}
    \end{subfigure}
    \hfill
    \begin{subfigure}[b]{0.30\textwidth}
        \centering
        \includegraphics[width=\textwidth]{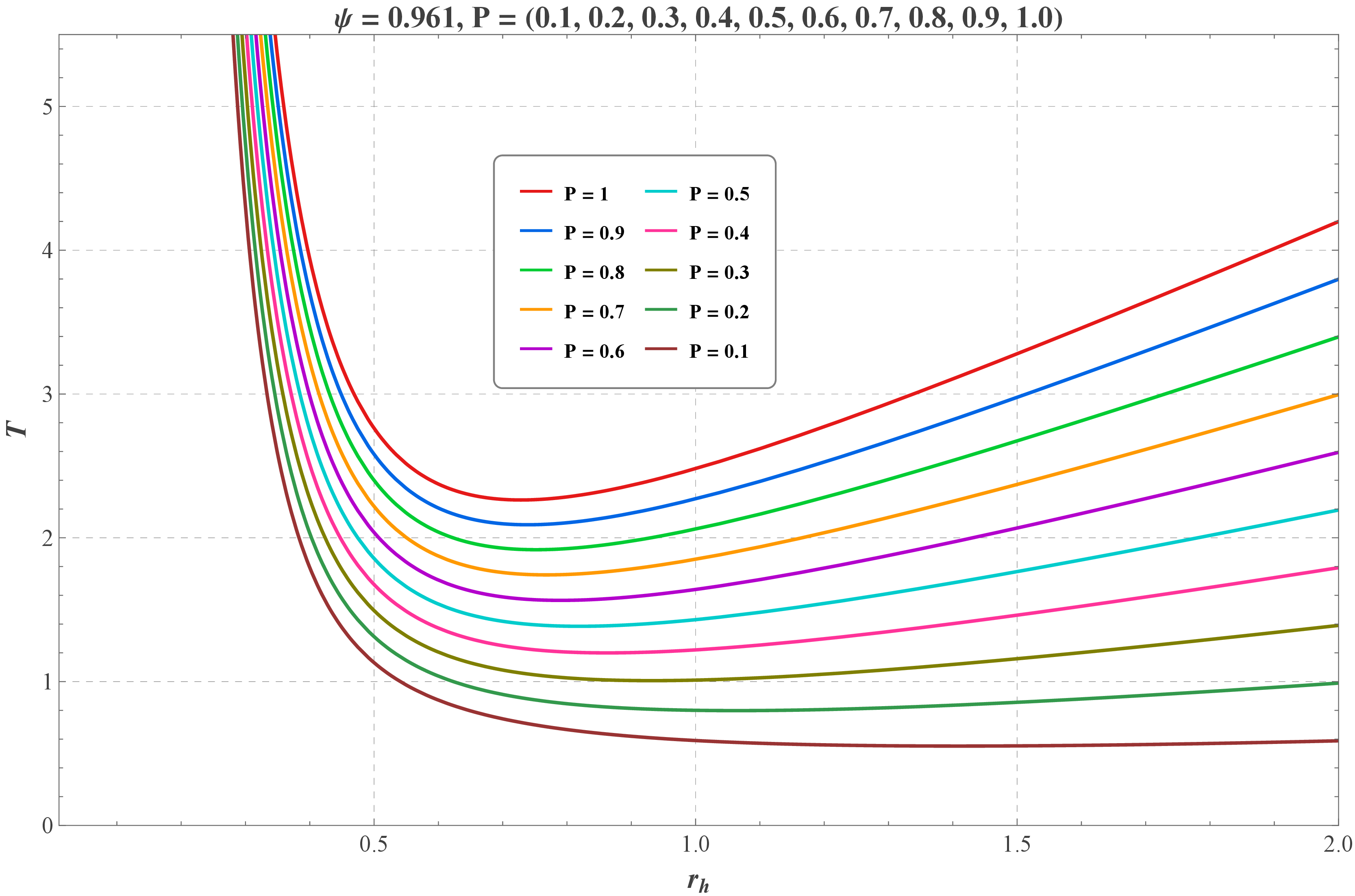}
        \caption{$\psi=0.961$}
        \label{fig:t14}
    \end{subfigure}

    \caption{Temperature as a function of the horizon radius for representative values of the electric potential in Region~VI. The plots are shown for $\psi=0.521$, $0.571$, $0.641$, $0.701$, $0.801$, and $0.961$, and different pressures respectively. As the electric potential increases, the oscillatory behavior gradually disappears and the temperature approaches a smooth curve with a single global minimum. In all panels, increasing the pressure shifts the minimum toward higher temperatures while preserving the overall shape.}
    \label{fig:t9-t14}
\end{figure}

Nevertheless, the absence of a critical point does not imply a trivial temperature profile. Figure~\ref{fig:t9-t14} shows that the temperature still exhibits a local maximum at small horizon radii and a local minimum at larger horizon radii for moderate values of the electric potential. As the electric potential is increased further, the oscillatory structure gradually disappears, and the temperature approaches a smooth hyperbola-like profile possessing only a single global minimum. 

\begin{figure}[H]
    \centering

    \begin{subfigure}[b]{0.40\textwidth}
        \centering
        \includegraphics[width=\textwidth]{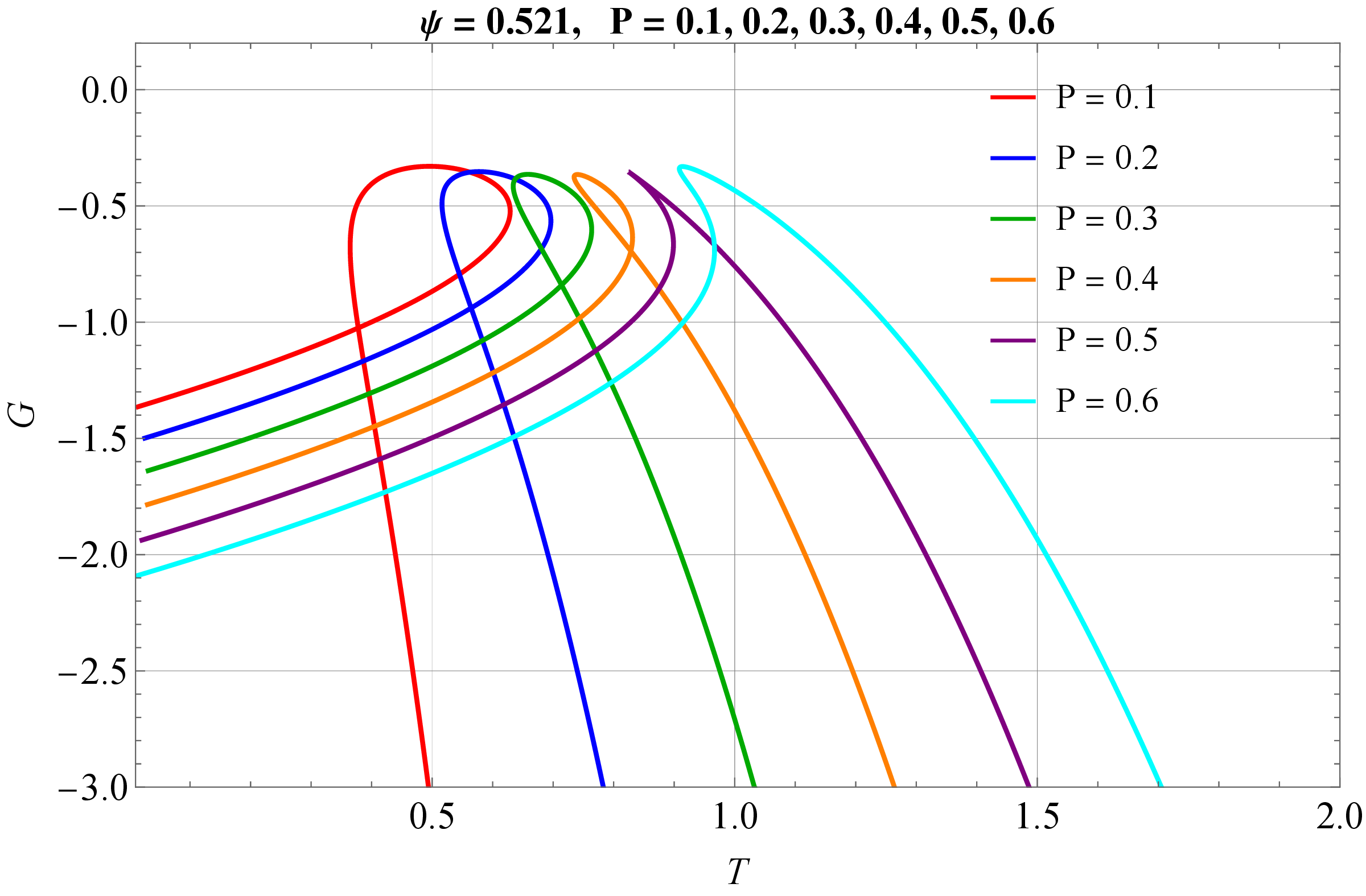}
        \caption{$\psi=0.521$}
        \label{fig:g7}
    \end{subfigure}
    \qquad
    \begin{subfigure}[b]{0.40\textwidth}
        \centering
        \includegraphics[width=\textwidth]{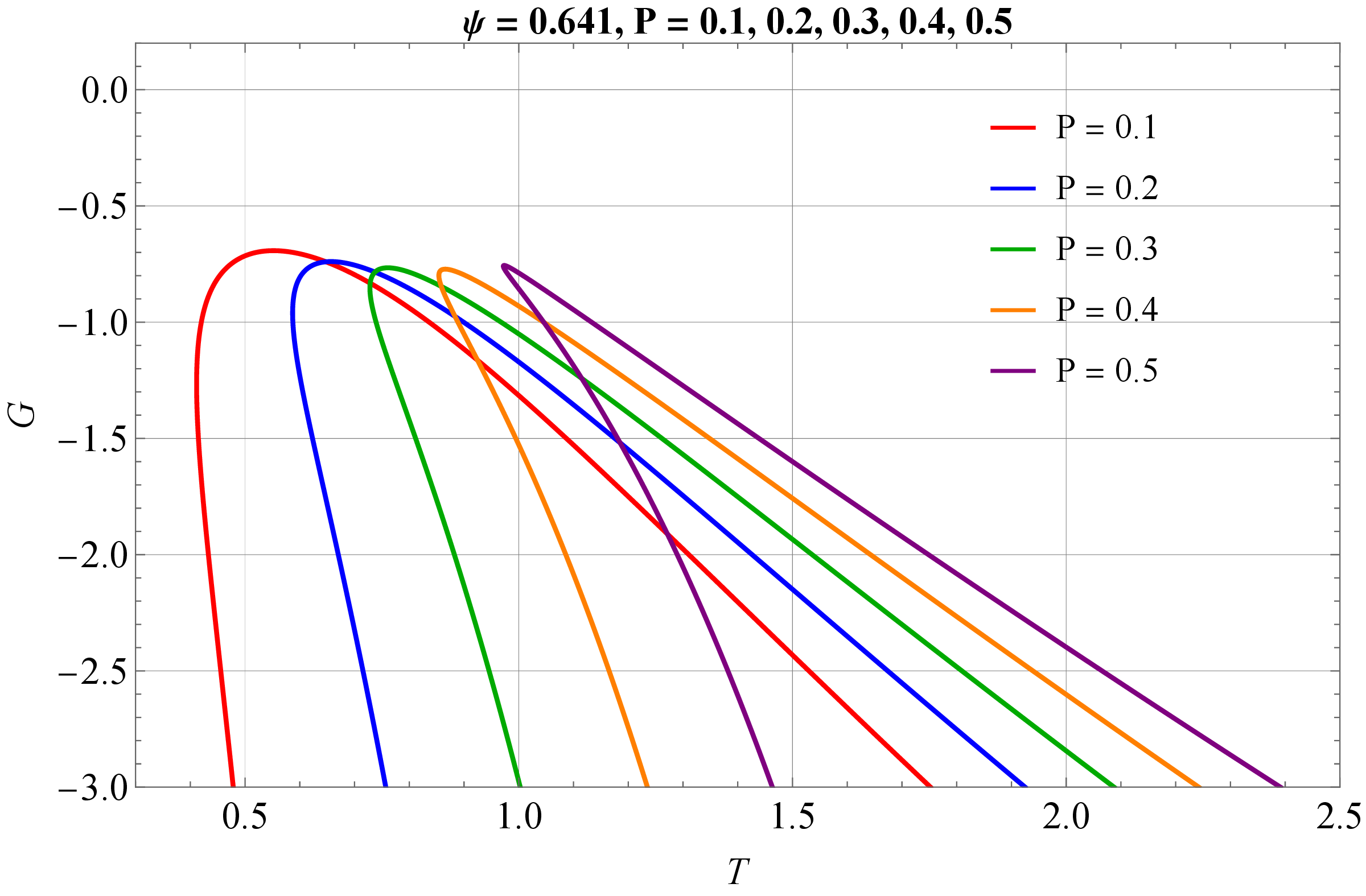}
        \caption{$\psi=0.641$}
        \label{fig:g8}
    \end{subfigure}

    \par\medskip

    \begin{subfigure}[b]{0.40\textwidth}
        \centering
        \includegraphics[width=\textwidth]{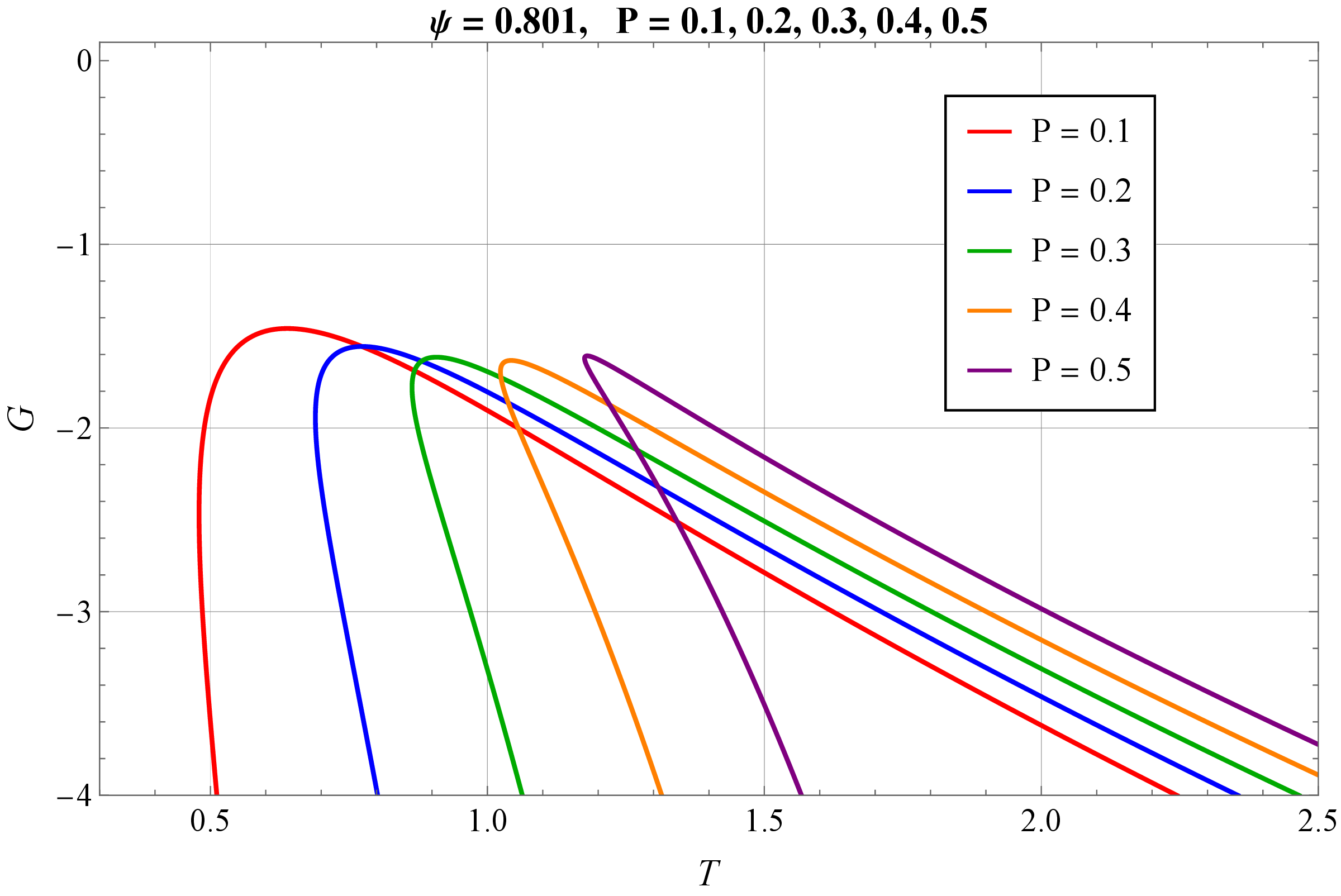}
        \caption{$\psi=0.801$}
        \label{fig:g9}
    \end{subfigure}
    \qquad
    \begin{subfigure}[b]{0.40\textwidth}
        \centering
        \includegraphics[width=\textwidth]{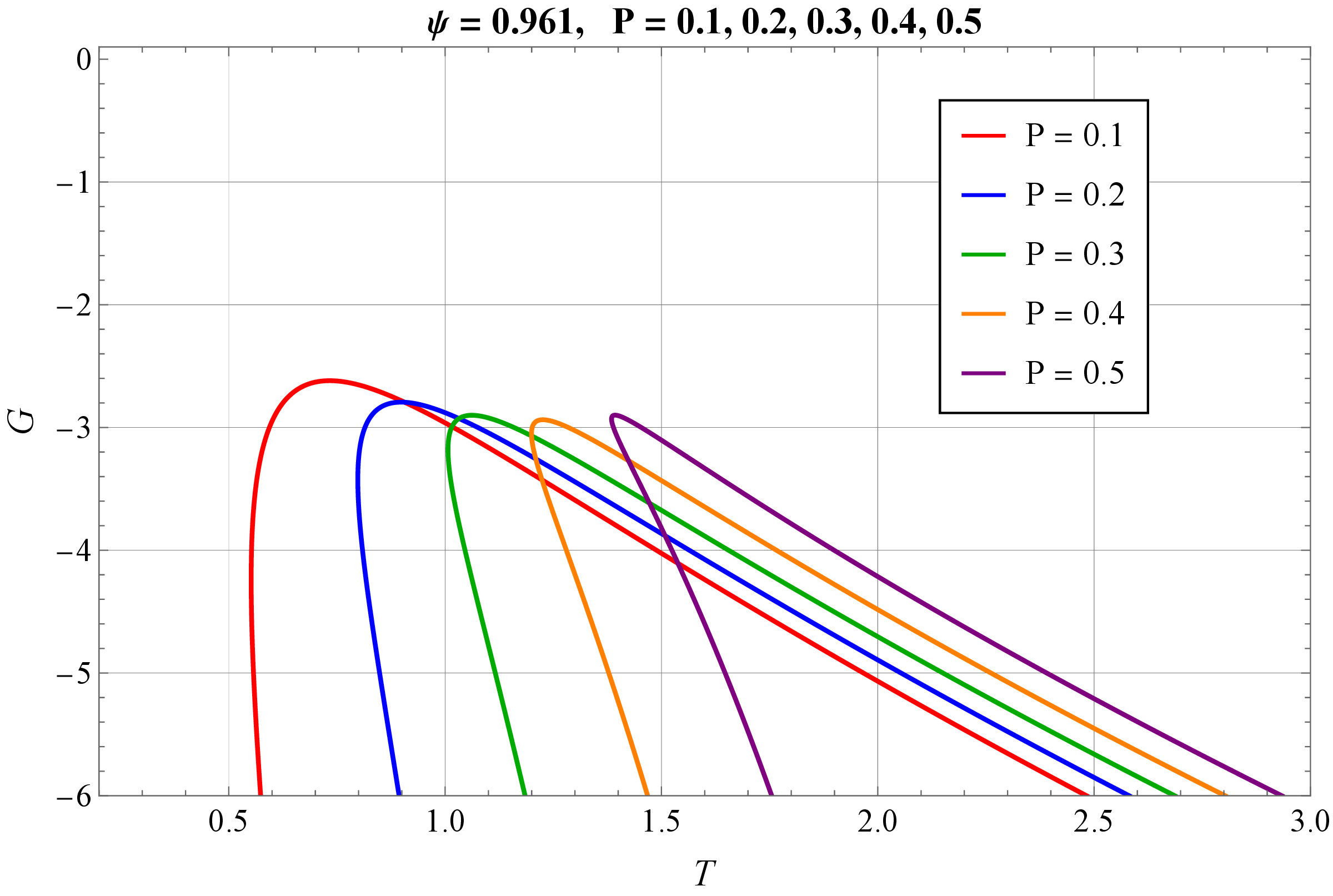}
        \caption{$\psi=0.961$}
        \label{fig:g10}
    \end{subfigure}

    \caption{Gibbs free energy as a function of temperature for representative values of the electric potential in Region~VI. The ribbon-like structure gradually contracts as $\psi$ increases, while both thermodynamic branches move toward more negative values of the Gibbs free energy. No swallowtail develops, confirming the absence of first-order phase transitions.}
    \label{fig:g7-g10}
\end{figure}

The Gibbs free energy provides additional confirmation of this picture. As shown in Fig.~\ref{fig:g7-g10}, the ribbon-like structure gradually shrinks as the electric potential increases. The two thermodynamic branches move closer together and eventually approach one another entirely within the negative Gibbs free-energy region. Consequently, no swallowtail structure develops, indicating the absence of a first-order phase transition throughout Region~VI. Moreover, as $\psi$ increases, the large-black-hole branch shifts progressively toward lower values of the Gibbs free energy, while the small-black-hole branch bends further toward the south-east in the $G$--$T$ plane, reflecting the gradual suppression of critical behavior. Although the conventional critical-point analysis does not fully resolve the detailed differences between the phase structures in these regions, the topological analysis presented in the next section reveals these distinctions more clearly.

%%%%%%%%%%%%%%%%%%%%%%%%%%%%%%%%%%%%%%%%%%%%%%%%%%%%%%%
\section{Topological formalism for phase transitions}
\label{sec5}

In this approach a generalized off-shell free energy that depends on both the horizon radius $r_h$ and an auxiliary inverse temperature parameter $\tau$, which originates from the cavity parameter in the Hawking--York Euclidean path integral approach is defined~\cite{York:1986it}. The standard on-shell free energy is recovered when $\tau = \beta = 1/T$.

A two-component vector field $\phi = (\phi^{r_h}, \phi^\Theta)$ is defined as
\begin{equation}
\phi^{r_h} = \frac{\partial \mathcal{F}}{\partial r_h}, \qquad 
\phi^\Theta = -\cot \Theta \csc \Theta,
\label{eq:vector_field}
\end{equation}
where $\mathcal{F}(r_h, \tau) = M(r_h) - S(r_h)/\tau$ is the generalized free energy for a black hole of mass $M$ (enthalpy in extended phase space) and entropy $S$ enclosed in a cavity at fixed temperature $1/\tau$, and $\Theta \in (0,\pi)$ is an auxiliary angular coordinate introduced to apply Duan's topological current $\phi$-mapping theory.
The zero points of $\phi^{r_h}$ (i.e., $\partial \mathcal{F}/\partial r_h = 0$) correspond to the black hole solutions. 

Using Duan's $\phi$-mapping topological current theory, a unit vector field
\begin{equation}
n^a=\frac{\phi^a}{\|\phi\|}, a=1,2
\end{equation}
is introduce where $n^{a}=(n^{r_h},n^{\Theta})$. A conserved topological current also is defined by
\begin{equation}
j^\mu=
\frac{1}{2\pi}
\epsilon^{\mu\nu \rho}\epsilon_{ab}
\partial_\nu n^a\partial_\rho n^b,
\qquad
\mu,\nu,\rho=0,1,
\label{eq:topocurrent}
\end{equation}
which satisfies $\partial_\mu j^\mu=0$. This current can be expressed as
\begin{equation}
j^\mu=
\delta^2(\phi)
J^\mu\left(\frac{\phi}{x}\right),
\label{eq:topocurrent2}
\end{equation}
showing that it is localized at the zero points of $\phi$, where $J^\mu(\phi/x)$ denotes the Jacobian vector.

A winding number associated with the $i$th zero point($ZP_{i}$) is given by
\begin{equation}\label{wi}
w_i=\int j^0\,d^2x=\beta_i\eta_i,
\end{equation}
where
\begin{equation}
j^0=
\beta_i\eta_i\,
\delta^2(\vec{x}-\vec{z}_i).
\end{equation}
Here, the positive integer $\beta_i$ is the Hopf index, which counts the number of times the mapping $\phi^a$ wraps around its internal space as $x^\mu$ encircles the zero point $z_i$, while
\begin{equation}
\eta_i=
\mathrm{sign}
\left[
J^0\left(\frac{\phi}{x}\right)_{z_i}
\right]
=\pm1
\end{equation}
is the Brouwer degree, determining the orientation of the vector field around the zero.

For isolated nondegenerate zero points the winding number associated with ith zero can be determined through the relation
\begin{equation}\label{nd}
w_i=\eta_i=
\mathrm{sign}
\left(
\frac{\partial^2\mathcal F}
{\partial r_h^2}
\right)
\Bigg|_{r_h=\mathrm{ZP}_i}.
\end{equation}
Here, "nondegenerate" means that the zero points of the vector field are isolated and the Jacobian determinant at each zero point is nonzero, ensuring that the winding number is well-defined and takes only the values $\pm 1$. Hence, a positive (negative) second derivative corresponds to a winding number $w_i=+1$ ($w_i=-1$), identifying a locally stable (unstable) black-hole branch.

For numerical calculations, the winding number may equivalently be obtained from the deflection angle of the normalized vector field along a closed contour $C$ enclosing a given zero point,
\begin{equation}\label{winding}
w=
\frac{1}{2\pi}
\int_0^{2\pi}
\epsilon_{ab}
n^a
\frac{\partial n^b}{\partial\vartheta}\,
d\vartheta, =\frac{\Omega(2\pi)}{2\pi}
\end{equation}
where $\vartheta$ parametrizes the contour. In practice, we choose the contour to be traversed counterclockwise by convention, with the parametrization
\begin{equation}
\label{eq:contour_param}
\begin{aligned}
r_h(\vartheta)&=a\cos\vartheta+z_0,\\
\Theta(\vartheta)&=b\sin\vartheta+\frac{\pi}{2},
\end{aligned}
\qquad
0\leq\vartheta\leq2\pi,
\end{equation}
where the parameters $a$ and $b$ are selected such that the contour encloses the desired zero point $z_0$.

The global topological number is obtained by summing over all isolated zero points,
\begin{equation}
W=\sum_{i=1}^{N}w_i,
\label{eq:totalwinding}
\end{equation}
where $N$ denotes the total number of zeros in the region of interest. During the phase transition $W$ remains invariant~\cite{Ahmed:2022kyv}.

\subsection{Known topological classes}

The established classification of black-hole thermodynamics begins with four
original topological classes,
\[
W^{1-},\qquad W^{0+},\qquad W^{0-},\qquad W^{1+}.
\]
They are distinguished by the asymptotic behavior of the inverse temperature
$\beta(r_h)=1/T(r_h)$ near the minimal horizon radius $r_m$ and as
$r_h\rightarrow\infty$:
\[
\begin{array}{lll}
W^{1-}: & \beta(r_m)=0, & \beta(\infty)=\infty,\\
W^{0+}: & \beta(r_m)=\infty, & \beta(\infty)=\infty,\\
W^{0-}: & \beta(r_m)=0, & \beta(\infty)=0,\\
W^{1+}: & \beta(r_m)=\infty, & \beta(\infty)=0.
\end{array}
\]
The corresponding global topological numbers are $W=-1$, $0$, $0$, and
$+1$, respectively.

The same classification can be visualized from the direction of the unit
vector field $\phi$ on the boundary of the parameter space. We consider the
closed contour
$C=I_1\cup I_2\cup I_3\cup I_4$ enclosing the parameter region
$(r_h,\Theta)$, with $r_h\in(r_m,\infty)$ and $\Theta\in(0,\pi)$:
\begin{align}
I_1 &= \{r_h=\infty,\; \Theta\in(0,\pi)\}, \nonumber\\
I_2 &= \{r_h\in(\infty,r_m),\; \Theta=\pi\}, \nonumber\\
I_3 &= \{r_h=r_m,\; \Theta\in(\pi,0)\}, \nonumber\\
I_4 &= \{r_h\in(r_m,\infty),\; \Theta=0\}. \nonumber
\end{align}
On $I_2$ and $I_4$, the vector field $\phi$ is orthogonal to the boundary by
construction. The relevant asymptotic behavior therefore occurs on $I_1$
and $I_3$. Along these boundaries, the radial component satisfies
\begin{equation}
\phi^{r_h}=\left(\frac{\partial S}{\partial r_h}\right)\left(\frac{1}{\beta}-\frac{1}{\tau}\right),
\end{equation}
where $\tau$ is the fixed inverse cavity temperature and we have assumed $\frac{\partial S}{\partial r_h}>0$. Hence, $\phi^{r_h}>0$ when
$\beta\rightarrow0$, whereas $\phi^{r_h}<0$ when
$\beta\rightarrow\infty$. The resulting boundary directions for the four original classes are represented in the table \ref{tab:topo_boundary}.
\begin{table}[H]
\centering
\caption{Boundary directions of the unit vector field for the four original
topological classes.}
\label{tab:topo_boundary}
\begin{tabular}{lccccc}
\toprule
Class & $I_1$ & $I_2$ & $I_3$ & $I_4$ & $W$ \\
\midrule
$W^{1-}$ & $\leftarrow$ & $\uparrow$ & $\rightarrow$ & $\downarrow$ & $-1$ \\
$W^{0+}$ & $\leftarrow$ & $\uparrow$ & $\leftarrow$ & $\downarrow$ & $0$ \\
$W^{0-}$ & $\rightarrow$ & $\uparrow$ & $\rightarrow$ & $\downarrow$ & $0$ \\
$W^{1+}$ & $\rightarrow$ & $\uparrow$ & $\leftarrow$ & $\downarrow$ & $+1$ \\
\bottomrule
\end{tabular}
\end{table}

Several black-hole solutions exhibit thermodynamic structures that cannot be fully described by these four classes. The extended classification therefore
includes the temperature-dependent class $W^{0-\leftrightarrow1+}$ and the subclasses
\[
\overline{W}^{1+},\qquad
\widehat{W}^{1+},\qquad
\widetilde{W}^{1+},\qquad
\ddot{W}^{1-}.
\]
These classes are distinguished not only by the global topological number
$W$, but also by the stability of the innermost and outermost branches and by the arrangement of stable and unstable states in the low- and high-temperature
limits. For more details and further reading, see Refs.~\cite{Wu:2025,Wu:2025EPJC,AiWu:2025,Chen:2025vsz,Babaei-Aghbolagh:2025qxm,Wu:2023xpq,Wu:2023fcw,Wu:2023meo,Chen:2024atr,Zhu:2024zcl,Liu:2025iyl,Chen:2025nto,Chen:2025fse,Wu:2025wpz,Tian:2026qnt,Wu:2026gqz}.

%%%%%%%%%%%%%%%%%%%%%%%%%%%%%%%%%%%%%%%%%%%%%%%%%%%%%%%%%%%
\section{Topological Phase Structure of $F^{\alpha \beta } F^{\gamma \lambda } R_{\alpha \gamma } R_{\beta \lambda }$ Coupling in the Grand Canonical Ensemble}\label{sec6}

In this section, we employ the topological approach to study the results obtained in Section 3 and develop the results. The off-shell Gibbs free energy and its associated $\phi^{r_h}$ are as follows:
\begin{equation}\label{offG}
\begin{aligned}
\mathcal{G} &= \mathcal{M} - \psi Q - S/\tau \\
&= \frac{4\pi P r_h^3}{3} - 2\psi^2 r_h + \frac{r_h}{2} \\
&\quad + \left( \frac{512\pi P \psi^4}{5 r_h} - \frac{1024\psi^6}{9 r_h^3} + \frac{64\pi P \psi^2}{r_h} + \frac{32\psi^4}{r_h^3} + 1024\pi^2 P^2 \psi^2 r_h \right) \epsilon \\
&\quad - \left( \pi r_h^2 + \left( 256\pi^2 P \psi^2 + \frac{128\pi \psi^4}{r_h^2} \right) \epsilon \right) \frac{1}{\tau}
\end{aligned}
\end{equation}

and 
\begin{equation}\label{vector}
\begin{aligned}
\phi^{r_h} &= \frac{\partial \mathcal{G}}{\partial r_h} 
= 4\pi P r_h^2 - 2\psi^2 + \frac{1}{2} \\
&\quad + \left( -\frac{512\pi P \psi^4}{5 r_h^2} + \frac{1024\psi^6}{3 r_h^4} - \frac{64\pi P \psi^2}{r_h^2} - \frac{96\psi^4}{r_h^4} + 1024\pi^2 P^2 \psi^2 \right) \epsilon \\
&\quad + \left( -2\pi r_h + \frac{256\pi \psi^4 \epsilon}{r_h^3} \right) \frac{1}{\tau}
\end{aligned}
\end{equation} 

To investigate the global thermodynamic behavior, we solve the equation $\phi^{r_h}=0$, which yields
\begin{equation}\label{tau}
\begin{split}
\tau &= \frac{N}{D},  \\
N &= 60\pi r_h^5-7680\pi r_h \psi^4 \epsilon, \\
D &= 120\pi P r_h^6 - 60\psi^2 r_h^4 + 15 r_h^4 \\
  &\quad + \left(30720\pi^2 P^2 \psi^2 r_h^4
  - 3072\pi P \psi^4 r_h^2
  - 1920\pi P \psi^2 r_h^2
  + 10240\psi^6
  - 2880\psi^4\right)\epsilon .
\end{split}
\end{equation}
This relation allows us to construct the $r_h$--$\tau$ diagrams and investigate the global thermodynamic behavior of the system.
For each region, we employ three complementary analyses. First, we examine the unit-vector field and its zeros, which determine the boundary orientations and associated topological charges. Second, we calculate the deflection angle around each zero to independently confirm the corresponding winding number. Finally, we analyze the $r_h$--$\tau$ diagrams, which reveal the off-shell thermodynamic branches and the turning points associated with the generation and annihilation of branches.

\subsection{Topological Phase Transition in Region I}

\begin{figure}[H]
\centering

\begin{subfigure}[b]{0.38\textwidth}
    \centering
    \includegraphics[width=\linewidth]{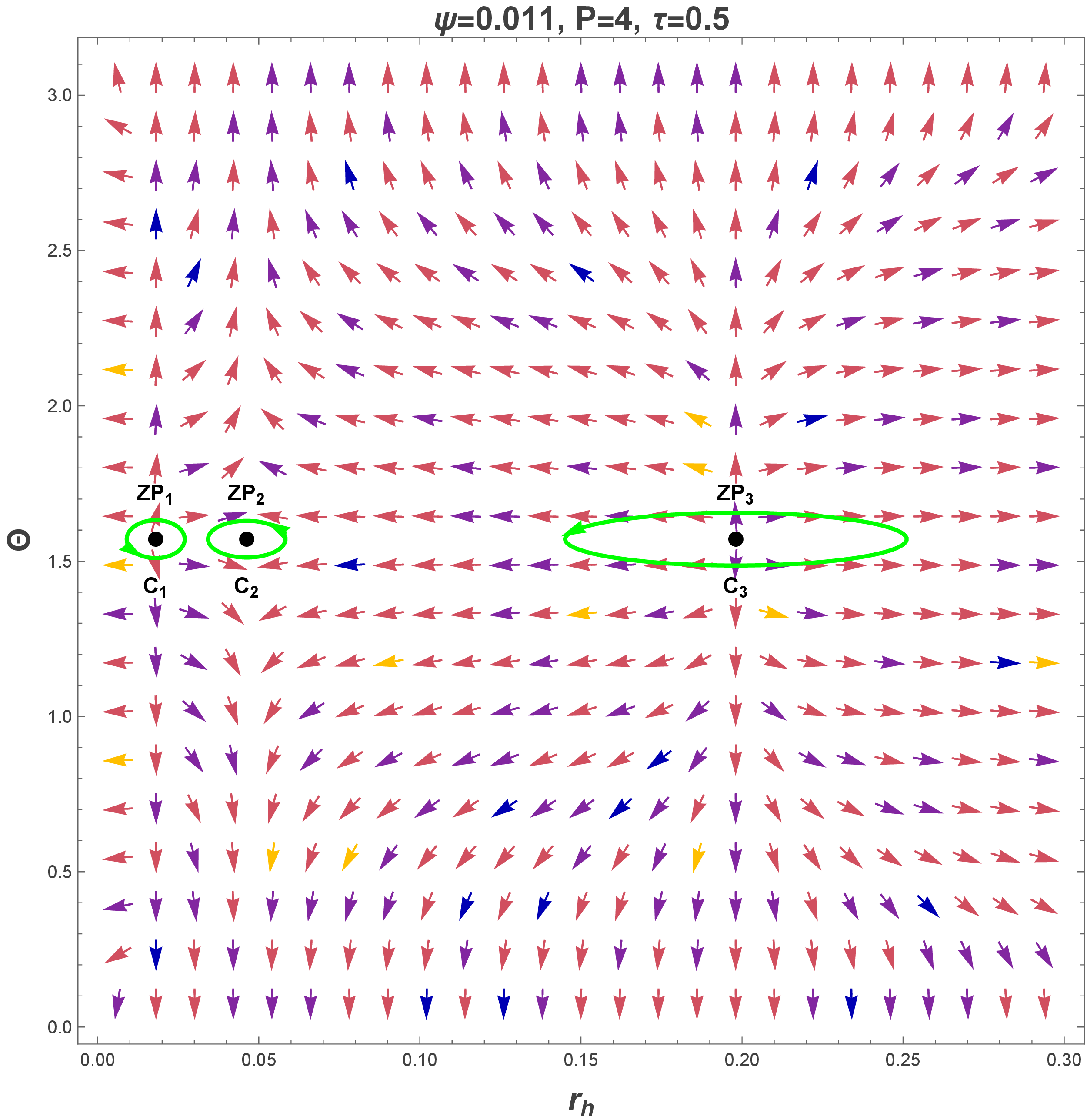}
    \caption{$\psi=0.011$, $P=4$, $\tau=0.5$.}
\end{subfigure}
\quad
\begin{subfigure}[b]{0.38\textwidth}
    \centering
    \includegraphics[width=\linewidth]{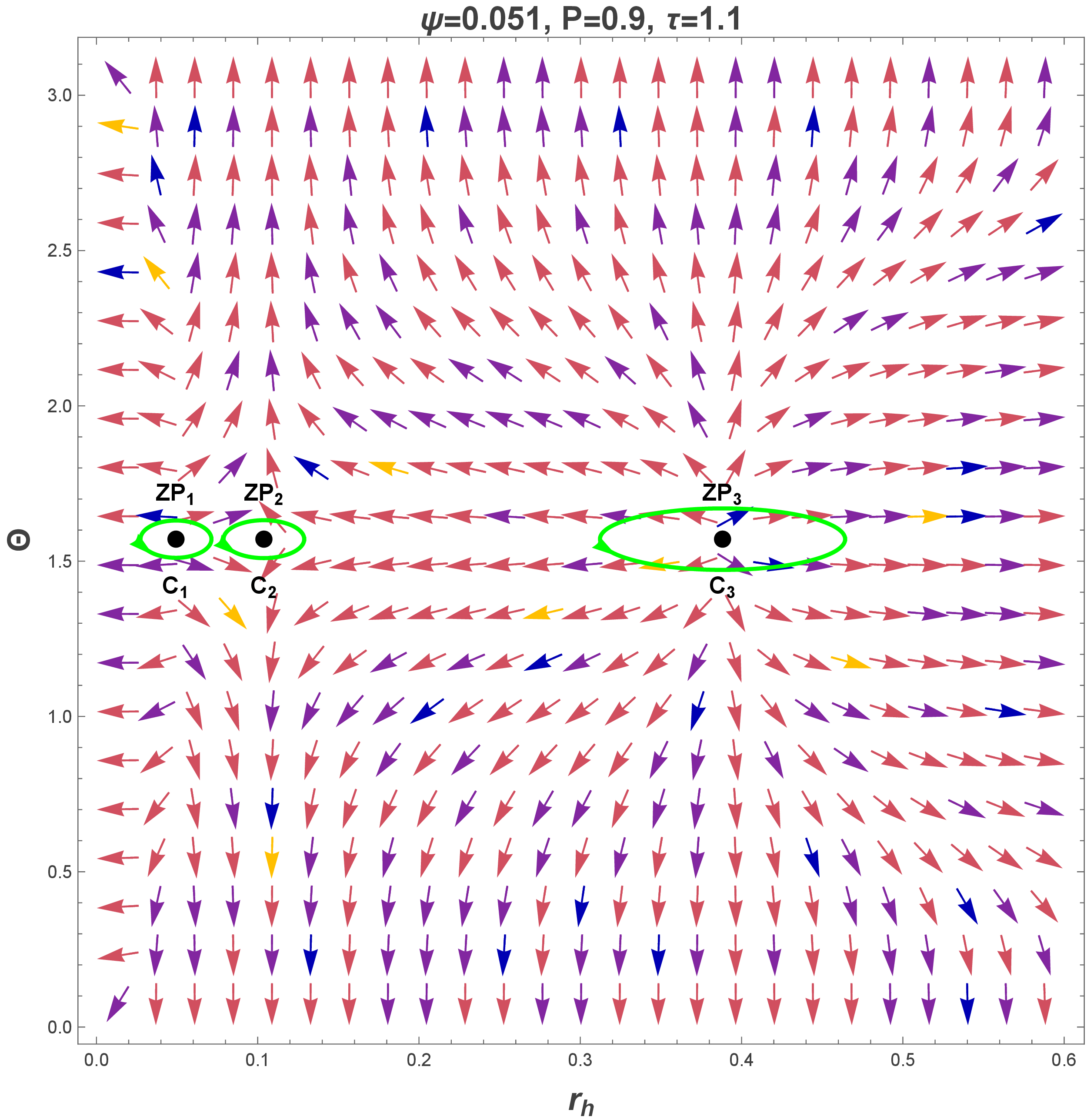}
    \caption{$\psi=0.051$, $P=0.9$, $\tau=1.1$.}
\end{subfigure}

\begin{subfigure}[b]{0.38\textwidth}
    \centering
    \includegraphics[width=\linewidth]{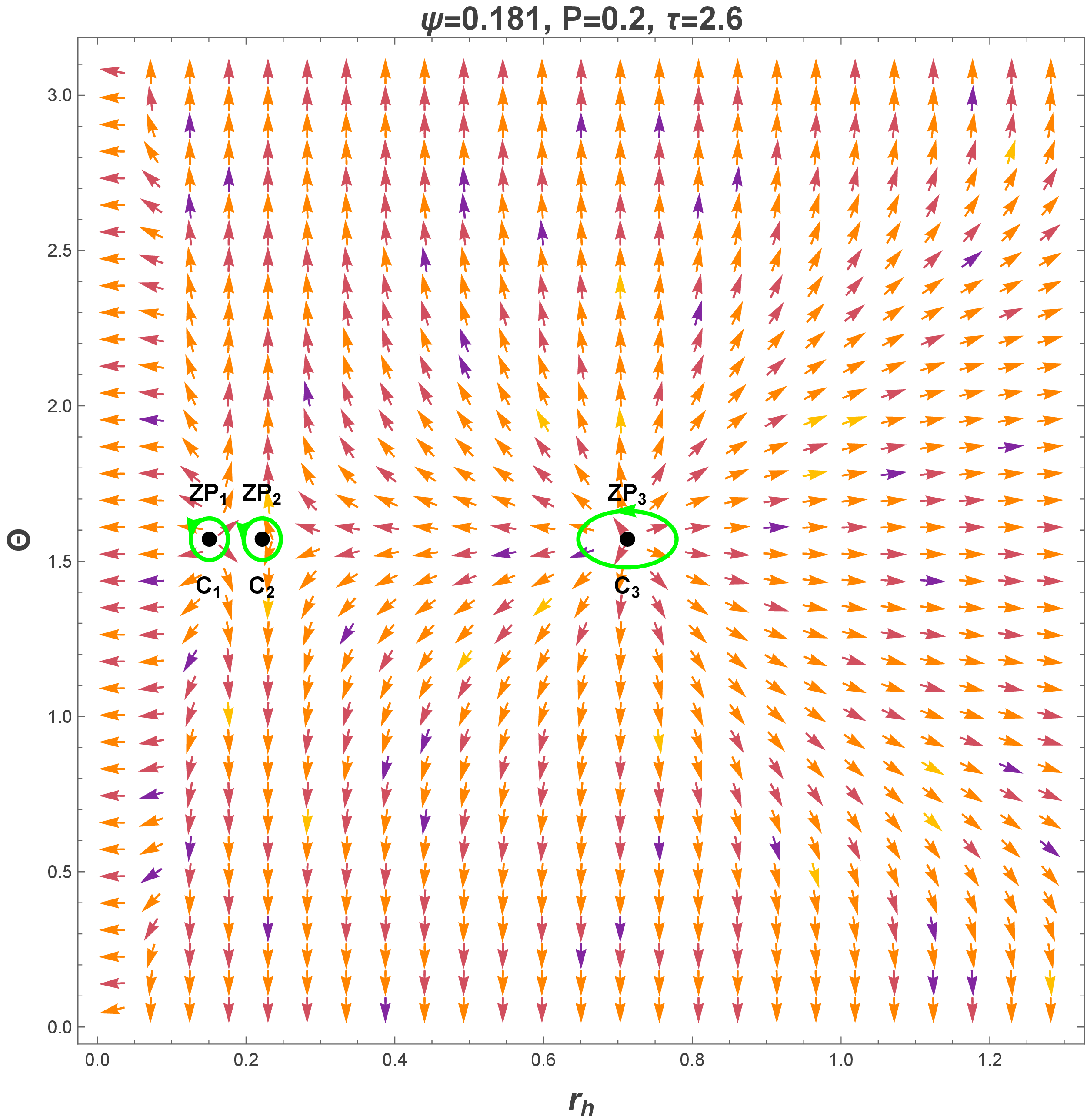}
    \caption{$\psi=0.181$, $P=0.2$, $\tau=2.6$.}
\end{subfigure}
\quad
\begin{subfigure}[b]{0.38\textwidth}
    \centering
    \includegraphics[width=\linewidth]{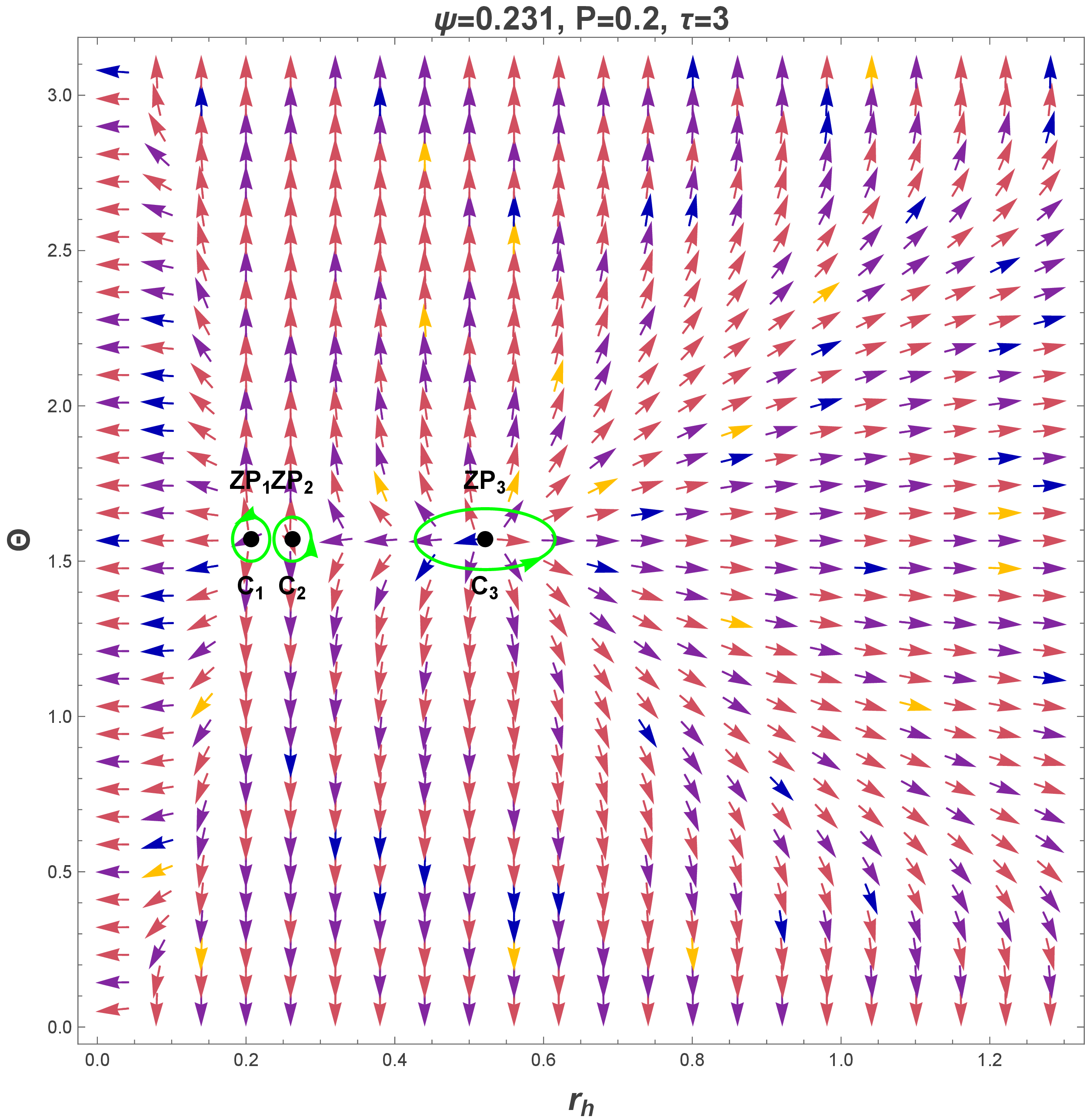}
    \caption{$\psi=0.231$, $P=0.2$, $\tau=3$.}
\end{subfigure}

\caption{
Unit vector fields in Region~I for four representative values of the electric potential. The black dots denote the zero points (topological defects) of the vector field, while the closed contours surrounding each defect are used to determine the corresponding local winding numbers. The asymptotic orientation of the unit vector field remains unchanged in all cases, indicating that the thermodynamic configurations throughout Region~I belong to the topological class with total winding number $W=+1$.
}
\label{fig:v0v1v2v3}
\end{figure}

The unit vector fields corresponding to $\psi=0.011$, $0.051$, $0.181$, and $0.231$ are shown in Fig.~\ref{fig:v0v1v2v3}. For each value of the electric potential, the zero points (topological defects) of the vector field are identified and enclosed by closed contours, allowing the local winding numbers to be determined. Although the locations of the defects vary with $\psi$, the asymptotic behavior of the unit vector field remains unchanged throughout the entire region.

The unit vector field points leftward at the left boundary ($I_3$) and rightward at the right boundary ($I_1$). This means the system in this region belongs to $W=1$ classification. For the $\tau$ values chosen in Figure~\ref{fig:v0v1v2v3}, we solved the equation $\phi^{r_h}=0$ and plotted $\tau$ against the horizon to estimate its range of variation.

\begin{figure}[H]
\centering

\begin{subfigure}[b]{0.3\textwidth}
    \centering
    \includegraphics[width=\linewidth]{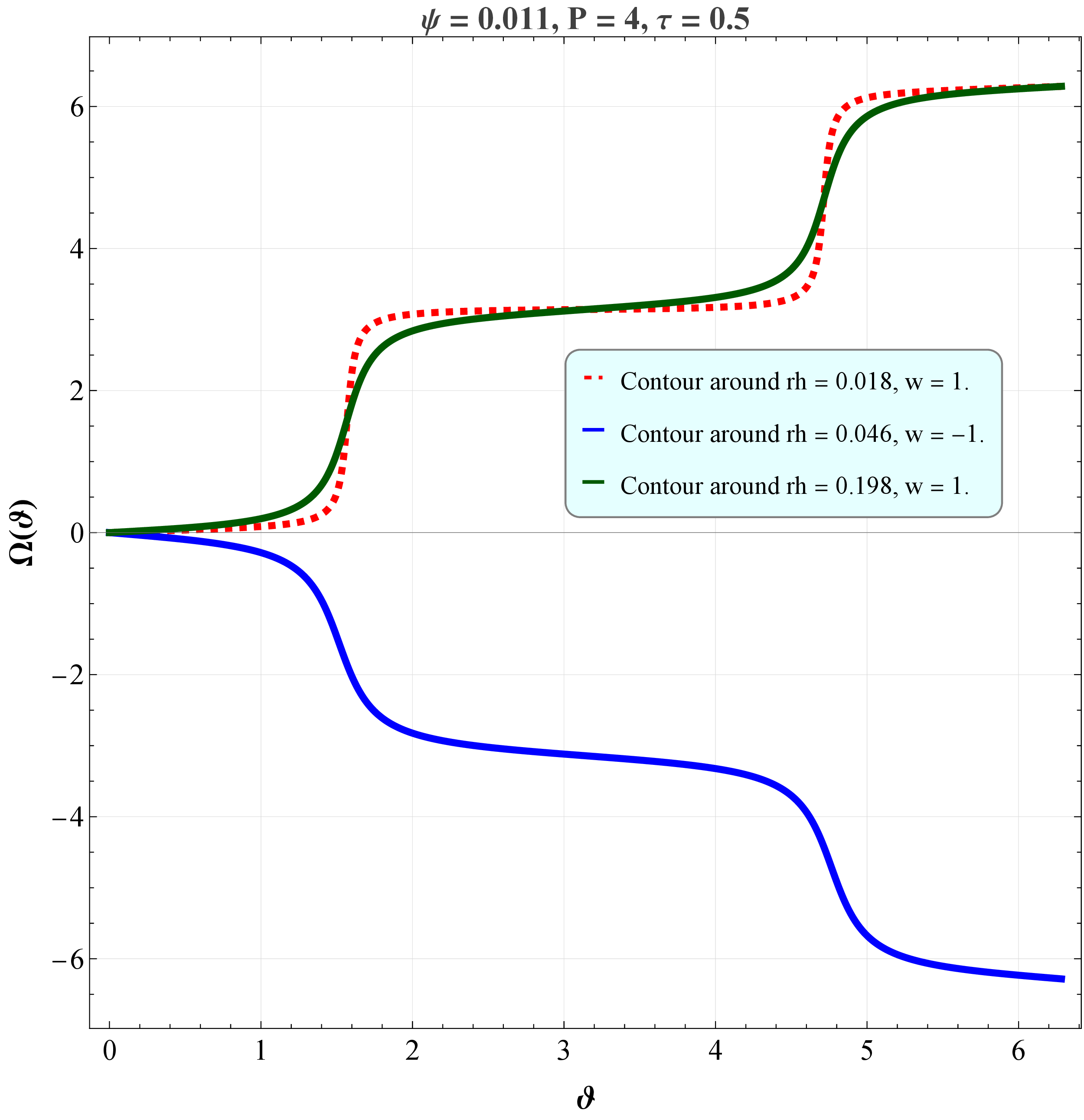}
    \caption{$\psi=0.011$, $P=4$, $\tau=0.5$. Deflection-angle diagram for a closed loop enclosing the three zero points located at $r_h=0.018$, $0.046$, and $0.198$.}
\end{subfigure}
\quad
\begin{subfigure}[b]{0.3\textwidth}
    \centering
    \includegraphics[width=\linewidth]{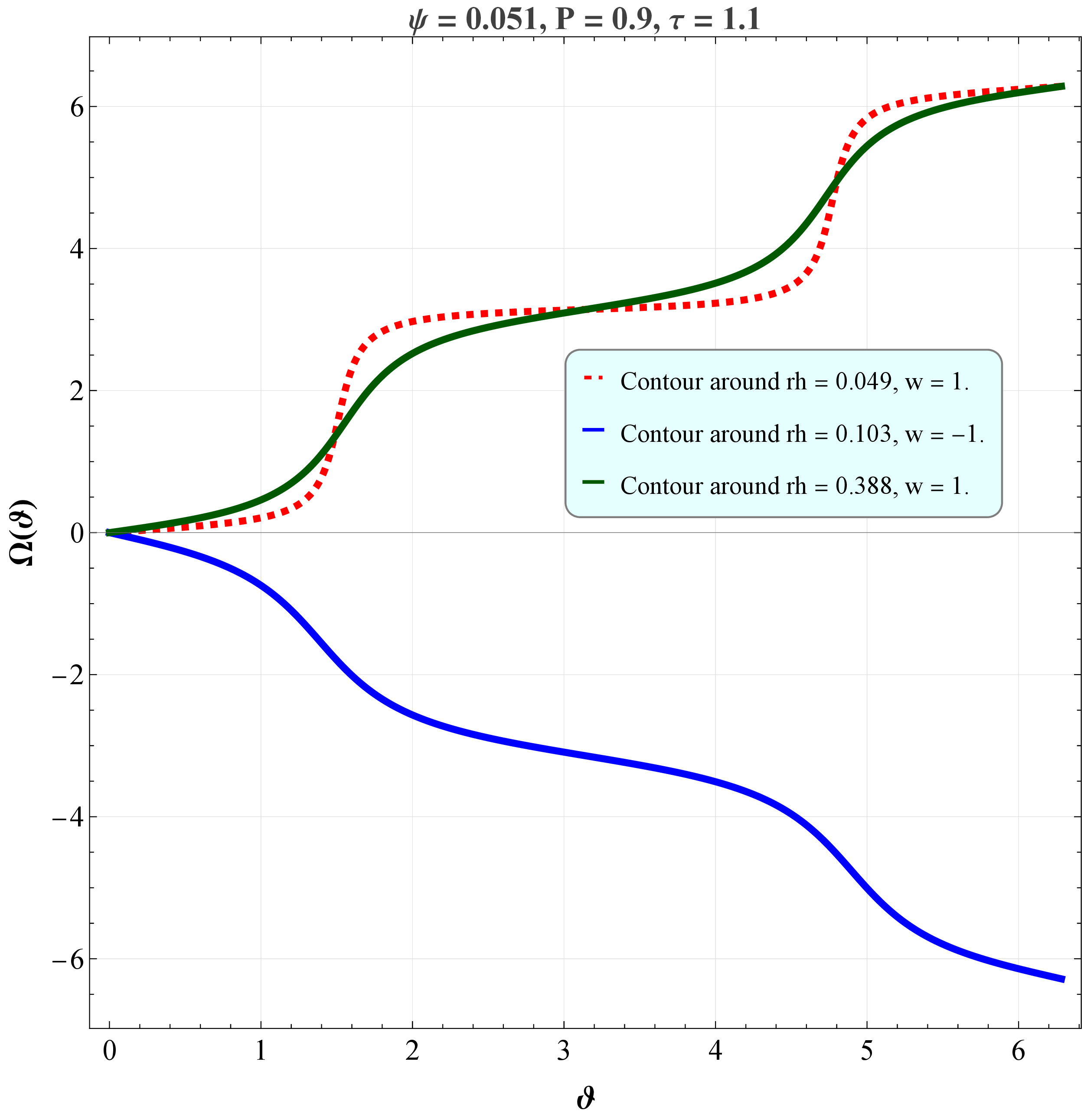}
    \caption{$\psi=0.051$, $P=0.9$, $\tau=1.1$. Deflection-angle diagram for a closed loop enclosing the three zero points located at $r_h=0.049$, $0.103$, and $0.388$.}
\end{subfigure}

\vspace{0.2cm}

\begin{subfigure}[b]{0.3\textwidth}
    \centering
    \includegraphics[width=\linewidth]{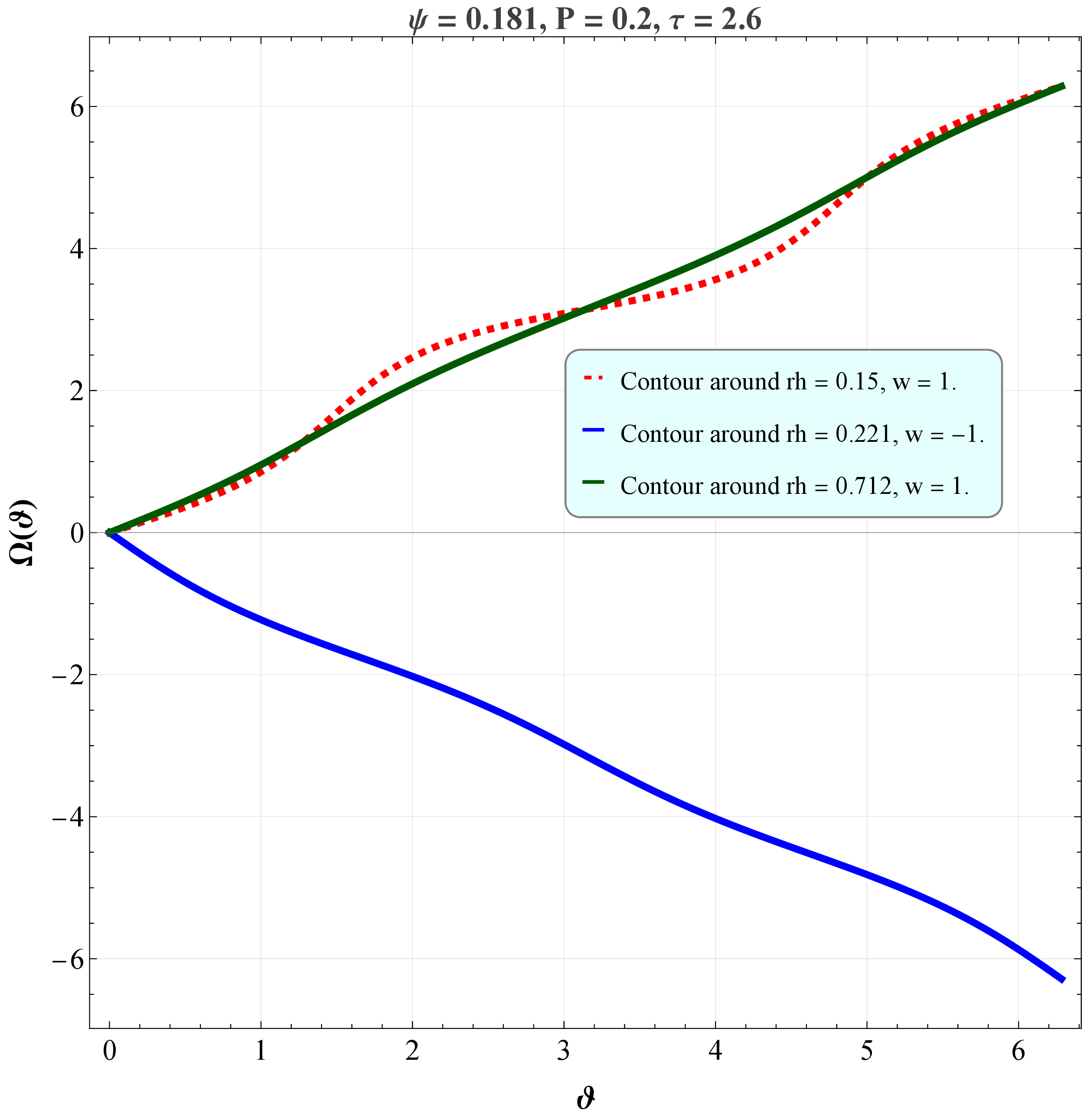}
    \caption{$\psi=0.181$, $P=0.2$, $\tau=2.6$. Deflection-angle diagram for a closed loop enclosing the three zero points located at $r_h=0.150$, $0.221$, and $0.712$.}
\end{subfigure}
\quad
\begin{subfigure}[b]{0.3\textwidth}
    \centering
    \includegraphics[width=\linewidth]{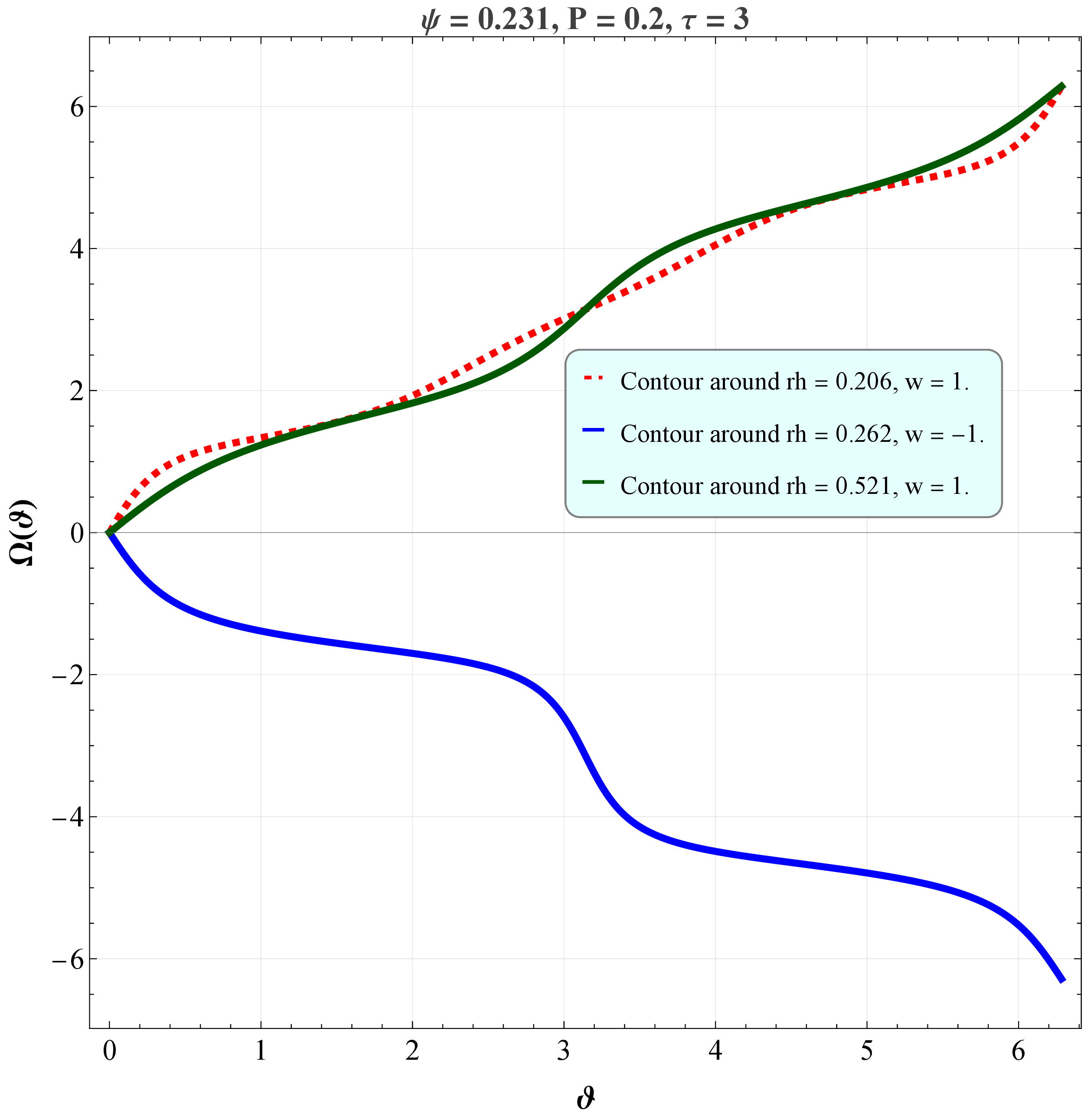}
    \caption{$\psi=0.231$, $P=0.2$, $\tau=3$. Deflection-angle diagram for a closed loop enclosing the three zero points located at $r_h=0.206$, $0.262$, and $0.521$.}
\end{subfigure}

\caption{
Representative deflection-angle diagrams in Region~I for several values of the electric potential $\psi$.  The winding-number sequence is $(+1,-1,+1)$. 
}\label{fig:d1-d4}
\end{figure}

Relation~\eqref{winding} can be used to calculate the winding number associated with each topological defect and thereby verify the classification obtained from the off-shell vector field. The corresponding deflection-angle diagrams for representative parameter sets in Region~I are shown in Fig.~\ref{fig:d1-d4}. In every case, the defects exhibit the winding-number sequence $(+1,-1,+1)$, confirming the expected topological structure of this region. Since the innermost zero, corresponding to the black hole, has a winding number of $+1$, the system in this region is classified as $W^{1_{+}}$.

To calculate the winding number, we first choose a value of $\tau$ within the relevant range and identify the corresponding zeros. We then select a contour enclosing each zero individually. The contours may be chosen as those shown in Figure~\ref{fig:v0v1v2v3} or as arbitrary closed contours. Using numerical methods, we then calculate and plot the deflection angle along a complete loop around each zero.

In the context of topological phase transitions in AdS black holes, the $r_h$-$\tau$ diagrams serve as fundamental tools that reveal the complete thermodynamic phase structure of the system. Each point on the curve represents an on-shell black hole solution satisfying the equation $\phi^{r_h}=0$, with the horizon radius $r_h$ plotted against the inverse temperature $\tau$, thereby mapping all possible black hole states for a given set of parameters. For a fixed value of $\tau$, the number of intersections with the curve directly indicates the number of coexisting black hole phases; for example, a single intersection corresponds to one stable phase, whereas three intersections signal the coexistence of small, intermediate, and large black hole branches. The extrema, or turning points, of the curve mark the critical inverse temperatures where phase transitions occur, and at these points pairs of branches typically one stable and one unstable are either generated or annihilated.

\begin{figure}[H]
\centering

\begin{subfigure}[b]{0.28\textwidth}
    \centering
    \includegraphics[width=\linewidth]{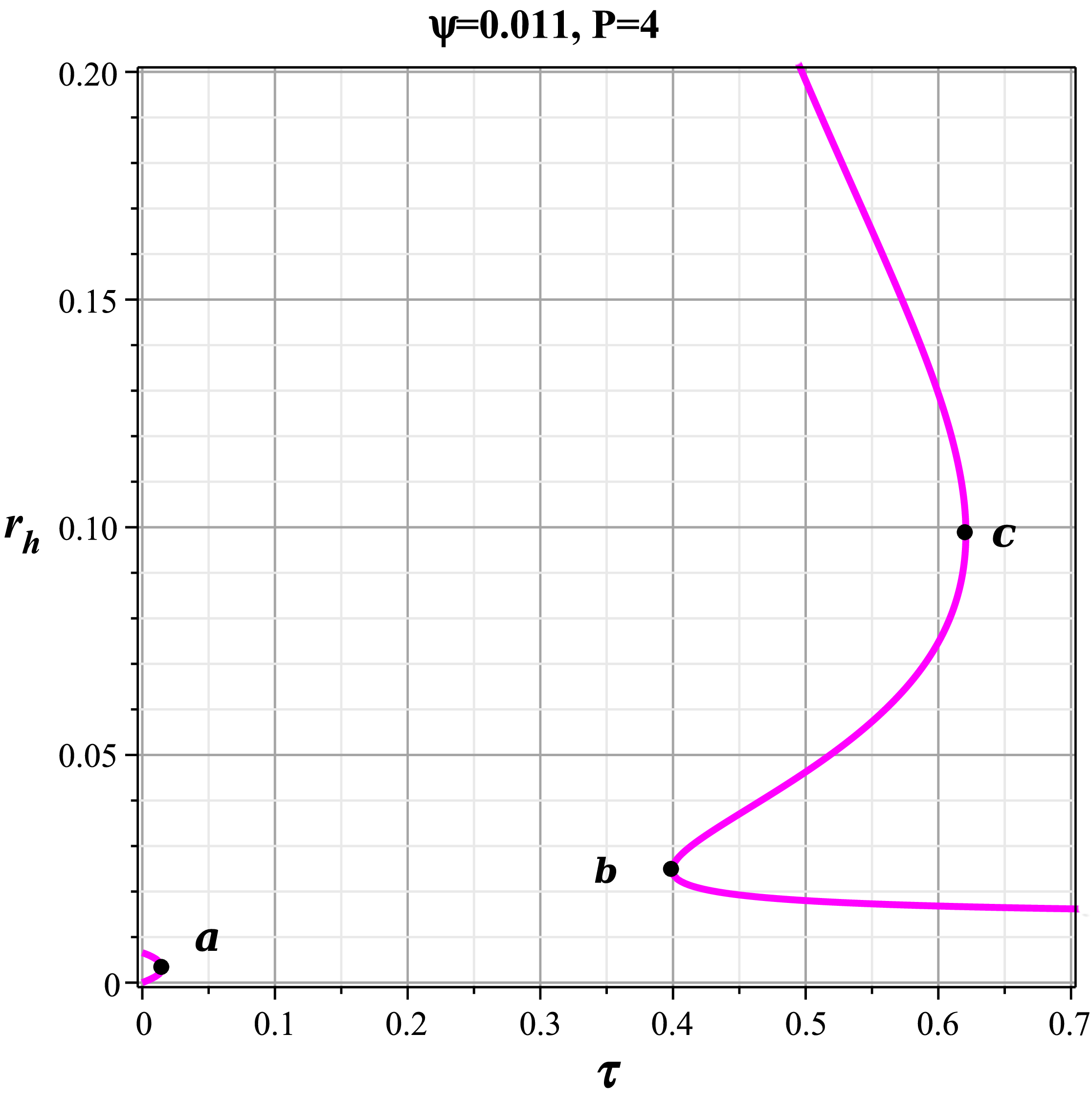}
    \caption{$r_h$--$\tau$ diagram for $\psi=0.011$ and $P=4$.}
\end{subfigure}
\quad
\begin{subfigure}[b]{0.28\textwidth}
    \centering
    \includegraphics[width=\linewidth]{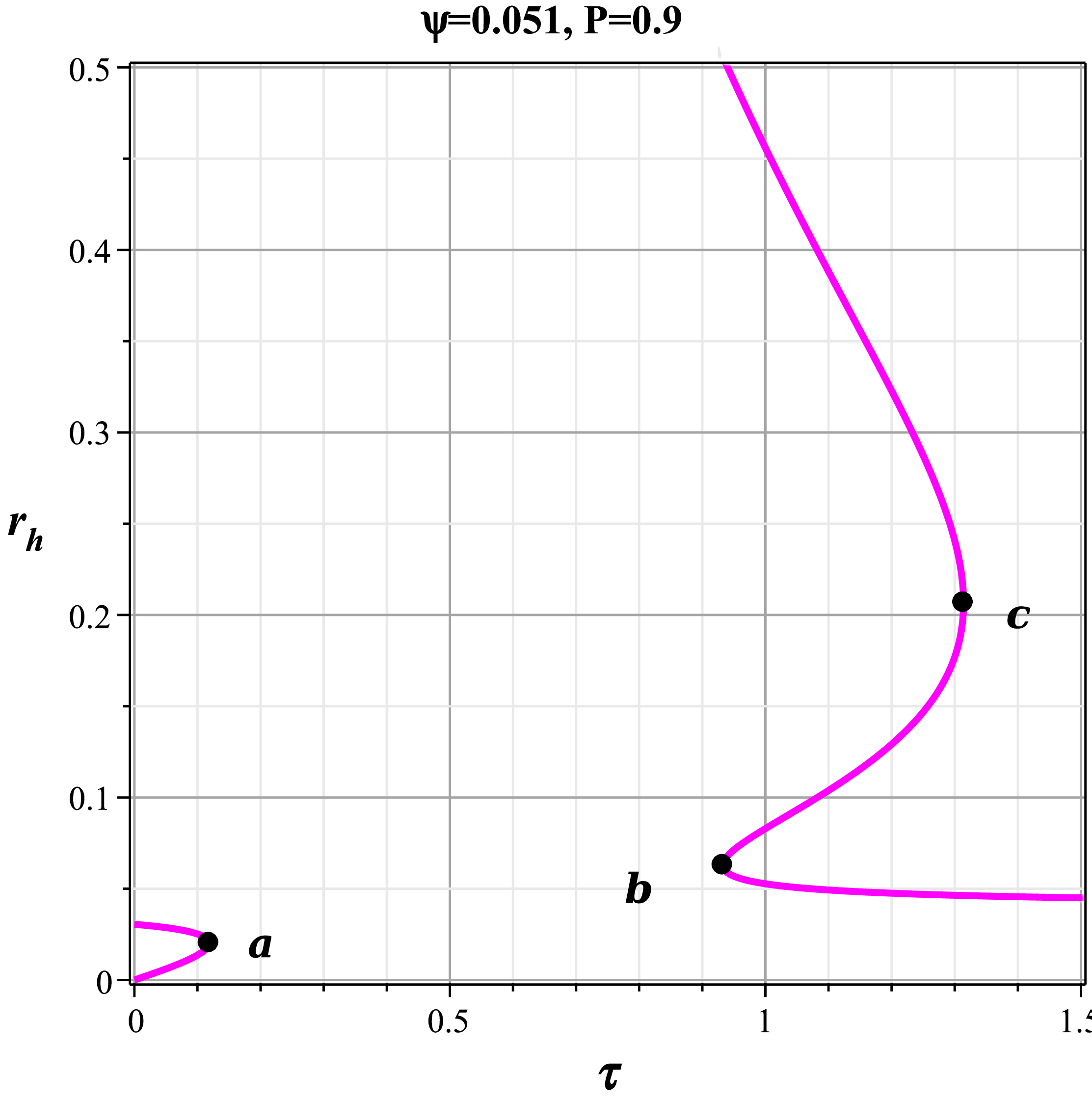}
    \caption{$r_h$--$\tau$ diagram for $\psi=0.051$ and $P=0.9$.}
\end{subfigure}

\vspace{0.2cm}

\begin{subfigure}[b]{0.28\textwidth}
    \centering
    \includegraphics[width=\linewidth]{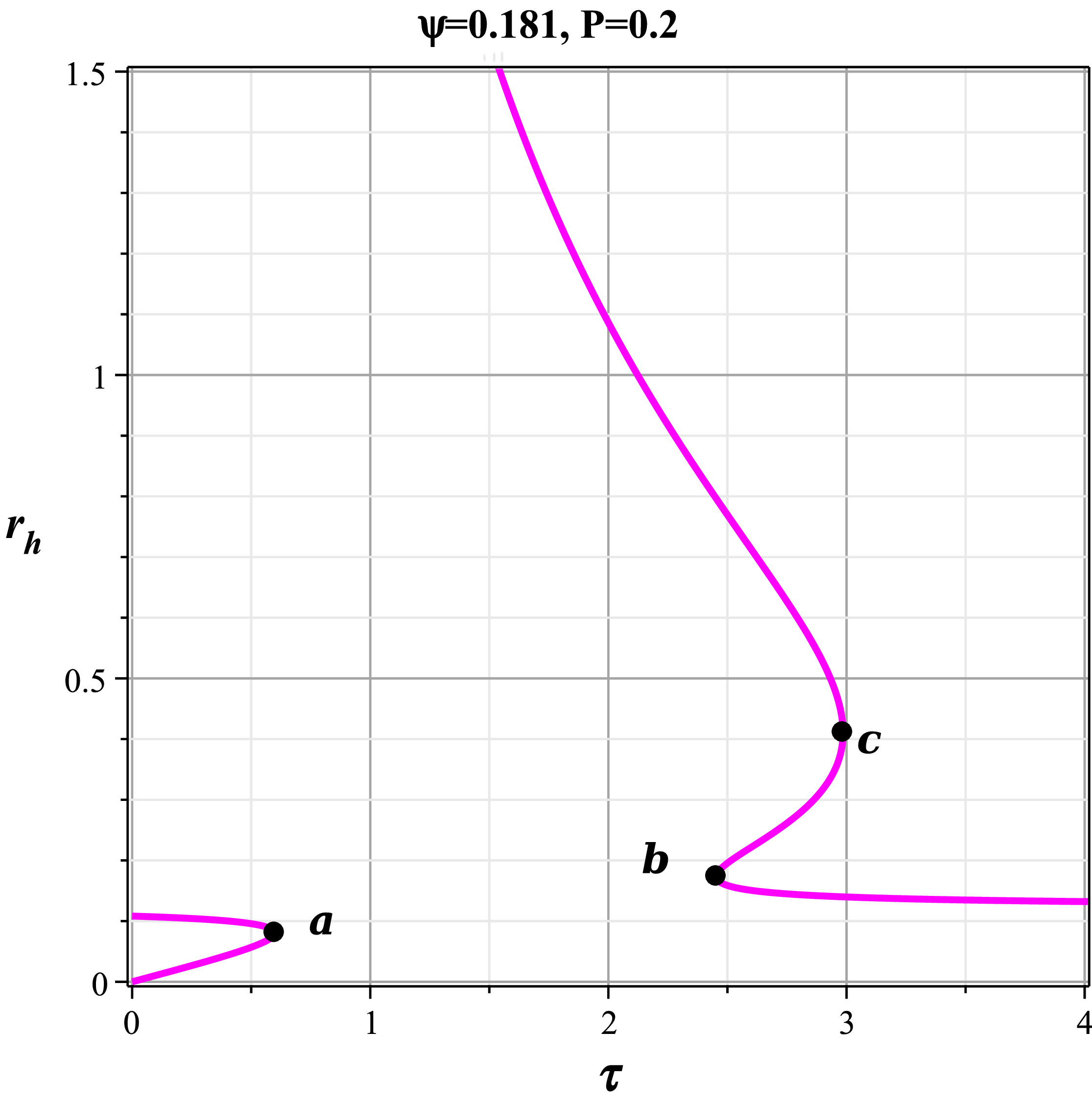}
    \caption{$r_h$--$\tau$ diagram for $\psi=0.181$ and $P=0.2$.}
\end{subfigure}
\quad
\begin{subfigure}[b]{0.28\textwidth}
    \centering
    \includegraphics[width=\linewidth]{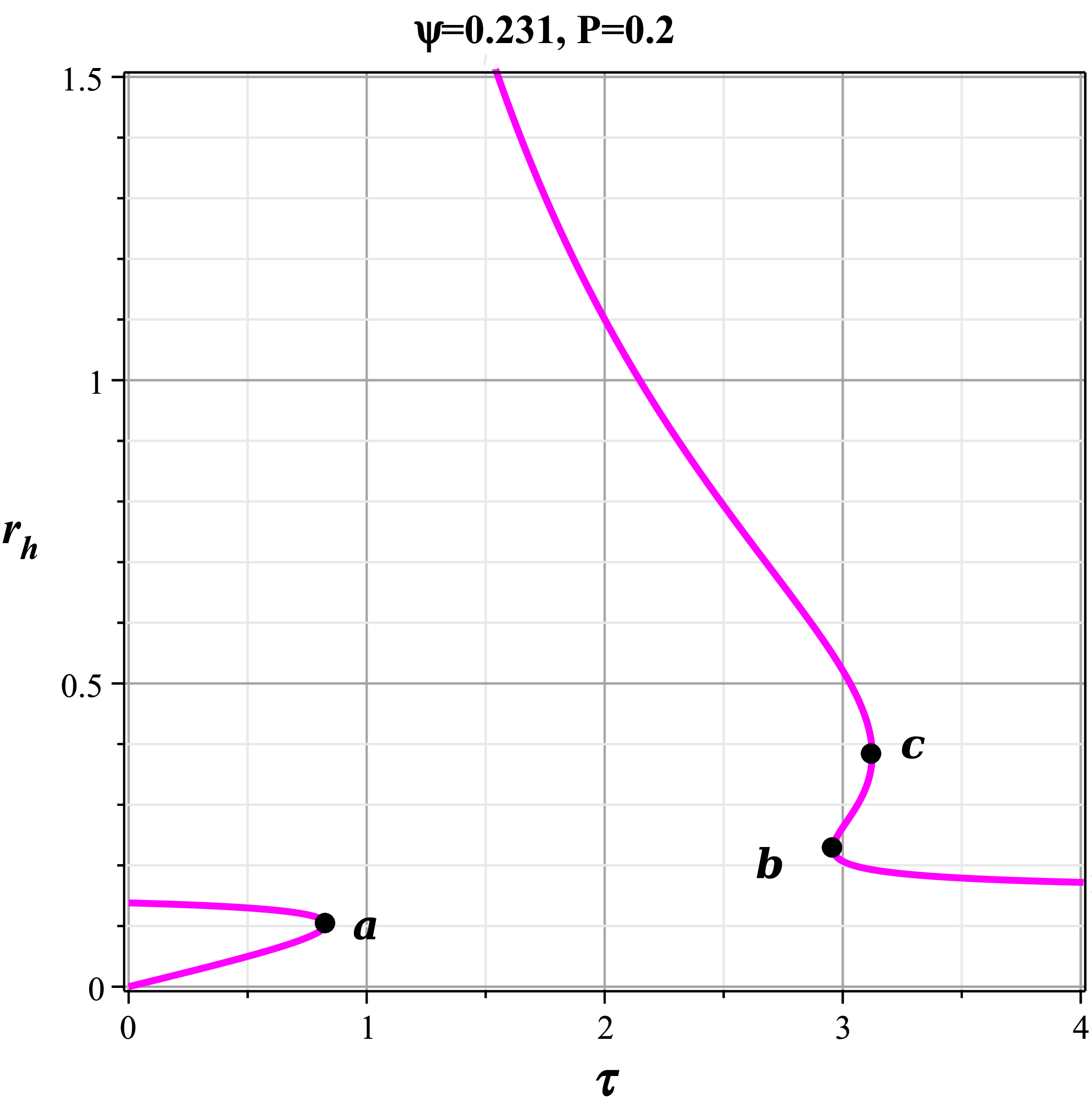}
    \caption{$r_h$--$\tau$ diagram for $\psi=0.231$ and $P=0.2$.}
\end{subfigure}

\caption{
Representative $r_h$--$\tau$ diagrams for Region~I. Although all parameter sets belong to the same topological class, with total winding number $W=+1$, their thermodynamic evolution differs from that of a conventional van der Waals system. In addition to the familiar van der Waals-like branch, an additional disconnected branch emerges at small horizon radii and small $\tau$ values (corresponding to the high-temperature regime). The characteristic values $\tau_a$, $\tau_b$, and $\tau_c$ correspond to successive annihilation, generation, and annihilation events of topological defects. 
}\label{fig:tau1-tau4}
\end{figure}

The $r_h$--$\tau$ diagrams shown in Fig.~\ref{fig:tau1-tau4} demonstrate that Region~I belongs to the $W=+1$ topological class, although its thermodynamic evolution differs significantly from that of the conventional Van der Waals system where we could not realized directly from the usual thermodynamic study. In addition to the familiar Van der Waals-like branch, an isolated branch (or a disconnected family of two branches) appears at small horizon radius, resulting in a modified sequence of topological-defect generation and annihilation while preserving the total winding number.

The evolution is naturally divided into four temperature intervals. For $0<\tau<\tau_a$, two disconnected families of black holes coexist: one consists of a stable small black hole and an unstable intermediate black hole, while the other contains a stable large black hole from the other family. The corresponding winding numbers are $(+1,-1,+1)$, giving a total winding number $W=+1$. At $\tau=\tau_a$, the stable small-black-hole branch annihilates with the unstable intermediate branch, leaving only the stable large-black-hole branch for $\tau_a<\tau<\tau_b$. For each branch, or black hole state, associated with a non-degenerate zero, one can immediately estimate the sign of the winding number from relation~\ref{nd} and thereby obtain the sequence of stabilities of the states. In this case, the sequence is $(+1,-1,+1)$.

At $\tau=\tau_b$, a new pair of topological defects is created. Consequently, for $\tau_b<\tau<\tau_c$, three black-hole branches coexist, namely a stable small black hole, an unstable intermediate black hole, and a stable large black hole, reproducing the characteristic Van der Waals branch structure. Finally, at $\tau=\tau_c$, the unstable intermediate branch annihilates with the stable large-black-hole branch, so that for $\tau>\tau_c$ only the stable small-black-hole branch survives.

Therefore, although the global topological invariant remains $W=+1$, the thermodynamic evolution is not uniquely determined by the total winding number. In particular, the same winding-number structure $(+1,-1,+1)$ can correspond to different arrangements and connectivities of thermodynamic branches in the $r_h$--$\tau$ plane. The presence of a disconnected branch changes the way the thermodynamic branches emerge, merge, and disappear as $\tau$ varies. This demonstrates that black-hole solutions belonging to the same topological class may nevertheless exhibit qualitatively different thermodynamic evolutions.

The $r_h-\tau$ diagrams provide a temperature-dependent perspective on the phase-transition structure; however, the topological sequence $(+1,-1,+1)$ remains unchanged. For example, in panel (d) of Fig.~\ref{fig:d1-d4}, for $\psi=0.231$, $P=0.2$, and $\tau=3$, the three black points correspond to the winding-number sequence $(+1,-1,+1)$. Similarly, if we consider $\tau=0.5$ in panel (d) of Fig.~\ref{fig:tau1-tau4}, which corresponds to a different family of states, we again obtain the same winding-number sequence, $(+1,-1,+1)$. We have verified this behavior for several values of $\tau$, confirming that the topological classification remains unchanged across these different temperature-dependent branches. The behavior in this region is more complicated than that of a conventional van der Waals system. This complexity originates from the nonminimal coupling, which leads to a more intricate form of the entropy.

The cusp-like family at small horizon radii  consists of two lower stable branches and an upper unstable branch. It approaches $\tau=0$ at $r_h=0$ and at another finite value of $r_h$ associated with the unstable branch. For example, for $\psi=0.181$, as $\tau\rightarrow\infty$, one branch approaches a finite horizon radius, $r_h\simeq0.123$. Meanwhile, the upper branch of the cusp intersects the $r_h$-axis at approximately $r_h=0.108$. The van der Waals-like family, on the other hand, approaches $\tau=0$ in the limit $r_h\rightarrow\infty$. These features demonstrate the richer structure of the phase space induced by the nonminimal coupling.

\subsection{Topological Phase Transition in Region II}

The unit vector field at the boundaries and the corresponding deflection-angle analysis in Region II are identical to those in Region I, yielding the winding-number sequence $(+1,-1,+1)$ and the total winding number $W=+1$. Since the topological charges remain unchanged, these plots are not repeated here.

\begin{figure}[H]
\centering

\begin{subfigure}[b]{0.32\textwidth}
    \centering
    \includegraphics[width=\textwidth]{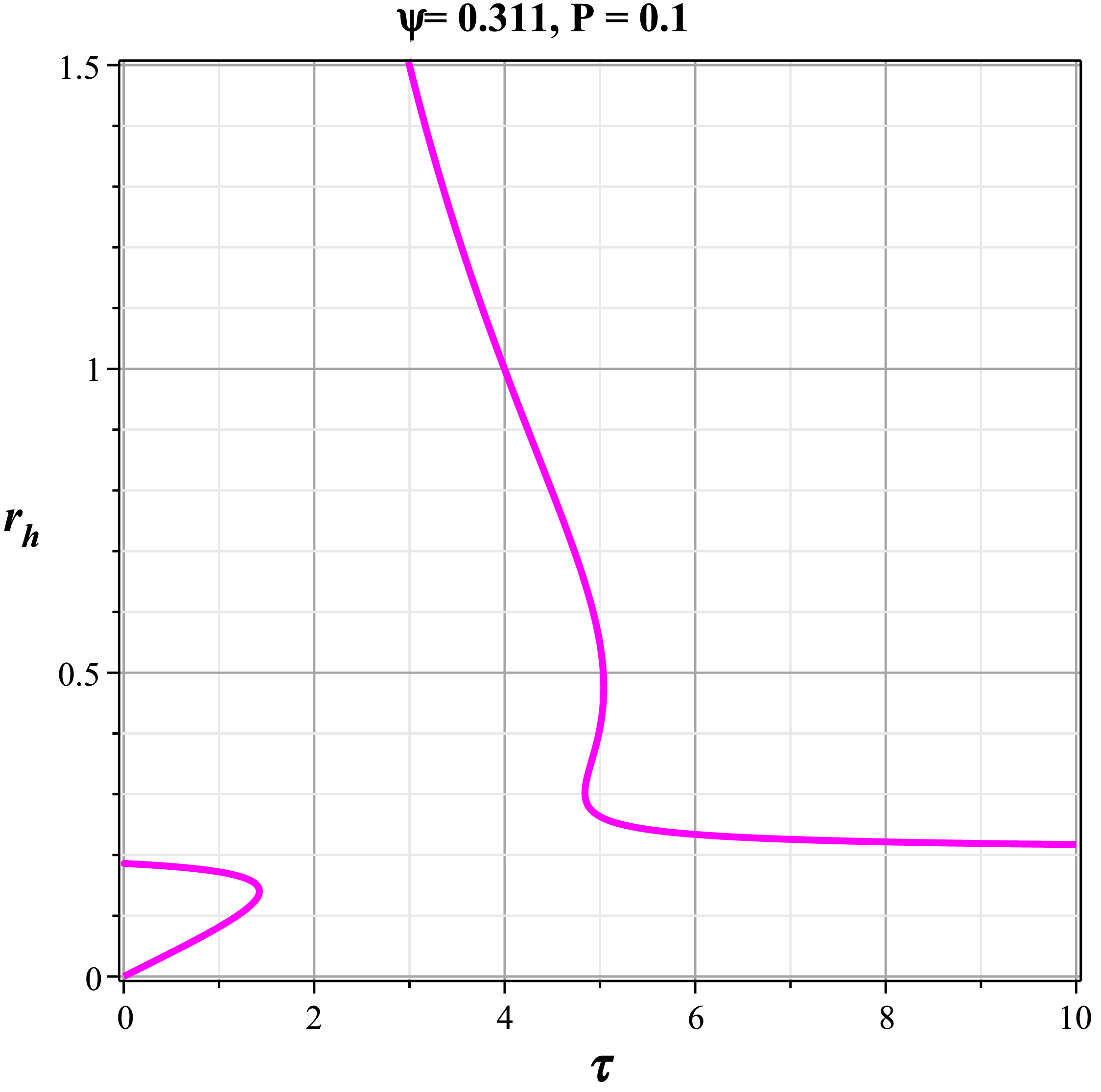}
    \caption{$P=0.1<P_c$}
    \label{fig:tau101}
\end{subfigure}
\hfill
\begin{subfigure}[b]{0.32\textwidth}
    \centering
    \includegraphics[width=\textwidth]{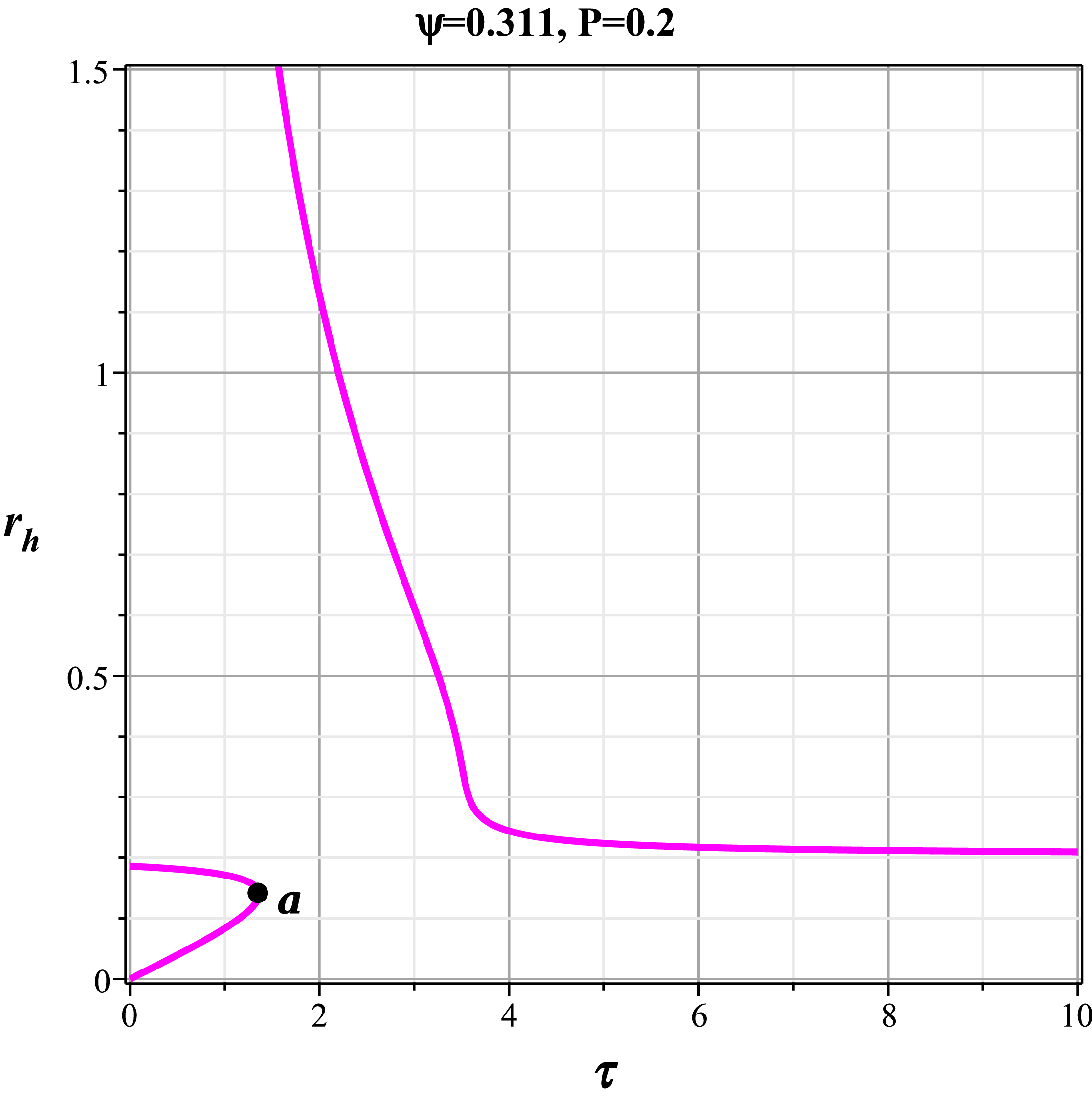}
    \caption{$P=0.2<P_c$}
    \label{fig:tau10}
\end{subfigure}
\hfill
\begin{subfigure}[b]{0.32\textwidth}
    \centering
    \includegraphics[width=\textwidth]{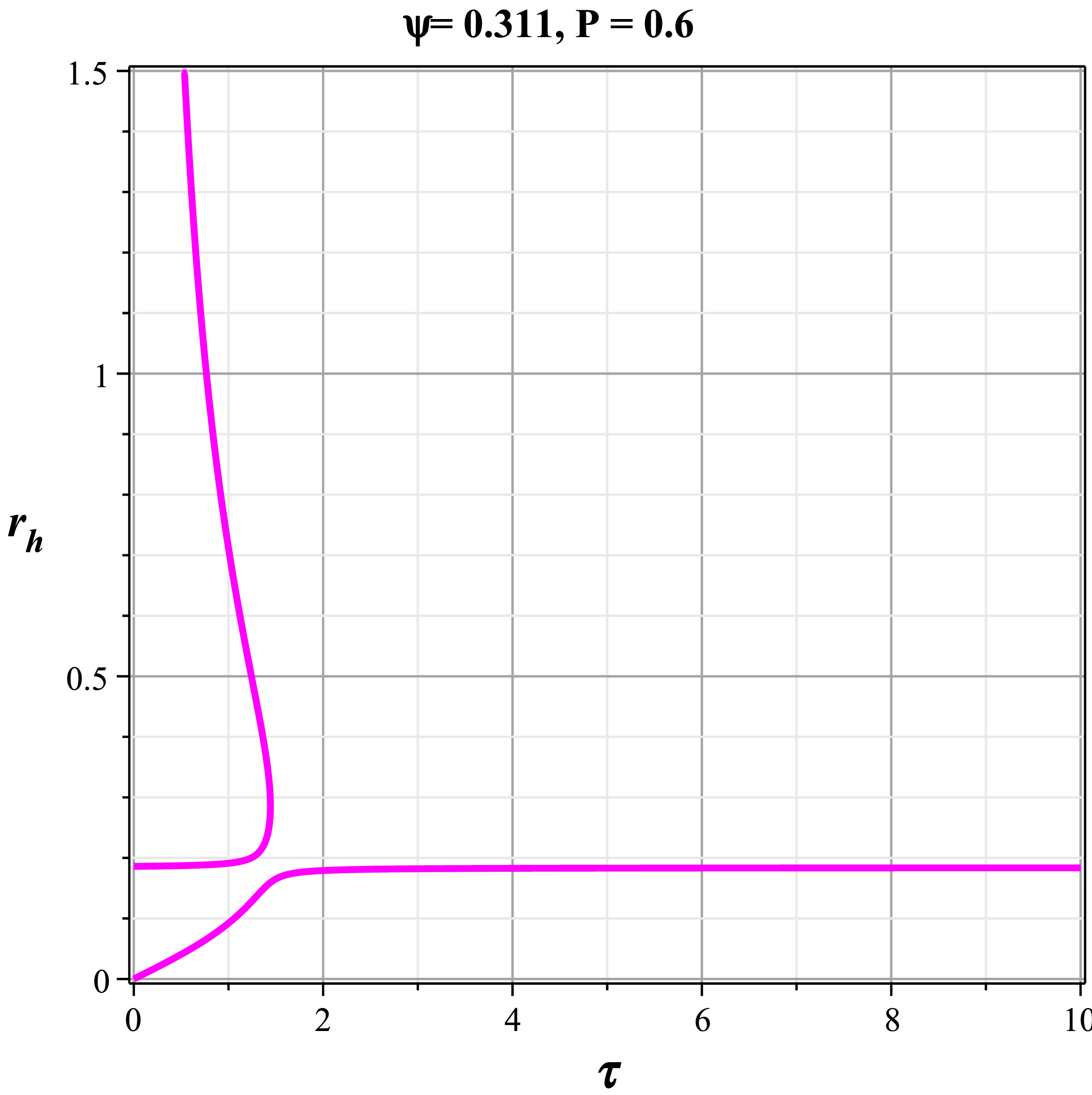}
    \caption{$P=0.6>P_c$}
    \label{fig:tau102}
\end{subfigure}

\caption{
The $r_h$--$\tau$ diagrams in Region II for $\psi=0.311$ at different pressures, with $P_c=0.376$. 
Panel (a), for $P=0.1<P_c$, exhibits a van der Waals-like family together with an additional disconnected cusp-like family at small $r_h$ and small $\tau$. 
In panel (b), for $P=0.2<P_c$, the oscillatory behavior of the van der Waals-like family disappears, while the cusp-like family persists for small $\tau$; for $\tau>\tau_a$, only the monotonic stable branch remains. 
In panel (c), for $P=0.6>P_c$, the cusp-like family extends to relatively large horizon radii at small $\tau$, corresponding to the high-temperature regime. Hence, the large-horizon black holes at high temperature belong to the cusp-like family rather than to the van der Waals-like family. 
}
\label{fig:tau1012}
\end{figure}

The distinctive feature of Region II is instead revealed by the $r_h$--$\tau$ diagrams shown in Fig.~\ref{fig:tau1012}.

For $\psi=0.311$ and $P=0.1$, which is below the critical pressure $P_c=0.376$, the left panel shows a van der Waals-like family accompanied by an additional disconnected cusp-like family, similar to that observed in Region I. 

As the pressure is increased to $P=0.2$, which remains below the critical pressure, the oscillatory behavior of the van der Waals-like family decreases, and a smoother behavior emerges, as shown in the middle panel. For $\tau<\tau_a$, three equilibrium black-hole solutions coexist, corresponding to a small stable black hole, an intermediate unstable black hole, and a large stable black hole. Their winding numbers remain $(+1,-1,+1)$, giving a total winding number $W=+1$. However, for $\tau>\tau_a$, the disconnected cusp-like family disappears, and only the monotonic stable branch remains. Consequently, the thermodynamic landscape is simplified to a single stable black-hole phase, although the total winding number remains unchanged.

For $P=0.6$, which is above the critical pressure, the oscillatory behavior of the van der Waals-like family disappears, and the corresponding branch becomes smooth. Interestingly, this smooth branch lies below the surviving cusp-like family in the $r_h$--$\tau$ diagram. 
Interestingly, this downward displacement of the smooth branch below the cusp-like family is not observed for pressures above the critical pressure in Region I. The cusp-like family itself persists and contains branches extending to relatively large horizon radii at small $\tau$, corresponding to the high-temperature regime. Thus, the large-horizon black holes at high temperature belong to the surviving cusp-like family rather than to the smooth branch. This behavior is qualitatively different from the conventional van der Waals picture and demonstrates the nontrivial modification of the equilibrium structure induced by the non-minimal coupling. At large $\tau$, corresponding to low temperatures, a relatively small-horizon stable black-hole branch survives.

Therefore, Region II provides an example in which the   variations of the pressure can eliminate the oscillatory behavior of the van der Waals-like family and modify the connectivity of the equilibrium branches without changing the total winding number, which remains $W=+1$.

\subsection{Topological Phase Structure in Region III}

As discussed in Sec.~\ref{sec4}, Region III is characterized by the simultaneous increase of the critical radius, pressure, and temperature. The off-shell topological analysis confirms that this region belongs to the same topological class as Regions I and II, with the total winding number $W=+1$. This behavior is observed throughout the entire parameter range of Region III. 

Similar to Region II, the equilibrium solutions  consist of two disconnected families. One family forms a monotonic branch extending over the entire range of $\tau$, while the second family exists only for $\tau<\tau_a$. Consequently, for $\tau>\tau_a$ only the monotonic branch survives, corresponding to a single thermodynamically stable black-hole phase. However, by decreasing the pressure, curves with two turning points may be observed above the cusp-like curves. 

\begin{figure}[H]
\centering
\begin{subfigure}[b]{0.38\textwidth}
\includegraphics[width=\linewidth]{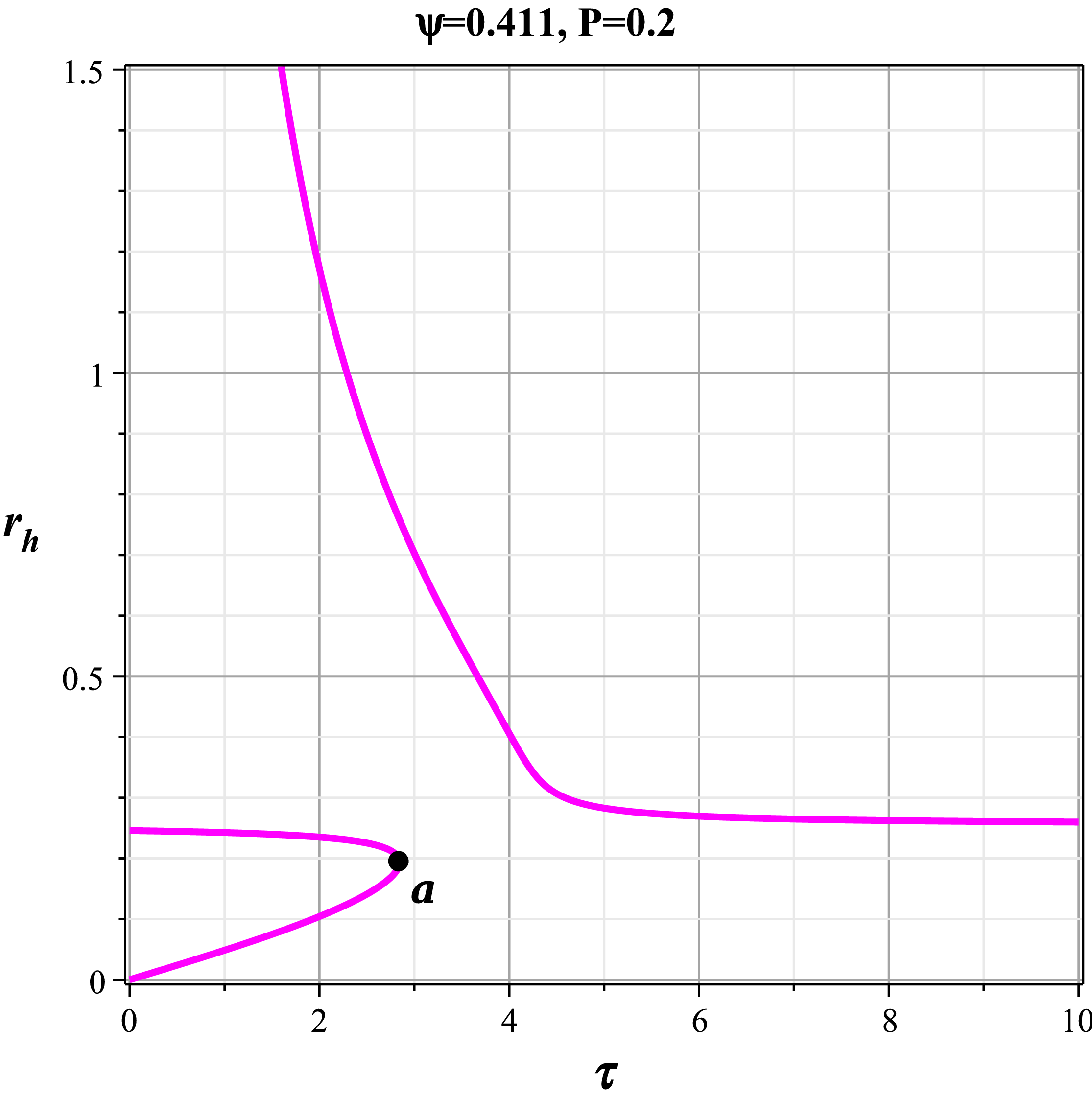}
\caption{$r_h$--$\tau$ diagram.}
\end{subfigure}
\quad
\begin{subfigure}[b]{0.38\textwidth}
\includegraphics[width=\linewidth]{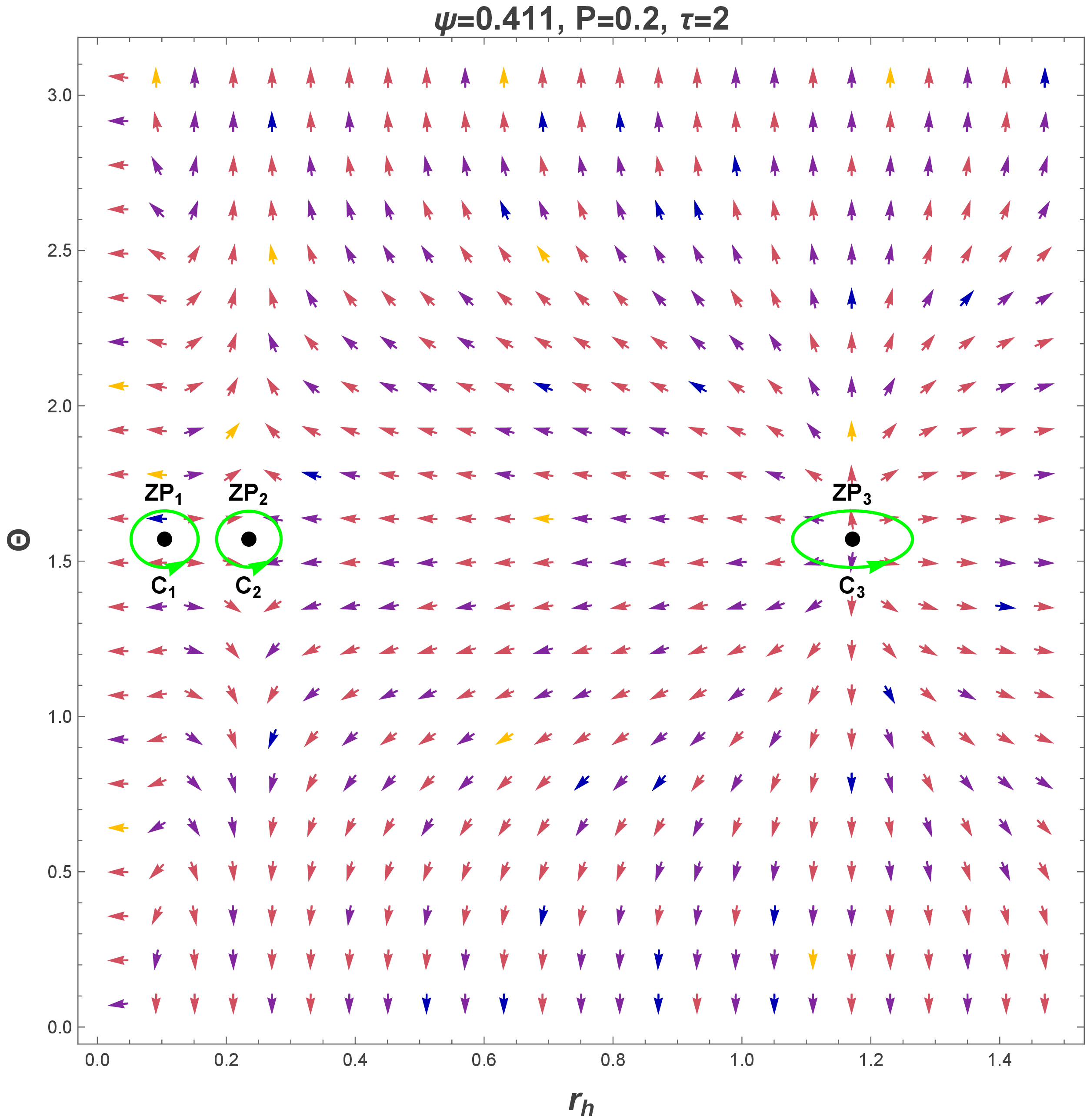}
\caption{Unit vector field for $\tau=2<\tau_a$.}
\end{subfigure}

\begin{subfigure}[b]{0.38\textwidth}
\includegraphics[width=\linewidth]{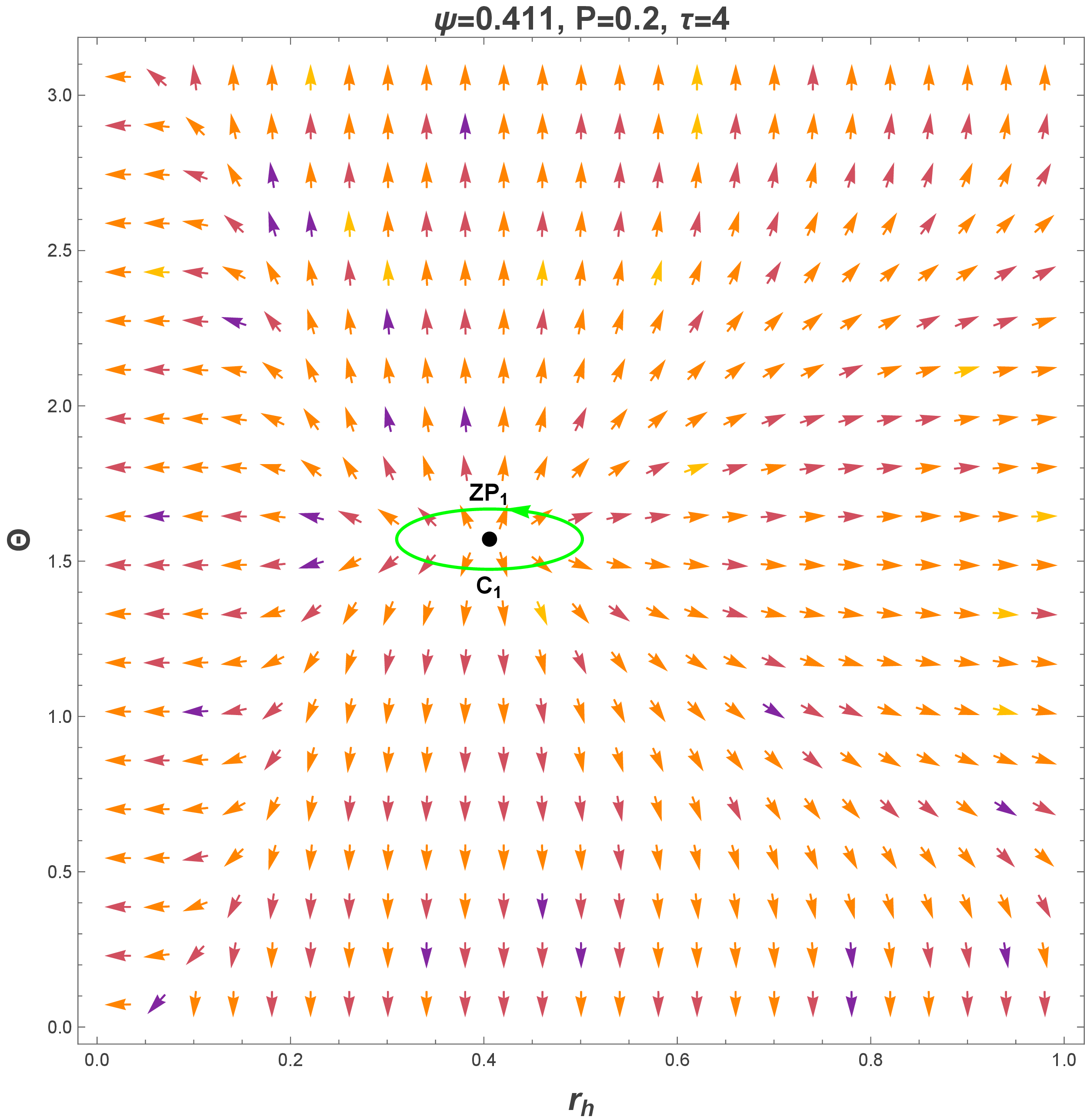}
\caption{Unit vector field for $\tau=4>\tau_a$.}
\end{subfigure}
\quad
\begin{subfigure}[b]{0.38\textwidth}
\includegraphics[width=\linewidth]{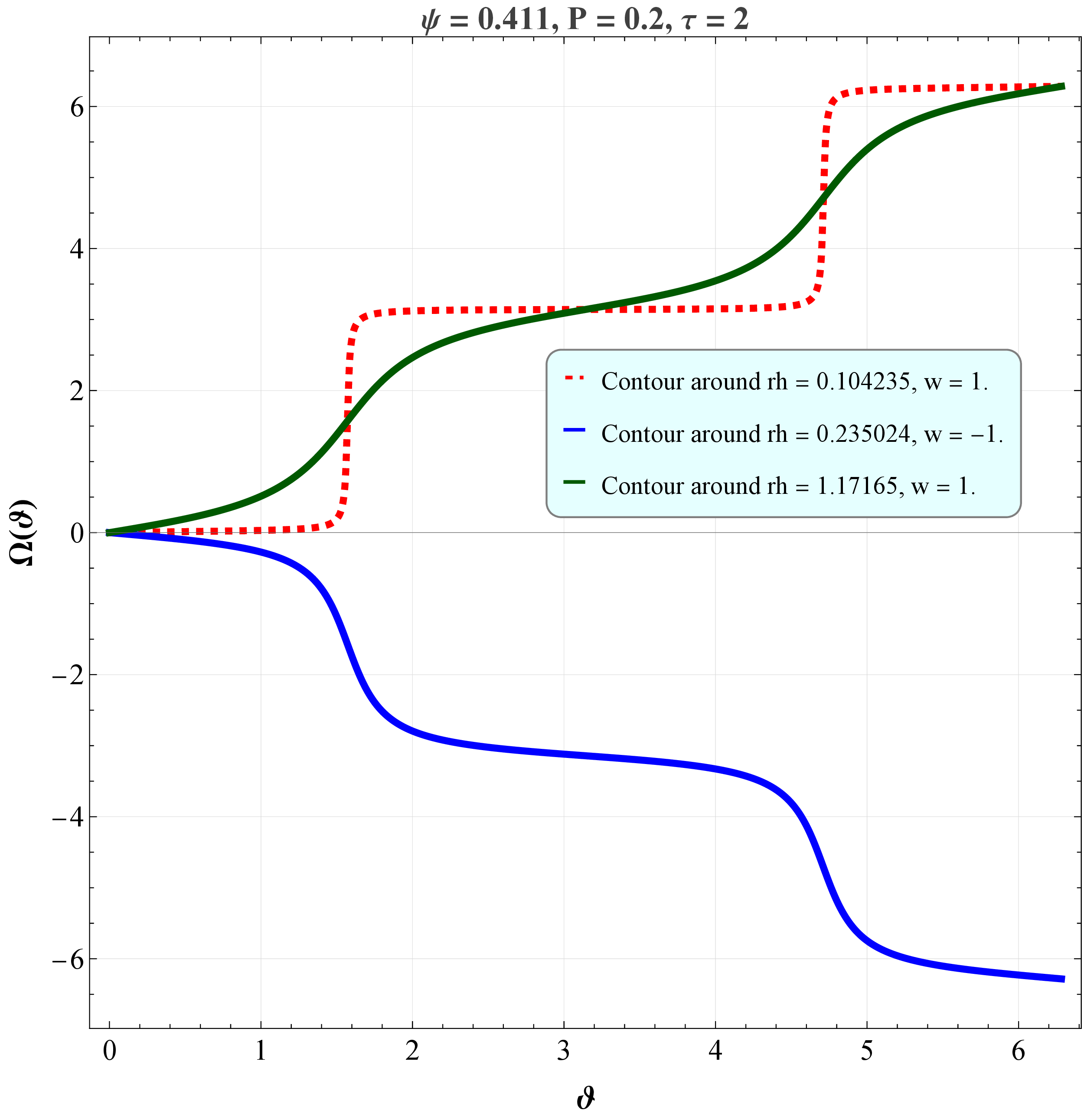}
\caption{Deflection angle for $\tau=2<\tau_a$.}
\end{subfigure}

\caption{Topological structure of Region III for $\psi=0.411$ and $P=0.2$. (a) The $r_h$--$\tau$ diagram consists of two disconnected equilibrium families. The secondary family disappears at $\tau=\tau_a$, leaving only the monotonic stable branch for $\tau>\tau_a$. (b) The unit vector field for $\tau<\tau_a$ exhibits three topological defects. (c) For $\tau>\tau_a$, only a single defect survives. (d) The deflection-angle analysis confirms the winding-number sequence $(+1,-1,+1)$, yielding the total winding number $W=+1$.}
\label{fig:regionIII}
\end{figure}

The corresponding unit vector fields are shown in Figs.~\ref{fig:regionIII}(b) and \ref{fig:regionIII}(c). For $\tau=2<\tau_a$, with $P=0.2$ and $\psi=0.411$, the off-shell vector field possesses three defects located at
\[
r_h=0.1042354492,\quad
r_h=0.2350238619,\quad
r_h=1.171650651,
\]
representing the small, intermediate, and large black-hole branches, respectively. In contrast, for $\tau=4>\tau_a$, only a single defect remains, located at$r_h=0.4058050180,$ consistent with the disappearance of the disconnected family in panel (a).

The corresponding deflection-angle analysis is presented in Fig.~\ref{fig:regionIII}(d) for $\tau=2<\tau_a$. The winding-number sequence is found to be $(+1,-1,+1)$, confirming that the total topological charge remains
$W=+1$
Therefore, although the number of equilibrium branches changes from three to one as $\tau$ crosses $\tau_a$, the global topological invariant is preserved. 

\begin{figure}[H]
\centering
\begin{subfigure}[b]{0.35\textwidth} 
    \centering
    \includegraphics[width=\linewidth]{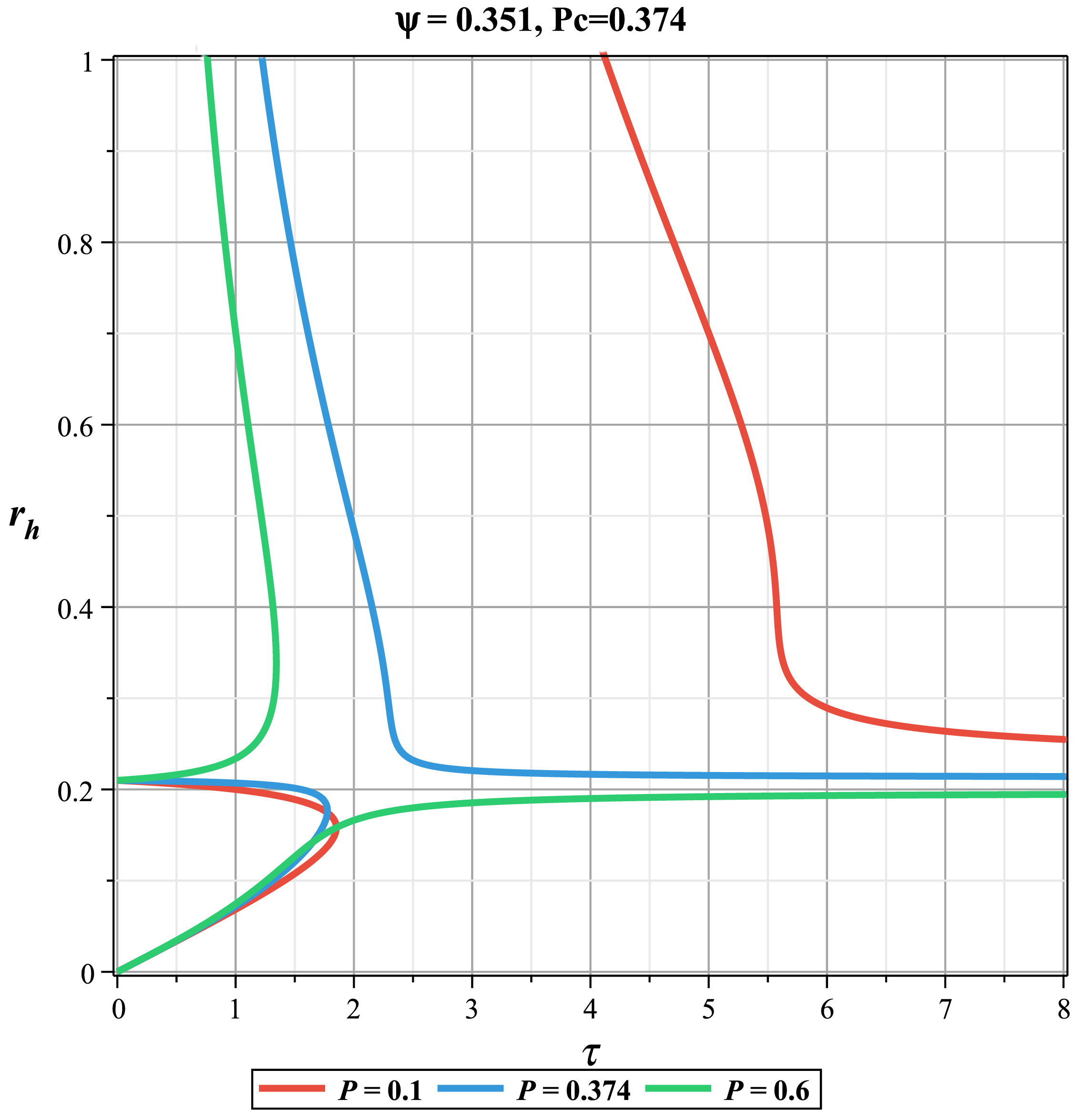}
    \caption{$\psi=0.351$, with $P=0.1$, $0.374$, and $0.6$.}
    \label{fig:tau42}
\end{subfigure}
\quad
\begin{subfigure}[b]{0.35\textwidth} 
    \centering
    \includegraphics[width=\linewidth]{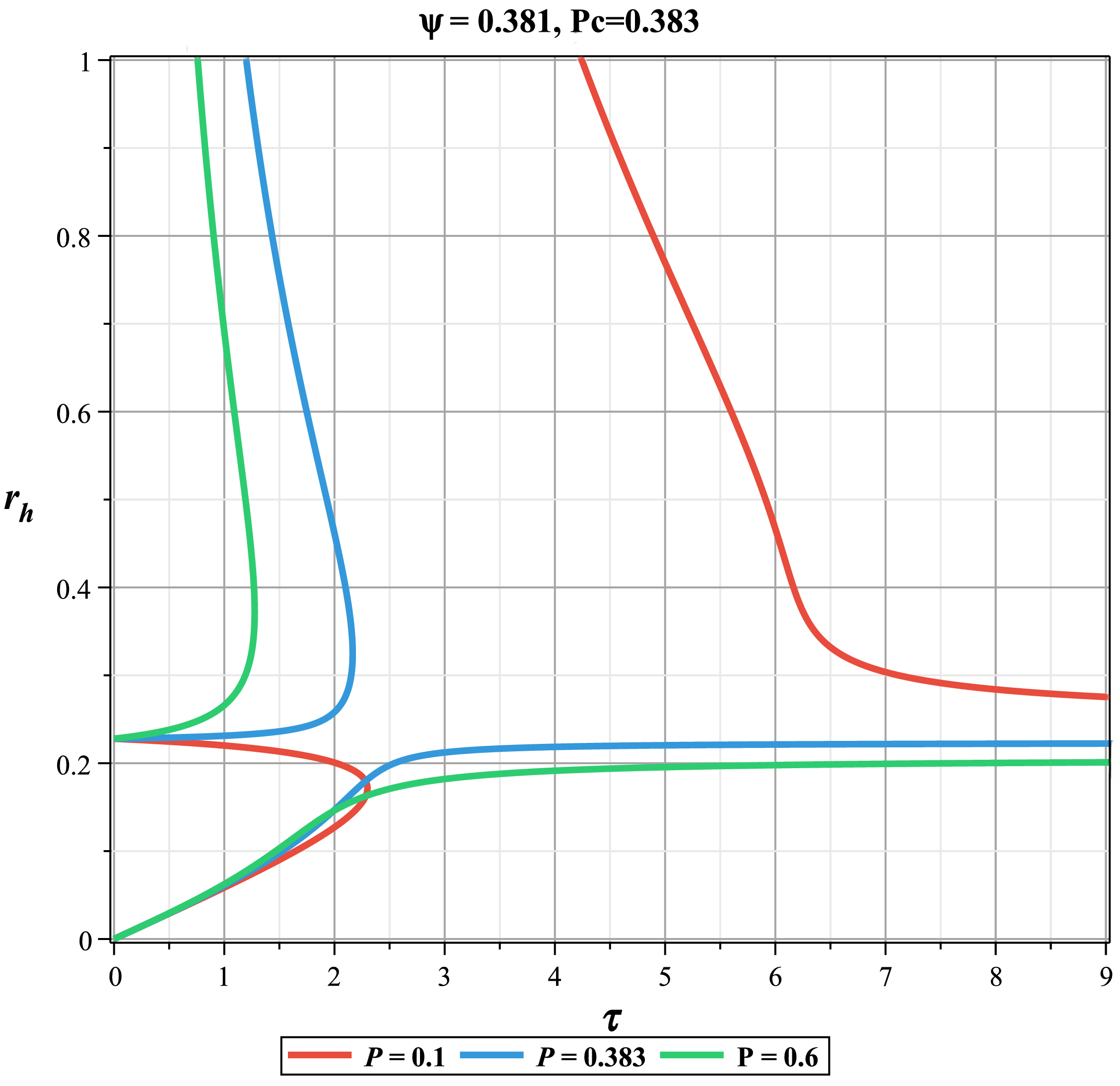}
    \caption{$\psi=0.381$, with $P=0.1$, $0.383$, and $0.6$.}
    \label{fig:tau43}
\end{subfigure}

\caption{The $r_h$--$\tau$ diagrams for two representative values of the electric potential in Region~III. For $\psi=0.351$ (left panel), the curves are plotted for $P=0.1$, $P=P_c=0.374$, and $P=0.6$. For $\psi=0.381$ (right panel), the corresponding pressures are $P=0.1$, $P=P_c=0.383$, and $P=0.6$. Compared with Region~II, the multiple-turning-point structure becomes less pronounced, and the branches progressively exhibit smoother behavior as $\psi$ increases.}
\label{fig:tau4243}
\end{figure}

For further confirmation, we plot the $r_h$--$\tau$ diagrams for $\psi=0.351$ and $\psi=0.381$, as shown in Fig.~\ref{fig:tau4243}. The overall behavior remains consistent with that observed in Region~II. However, as the parameters move into this region, the tendency of the $r_h$--$\tau$ curves to develop multiple turning points becomes less pronounced. Instead, the branches become progressively smoother, particularly for the curves corresponding to the larger values of $\psi$. 

Interestingly, despite the distinct morphology of these branches, the total topological charge remains $W=+1$. This demonstrates that the topological classification is not uniquely tied to the detailed geometric morphology of the thermodynamic branches: different branch structures can belong to the same topological class.

\subsection{Topological Phase Structure in Region IV}

As discussed in Sec.~\ref{sec4}, Region IV is characterized by a decrease in the critical horizon radius, while both the critical pressure and critical temperature continue to increase. The off-shell topological analysis confirms that this region remains in the same topological class as the previous regions, with the total winding number $W=+1$.

\begin{figure}[H]
\centering
\begin{subfigure}[b]{0.38\textwidth}
\centering
\includegraphics[width=\linewidth]{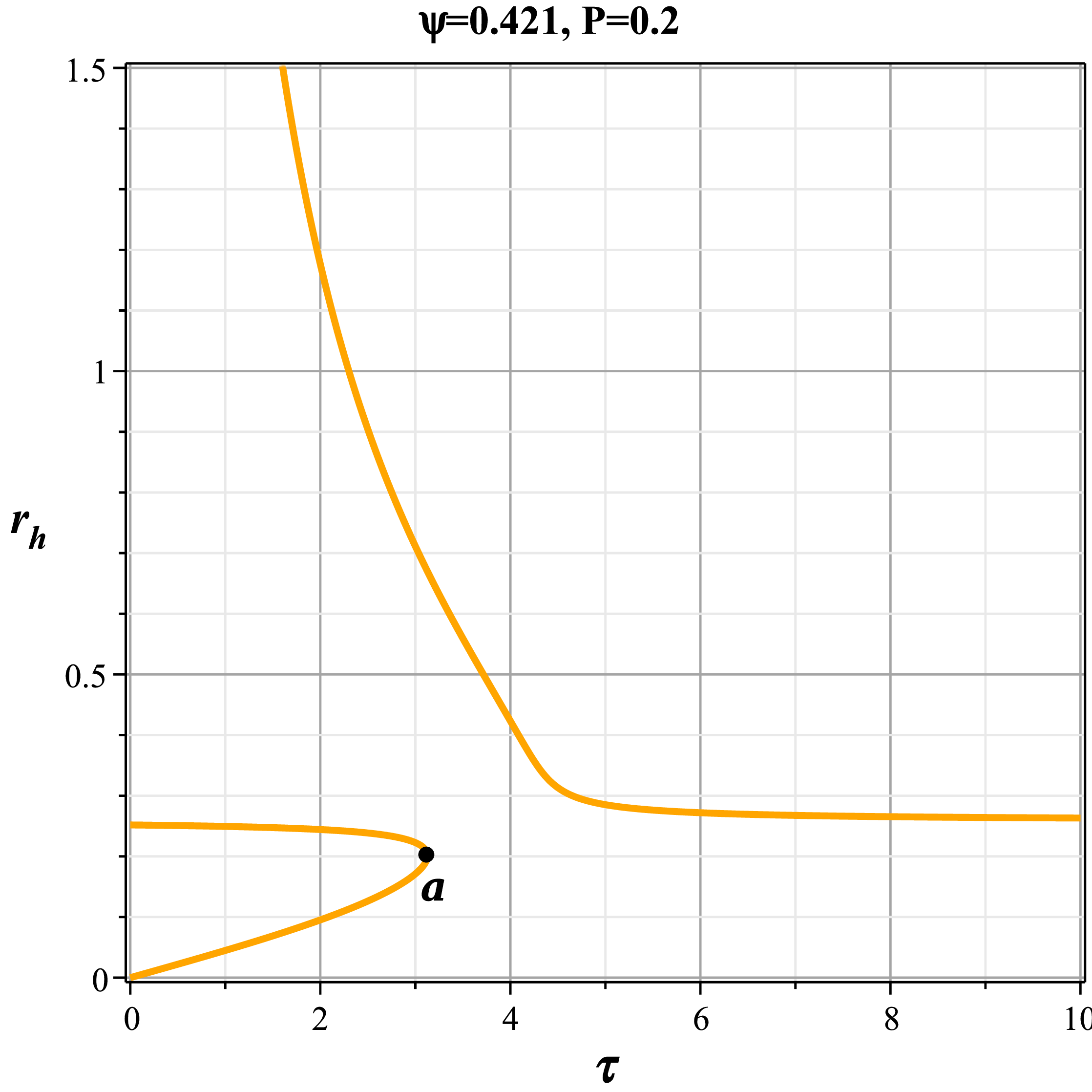}
\caption{$\psi=0.421$ and $P=0.2$.}
\end{subfigure}
\quad
\begin{subfigure}[b]{0.38\textwidth}
\centering
\includegraphics[width=\linewidth]{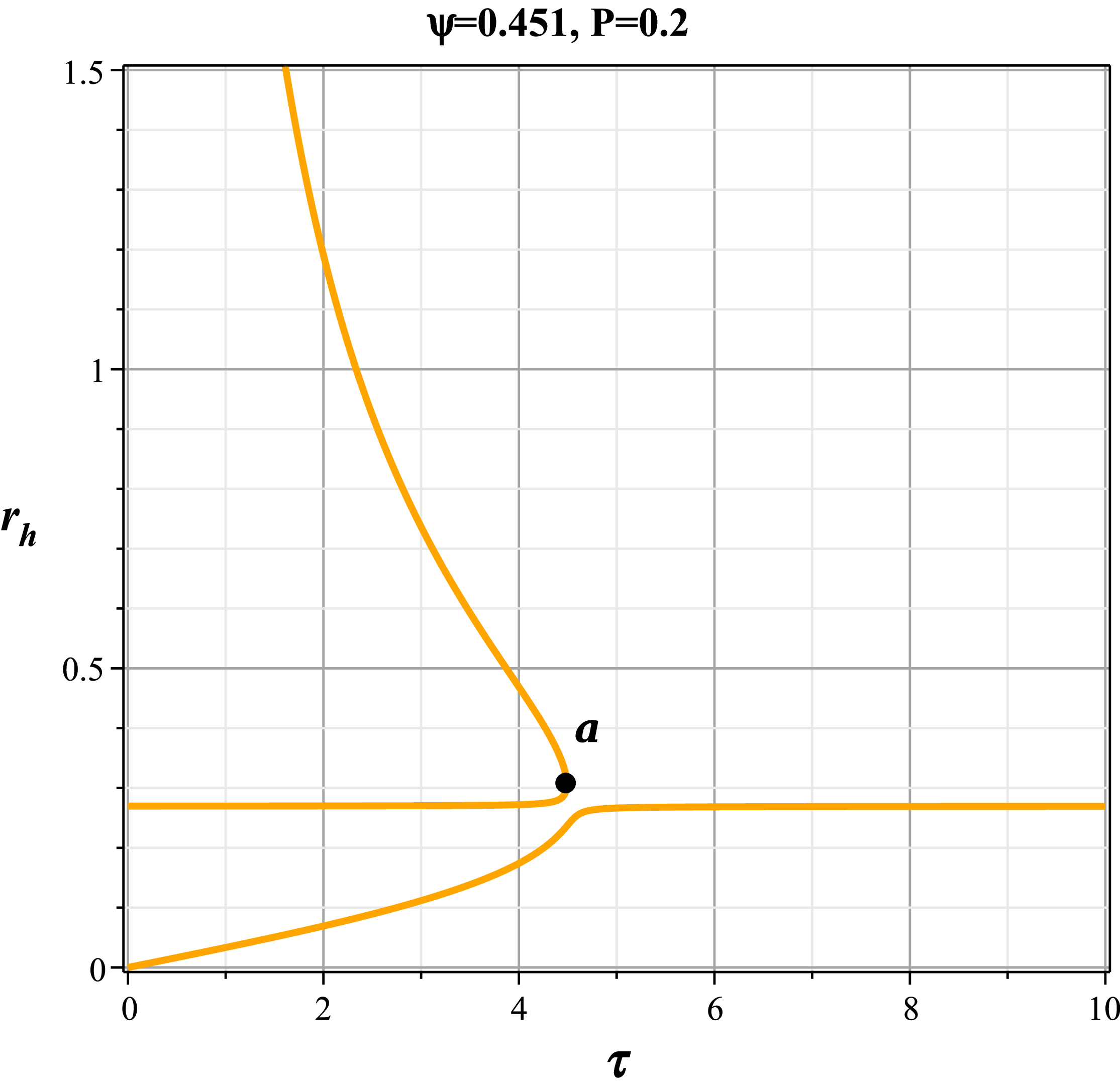}
\caption{$\psi=0.451$ and $P=0.2$.}
\end{subfigure}
\caption{
Representative $r_h$--$\tau$ diagrams in Region IV for $P=0.2$. Panel (a) ($\psi=0.421$) exhibits the same topology as Region III, where the monotonic stable branch lies above the disconnected family. Panel (b) ($\psi=0.451$) shows a qualitative rearrangement of the equilibrium branches: the monotonic stable branch moves below the disconnected family, while the total winding number remains $W=+1$.
}
\label{fig:regionIV}
\end{figure}

To illustrate the evolution of the equilibrium branches, Fig.~\ref{fig:regionIV} presents the $r_h$--$\tau$ diagrams for two representative values of the electric potential, $\psi=0.421$ and $\psi=0.451$, at $P=0.2$, which is below the corresponding critical pressure.

For $\psi=0.421$, shown in Fig.~\ref{fig:regionIV}(a), the $r_h$--$\tau$ diagram is qualitatively identical to that of Region III. The equilibrium solutions consist of two disconnected families: a monotonic branch extending over the entire range of $\tau$, together with a disconnected family that exists only for $\tau<\tau_a$. Consequently, for $\tau>\tau_a$ only the monotonic stable branch survives.

A qualitative change occurs at $\psi=0.451$, as illustrated in Fig.~\ref{fig:regionIV}(b). Although the equilibrium solutions are still composed of two disconnected families, their relative positions are reversed. The monotonic stable branch now lies below the disconnected family and therefore becomes the preferred equilibrium solution for smaller horizon radii over the entire range of $\tau$. The disconnected family, consisting of an unstable and a stable branch, exists only for $\tau<\tau_a$ and disappears beyond the critical point.

\begin{figure}[H]
\centering
\begin{subfigure}[b]{0.38\textwidth}
\centering
\includegraphics[width=\linewidth]{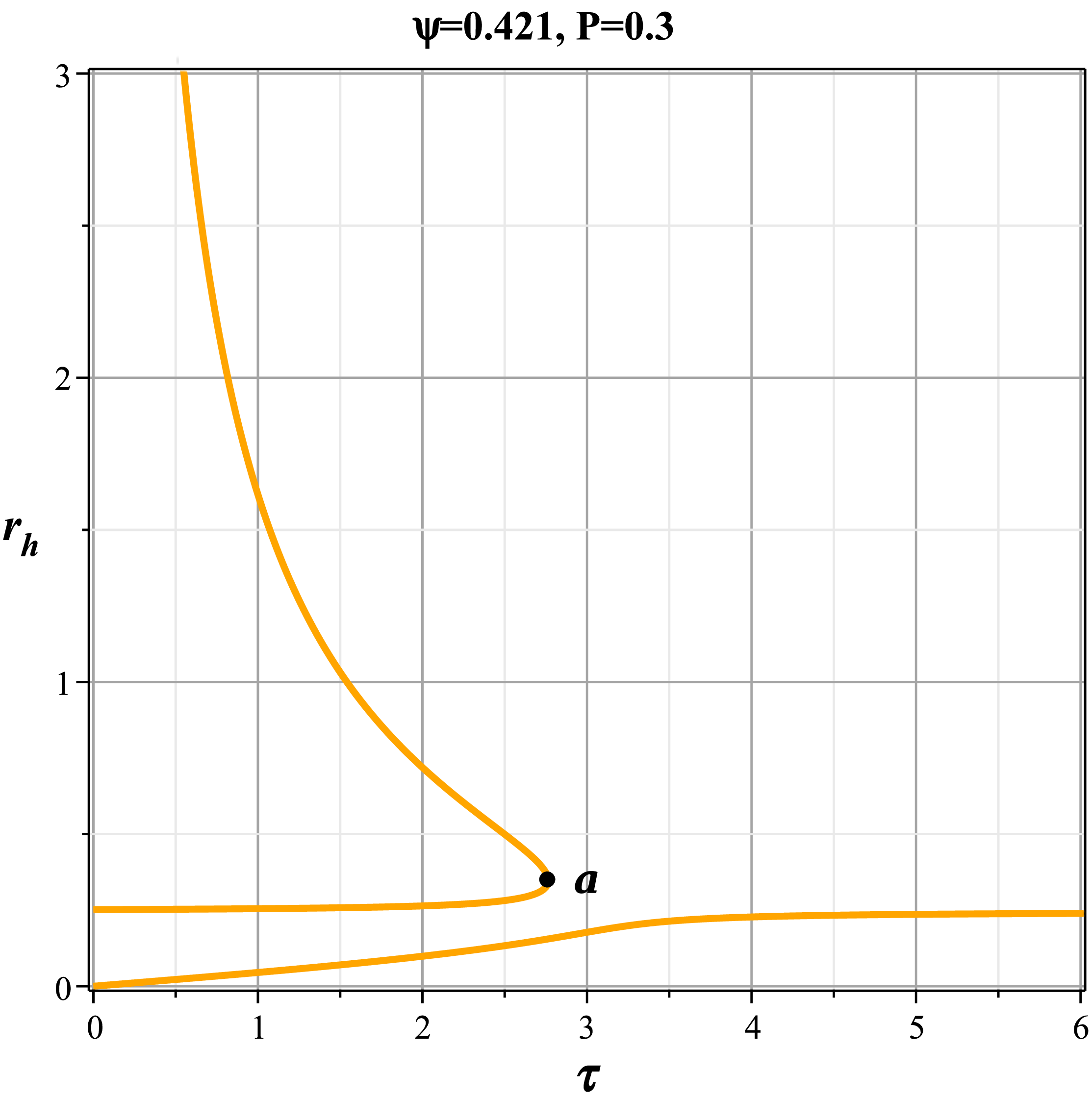}
\caption{$\psi=0.421,\;P=0.3$.}
\end{subfigure}
\quad
\begin{subfigure}[b]{0.38\textwidth}
\centering
\includegraphics[width=\linewidth]{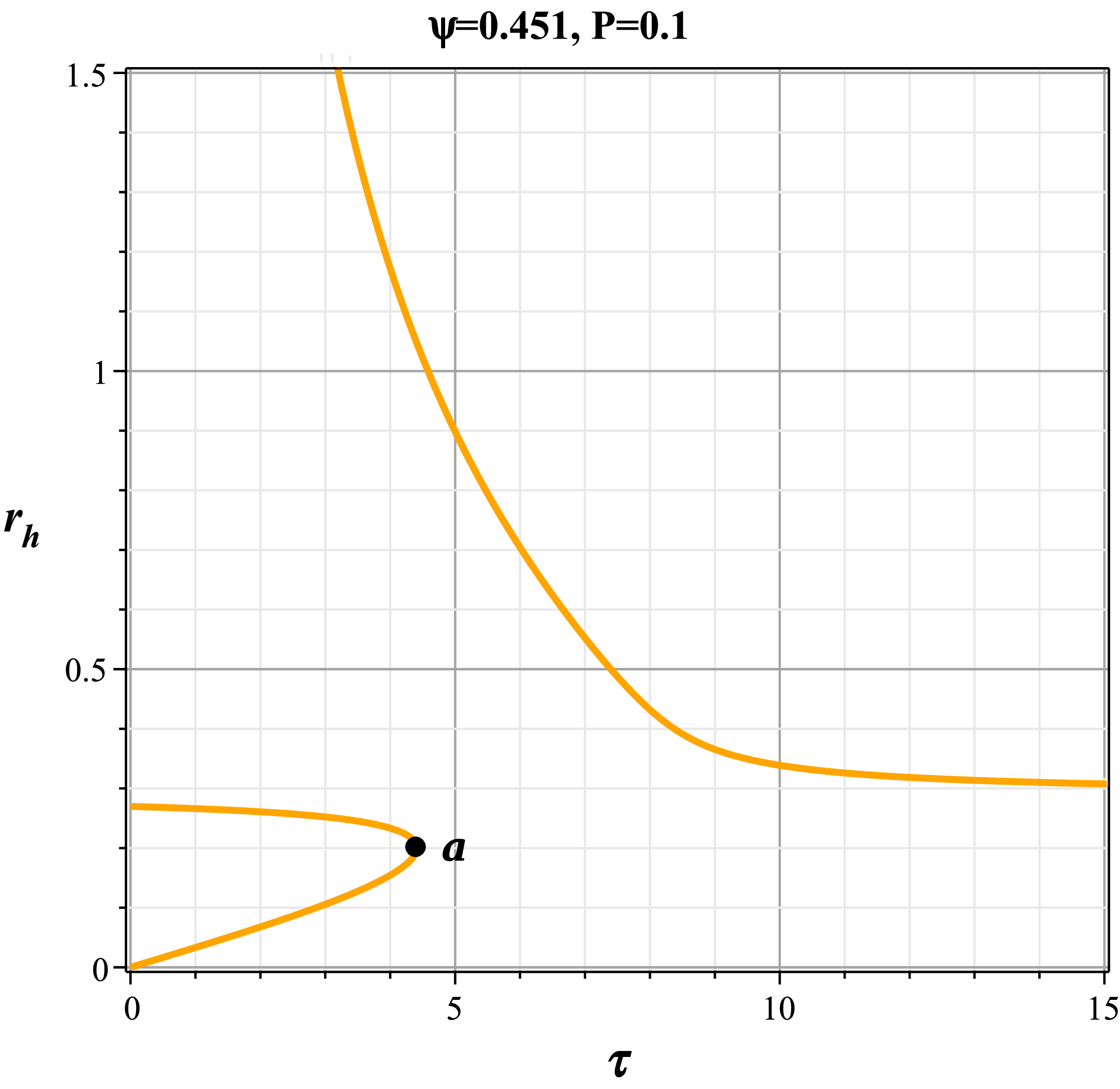}
\caption{$\psi=0.451,\;P=0.1$.}
\end{subfigure}
\caption{
Pressure-induced rearrangement of the off-shell equilibrium branches in Region IV. (a) For $\psi=0.421$ and $P=0.3$ reverses the ordering of the two disconnected families compared with Fig.~\ref{fig:regionIV}(a). (b) For $\psi=0.451$, and $P=0.1$ restores the opposite ordering, with the monotonic branch lying above the disconnected family. 
}
\label{fig:regionIVP}
\end{figure}

As shown in Fig.~\ref{fig:regionIVP}, by variying the pressure the relative position of the curves is changed. Despite these qualitative rearrangements of the equilibrium branches, the winding-number sequence remains $(+1,-1,+1)$ in all cases, and consequently the total winding number is unchanged ($W=+1$). Therefore, the rearrangement of the equilibrium manifold is a geometric property of the off-shell solutions rather than a change of the topological classification.

\begin{figure}[H]
\centering
\begin{subfigure}[b]{0.49\textwidth}
    \centering
    \includegraphics[width=\textwidth]{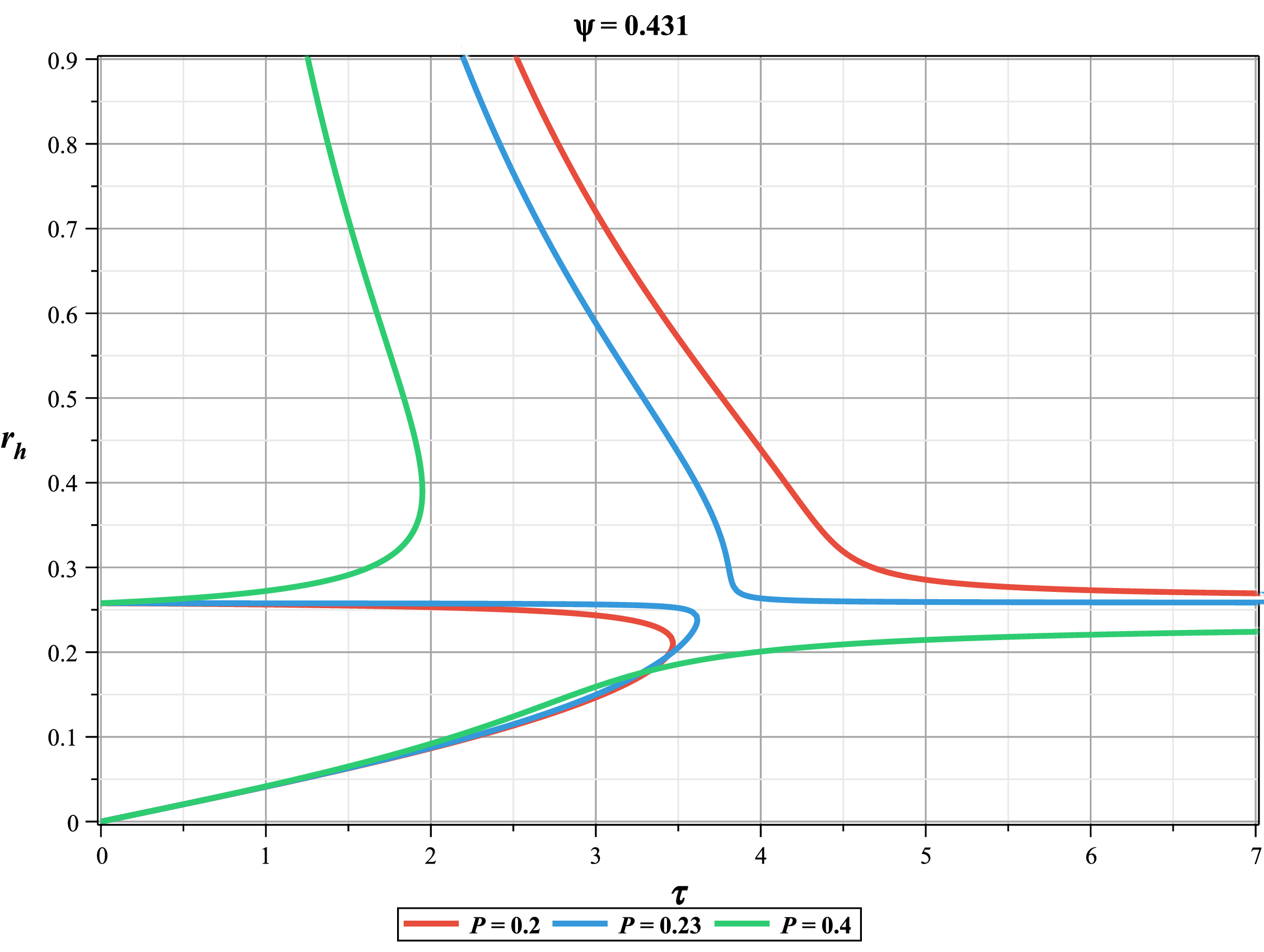}
    \caption{$\psi=0.431$.}
    \label{fig:tau52}
\end{subfigure}
\hfill
\begin{subfigure}[b]{0.49\textwidth}
    \centering
    \includegraphics[width=\textwidth]{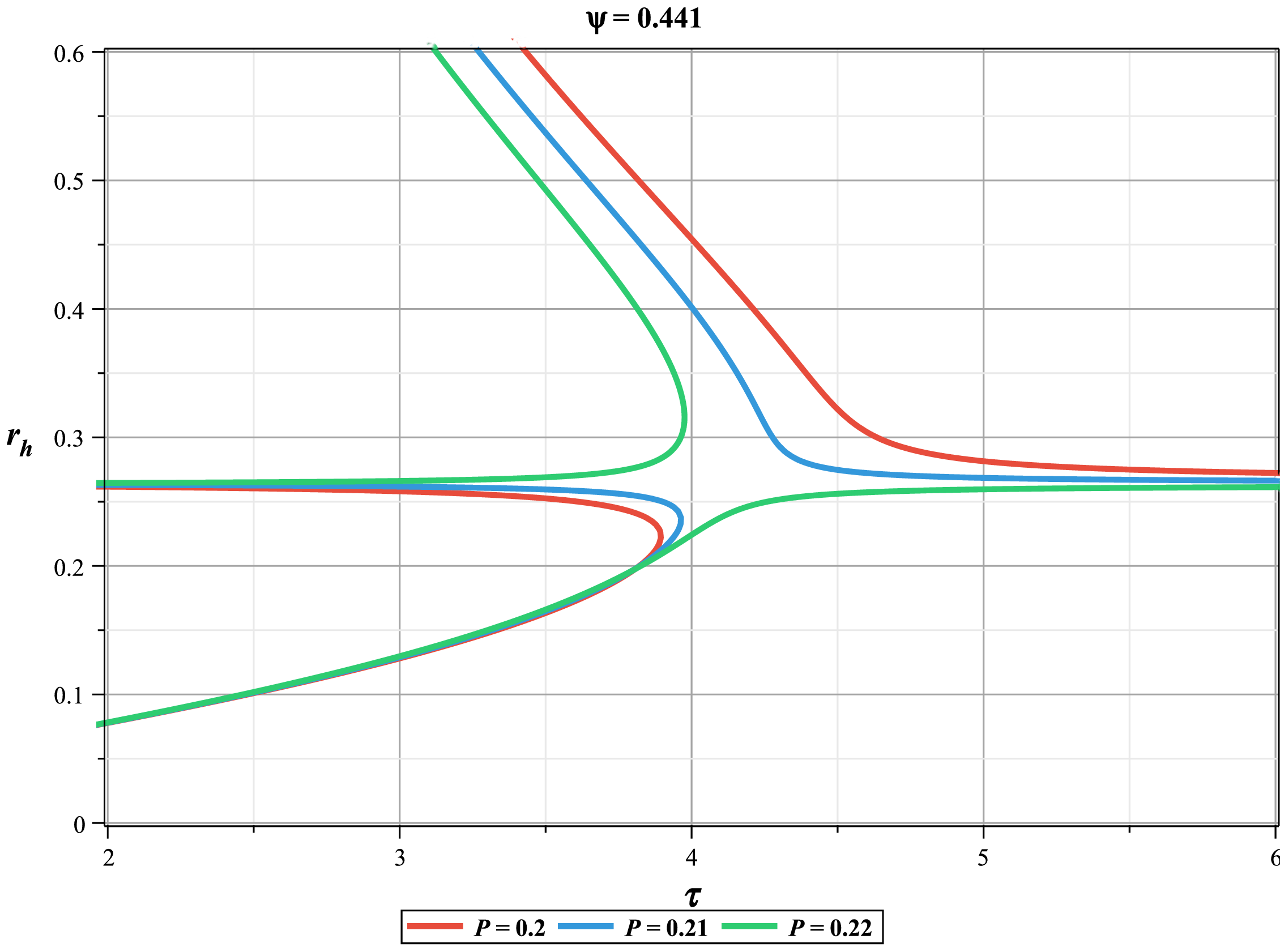}
    \caption{$\psi=0.441$.}
    \label{fig:tau53}
\end{subfigure}

\caption{The $r_h$--$\tau$ diagrams for $\psi=0.431$ and $\psi=0.441$ at several representative values of the pressure. The figures demonstrate the evolution of the horizon-radius branches and their turning-point structures as the pressure is varied.}
\label{fig:tau5253}
\end{figure}

For further confirmation, we illustrate the $r_h$--$\tau$ diagrams for the potential values $\psi=0.431$ and $\psi=0.441$ at several representative values of the pressure. Figure~\ref{fig:tau5253} demonstrates how the structure of the thermodynamic branches and the associated phase-transition behavior evolves as the pressure is varied. In particular, the diagrams reveal that changing the pressure can significantly modify the structure and the morphology of the $r_h$--$\tau$ branches, even for a fixed value of the electric potential. 

\subsection{Topological Phase Transition in Region V}

The topological analysis confirms that the system remains in the topological class $W=+1$ throughout Region~V, as indicated by the orientation of the unit vector field at the boundaries. To illustrate this behavior, we investigate two representative values, $\psi=0.461$ and $\psi=0.511$.

\begin{figure}[H]
\centering

\begin{subfigure}[b]{0.28\textwidth}
    \centering
    \includegraphics[width=\linewidth]{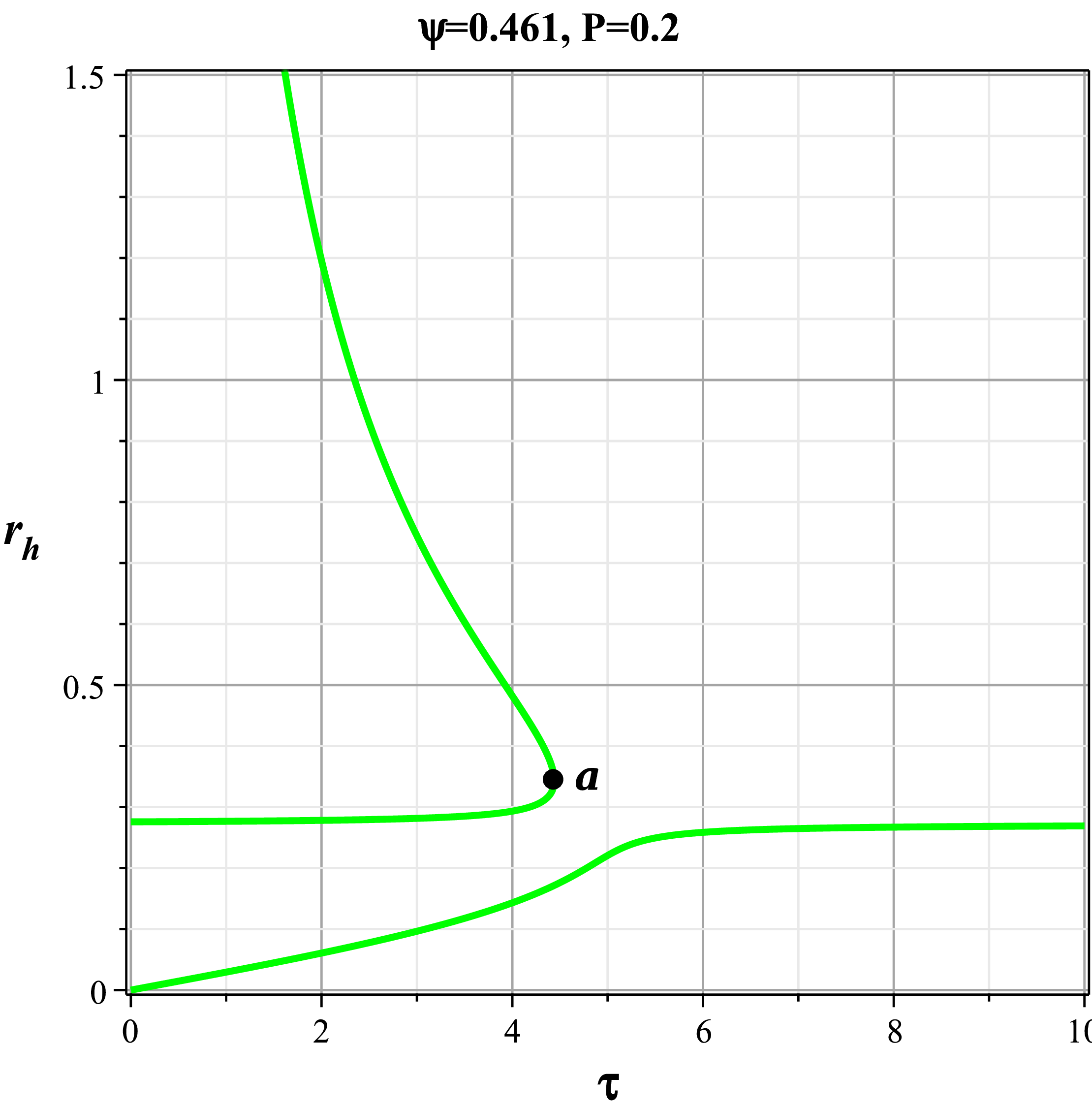}
    \caption{$\psi=0.461,\;P=0.15$}
\end{subfigure}
\quad
\begin{subfigure}[b]{0.28\textwidth}
    \centering
    \includegraphics[width=\linewidth]{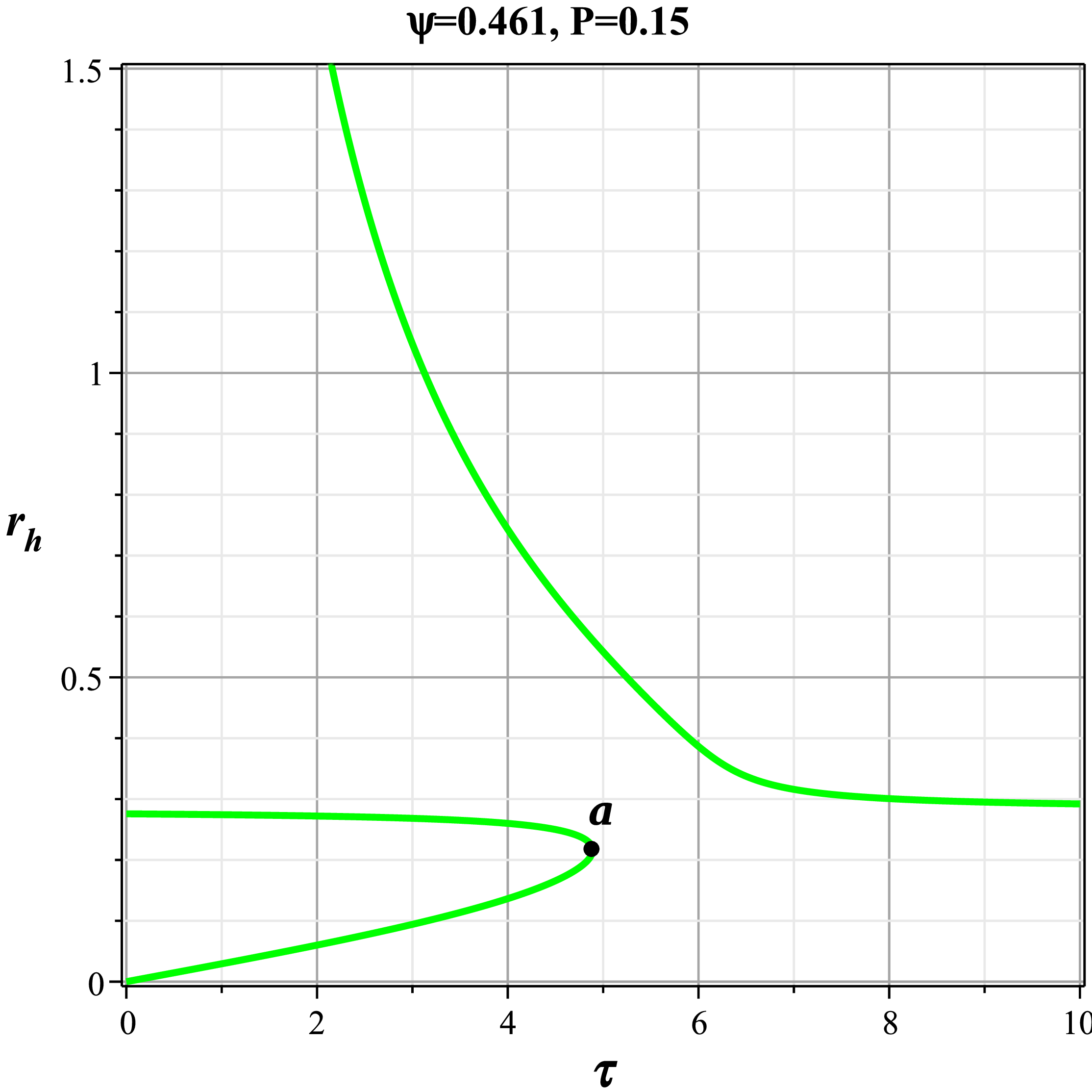}
    \caption{$\psi=0.461,\;P=0.2$}
\end{subfigure}

\begin{subfigure}[b]{0.28\textwidth}
    \centering
    \includegraphics[width=\linewidth]{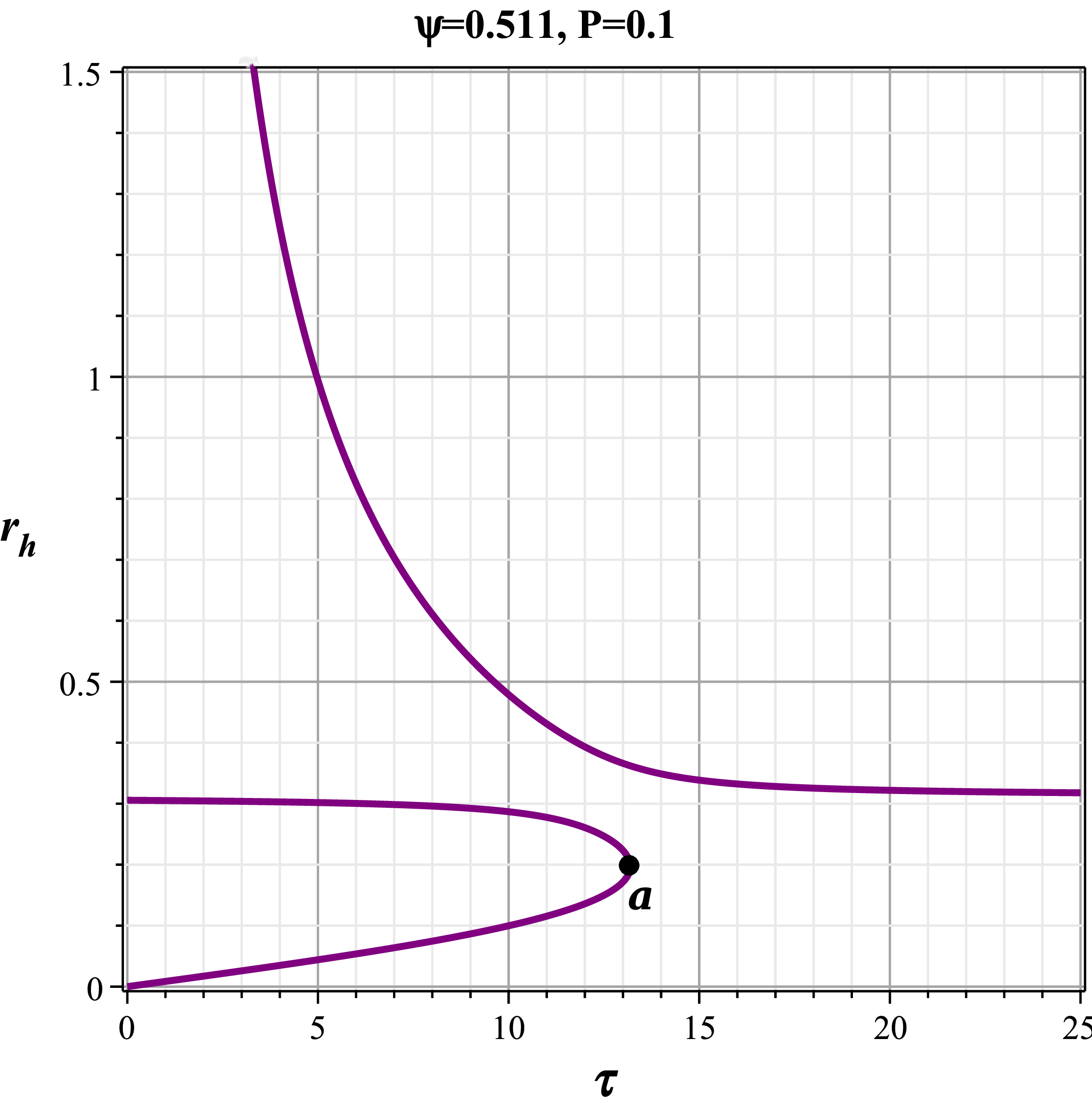}
    \caption{$\psi=0.511,\;P=0.1$}
\end{subfigure}
\quad
\begin{subfigure}[b]{0.28\textwidth}
    \centering
    \includegraphics[width=\linewidth]{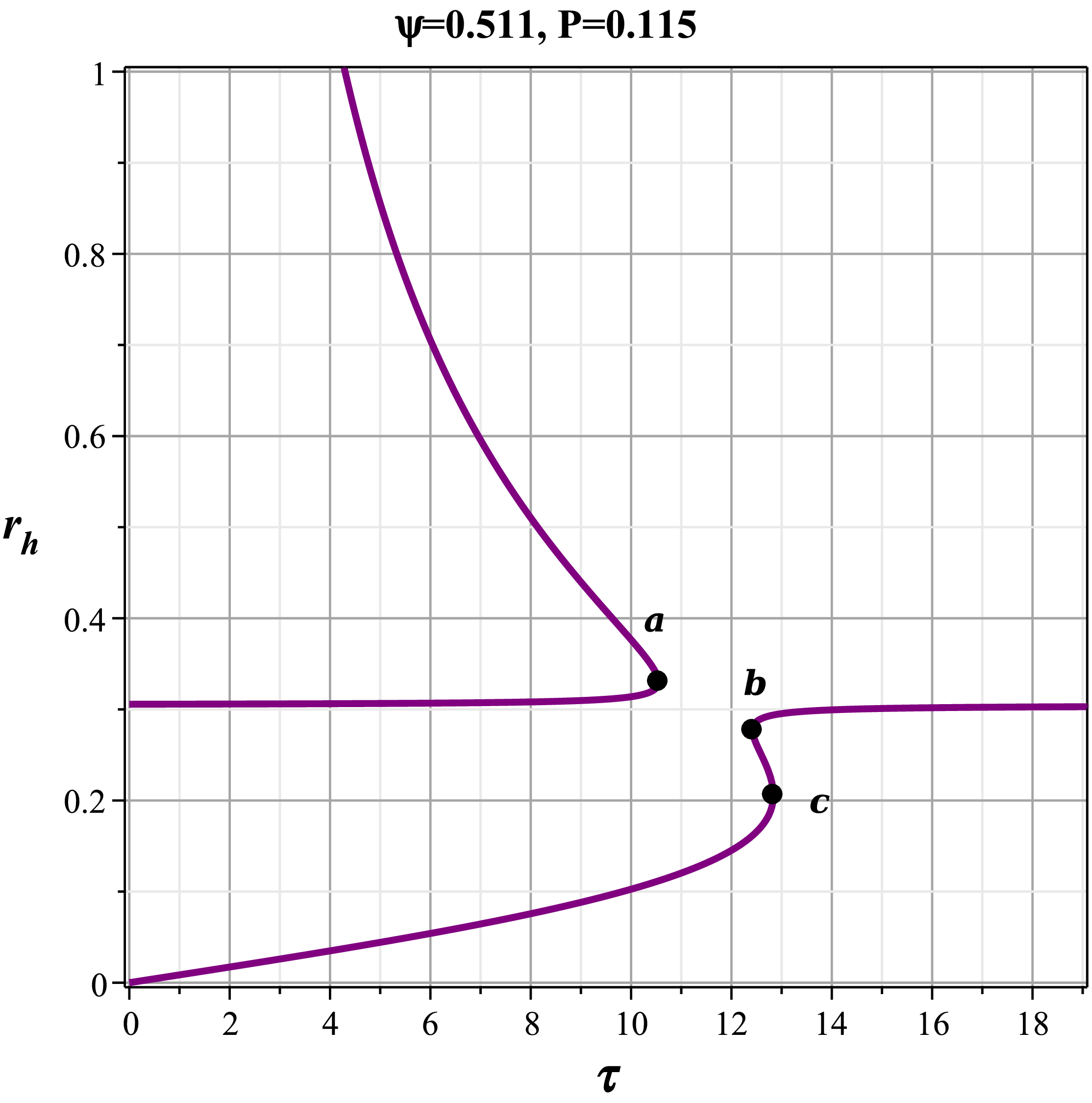}
    \caption{$\psi=0.511,\;P=0.115$}
\end{subfigure}

\begin{subfigure}[b]{0.28\textwidth}
    \centering
    \includegraphics[width=\linewidth]{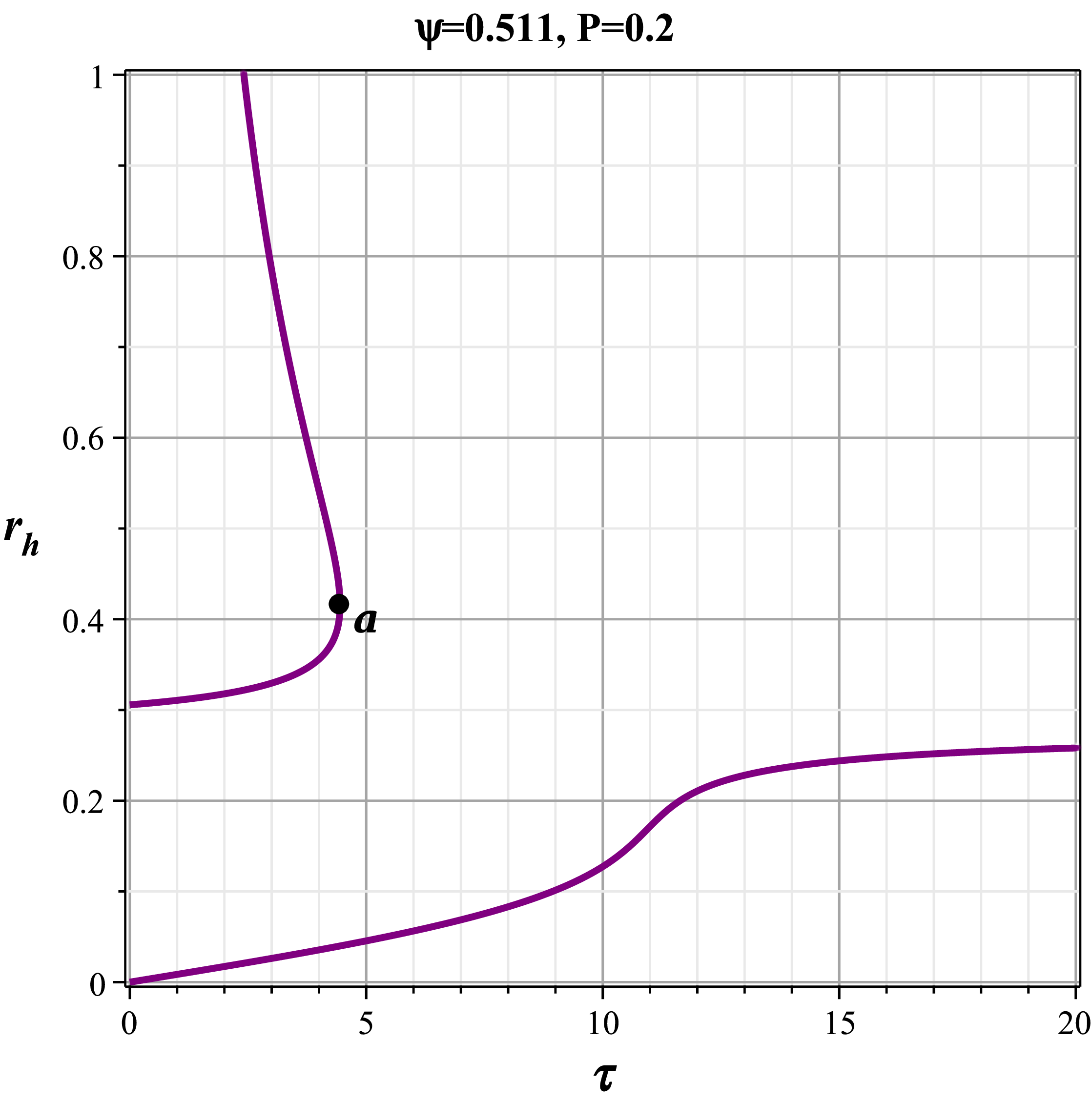}
    \caption{$\psi=0.511,\;P=0.2$}
\end{subfigure}

\caption{Representative $r_h$--$\tau$ diagrams illustrating the pressure dependence of the off-shell free-energy structure in Region~V. Panels (a) and (b) correspond to $\psi=0.461$, while panels (c)--(e) correspond to $\psi=0.511$. Increasing the pressure changes the relative ordering of the monotonic and disconnected branches and, at $P=0.115$, produces an intermediate S-shaped branch. Despite these structural rearrangements, the total winding number remains $W=+1$.}
\label{fig:tau16-tau20}
\end{figure}

In this region also the relative ordering of the disconnected branches depends sensitively on the pressure. For $\psi=0.461$, the monotonic branch lies below the disconnected two-branched family at $P=0.15$, whereas it moves above the disconnected family when the pressure is increased to $P=0.2$. A similar pressure-induced rearrangement is observed for $\psi=0.511$. At $P=0.1$, the monotonic branch is located above the disconnected family. Increasing the pressure to $P=0.115$ produces a more intricate structure: the lower branch is no longer globally monotonic but develops an intermediate S-shaped segment, while the disconnected two-branched family appears above it. Upon further increasing the pressure to $P=0.2$, the lower branch becomes monotonic again, whereas the upper branch retains its disconnected two-branched structure. These representative cases are shown in Fig.~\ref{fig:tau16-tau20}.

It is noteworthy that these pressure-induced rearrangements do not modify the global topological classification. Although the ordering of the branches changes, the winding-number sequence associated with the defects remains unchanged, and the total winding number is preserved as $W=+1$. 

\begin{figure}[H]
\centering

\begin{subfigure}[b]{0.37\textwidth}
    \centering
    \includegraphics[width=\linewidth]{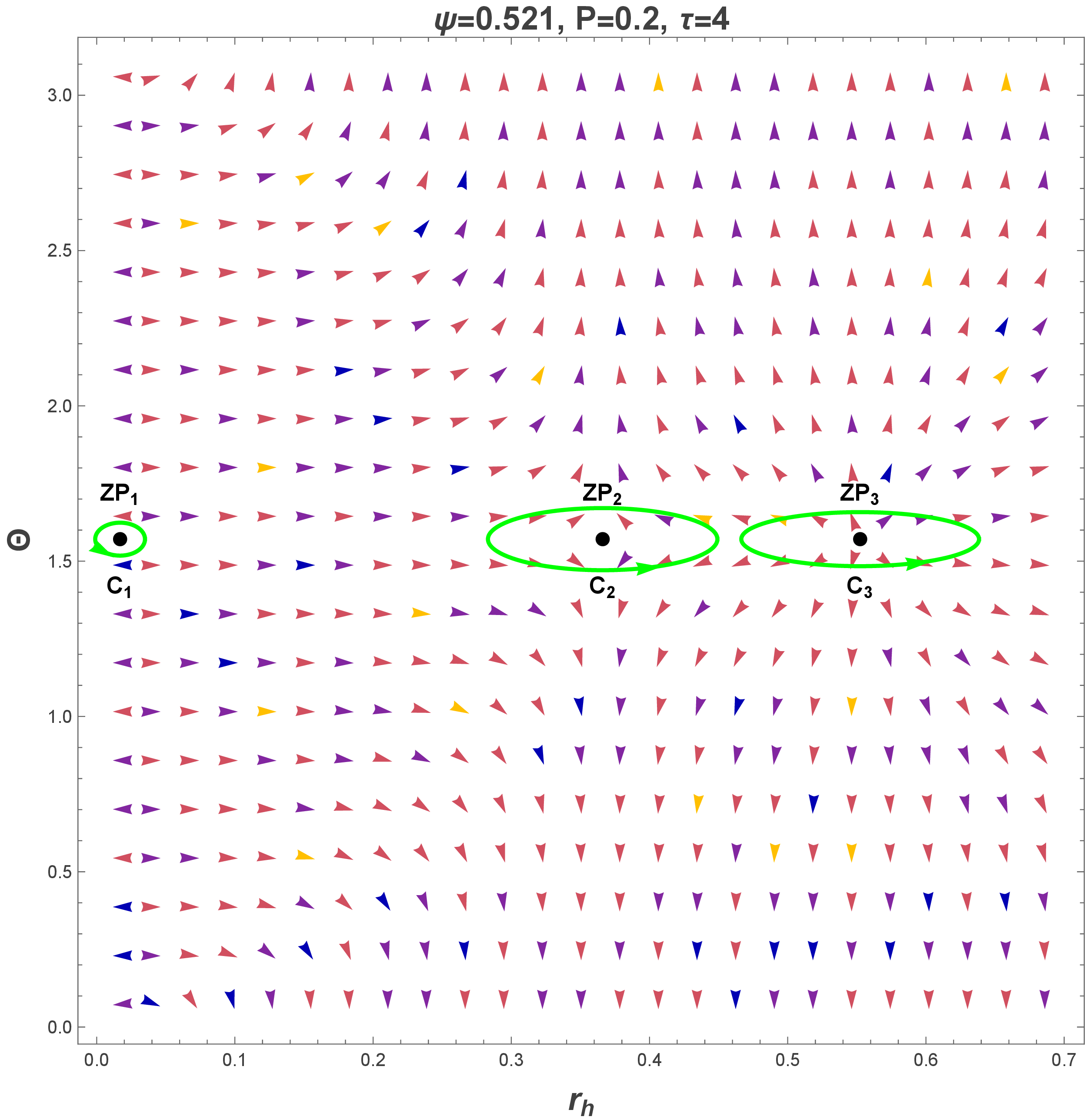}
    \caption{Unit vector field for $\psi=0.521$, $P=0.2$, and $\tau=4$.}
\end{subfigure}
\quad
\begin{subfigure}[b]{0.40\textwidth}
    \centering
    \includegraphics[width=\linewidth]{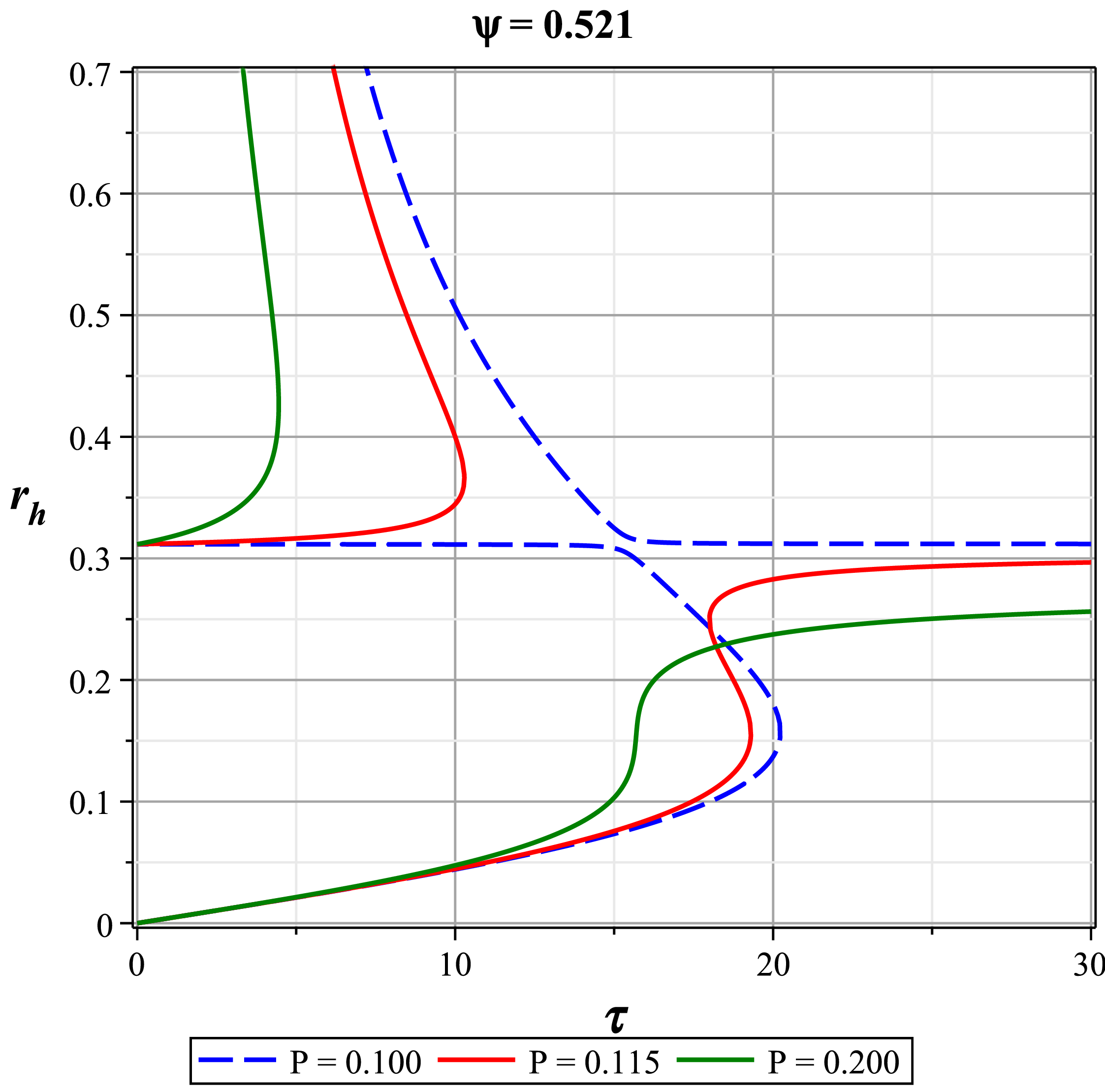}
    \caption{$r_h$--$\tau$ diagrams for $\psi=0.521$ at different pressures.}
\end{subfigure}

\caption{Topological characterization of the case $\psi=0.521$. The unit vector field confirms the boundary orientation corresponding to $W=+1$, while the $r_h$--$\tau$ diagrams show that the system remains in the same topological class for different pressures despite the absence of conventional criticality.}
\label{fig:v13tau21}
\end{figure}

An even more interesting observation emerges at $\psi=0.521$. Based solely on the conventional local thermodynamic analysis, this case was assigned to Region~VI because the criticality conditions are not satisfied, despite the oscillatory behavior of the  temperature for small horizon radii and the ribbon-shaped Gibbs free-energy curves that resemble a small/large black-hole transition. From the topological perspective, however, the picture becomes clearer. The $r_h$--$\tau$ diagrams for several pressures consistently yield a total winding number of $W=+1$, while the unit vector field exhibits the same boundary orientation as in Region~V (see Fig.~\ref{fig:v13tau21}). This demonstrates that, although the system does not possess a conventional van der Waals critical point, it still belongs to the same topological class associated with small/large black-hole phase transitions.

\emph{These results suggest that the topological approach provides information complementary to conventional local thermodynamic criteria.}  In particular, the global topological classification remains robust even in parameter regions where the usual criticality conditions fail, revealing a continuous topological connection between Regions~V and VI.

\subsection{Topological Phase Transition in Region VI}

\begin{figure}[H]
\centering

\begin{subfigure}[b]{0.38\textwidth}
    \centering
    \includegraphics[width=\linewidth]{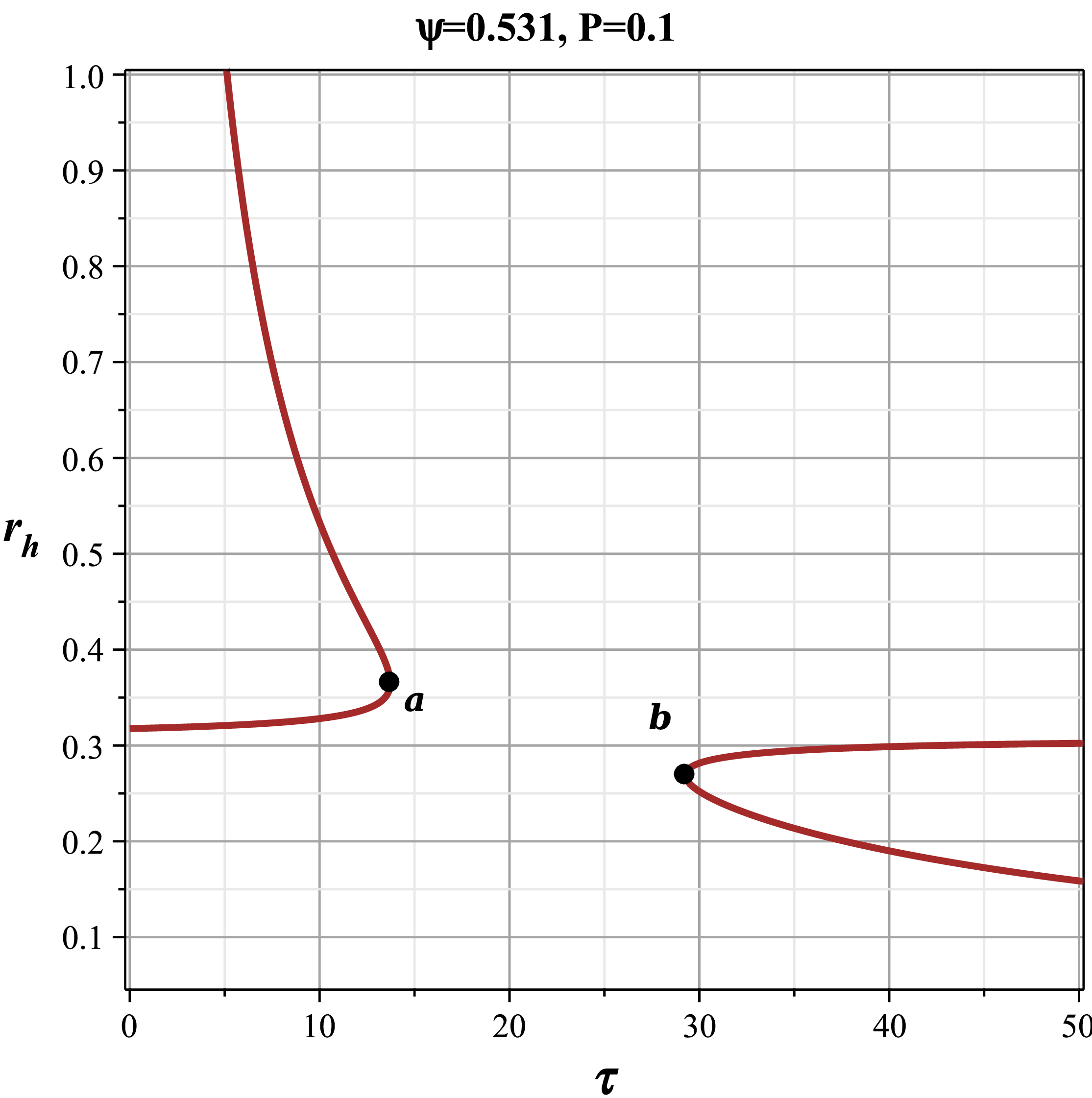}
    \caption{$\psi=0.531,\;P=0.1$.}
\end{subfigure}
\quad
\begin{subfigure}[b]{0.38\textwidth}
    \centering
    \includegraphics[width=\linewidth]{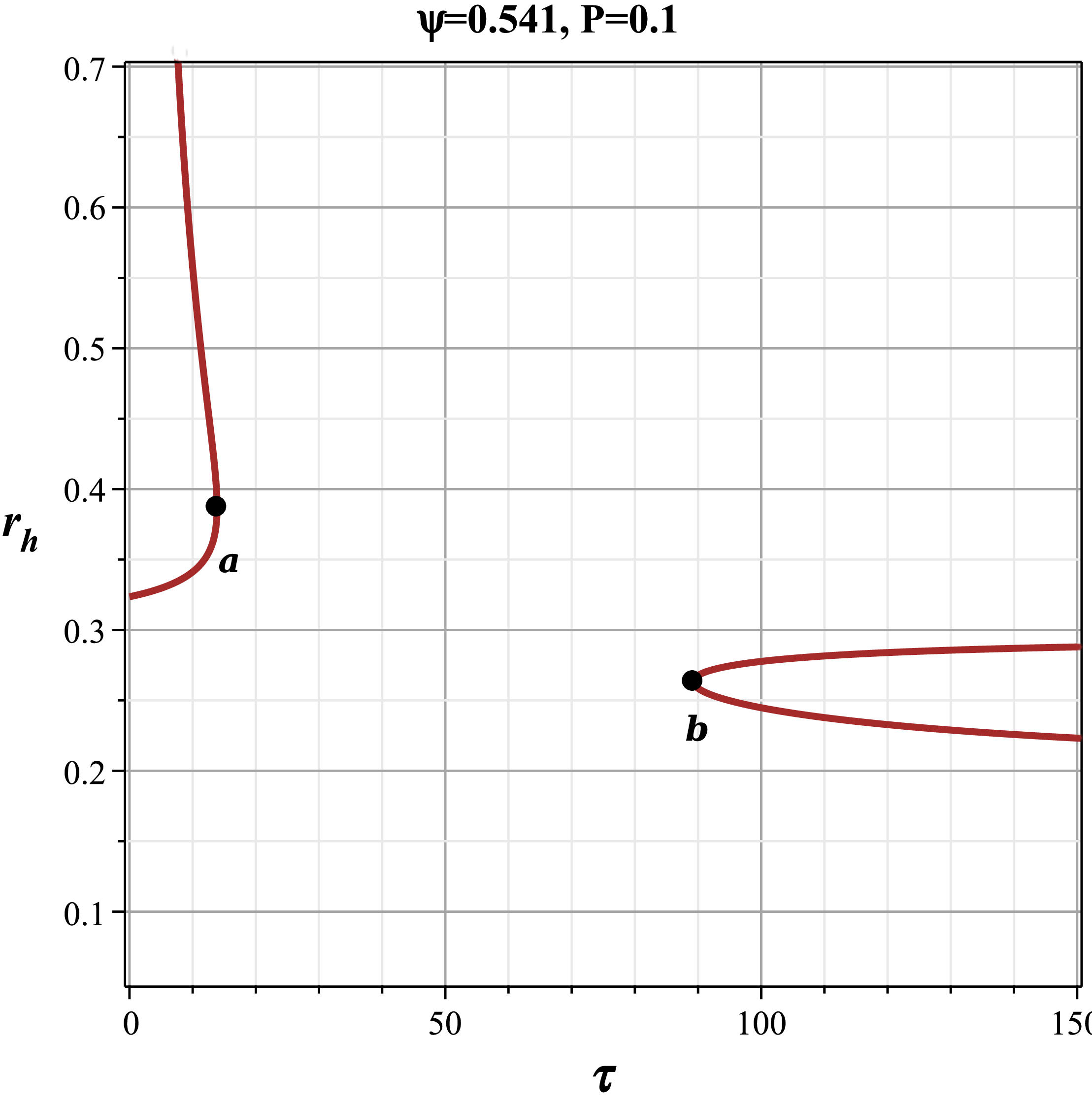}
    \caption{$\psi=0.541,\;P=0.1$.}
\end{subfigure}

\caption{Representative $r_h$--$\tau$ diagrams in Region~VI. Two disconnected V-shaped branches appear: the left branch opens toward smaller $\tau$, whereas the right branch opens toward larger $\tau$. The interval between the turning points $\tau_a$ and $\tau_b$ contains no black-hole solutions. Each disconnected branch contains a pair of defects with winding numbers $(-1,+1)$, yielding a total winding number $W=0$.}
\label{fig:tau22tau23}
\end{figure}

Beginning with $\psi=0.531$, the system enters the topological class $W=0$. See Fig.~\ref{fig:tau22tau23}. The $r_h$--$\tau$ diagrams exhibit a qualitatively different structure from those found in the previous regions. Instead of a monotonic or oscillatory branches  together with a disconnected two-branched family, the solutions split into two disconnected V-shaped  or cusp-liked curves. The left curve opens toward smaller values of $\tau$, whereas the right curve opens toward larger values of $\tau$, leaving an empty interval bounded by two turning points, $\tau_a$ and $\tau_b$, in which no black-hole solution exists.

For $\tau<\tau_a$, the left V-shaped branch contains two defects with winding numbers $(-1,+1)$, yielding a total winding number $W=(-1)+(+1)=0$.
Similarly, for $\tau>\tau_b$, the right V-shaped branch also contains two defects with winding numbers $(-1,+1)$, again giving $W=(-1)+(+1)=0$. Between the two turning points,$
\tau_a<\tau<\tau_b$,
no black-hole states exist because the $r_h$--$\tau$ diagram contains no solutions in this interval. See Fig.~\ref{fig:tau22tau23}

As $\psi$ increases further for a fixed pressure, the forbidden interval expands considerably, with the separation between the two disconnected V-shaped branches becoming much more pronounced. This behavior marks a clear topological distinction between Region~VI and the preceding $W=+1$ regions. Unlike Regions~I--V, where at least one continuous branch connected the small and large-horizon solutions, Region~VI consists of two completely disconnected topological sectors separated by a range of $\tau$ for which no black-hole configurations exist.

\begin{figure}[t]
\centering

\begin{subfigure}[b]{0.38\textwidth}
    \centering
    \includegraphics[width=\linewidth]{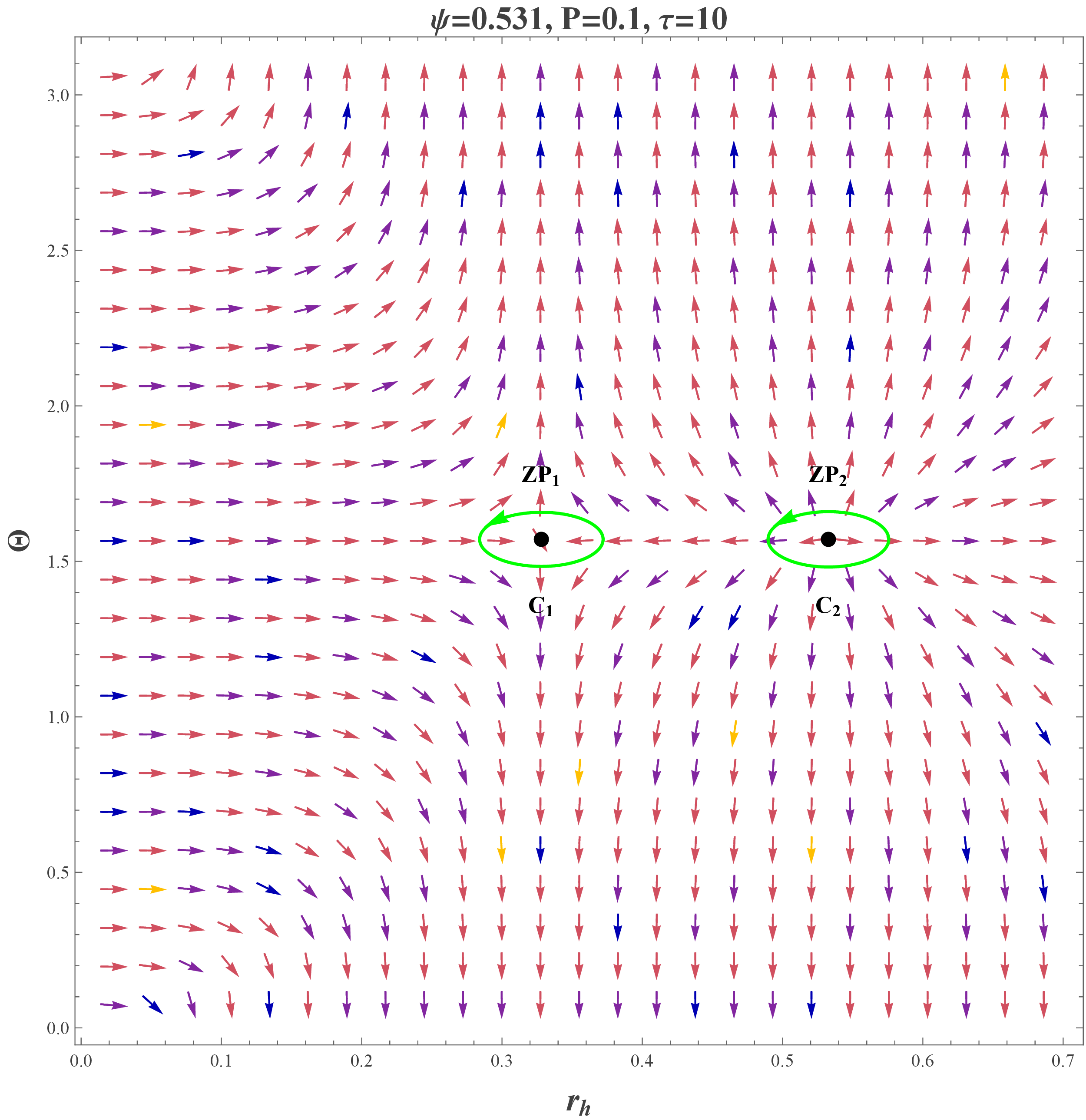}
    \caption{$\psi=0.531$, $P=0.1$, $\tau=10$, $ZP_{1}=0.328, ZP_{2}=0.532$ .}
\end{subfigure}
\quad
\begin{subfigure}[b]{0.38\textwidth}
    \centering
    \includegraphics[width=\linewidth]{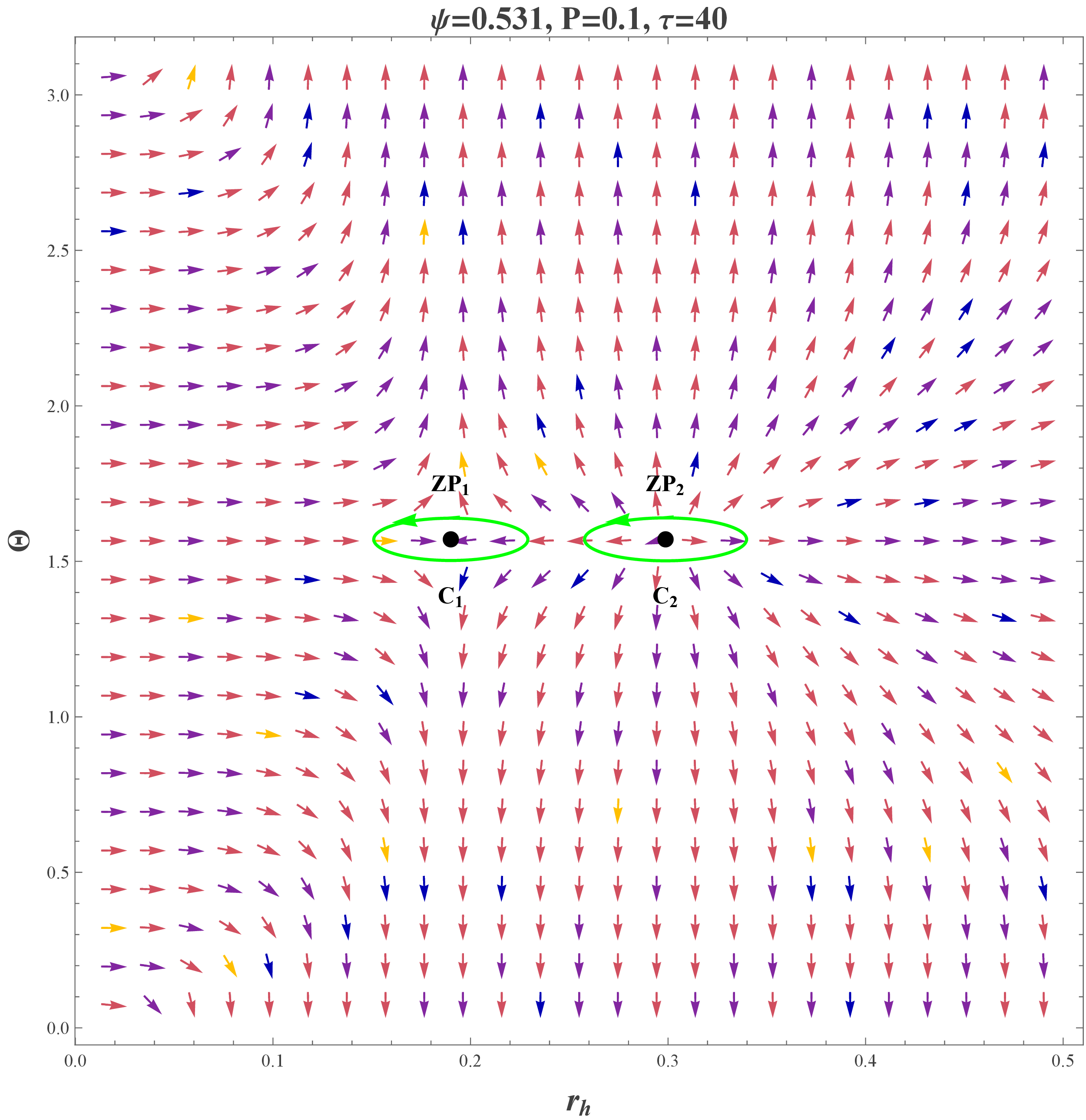}
    \caption{$\psi=0.531$, $P=0.1$, $\tau=40$, $ZP_{1}=0.190, ZP_{2}=0.298$ .}
\end{subfigure}

\caption{Unit vector field for $\psi=0.531$ and $P=0.1$ at two representative values of the off-shell parameter $\tau$. In both cases the vector field contains two topological defects with opposite winding numbers, $w=-1$ and $w=+1$, yielding a total winding number $W=0$. Although the defect positions shift as $\tau$ varies, the boundary orientation and total topological charge remain unchanged, confirming that the system stays within the same topological phase.}
\label{fig:v14v15}
\end{figure}

The unit vector field provides an independent confirmation of the topological classification. Figure~\ref{fig:v14v15} displays the unit vector field for $\psi=0.531$ and $P=0.1$ at two representative values, $\tau=10$ and $\tau=40$. Although the locations of the two defects change with $\tau$, the total winding number is $W=(-1)+(+1)=0$,
which agrees with the topology inferred from the corresponding $r_h$--$\tau$ diagrams. The boundary orientation of the unit vector field is unchanged between the two cases, demonstrating that the entire parameter region belongs to the same topological sector with $W=0$. 

\begin{figure}[H]
\centering

\begin{subfigure}[b]{0.38\textwidth}
    \centering
    \includegraphics[width=\linewidth]{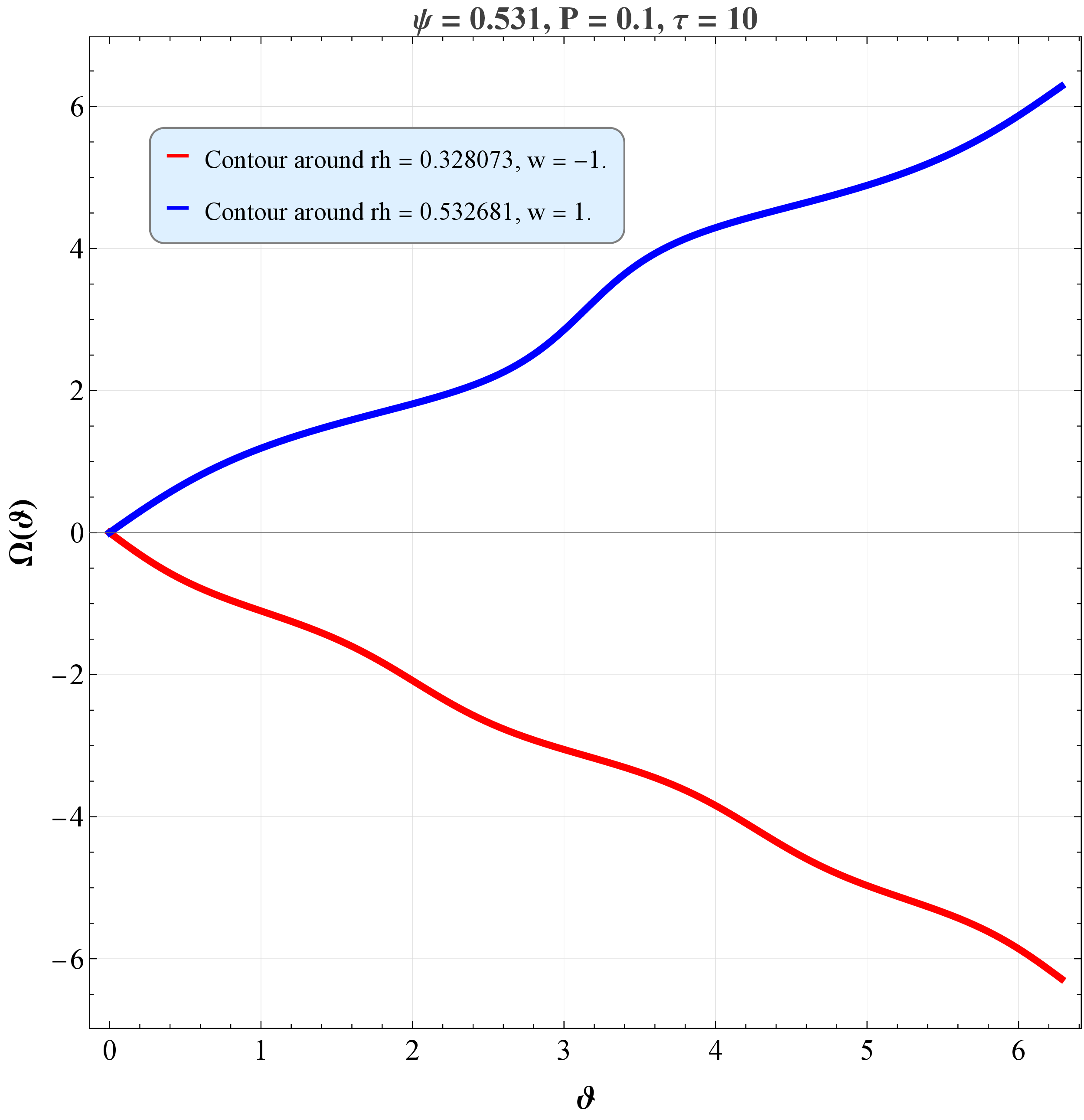}
    \caption{$\psi=0.531$, $P=0.1$, $\tau=10$.}
\end{subfigure}
\quad
\begin{subfigure}[b]{0.38\textwidth}
    \centering
    \includegraphics[width=\linewidth]{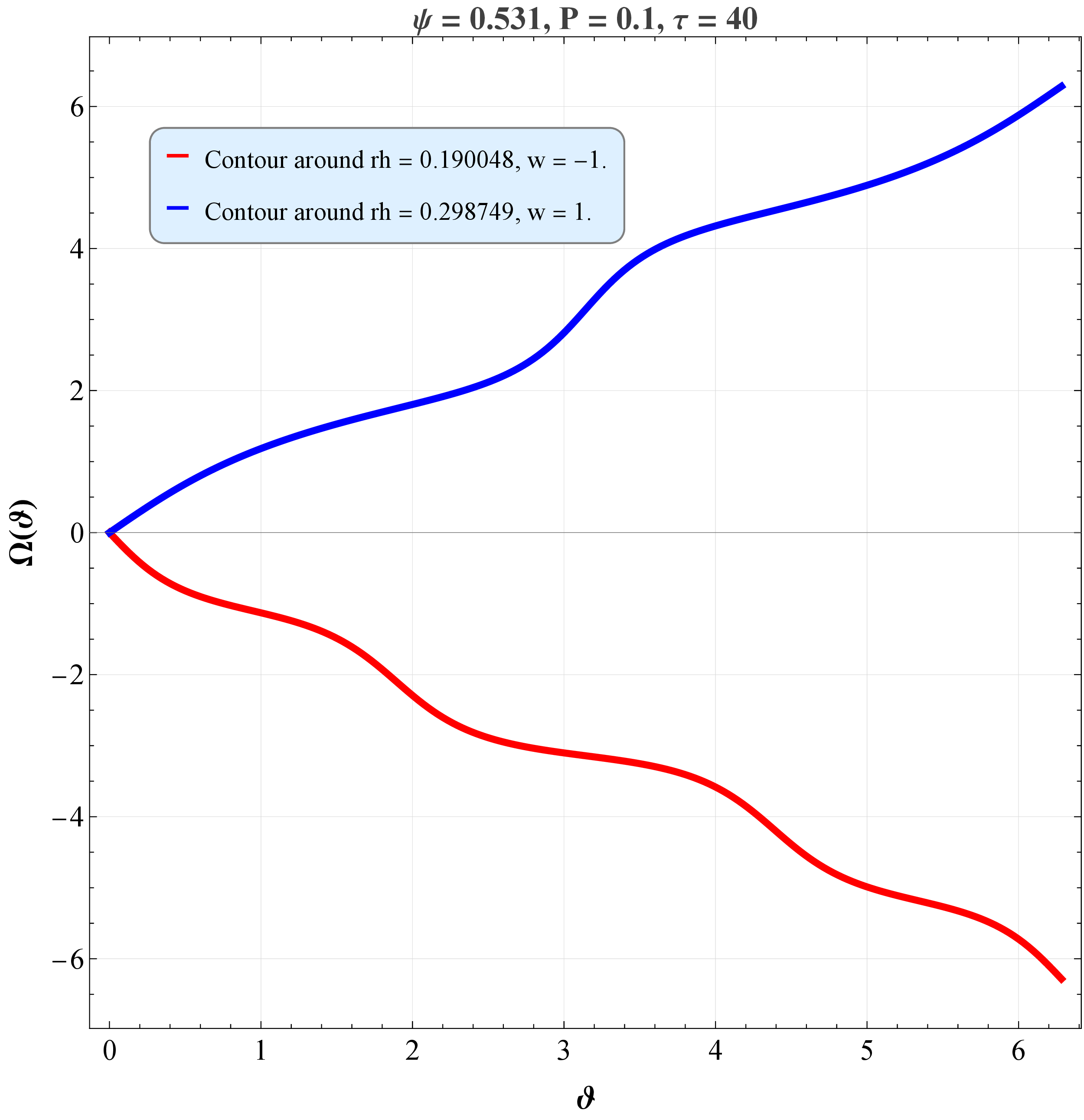}
    \caption{$\psi=0.531$, $P=0.1$, $\tau=40$.}
\end{subfigure}

\caption{Deflection angle $\Omega(\vartheta)$ associated with the two topological defects for $\psi=0.531$ and $P=0.1$. Panel (a) corresponds to $\tau=10$, while panel (b) corresponds to $\tau=40$. In both cases, the two defects possess winding numbers $w=-1$ and $w=+1$, yielding a conserved total winding number $W=0$.}
\label{fig:d11d12}
\end{figure}

The topological classification is further corroborated by the deflection angle analysis. Figure~\ref{fig:d11d12} displays the deflection angle $\Omega(\vartheta)$ for the two defects at $\psi=0.531$, $P=0.1$, and $\tau=10$. The contour surrounding the left defect yields a winding number $w=-1$, whereas the contour enclosing the right defect gives $w=+1$. Consequently, the total winding number is $W=(-1)+(+1)=0$, in complete agreement with the unit vector field and the $r_h$--$\tau$ diagram. The opposite signs of the two defects demonstrate that the system contains a topologically neutral pair whose total charge remains conserved. Thus, despite the unusual disconnected structure of the $r_h$--$\tau$ diagram and the existence of a forbidden interval in $\tau$, the system remains in the $W=0$ topological sector.

For $\psi \geq 0.551$, the structure of the solution branches changes
qualitatively, and the system is characterized by a single V-shaped family of curves. Beyond the turning point, no black-hole
state is present. This behavior is also consistent with the
corresponding $G-T$ diagrams, where the relevant branch lies in the region of negative Gibbs free energy. Here, we focus on characterizing the behavior of the system rather than determining the precise value of $\psi$ at which this change of structure occurs. As $\psi$ is increased, the system progressively moves away from the more intricate multi-branch structures associated with the nonminimal coupling, leading to the simpler V-shaped configuration observed in this region.

\begin{figure}[H]
\centering

\begin{subfigure}[b]{0.38\textwidth}
    \centering
    \includegraphics[width=\textwidth]{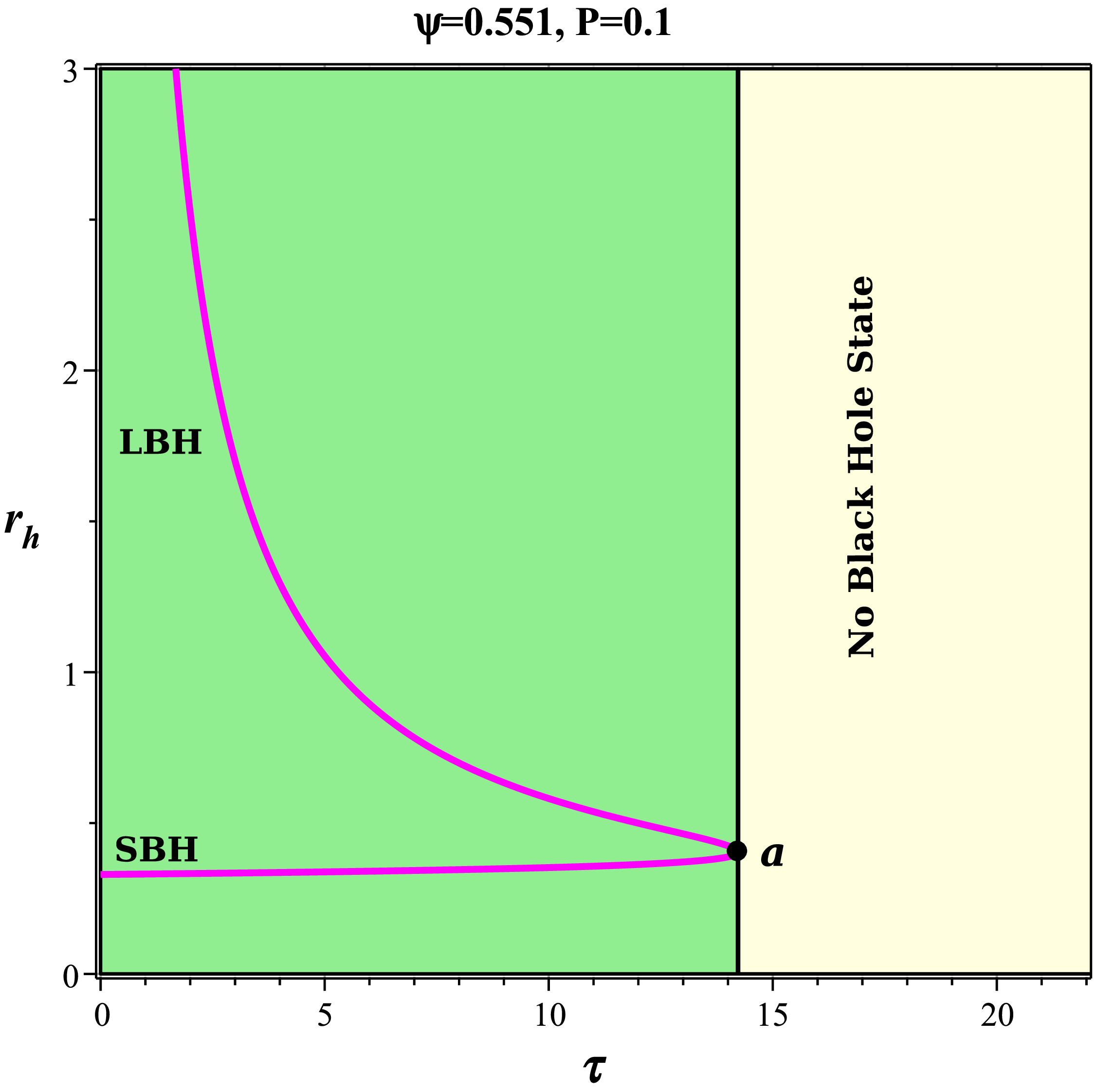}
    \caption{$r_h-\tau$ diagram for $\psi=0.551$ and $P=0.1$. }
    \label{fig:tau24}
\end{subfigure}
\quad
\begin{subfigure}[b]{0.38\textwidth}
    \centering
    \includegraphics[width=\textwidth]{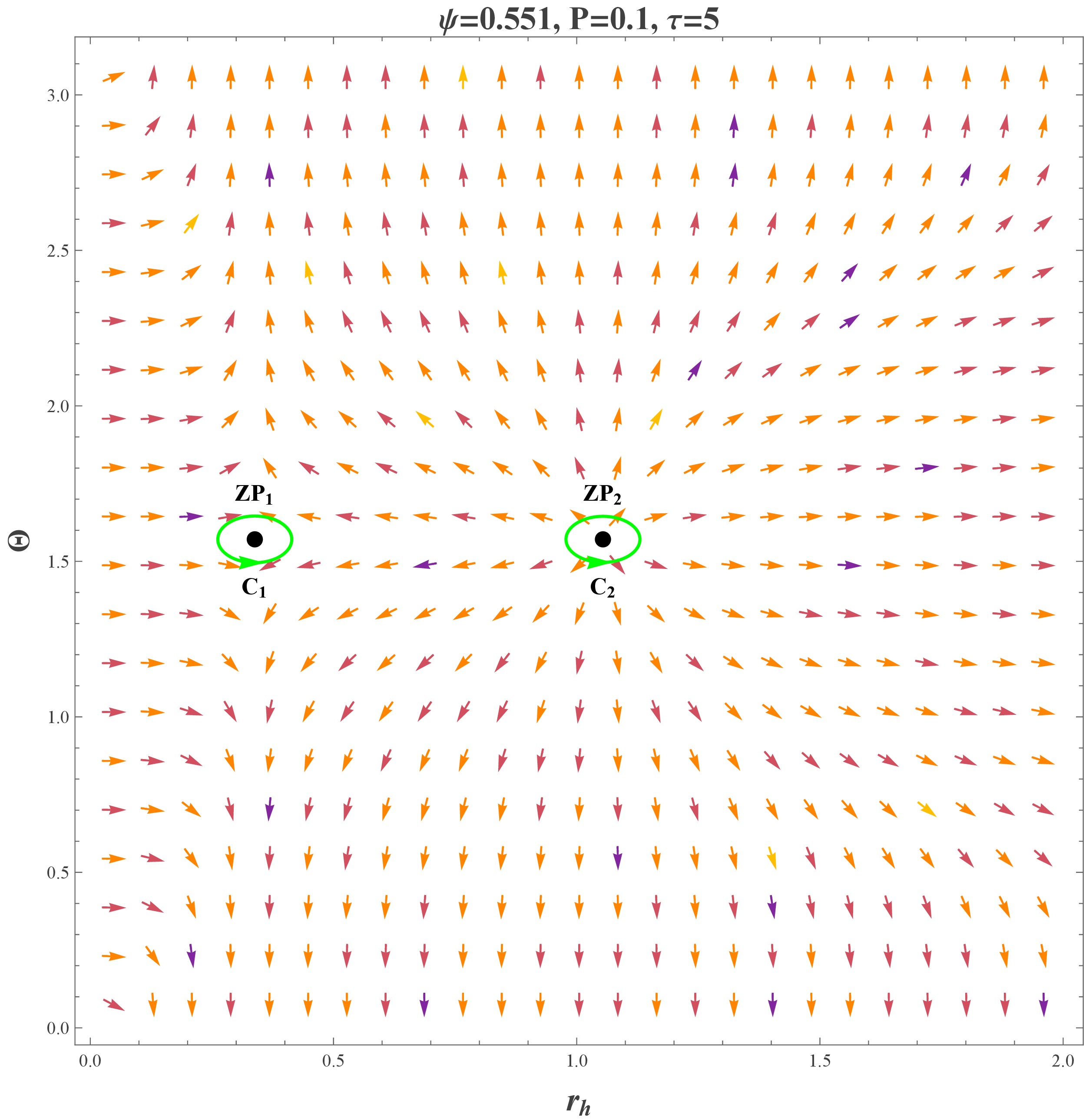}
    \caption{Unit vector field for $\psi=0.551, P=0.1$ and $\tau=5$.}
    \label{fig:v16}
\end{subfigure}

\caption{
Topological structure for $\psi=0.551$ and $P=0.1$.
Panel (a) shows the $r_h-\tau$ diagram, exhibiting a V-shaped family of black-hole solutions that terminates at a turning point. Panel (b)
shows the corresponding unit vector field for $\tau=5$, with the associated topological defects. The boundary orientation gives $W=0$.
}
\label{fig:tau24v16}
\end{figure}

The left panel of Fig.~\ref{fig:tau24v16} shows the $r_h-\tau$ behavior for $\psi=0.551$ and $P=0.1$. The two branches merge at a turning point, beyond which no black-hole solution is present. The
right panel shows the corresponding unit vector field for
$\psi=0.551$, $P=0.1$, and $\tau=5$, together with the associated
topological defects. The orientation of the unit vector field at the
boundaries confirms the topological classification
$W=0$. The same diagrams are obtained for other pressure values, but we only focused on this.

\begin{figure}[H]
\centering

\begin{subfigure}[b]{0.38\textwidth}
    \centering
    \includegraphics[width=\textwidth]{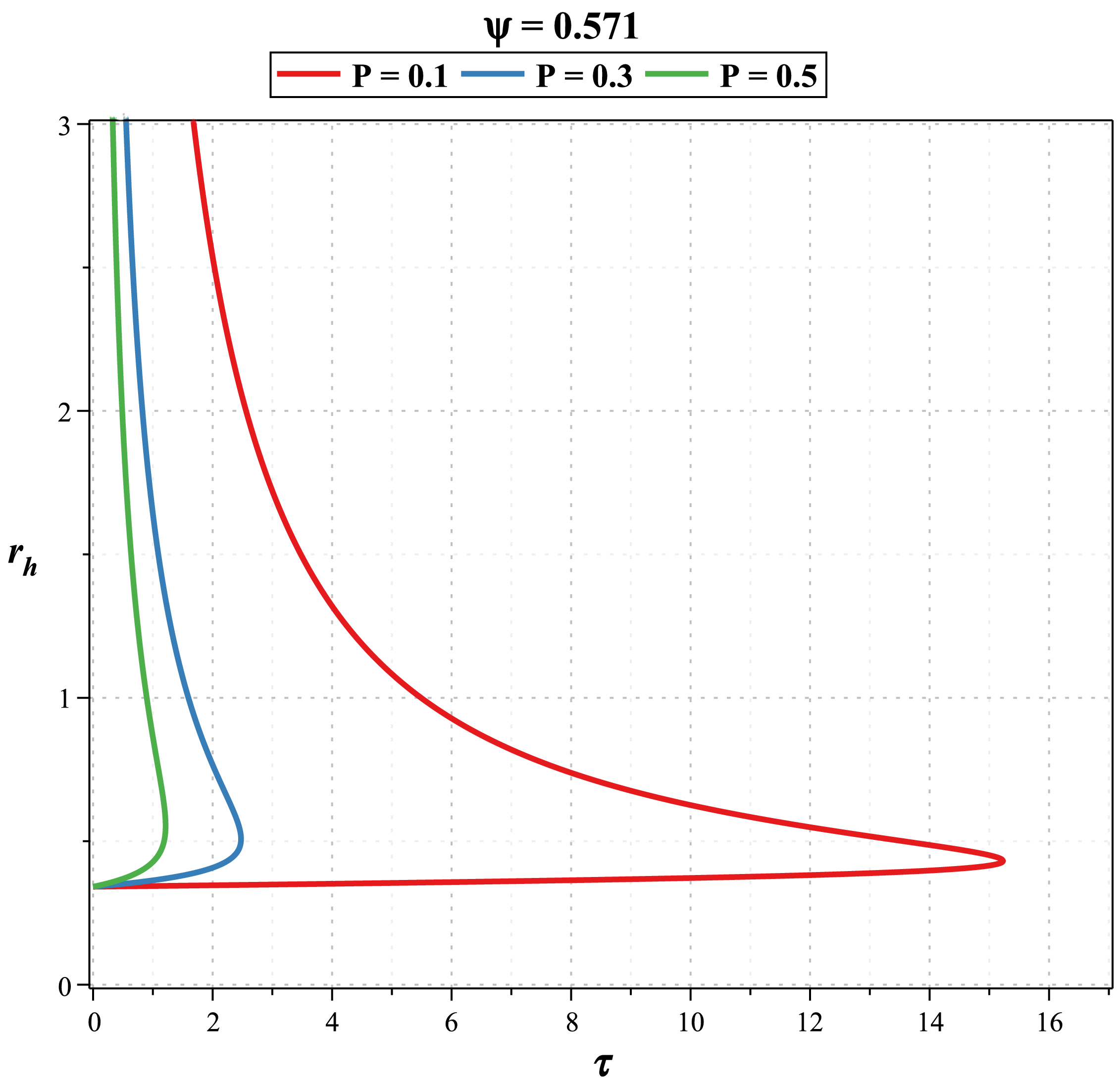}
    \caption{$r_h-\tau$ diagrams for $\psi=0.571$ and
    $P=0.1,\,0.3,\,0.5$.}
    \label{fig:tau25}
\end{subfigure}
\quad
\begin{subfigure}[b]{0.38\textwidth}
    \centering
    \includegraphics[width=\textwidth]{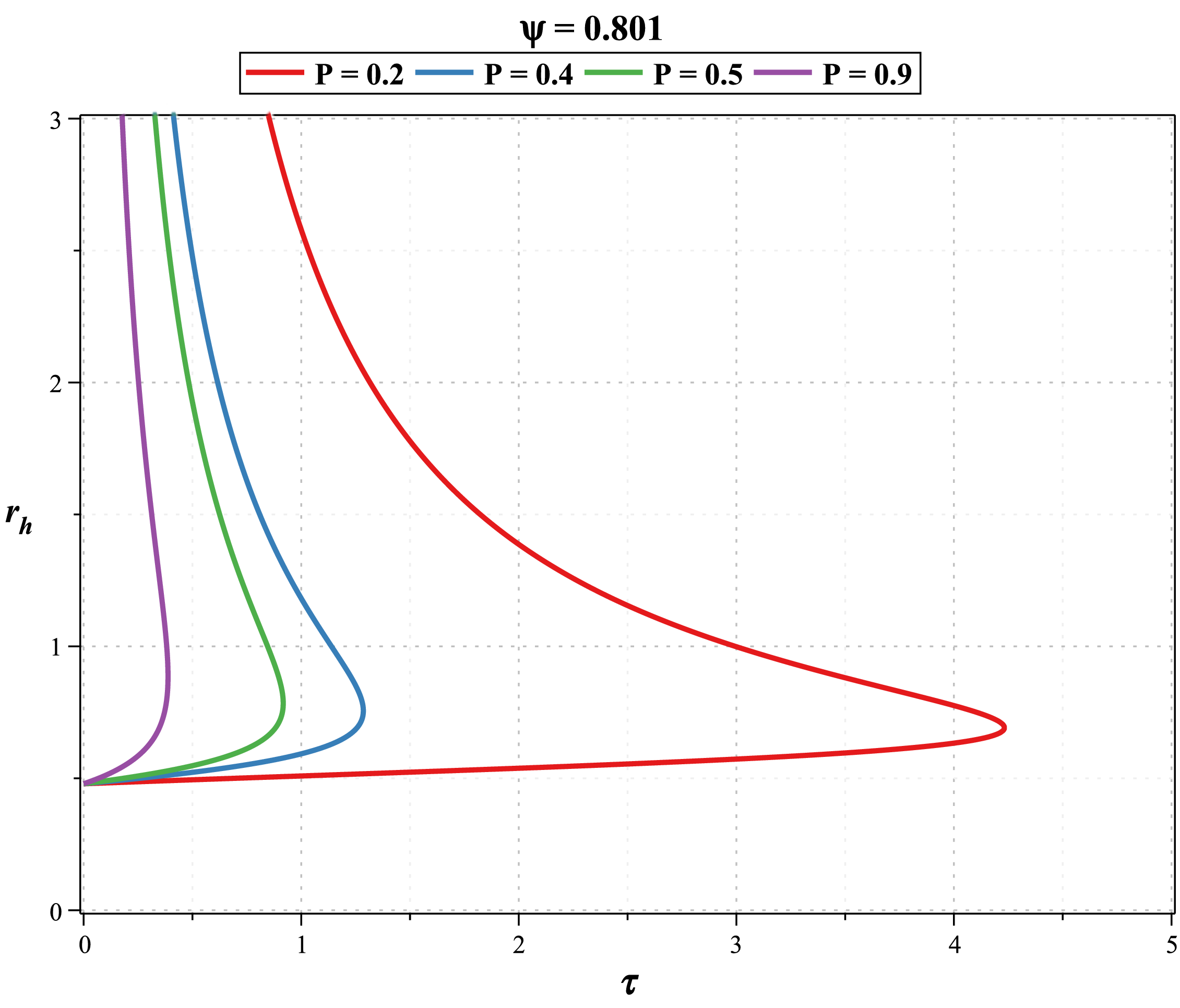}
    \caption{$r_h-\tau$ diagrams for $\psi=0.801$ and
    $P=0.2,\,0.4,\,0.5,\,0.9$.}
    \label{fig:tau26}
\end{subfigure}

\caption{
The V-shaped behavior of the $r_h-\tau$ diagrams for different values of $\psi$ and $P$. Panel (a) corresponds to $\psi=0.571$ with $P=0.1,\,0.3,\,0.5$, while panel (b) corresponds to $\psi=0.801$ with $P=0.2,\,0.4,\,0.5,\,0.9$. The V-shaped structure
persists for both parameter sets.}
\label{fig:tau25tau26}
\end{figure}

For further confirmation, we illustrate the $r_h-\tau$ diagrams, which provide a complementary view of the thermodynamic and topological behavior, for the parameter sets
$(\psi=0.571,\; P=0.1,\,0.3,\,0.5)$ and
$(\psi=0.801,\; P=0.2,\,0.4,\,0.5,\,0.9)$ in panels (a) and (b), respectively, of Fig.~\ref{fig:tau25tau26}. In both cases, the solution branches exhibit the characteristic V-shaped behavior identified in this region.
As the potential value increases, the $r_h$--$\tau$ diagrams exhibit a simpler structure characterized by a V-shaped family of curves.

%%%%%%%%%%%%%%%%%%%%%%%%%%%%%%%%%%%%%%%%%%%%%%%%%%%%%%%%%%

\section{Conclusions}\label{sec7}

In this work, we have investigated the thermodynamic and topological phase structure of charged AdS black holes with the nonminimal gauge--curvature coupling
$
F^{\alpha\beta}F^{\gamma\lambda}R_{\alpha\gamma}R_{\beta\lambda}
$
in the grand canonical ensemble. Working perturbatively to first order in the nonminimal coupling, we first derived the thermodynamic quantities and established the extended first law, including the contribution conjugate to the nonminimal coupling parameter in the canonical ensemble. We then examined the critical behavior, Gibbs free energy, and off-shell thermodynamic structure as functions of the electric potential in the grand canonical ensemble. It is important to emphasize that, in the minimal-coupling limit $\epsilon=0$, the grand canonical system reduces to the RN--AdS case, which exhibits Hawking--Page-like behavior but no van der Waals-like criticality. Therefore, the critical structures found for $\epsilon\neq0$ are generated by the nonminimal gauge--curvature interaction rather than representing perturbations of a pre-existing grand canonical critical point.

The conventional thermodynamic analysis reveals a remarkably diverse phase structure. According to the evolution of the critical quantities, the parameter space can be divided into six regions. The first region exhibits the van der Waals-like or small/large black-hole phase transition, while the subsequent regions display qualitatively different scaling behaviors of the critical horizon radius, pressure, and temperature. In particular, we find regimes in which the critical temperature increases while the critical pressure decreases, a regime in which all three critical quantities increase simultaneously, and a regime in which the critical horizon radius decreases while the critical pressure and temperature increase. In the double-critical region, two mathematical branches of critical points coexist. However, a detailed examination of the temperature and Gibbs free-energy profiles shows that only one branch exhibits the standard thermodynamic signatures of a physical van der Waals-like critical point. Thus, the existence of a solution to the local criticality conditions alone is insufficient to establish its thermodynamic relevance. For sufficiently large electric potential, the real critical solutions disappear and conventional van der Waals criticality is lost.

The off-shell $r_h-\tau$ analysis provides a broader view of the equilibrium solution space. We find that the number, ordering, connectivity, and morphology of the black-hole branches depend sensitively on both the pressure and the off-shell inverse temperature. Monotonic and disconnected branches can exchange their relative positions as the pressure changes, while S-shaped and cusp-like structures can emerge in different parameter regimes. Importantly, these geometric rearrangements do not necessarily correspond to changes in the global topological class. Hence, the detailed geometry of the equilibrium manifold and its global winding number encode complementary information about the thermodynamic system.

The topological analysis shows that the systems investigated in Regions I--V remain in the $W=+1$ class despite the substantial changes in their conventional critical behavior and off-shell branch structure. The winding-number sequence remains unchanged under continuous rearrangements of the equilibrium branches, demonstrating the robustness of the global topological invariant. Most importantly, this topological characterization persists beyond the disappearance of conventional criticality. At $\psi=0.521$, for example, the local criticality conditions no longer yield a conventional critical point, while the topological analysis still gives $W=+1$. Nontrivial thermodynamic structures also persist in this regime. This demonstrates that the disappearance of a conventional critical point does not, by itself, constitute a topological phase transition.

For larger electric potential, the system eventually enters a $W=0$ sector. In this regime, the equilibrium solution space may split into disconnected families containing defects with opposite winding numbers, whose contributions cancel to give $W=0$. With further increase of the electric potential, these structures evolve toward simpler cusp-like configurations while the total winding number remains unchanged. Thus, the transition from $W=+1$ to $W=0$ is distinct from the disappearance of conventional criticality and reflects a change in the global topology of the thermodynamic solution space.

Overall, our results demonstrate that conventional thermodynamic criticality and thermodynamic topology describe different but complementary aspects of black-hole phase structure. Local criticality conditions characterize the appearance or disappearance of conventional critical points, whereas the winding number captures global properties of the equilibrium solution space. The persistence of $W=+1$ beyond the range of conventional criticality, followed by the eventual emergence of a $W=0$ sector, highlights the ability of the topological approach to reveal global structural information that is not accessible from local criticality conditions alone. At the same time, the sensitivity of the $r_h-\tau$ branch geometry to pressure and off-shell temperature demonstrates that the global winding number does not uniquely determine the detailed thermodynamic structure. The combined use of conventional thermodynamics, off-shell solution geometry, and topological analysis therefore provides a more complete characterization of the phase structure of nonminimally coupled AdS black holes.

\vspace{1cm}
\noindent \textbf{Data Availability Statement:} No data were generated or analyzed in this study; therefore, data sharing is not applicable.\\

%%%%%%%%%%%%%%%%%%%%%%%%%%%%%%%%%%%%%%%%%%%%%%%%5
\appendix

\section{Explicit form of \(T^{(I)}_{\mu\nu}\)}
\label{app:B}

The energy-momentum tensor arising from the nonminimal coupling is
\begin{align}
	T^{(I)}_{\mu \nu }
	&= \tfrac{1}{2} F^{ \alpha \beta } F^{\gamma \lambda } g_{\mu \nu } R_{\alpha \gamma } R_{\beta \lambda }
	- 2 F^{\beta \gamma } F_{\nu }{}^{\alpha } R_{\alpha \gamma } R_{\mu \beta }
	- 2 F^{\beta \gamma } F_{\mu }{}^{\alpha } R_{\alpha \gamma } R_{\nu \beta }
	- F^{\alpha \beta } g_{\mu \nu } R_{\alpha \gamma } \nabla_{\beta }\nabla_{\lambda }F^{\gamma \lambda } \nonumber \\
	&\quad - F^{\alpha \beta } R_{\alpha \gamma } \nabla_{\beta }\nabla_{\mu }F_{\nu }{}^{\gamma }
	- F^{\alpha \beta } R_{\alpha \gamma } \nabla_{\beta }\nabla_{\nu }F_{\mu }{}^{\gamma }
	+ 2 F^{\alpha \beta } g_{\mu \nu } \nabla_{\beta }R_{\alpha }{}^{\lambda } \nabla_{\gamma }F^{\gamma }{}_{\lambda }
	- F_{\nu }{}^{\alpha } R_{\alpha \beta } \nabla_{\gamma }\nabla^{\gamma }F_{\mu }{}^{\beta } \nonumber \\
	&\quad - F_{\mu }{}^{\alpha } R_{\alpha \beta } \nabla_{\gamma }\nabla^{\gamma }F_{\nu }{}^{\beta }
	- F_{\mu }{}^{\alpha } F_{\nu }{}^{\beta } \nabla_{\gamma }\nabla^{\gamma }R_{\alpha \beta }
	- F_{\nu }{}^{\alpha } R_{\alpha \beta } \nabla_{\gamma }\nabla_{\mu }F^{\beta \gamma }
	- F^{\beta \gamma } F_{\nu }{}^{\alpha } \nabla_{\gamma }\nabla_{\mu }R_{\alpha \beta } \nonumber \\
	&\quad - F_{\mu }{}^{\alpha } R_{\alpha \beta } \nabla_{\gamma }\nabla_{\nu }F^{\beta \gamma }
	- F^{\beta \gamma } F_{\mu }{}^{\alpha } \nabla_{\gamma }\nabla_{\nu }R_{\alpha \beta }
	- 2 R_{\alpha \beta } \nabla_{\gamma }F_{\mu }{}^{\alpha } \nabla^{\gamma }F_{\nu }{}^{\beta }
	- 2 F_{\nu }{}^{\alpha } \nabla_{\gamma }F_{\mu }{}^{\beta } \nabla^{\gamma }R_{\alpha \beta } \nonumber \\
	&\quad - 2 F_{\mu }{}^{\alpha } \nabla_{\gamma }F_{\nu }{}^{\beta } \nabla^{\gamma }R_{\alpha \beta }
	- g_{\mu \nu } R_{\alpha \beta } \nabla_{\gamma }F^{\beta \lambda } \nabla_{\lambda }F^{\alpha \gamma }
	- g_{\mu \nu } R_{\alpha \beta } \nabla_{\gamma }F^{\alpha \gamma } \nabla_{\lambda }F^{\beta \lambda }
	+ 2 F^{\alpha \beta } g_{\mu \nu } \nabla_{\alpha }F^{\gamma \lambda } \nabla_{\lambda }R_{\beta \gamma } \nonumber \\
	&\quad - F^{\alpha \beta } g_{\mu \nu } R_{\alpha \gamma } \nabla_{\lambda }\nabla_{\beta }F^{\gamma \lambda }
	- F^{\alpha \beta } F^{\gamma \lambda } g_{\mu \nu } \nabla_{\lambda }\nabla_{\beta }R_{\alpha \gamma }
	- F_{\nu }{}^{\alpha } \nabla_{\gamma }R_{\alpha }{}^{\beta } \nabla_{\mu }F_{\beta }{}^{\gamma }
	- R_{\alpha \beta } \nabla_{\gamma }F_{\nu }{}^{\alpha } \nabla_{\mu }F^{\beta \gamma } \nonumber \\
	&\quad - R_{\alpha \beta } \nabla_{\gamma }F^{\beta \gamma } \nabla_{\mu }F_{\nu }{}^{\alpha }
	- F^{\alpha \beta } \nabla_{\beta }R_{\alpha }{}^{\gamma } \nabla_{\mu }F_{\nu \gamma }
	+ F_{\nu }{}^{\alpha } \nabla_{\beta }F^{\beta \gamma } \nabla_{\mu }R_{\alpha \gamma }
	+ F^{\alpha \beta } \nabla_{\alpha }F_{\nu }{}^{\gamma } \nabla_{\mu }R_{\beta \gamma } \nonumber \\
	&\quad - F_{\mu }{}^{\alpha } \nabla_{\gamma }R_{\alpha }{}^{\beta } \nabla_{\nu }F_{\beta }{}^{\gamma }
	- R_{\alpha \beta } \nabla_{\gamma }F_{\mu }{}^{\alpha } \nabla_{\nu }F^{\beta \gamma }
	- R_{\alpha \beta } \nabla_{\gamma }F^{\beta \gamma } \nabla_{\nu }F_{\mu }{}^{\alpha }
	- F^{\alpha \beta } \nabla_{\beta }R_{\alpha }{}^{\gamma } \nabla_{\nu }F_{\mu \gamma } \nonumber \\
	&\quad + F_{\mu }{}^{\alpha } \nabla_{\beta }F^{\beta \gamma } \nabla_{\nu }R_{\alpha \gamma }
	+ F^{\alpha \beta } \nabla_{\alpha }F_{\mu }{}^{\gamma } \nabla_{\nu }R_{\beta \gamma }.
\end{align}

This tensor contributes to the gravitational field equations and is obtained from the variation of the nonminimal interaction term $ \epsilon F^{\alpha\beta}F^{\gamma\lambda}R_{\alpha\gamma}R_{\beta\lambda}$ with respect to the metric \(g^{\mu\nu}\).

\section{Explicit Form of the Field Equations at First Order}
\label{app:C}

In this appendix, we present the explicit form of the \(tt\) and \(rr\) components of the gravitational field equations and Maxwell equations at first order in the nonminimal coupling parameter \(\epsilon\). 

The \(tt\) component of Eq.~(\ref{EOM1}) is:
\begin{equation}\label{tteq}
4 r f_0' + 4 f_0 + \kappa \alpha r^2 e^{2H_0} h_0'^2 + 4\Lambda r^2  - \epsilon B_1(r) = 0,
\end{equation}
where \(B_1(r)\) is defined as:
\begin{equation}
\begin{aligned}
B_1(r) &= 3 \mathrm{e}^{2H(r)} \kappa r^2 h'(r)^2 f''(r)^2 \\
&\quad + \mathrm{e}^{2H(r)} \kappa f'(r)^2 h'(r) \Big[ 4 r \big(-2 + 3 r H'(r)\big) h''(r) + h'(r) \big( 8 - 28 r H'(r) + 35 r^2 H'(r)^2 + 10 r^2 H''(r) \big) \Big] \\
&\quad - 2 r f'(r) \Big[ 2 \mathrm{e}^{2H(r)} \kappa r h'(r) f''(r) h''(r) + \mathrm{e}^{2H(r)} \kappa h'(r)^2 \big( (-2 + 8 r H'(r)) f''(r) + r f^{(3)}(r) \big) \Big] \\
&\quad - 4 f(r) \Bigg[ 2 \mathrm{e}^{2H(r)} \kappa r^2 f''(r) h''(r)^2 \\
&\qquad - 2 \mathrm{e}^{2H(r)} \kappa r h'(r) \Big( -2 r h''(r) f^{(3)}(r) + f''(r) \big( (-7 + 2 r H'(r)) h''(r) - r h^{(3)}(r) \big) \Big) \\
&\qquad + \mathrm{e}^{2H(r)} \kappa f'(r) \Big( 2 r (2 - 3 r H'(r)) h''(r)^2 \\
&\qquad\quad - 2 h'(r) \Big( h''(r) \big( -2 + 6 r H'(r) + 7 r^2 H'(r)^2 + 11 r^2 H''(r) \big) + r (-2 + 3 r H'(r)) h^{(3)}(r) \Big) \\
&\qquad\quad + r h'(r)^2 \big( -27 H'(r)^2 + 7 r H'(r)^3 - 12 H''(r) - 27 r H'(r) H''(r) - 8 r H^{(3)}(r) \big) \Big) \\
&\qquad - \mathrm{e}^{2H(r)} \kappa h'(r)^2 \Big( 3 f''(r) \big( -1 - 2 r H'(r) + 3 r^2 H'(r)^2 + r^2 H''(r) \big) \\
&\qquad\quad - r \big( (5 + r H'(r)) f^{(3)}(r) + r f^{(4)}(r) \big) \Big) \Bigg] \\
&\quad - 4 \mathrm{e}^{2H(r)} \kappa f(r)^2 \Bigg[ 4 r h''(r)^2 \big( -H'(r) + r H'(r)^2 - r H''(r) \big) \\
&\qquad + 4 h'(r) \Big( 4 r^2 H'(r)^3 h''(r) + r H'(r)^2 \big( -h''(r) + r h^{(3)}(r) \big) \\
&\qquad\quad - H'(r) \big( 2 h''(r) + r h^{(3)}(r) \big) - r \big( r H''(r) h^{(3)}(r) + h''(r) (5 H''(r) + 2 r H^{(3)}(r)) \big) \Big) \\
&\qquad + h'(r)^2 \Big( 10 r H'(r)^3 + 5 r^2 H'(r)^4 - 6 H''(r) - 3 r^2 H''(r)^2 \\
&\qquad\quad + 6 H'(r)^2 \big( -1 + 3 r^2 H''(r) \big) - 2 r H'(r) \big( 9 H''(r) + 2 r H^{(3)}(r) \big) \\
&\qquad\quad - 2 r \big( 4 H^{(3)}(r) + r H^{(4)}(r) \big) \Big) \Bigg].
\end{aligned}
\end{equation}

The \(rr\) component of Eq.~(\ref{EOM1}) is:

\begin{equation}\label{rreq}
4 r f_0' + 4 f_0 - 8 r f_0 H_0' + \kappa \alpha r^2 e^{2H_0} h_0'^2 + 4\Lambda r^2  + \epsilon B_2(r) = 0,
\end{equation}

where \(B_2(r)\) is defined as:
\begin{equation}\label{B2}
\begin{aligned}
B_2(r) = &-3 e^{2H(r)} \kappa r^{2} \bigl( h'(r) \bigr)^{2} \bigl( f''(r) \bigr)^{2} \\
& - e^{2H(r)} \kappa \bigl( f'(r) \bigr)^{2} h'(r) \Bigl[ 4r \bigl( -2 + 3r H'(r) \bigr) h''(r) \\
&\qquad + h'(r) \Bigl( 8 - 28r H'(r) + 35 r^{2} \bigl( H'(r) \bigr)^{2} + 10 r^{2} H''(r) \Bigr) \Bigr] \\
& + 2r f'(r) \Bigl[ 2 e^{2H(r)} \kappa r h'(r) f''(r) h''(r) \\
&\qquad + e^{2H(r)} \kappa \bigl( h'(r) \bigr)^{2} \Bigl( \bigl( -2 + 8r H'(r) \bigr) f''(r) + r f'''(r) \Bigr) \Bigr] \\
& -4 e^{2H(r)} \kappa f(r)^{2} h'(r) \Biggl[ 4 h''(r) \Bigl( -2r \bigl( H'(r) \bigr)^{2} + r^{2} \bigl( H'(r) \bigr)^{3} \\
&\qquad + r H''(r) + H'(r) \bigl( 2 - r^{2} H''(r) \bigr) \Bigr) \\
&\qquad + h'(r) \Bigl( -10r \bigl( H'(r) \bigr)^{3} + 7 r^{2} \bigl( H'(r) \bigr)^{4} + 6 H''(r) \\
&\qquad - 6 r^{2} \bigl( H'(r) \bigr)^{2} H''(r) + 3 r^{2} \bigl( H''(r) \bigr)^{2} + 2r H'''(r) - 2 r^{2} H'(r) H'''(r) \Bigr) \Biggr] \\
& +4 f(r) \Biggl[ 15 e^{2H(r)} \kappa r^{2} f'(r) \bigl( h'(r) \bigr)^{2} \bigl( H'(r) \bigr)^{3} \\
& \qquad + e^{2H(r)} \kappa r h'(r) \bigl( H'(r) \bigr)^{2} \Bigl( -2r h'(r) f''(r) \\
&\qquad\quad + f'(r) \bigl( -15 h'(r) + 8r h''(r) \bigr) \Bigr) \\
& \qquad -2 e^{2H(r)} \kappa h'(r) h''(r) \Bigl( -r f''(r) + f'(r) \bigl( -2 + r^{2} H''(r) \bigr) \Bigr) \\
& \qquad - e^{2H(r)} \kappa h'(r) H'(r) \Bigl( 2 f'(r) h'(r) + 2r h'(r) f''(r) + 12r f'(r) h''(r) \\
&\qquad\quad + 2r^{2} f''(r) h''(r) + 4 r^{2} f'(r) h'(r) H''(r) + r^{2} h'(r) f'''(r) \Bigr) \\
& \qquad + e^{2H(r)} \kappa \bigl( h'(r) \bigr)^{2} \Bigl( 3 f''(r) \bigl( 1 + r^{2} H''(r) \bigr) \\
&\qquad\quad + r \Bigl( f'''(r) - f'(r) \bigl( 2 H''(r) + r H'''(r) \bigr) \Bigr) \Bigr) \Biggr] = 0.
\end{aligned}
\end{equation}

The functions \(B_1(r)\) and \(B_2(r)\) encode the contributions arising from the nonminimal interaction term \(F^{\alpha\beta}F^{\gamma\lambda}R_{\alpha\gamma}R_{\beta\lambda}\) at first order in \(\epsilon\). The explicit expressions above were obtained by substituting the perturbative expansions of the metric functions and gauge potential into the full field equations, and then extracting the terms linear in \(\epsilon\). These equations, together with the Maxwell equations at first order, determine the corrections to the metric and gauge potential presented in Sec.~\ref{sec2}.

The Maxwell equation \eqref{EOM-YM}, using the metric Eq.(\ref{metric}), leads to the following integral for the gauge function $h(r)$,

\begin{equation}
h(r)=\int ^{r}\frac{ e^{-H(u)}}{-\alpha u^2+\epsilon B_3(u)}du+C,
\end{equation}
where $B_3(u)$ is as, 
\begin{equation}
\begin{aligned}
B_3(u) &= 4 f'(u)^2 - 8 f(u) f'(u) H'(u) - 12 u f'(u)^2 H'(u) + 20 u f(u) f'(u) H'(u)^2 \\
&\quad + 9 u^2 f'(u)^2 H'(u)^2 - 8 u f(u)^2 H'(u)^3 - 12 u^2 f(u) f'(u) H'(u)^3 + 4 u^2 f(u)^2 H'(u)^4 \\
&\quad + 4 u f'(u) f''(u) - 4 u f(u) H'(u) f''(u) - 6 u^2 f'(u) H'(u) f''(u) + 4 u^2 f(u) H'(u)^2 f''(u) +\\
&\quad u^2 f''(u)^2  - 8 u f(u) f'(u) H''(u) + 8 u f(u)^2 H'(u) H''(u) + 12 u^2 f(u) f'(u) H'(u) H''(u) \\
&\quad - 8 u^2 f(u)^2 H'(u)^2 H''(u) - 4 u^2 f(u) f''(u) H''(u) + 4 u^2 f(u)^2 H''(u)^2.
\end{aligned}
\end{equation}

Substituting the zeroth-order solutions \(f_0(r)\), \(h_0(r)\), and \(H_0(r)=0\) into the above expressions, and using the matching condition (\ref{con}) to fix the integration constants, yields the first-order corrections \(f_1(r)\), \(h_1(r)\), and \(H_1(r)\) given in Eqs.~(\ref{f1}), (\ref{h1}), and (\ref{H1}), respectively.

%%%%%%%%%%%%%%%%%%%%%%%%%%%%%%%%%%%%%%%%%%%%


\begin{thebibliography}{99}

\bibitem{Bardeen:1973gs}
J.~M.~Bardeen, B.~Carter and S.~W.~Hawking,
``The Four laws of black hole mechanics,''
Commun. Math. Phys. \textbf{31}, 161-170 (1973)

\bibitem{Hawking:1975vcx}
S.~W.~Hawking,
``Particle Creation by Black Holes,''
Commun. Math. Phys. \textbf{43}, 199-220 (1975)

\bibitem{Hawking:1982dh}
S.~W.~Hawking and D.~N.~Page,
``Thermodynamics of Black Holes in anti-De Sitter Space,''
Commun. Math. Phys. \textbf{87}, 577 (1983)

\bibitem{Witten:1998qj}
E.~Witten,
``Anti de Sitter space and holography,''
Adv. Theor. Math. Phys. \textbf{2}, 253-291 (1998)

\bibitem{Kastor:2009wy}
D.~Kastor, S.~Ray and J.~Traschen,
``Enthalpy and the Mechanics of AdS Black Holes,''
Class. Quant. Grav. \textbf{26}, 195011 (2009)

\bibitem{Dolan2011}
B.~P.~Dolan, ``The Cosmological Constant and the Black Hole Equation of State,'' Class.Quant.Grav.\textbf{28} (2011) 125020.

\bibitem{Kubiznak:2012wp}
D.~Kubiznak and R.~B.~Mann,
``P-V criticality of charged AdS black holes,''
JHEP \textbf{07}, 033 (2012)

\bibitem{Belhaj:2012bg}
A.~Belhaj, M.~Chabab, H.~El Moumni and M.~B.~Sedra,
``On Thermodynamics of AdS Black Holes in Arbitrary Dimensions,''Chin. Phys. Lett. \textbf{29}, 100401 (2012)


\bibitem{Chemissany:2008fy}
W.~A.~Chemissany, M.~de Roo and S.~Panda,
``Thermodynamics of Born-Infeld Black Holes,''
Class. Quant. Grav. \textbf{25}, 225009 (2008)

\bibitem{Banerjee:2011cz}
R.~Banerjee and D.~Roychowdhury,
``Critical phenomena in Born-Infeld AdS black holes,'' Phys. Rev. D \textbf{85}, 044040 (2012)


\bibitem{Fernando:2006gh}
S.~Fernando, ``Thermodynamics of Born-Infeld-anti-de Sitter black holes in the grand canonical ensemble,''
Phys. Rev. D \textbf{74}, 104032 (2006)

\bibitem{Cai:2001dz}
R.~G.~Cai, ``Gauss-Bonnet black holes in AdS spaces,'' Phys. Rev. D \textbf{65}, 084014 (2002)


\bibitem{Frassino:2014pha}
A.~M.~Frassino, D.~Kubiznak, R.~B.~Mann and F.~Simovic, ``Multiple Reentrant Phase Transitions and Triple Points in Lovelock Thermodynamics,''
JHEP \textbf{09}, 080 (2014)


\bibitem{Cai:2014znn}
R.~G.~Cai, Y.~P.~Hu, Q.~Y.~Pan and Y.~L.~Zhang,
``Thermodynamics of Black Holes in Massive Gravity,''
Phys. Rev. D \textbf{91}, no.2, 024032 (2015)


\bibitem{Hendi:2017fxp}
S.~H.~Hendi, R.~B.~Mann, S.~Panahiyan and B.~Eslam Panah, ``Van der Waals like behavior of topological AdS black holes in massive gravity,''
Phys. Rev. D \textbf{95}, no.2, 021501 (2017)

\bibitem{Anabalon:2013oea}
A.~Anabalon, A.~Cisterna and J.~Oliva,
``Asymptotically locally AdS and flat black holes in Horndeski theory,''
Phys. Rev. D \textbf{89}, 084050 (2014)


\bibitem{Miao:2016aol}
Y.~G.~Miao and Z.~M.~Xu,
``Thermodynamics of Horndeski black holes with non-minimal derivative coupling,''
Eur. Phys. J. C \textbf{76}, no.11, 638 (2016)


\bibitem{Balakin:2005fu}
A.~B.~Balakin and J.~P.~S.~Lemos,
``Non-minimal coupling for the gravitational and electromagnetic fields: A General system of equations,'' Class. Quant. Grav. \textbf{22}, 1867-1880 (2005)

\bibitem{Balakin2010}
A.~B.~Balakin, J.~P.~S.~Lemos, and A.~E.~Zayats,
``Nonminimal coupling for the gravitational and electromagnetic fields: Traversable electric wormholes,'' Phys.\ Rev.\ D \textbf{81}, 084015 (2010).



\bibitem{Guo:2021ere}
G.~Guo, P.~Wang, H.~Wu and H.~Yang,
``Thermodynamics and phase structure of an Einstein-Maxwell-scalar model in extended phase space,''
Phys. Rev. D \textbf{105}, no.6, 064069 (2022)
 


\bibitem{Duan1984}
Y.~Duan, ``THE STRUCTURE OF THE TOPOLOGICAL CURRENT,''
SLAC-PUB-3301.

\bibitem{Duan1979}
Y.~S.~Duan and M.~L.~Ge, ``SU(2) Gauge Theory and Electrodynamics with N Magnetic Monopoles,''
Sci. Sin. \textbf{9}, no.11, 1072 (1979)

\bibitem{Wei:2021vdx}
S.~W.~Wei and Y.~X.~Liu,
``Topology of black hole thermodynamics,''
Phys. Rev. D \textbf{105}, no.10, 104003 (2022)
doi:10.1103/PhysRevD.105.104003


\bibitem{WeiLiu2026}
S.-W.~Wei and Y.-X.~Liu,
``Topology of black hole thermodynamics: A brief review,''
\textit{Sci. China-Phys. Mech. Astron.} \textbf{69}, 260401 (2026),
doi:10.1007/s11433-025-2923-3.


\bibitem{Ali:2023jox}
M.~S.~Ali, H.~El Moumni, J.~Khalloufi and K.~Masmar,
``Topology of Born{\textendash}Infeld-AdS black hole phase transitions: Bulk and CFT sides,''
Annals Phys. \textbf{465}, 169679 (2024)

\bibitem{Gogoi:2025ied}
N.~J.~Gogoi, D.~J.~Gogoi and J.~Bora,
``Topology of 5-dimensional Einstein{\textendash}Gauss{\textendash}Bonnet AdS black hole thermodynamics surrounded by a cloud of Strings,''
Phys. Dark Univ. \textbf{50}, 102099 (2025)

\bibitem{Yerra:2022eov}
P.~K.~Yerra and C.~Bhamidipati,
``Topology of Born-Infeld AdS black holes in 4D novel Einstein-Gauss-Bonnet gravity,''
Phys. Lett. B \textbf{835}, 137591 (2022)

\bibitem{EslamPanah:2024fls}
B.~Eslam Panah, B.~Hazarika and P.~Phukon,
``Thermodynamic Topology of Topological Black Hole in F(R)-ModMax Gravity{\textquoteright}s Rainbow,''
PTEP \textbf{2024}, no.8, 083E02 (2024)


\bibitem{Hazarika:2024xar}
B.~Hazarika, B.~Eslam Panah and P.~Phukon,
``Thermodynamic topology of topological charged dilatonic black holes,''
Eur. Phys. J. C \textbf{84}, no.11, 1204 (2024)

\bibitem{WeiLiuMann2022}
Shao-Wen Wei, Yu-Xiao Liu, and Robert B. Mann. Black Hole Solutions as Topological Thermodynamic Defects. Phys. Rev. Lett., 129(19):191101, 2022.


\bibitem{Wei:2024gfz}
S.~W.~Wei, Y.~X.~Liu and R.~B.~Mann,``Universal topological classifications of black hole thermodynamics,''
Phys. Rev. D \textbf{110}, no.8, L081501 (2024)


\bibitem{Wu:2022whe}
D.~Wu, ``Topological classes of rotating black holes,''
Phys. Rev. D \textbf{107}, no.2, 024024 (2023)


\bibitem{ZhuWu2024}
X.-D. Zhu, D. Wu, and D. Wen, ``Topological classes of thermodynamics of the rotating charged AdS black holes in gauged supergravities,'' \textit{Physics Letters B} \textbf{856} (2024) 138919.

\bibitem{WuGu2024}
Wu, D., Gu, SY., Zhu, XD. et al. Topological classes of thermodynamics of the static multi-charge AdS black holes in gauged supergravities: novel temperature-dependent thermodynamic topological phase transition. J. High Energ. Phys. 2024, 213 (2024).


\bibitem{Chen:2024sow}
Z.~Q.~Chen and S.~W.~Wei,
``Thermodynamical topology with multiple defect curves for dyonic AdS black holes,''
Eur. Phys. J. C \textbf{84}, no.12, 1294 (2024)



\bibitem{Zhang:2023uay}
M.~Zhang and J.~Jiang,
``Bulk-boundary thermodynamic equivalence: a topology viewpoint,''
JHEP \textbf{06}, 115 (2023)

\bibitem{Liu:2022aqt}
C.~Liu and J.~Wang,
``Topological natures of the Gauss-Bonnet black hole in AdS space,'' Phys. Rev. D \textbf{107}, no.6, 064023 (2023)


\bibitem{Ali:2024}
M.~S. Ali, H.~El Moumni, J.~Khalloufi, and K.~Masmar,
``Topology of Born--Infeld-AdS black hole phase transitions: Bulk and CFT sides,''
Ann. Phys. \textbf{465}, 169679 (2024).

\bibitem{Sadeghi:2026jic}
M.~Sadeghi and F.~Rahmani,
``Thermodynamic topology of 4D charged AdS black holes with $F^{\alpha \beta }F^{\gamma \lambda }R_{\alpha \gamma \beta \lambda }$ coupling,''
Eur. Phys. J. C \textbf{86}, no.5, 581 (2026)


\bibitem{Rahmani:2025iks}
F.~Rahmani and M.~Sadeghi,
``Topological classification of a 4D ads black hole with non-minimal maxwell coupling,''
Class. Quant. Grav. \textbf{43}, no.10, 105011 (2026)

\bibitem{Sadeghi:2026fyu}
M.~Sadeghi and F.~Rahmani,
``Thermodynamic and Topological Phase Transitions of AdS Black Holes with Nonminimal $F^{\alpha\beta}F^{\gamma \lambda}R_{\alpha \gamma}R_{\beta \lambda}$ Coupling,''
[arXiv:2606.22204 [hep-th]].


\bibitem{Drummond:1979pp}
I.~T.~Drummond and S.~J.~Hathrell,
``QED Vacuum Polarization in a Background Gravitational Field and Its Effect on the Velocity of Photons,'' Phys. Rev. D \textbf{22}, 343 (1980)



\bibitem{York:1986it}
J.~W.~York, Jr.,``Black hole thermodynamics and the Euclidean Einstein action,''
Phys. Rev. D \textbf{33}, 2092-2099 (1986)


\bibitem{Ahmed:2022kyv}
M.~B.~Ahmed, D.~Kubiznak and R.~B.~Mann,
``Vortex-antivortex pair creation in black hole thermodynamics,'' Phys. Rev. D \textbf{107}, no.4, 046013 (2023)


\bibitem{Wu:2025}
D.~Wu, W.~Liu, S.-Q.~Wu, and R.~B.~Mann,
``Novel topological classes in black hole thermodynamics,'' Phys. Rev. D \textbf{111}, L061501 (2025).


\bibitem{Wu:2025EPJC}
S.-P.~Wu, S.-J.~Yang, and S.-W.~Wei,
``Extended thermodynamical topology of black hole,''
Eur. Phys. J. C \textbf{85}, 1372 (2025).

\bibitem{AiWu:2025}
W.~Ai and D.~Wu,
``$\widetilde{W}^{1+}$ subclass: Extending the topological classification of black hole thermodynamics,'' Phys.~Rev.~D \textbf{112}, 124024 (2025).



\bibitem{Chen:2025vsz}
H.~Chen, M.~Y.~Zhang, H.~Hassanabadi, Q.~Huang and Z.~W.~Long,
``Novel topological subclass in Ho\v{r}ava{\textendash}Lifshitz black holes,''
Eur. Phys. J. C \textbf{85}, no.12, 1386 (2025)


\bibitem{Babaei-Aghbolagh:2025qxm}
H.~Babaei-Aghbolagh, H.~Esmaili, S.~He and H.~Mohammadzadeh,
``Thermodynamic topology of Einstein{\textendash}Maxwell-dilaton theories,''
Eur. Phys. J. C \textbf{86}, no.1, 78 (2026)

\bibitem{Wu:2023xpq}
D.~Wu,
``Classifying topology of consistent thermodynamics of the four-dimensional neutral Lorentzian NUT-charged spacetimes,''
Eur. Phys. J. C \textbf{83}, no.5, 365 (2023)




\bibitem{Wu:2023fcw}
D.~Wu,
``Consistent thermodynamics and topological classes for the four-dimensional Lorentzian charged Taub-NUT spacetimes,''
Eur. Phys. J. C \textbf{83}, no.7, 589 (2023)


\bibitem{Wu:2023meo}
D.~Wu,
``Topological classes of thermodynamics of the four-dimensional static accelerating black holes,''
Phys. Rev. D \textbf{108}, no.8, 084041 (2023)


\bibitem{Chen:2024atr}
H.~Chen, D.~Wu, M.~Y.~Zhang, H.~Hassanabadi and Z.~W.~Long,
``Thermodynamic topology of phantom AdS black holes in massive gravity,''
Phys. Dark Univ. \textbf{46}, 101617 (2024)

\bibitem{Zhu:2024zcl}
X.~D.~Zhu, W.~Liu and D.~Wu,
``Universal thermodynamic topological classes of rotating black holes,''
Phys. Lett. B \textbf{860}, 139163 (2025)


\bibitem{Liu:2025iyl}
W.~Liu, L.~Zhang, D.~Wu and J.~Wang,
``Thermodynamic topological classes of the rotating, accelerating black holes,''
Class. Quant. Grav. \textbf{42}, 125007 (2025)



\bibitem{Chen:2025nto}
Y.~Chen, X.~D.~Zhu and D.~Wu,
``Universal thermodynamic topological classes of three-dimensional BTZ black holes,''
Phys. Lett. B \textbf{865}, 139482 (2025)


\bibitem{Chen:2025fse}
H.~Chen, D.~Wu, M.~Y.~Zhang, S.~Zare, H.~Hassanabadi, B.~C.~L{\"u}tf{\"u}o{\u{g}}lu and Z.~W.~Long,
``Universal thermodynamic topological classes of static black holes in Conformal Killing Gravity,''
Eur. Phys. J. C \textbf{85}, no.8, 828 (2025)


\bibitem{Wu:2025wpz}
D.~Wu and S.~Q.~Wu,
``Thermodynamics and topological classifications of static non-extremal four-charge AdS black hole in the five-dimensional $\mathcal {N}= 2$, $STU-W^2U$ gauged supergravity,''
Eur. Phys. J. C \textbf{86}, no.2, 187 (2026)



\bibitem{Tian:2026qnt}
M.~Tian, Y.~Chen and D.~Wu,
``Dimensional structure of thermodynamic topology in ultraspinning Kerr-AdS black holes,''
Phys. Lett. B \textbf{879}, 140658 (2026)


\bibitem{Wu:2026gqz}
D.~Wu,
``Topological changes and response singularities in black hole thermodynamic branch structure,''
[arXiv:2607.07364 [hep-th]].










\end{thebibliography}
\end{document}